\documentclass[12pt,twoside,a4paper,openright]{book}

\usepackage{floatflt} % allow floating figures
\usepackage{pdfsync} % allow easy editing with pdf files
\usepackage[bookmarks=true, pdfpagelabels=true, colorlinks=true, linkcolor=black, urlcolor=black, citecolor=blue, anchorcolor=blue]{hyperref}

\usepackage{tabularx}
\usepackage{booktabs}
\usepackage[utf8]{inputenc}
\usepackage[T1]{fontenc}

\usepackage{amsmath,amssymb,bm}	  % math options
\usepackage{cancel}
\usepackage{graphics,xcolor}   % graphic and colors
\usepackage{times}
\usepackage{graphicx}
\usepackage{emptypage}
\usepackage{cite}
\usepackage[export]{adjustbox}
\usepackage{rotating}
\usepackage{makecell}

\usepackage{etoolbox}% http://ctan.org/pkg/etoolbox

\usepackage[font=small,labelfont=bf]{caption}

\usepackage[a4paper,total={16cm,24cm},centering,headsep=.5cm]{geometry}
\usepackage{fancyhdr}
\usepackage{sectsty}
\sectionfont{\fontsize{16}{18}\usefont{OT1}{phv}{m}{n}\selectfont}
\subsectionfont{\fontsize{13}{15}\usefont{OT1}{phv}{m}{n}\selectfont}
\subsubsectionfont{\fontsize{11}{13}\usefont{OT1}{phv}{m}{n}\selectfont}

\usepackage[Lenny]{fncychap}
\ChNameVar{\fontsize{14}{16}\usefont{OT1}{phv}{m}{n}\selectfont}
\ChNumVar{\fontsize{60}{62}\usefont{OT1}{phv}{m}{n}\selectfont}
\ChTitleVar{\fontsize{24}{26}\usefont{OT1}{phv}{m}{n}\selectfont}

\makeatletter
\renewcommand\ps@plain{\let\@mkboth\@gobbletwo
     \let\@oddhead\@empty
     \def\@oddfoot{\reset@font\hfil}
     \let\@evenhead\@empty\let\@evenfoot\@oddfoot}

\patchcmd{\@makechapterhead}{\vspace*{50\p@}}{}{}{}% Removes space above \chapter head
\patchcmd{\@makeschapterhead}{\vspace*{50\p@}}{}{}{}% Removes space above \chapter* head

\g@addto@macro\normalsize{%
  \setlength\abovedisplayskip{10pt}
  \setlength\belowdisplayskip{10pt}
  \setlength\abovedisplayshortskip{10pt}
  \setlength\belowdisplayshortskip{10pt}
}

\makeatother

\usepackage{makeidx} 
\makeindex

\newcommand{\ie}{\textit{i.e.}}
\newcommand{\eg}{\textit{e.g.}}
\newcommand{\amc}{{\sc MadGraph5}\_a{\sc MC@NLO}}

\newcommand{\nloct}{{\sc NloCT}}

\newcommand{\pyth}{{\sc Pythia8}}
\newcommand{\hw}{{\sc Herwig7}}
\newcommand{\fj}{{\sc FastJet}}

\newcommand{\fa}{{\sc FeynArts}}
\newcommand{\fc}{{\sc FormCalc}}

\begin{document}  %%%%%%%%%%%%%%%%%%%%%%%%%%%%%%%%%%%%%%

% Title, acknowledgment, table of contents...
%%%%%%%%%%%%%%%%%%%%%%%%%%%%%%%%%%%%%%%%%%%%%%%%%%%%
%
%      First pages ...
%
%
%%%%%%%%%%%%%%%%%%%%%%%%%%%%%%%%%%%%%%%%%%%%%%%%%%%

\begin{figure}[t!]
\centering
\includegraphics[scale=0.3]{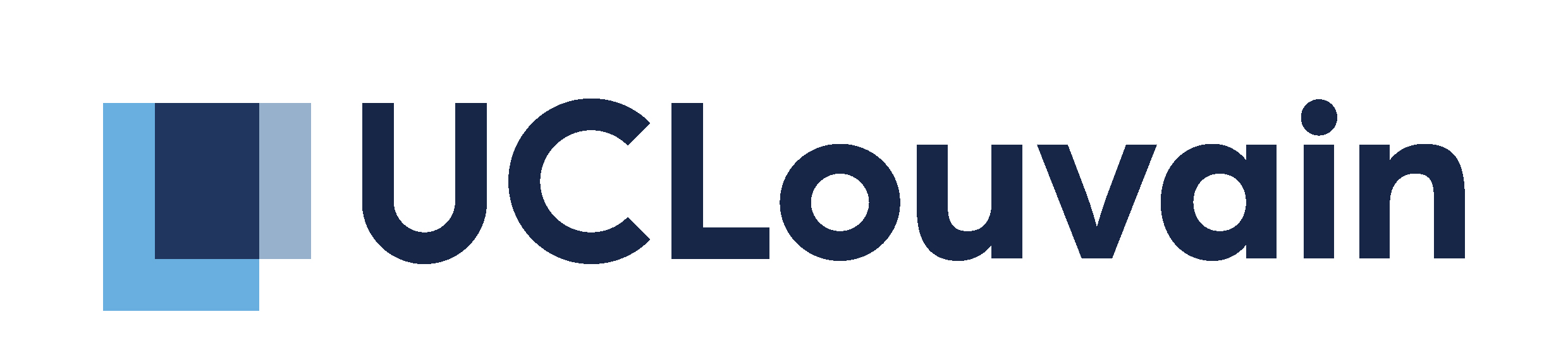}
\hspace{2.9cm}
\includegraphics[scale=0.3]{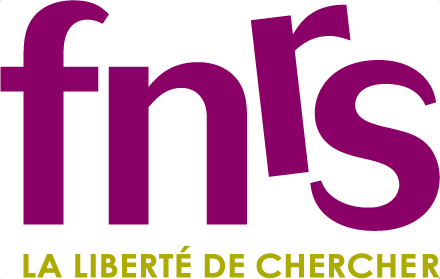}
\end{figure}

% \parbox[c][][c]{0.799\textwidth}{\vspace{0.05cm}
% \begin{flushright}
% \large Universit{\'e} catholique de Louvain\\[0.2\baselineskip]
% \normalsize Secteur des Sciences et Technologies\\[0.1\baselineskip] 
% Institut de Recherche en Math{\'e}matique et Physique\\[0.1\baselineskip]
% Center for Cosmology, Particle Physics and Phenomenology
% \end{flushright}
% }
\vspace{1.5cm}
\begin{center}
% \vspace{2cm}
\parbox{0.95\textwidth}{\fontsize{22}{30}\selectfont\centering{Resurrecting the Standard Model Effective Field Theory interferences at colliders}}
% \vspace{0.3cm}
\end{center}
\vspace{0.6cm}
\begin{center}
Doctoral dissertation presented by \\
\vspace{2mm}
{\Large Matteo Maltoni}\\
\vspace{2mm}
in fulfilment of the requirements for the degree of Doctor in Sciences
\end{center}
\vspace{\fill}
\begin{center}
\vfill
\begin{tabular}{llc} 
\multicolumn{3}{c}{\large Thesis Jury}                                  \\
                    &               &                           \\
\toprule 
Prof. Fabio Maltoni       & President     & Université Catholique de Louvain  \\
Prof. Céline Degrande     & Supervisor    & Université Catholique de Louvain  \\
Prof. Vincent Lemaitre    & Secretary     & Université Catholique de Louvain  \\
Dr. Ken Mimasu            &               & University of Southampton  \\
Dr. Ilaria Brivio         &               & Università di Bologna  \\
\bottomrule
\end{tabular}

\vspace*{0.5cm}
\textsl{November 14$^{th}$, 2025}\\[1pt]
\end{center}
\thispagestyle{empty}

%%% acknowledgment

\hfill
\newenvironment{acknowledgements}%
    {\cleardoublepage\thispagestyle{empty}\null\vfill\begin{center}%
    \bfseries Acknowledgements\end{center}}%
    {\vfill\null}
        \begin{acknowledgements}
           \centering
           \includegraphics[width=.9\textwidth]{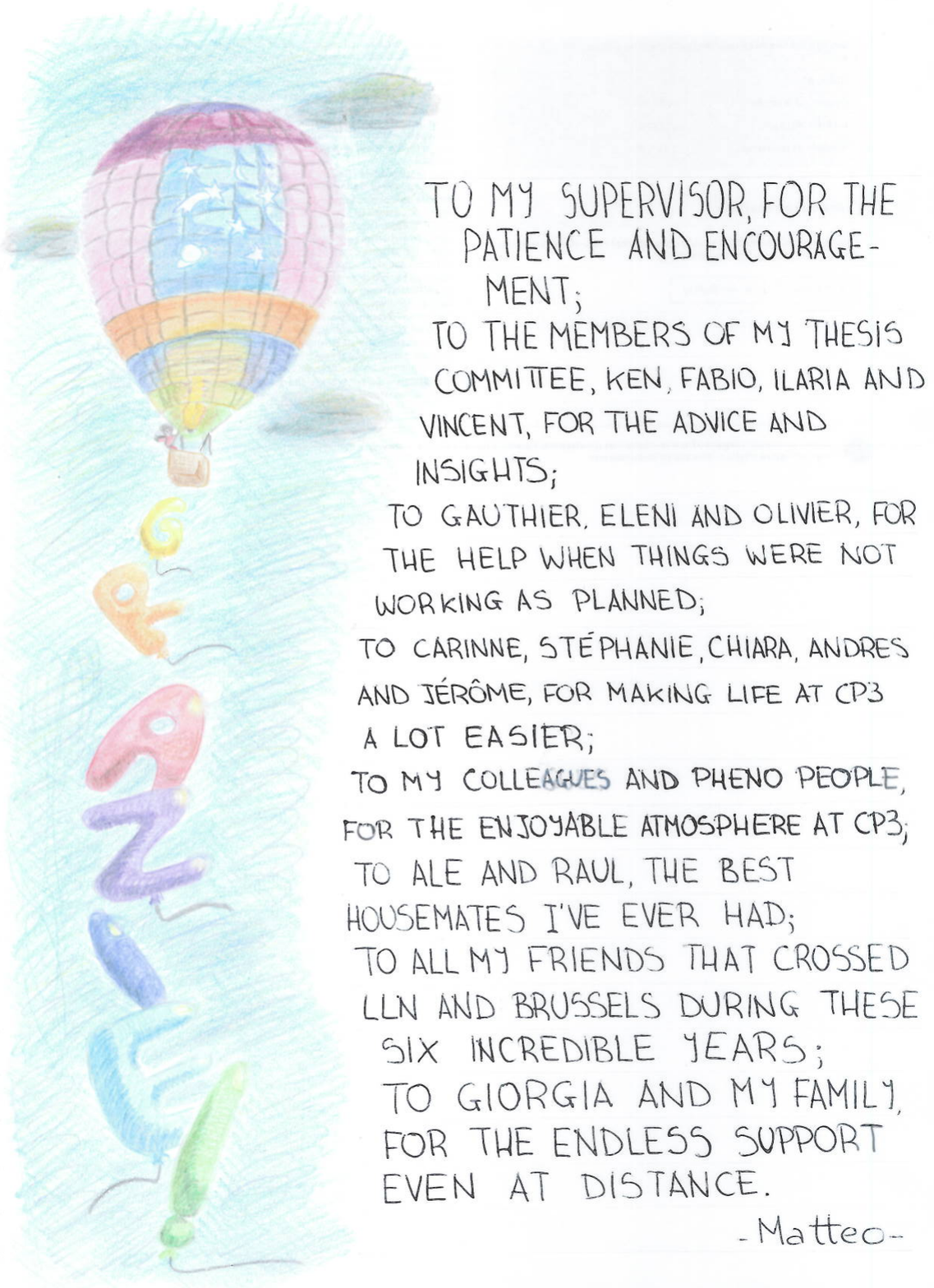}
        \end{acknowledgements}

%%% abstract

\hfill
\newenvironment{Abstract}%
    {\cleardoublepage\thispagestyle{empty}\null\vfill\begin{center}%
    \bfseries Abstract\end{center}}%
    {\vfill\null}
        \begin{Abstract}
           Even if some experimental evidence suggests the existence of physics beyond the SM, no clues of new resonances can be found in the data. In the case their masses are much larger than the energies of current experiments, the SMEFT formalism can be used to introduce new operators that parametrise small deviations from the SM predictions at the LHC, induced by interactions between the known and new states. This thesis focuses on some operators and processes for which the leading correction to the SM, namely its interference with dimension-6 operators, is suppressed, either because it is small all over the phase space as a result of a helicity mismatch in the SM and SMEFT amplitudes, or because a cancellation between large cross-section contributions with opposite sign occurs. Several useful quantities are introduced to distinguish among these two cases, and a phenomenological strategy to revive the interferences is developed. They are applied to the cases of the $O_G$ and $O_W$ operators, respectively in three-jet production and EW processes like VBF $Zjj$, $WZ$ and $W\gamma$. The comparison among them highlights how different procedures can be followed to restore the interference. The quantities introduced in this thesis can be used to find simple kinematic observables that are sensitive to the suppression and can yield competitive bounds on the coefficients of the operators, even outside the SMEFT validity region.
           In the last chapter, a study of ten four-light quark operators is presented at LO matched to parton shower. Individual and marginalised limits on them are obtained through multijet production and processes where the jets are generated together with EW bosons, like $Z,W,\gamma +$jets. Almost no interference suppression happens for these operators, but they can virtually affect any process at NLO.
        \end{Abstract}

\clearpage

\begin{center}

\vspace{121pt}

{\LARGE Publications covered in this thesis:}
\vspace{22pt}
\begin{description}
\item[\cite{interf_rev_OG}] 
C. Degrande, M.M., {\it “Reviving the interference: Framework and proof-of-principle for the anomalous gluon self-interaction in the SMEFT”}, Physical Review D 103 (2021), arXiv:2012.06595[hep-ph]
\item[\cite{interf_rev_OW}]
C. Degrande, M.M., {\it “EFT observable stability under NLO corrections through interference revival”}, Phys. Lett. B 856, 138970 (2024), arXiv:2403.16894[hep-ph]
\item[\cite{4lq_paper}]
C. Degrande, M.M., {\it “Constraining the four-light quark operators in the SMEFT with multijet and VBF processes at linear level”}, arXiv:2511.04517[hep-ph]
\end{description}

\vspace{44pt}

During my PhD, I published other works of research that are not presented in this manuscript:
\vspace{22pt}
\begin{description}
\item[\cite{4t_letter}] 
L. Darmé, B. Fuks, H.-L. Li, M.M., O. Mattelaer, J. Touchèque, {\it “Novel approach to probing top-philic resonances with boosted four-top tagging”}, Physical Review D 111 (2025), arXiv:2404.14482[hep-ph]
\item[\cite{4t_paper}]
L. Darmé, B. Fuks, H.-L. Li, M.M., J. Touchèque, {\it “Searching for top-philic heavy resonances in boosted four-top final states”}, submitted to JHEP, arXiv:2507.05334[hep-ph]

\end{description}
\end{center}

%%% table of content
\pagestyle{empty}
\tableofcontents

% Introduction
\addcontentsline{toc}{chapter}{Introduction}
%!TEX root = main.tex

%%%%%%%%%%%%%%%%%%%%%%%%%%%%%%%%%%%%%%%%%%%%%%%%%%%%
%
%      Introduction :
%
%
%%%%%%%%%%%%%%%%%%%%%%%%%%%%%%%%%%%%%%%%%%%%%%%%%%%

\chapter*{Introduction}
\label{chap:intro}
\pagestyle{fancy}

Despite the great achievements of accelerator physics in the past decades, no evidence for new resonances can be found in the data, and all the measurements so far seem to agree with the SM. The SMEFT provides a framework to parametrise eventual small deviations from the SM predictions at the experimental energies, induced by interactions among very heavy new states and the known ones. Complete sets of operators with mass-dimension larger than four are added to the SM Lagrangian, and the goal is to place bounds over the values of their multiplicative coefficients through comparison with the experimental data.

The leading corrections to the SM at colliders normally come from its interferences with the dimension-6 operators, but there are cases where these terms are suppressed. For certain processes and operators at a given collider, the interference can be small over the entire phase space, for example if the helicities of the final states produced by the SM and the dimension-6 amplitudes do not match. Alternatively, since the interferences are not positive-definite, a cancellation between large cross-section contributions with opposite sign can occur.

In this thesis, I will introduce new useful quantities to identify a suppression of the interference and, in that case, to distinguish between the two scenarios above. I will also review the strategies to revive this term and show their application to the notable examples of the $O_G$ and $O_W$ operators, for which the cancellation of the interference contribution is well known in dijet and diboson processes, respectively. In both cases, I will also describe how the quantities we introduced can be used to find simple observables, built out of kinematic ones, that are sensitive to the suppression and can set competitive bounds over the coefficients of the operators at the LHC.

Additionally to this, I will illustrate our study of ten four-light quark operators, that can virtually affect any process at NLO. We placed bounds on their coefficients by investigating their LO interference contributions to multijet production and VBF processes like $Z,W,\gamma +$jets: the different EW bosons are sensitive to different quantum numbers of the quark fields and can thus constrain certain operators more than others.

This thesis is structured as follows: Chapters \ref{chap:sm} and \ref{chap:smeft} introduce some concepts about the SM and SMEFT that will be useful in the following. Chapter \ref{chap:interf_OG} explains the issue of interference suppression and some strategies to revive it, and shows their application to the case of the $O_G$ operator in three-jet production; in Chapter \ref{chap:interf_OW}, the same principles are applied to restore the $O_W$ interference contribution to $Zjj$, $WZ$ and $W\gamma$. Finally, Chapter \ref{chap:fourLQ} is about the four-light quark operators.

% Chapters
%!TEX root = main.tex

%%%%%%%%%%%%%%%%%%%%%%%%%%%%%%%%%%%%%%%%%%%%%%%%%%%%
%
%      Chapter I :
%
%
%%%%%%%%%%%%%%%%%%%%%%%%%%%%%%%%%%%%%%%%%%%%%%%%%%%

\chapter{An introduction to the Standard Model of particle physics}
\label{chap:sm}
\pagestyle{fancy}

\hfill
\begin{minipage}{\textwidth}
   \centering
   \includegraphics[width=.7\textwidth]{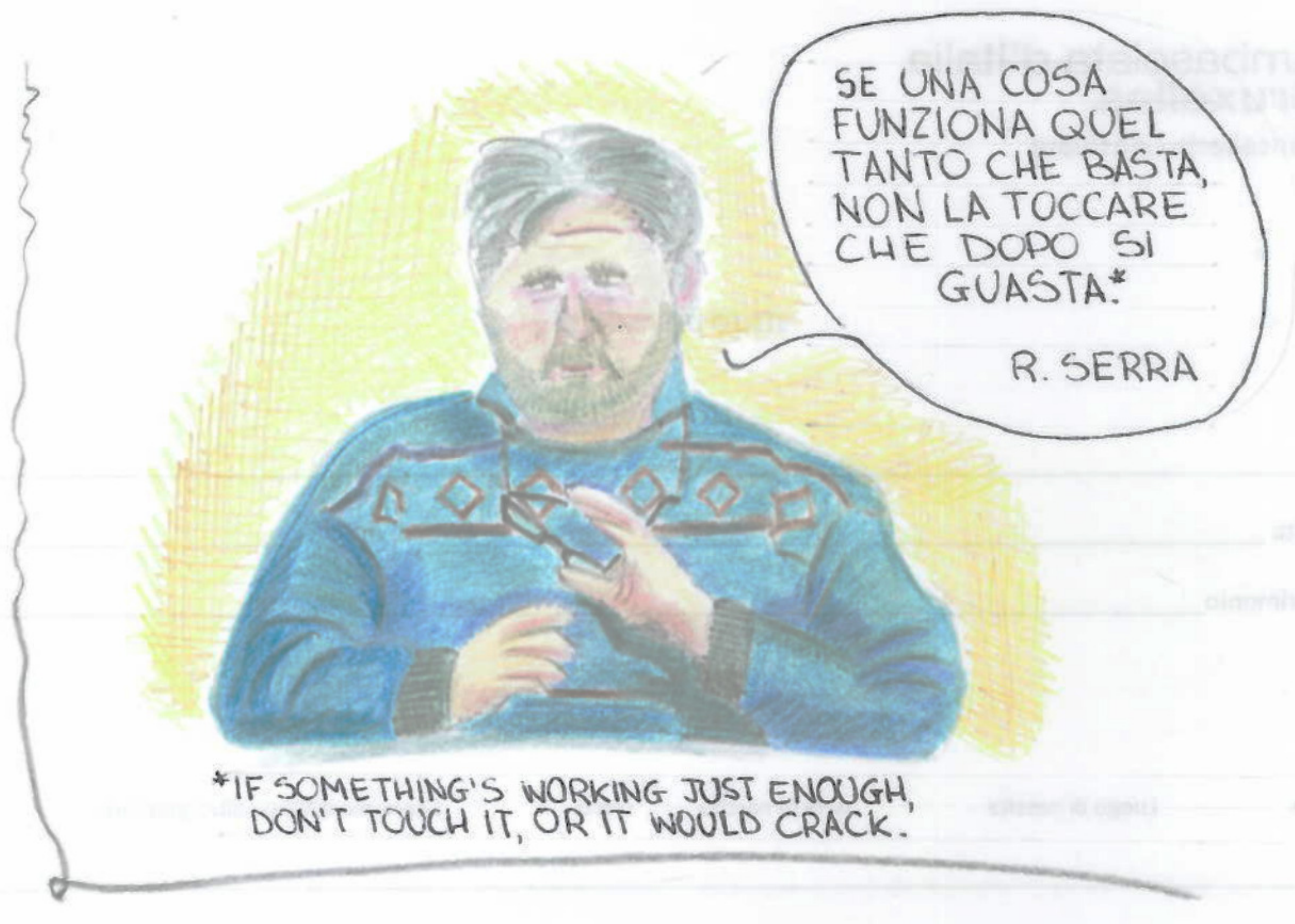}
\end{minipage}

\vspace{0.5cm}

The Standard Model (SM) describes the interactions among the fundamental constituents of matter.

In this framework, an object is ``fundamental'' if there is no scale associated to its size. The interactions are mediated by the exchange of particles grouped under the name of ``bosons'': the distribution of their masses follows a strange pattern for which we know no rule by now, and they can be both scalars and vectors. For each interaction, the charge is the strength of the coupling of a particle with the mediators. The only spin-zero boson that we have discovered is the Higgs $h$, that is a real scalar with invariant mass close to 125 GeV. Among the vectors, we have inferred the existence of the $W^\pm$ and $Z$ bosons, the gluon $g$ and the photon $\gamma$, all with spin $s=1$; the $\pm$ sign stands for the electric charge. The first two have masses of the order of 100 GeV, while the last two are supposed to be massless. For the photon, this is a theoretical requirement to ensure that a symmetry called ``gauge invariance'' is respected, but we will get to that later. On the gluon side, the argument about the null mass and the range of its interaction is more complicated, and we will get to that later on as well. The photon vehiculates the electromagnetic (EM) interaction, but it does not carry an EM charge and so it does not feel the respective field: for this reason, it is called an ``Abelian carrier''. These phenomena are described by the theory of Quantum Electrodynamics (QED). The $W$, $Z$ and $g$ bosons, on the other hand, are non-Abelian carriers as they present a charge for the forces that they mediate: the weak one for the first two and the strong one for the last, also called ``colour charge''. For this reason, they can self-interact with their own field. The framework that describes strong interactions is called Quantum Chromodynamics (QCD).

The matter constituents are grouped in a class of particles called ``fermions'': they always have $s = 1/2$ and are further divided into ``quarks'' and ``leptons''. The first subgroup feels the strong, EM and weak interactions (with the last two usually combined in the electroweak interaction, abbreviated as EW), while the second class only feels the EW one. Since they all are affected by the EW force, they are in the fundamental representation of its group $SU(2)_L$ and are organised in doublets. This happens to their left-handed components only, as the right-handed ones had been observed to be singlets under this symmetry at experiments, and in particular the one conducted in 1957 by Wu, Ambler, Hayward, Hoppes and Hudson on the decay of cobalt-60 \cite{wu_experiment}. The SM is thus chiral and parity is maximally violated. The left-handed doublets are split as
\begin{equation}
   \begin{array}{cccc}
       & 1^\text{st} \text{ gen.} & 2^\text{nd} \text{ gen.} & 3^\text{rd} \text{ gen.} \\[6pt]
      \text{Quarks } q & \dbinom{u}{d}_L & \dbinom{c}{s}_L & \dbinom{t}{b}_L \\[9pt]
      \text{Leptons } \ell & \dbinom{\nu_e}{e}_L & \dbinom{\nu_\mu}{\mu}_L & \dbinom{\nu_\tau}{\tau}_L \\
   \end{array} \label{su2_doublets}
\end{equation}
and an analogous distinction holds for the respective antiparticles. They are spread over three flavours (also called generations) with the same structure and the increasing mass as the only difference; ordinary matter is composed of the particles in the first family only, the lightest. Flavour-mixing can occur between the doublets.

The upper components of the lepton doublets are called ``neutrinos'' and can only interact through the weak force; this makes them extremely difficult to detect.

Because of the effect of the strong interaction, no asymptotic states seem to be possible for the quarks: they can only exist confined in singlets under the respective group $SU(3)_c$, called ``hadrons''. Among them, structures formed by quark-antiquark pairs, called ``mesons'', and others composed by three quarks or antiquarks, called ``baryons'', can be further identified. Some examples of mesons are the pions and kaons, while protons and neutrons are baryons.

\section{Symmetries and gauge invariance}
In the SM, symmetries determine the structure and properties of the theory. As a matter of fact, a theorem by N\"{o}ther states that for every continuous global transformation that leaves $\mathcal{L}$ invariant, there exists a conserved quantity. The necessity of the Lagrangian density $\mathcal{L}$ to be invariant under Lorentz transformations, also, introduces redundancies in the formalism: vectors need to present four components, but only have two degrees of freedom (d.o.f.) if massless and three if massive.

\subsection{Abelian theories}
In the following, the Dirac spinor will be identified with the symbol $\psi$, and $\bar \psi = \psi^\dagger \gamma^0$ will be its Dirac adjoint. The Dirac Lagrangian has the form
\begin{equation}
   \mathcal{L} = \bar \psi \left( i \gamma^\mu \partial_\mu -m \right) \psi,
\end{equation}
where $\gamma^\mu$ are the Dirac matrices and the parameter $m>0$ can be interpreted as the mass.
$\mathcal{L}$ shows a global phase symmetry under transformations $\psi \rightarrow e^{i\alpha}\psi$ and the conserved current is the EM one, $J^\nu = \bar \psi \gamma^\nu \psi$. The phase $\alpha$ here is not physical, as it cannot enter in any physical-quantity ecpression. In order to build a local symmetry, the parameter should depend on the coordinates as $\alpha(x)$. This produces a translation when the derivative $\partial_\mu$ is applied to the transformed field,
\begin{equation}
   \partial_\mu \left( e^{i\alpha(x)} \psi (x) \right) = e^{i\alpha(x)} \left[ \partial_\mu +i \partial_\mu \alpha(x) \right] \psi(x).
\end{equation}
The term $\bar \psi \gamma^\mu \partial_\mu \psi$ in $\mathcal{L}$ would no longer be invariant, unless a covariant derivative is defined as
\begin{equation}
   D_\mu = \partial_\mu +i e A_\mu,
\end{equation}
and used in place of the usual one.
In this expression, the electric charge $e$ is included for dimensional reasons and $A_\mu$ is called the ``gauge field''. If we require the new derivative to transform as the spinor, \ie\ $D_\mu \psi(x) \rightarrow e^{i\alpha(x)} D_\mu \psi(x)$, we obtain a rule for the transformation of the gauge field as well, namely 
\begin{equation}
   A_\mu (x) \rightarrow A_\mu (x) +\frac{1}{e} \partial_\mu \alpha (x). \label{gauge_fix}
\end{equation}
The term $\bar \psi \gamma^\mu D_\mu \psi$ is now invariant under local phase transformations and the global symmetry has been translated into a local one, but a new derivative had to be introduced together with a gauge field that transforms consistently with the spinors. 

$A_\mu$ is not a physical quantity, as we can only measure its components, the electric and magnetic fields, but it appears in the Lagrangian density as a field to which the spinors couple with a strength $e$: it can be associated to the EM field. $A_\mu$ has four components but only two or three d.o.f., depending if it is massless or not. This means that it is locally redundant, since it contains non-physical states mixed together inside it. It is then possible to apply changes that act on these states without modifying the physics, called ``gauge transformations''. %and Eq. \eqref{gauge_fix} is one of the possible ways to fix the gauge to perform the calculations.

Thanks to the covariant derivative, it is possible to build a Lagrangian density for QED,
\begin{equation}
   \mathcal{L} = -\frac{1}{4} F_{\mu\nu} F^{\mu\nu} + \bar \psi \left( i \cancel{D}-m \right) \psi, \label{l_abelian}
\end{equation}
that is locally invariant under the simultaneous transformations $\psi(x) \rightarrow e^{i\alpha(x)} \psi(x)$ and $A_\mu (x) \rightarrow A_\mu(x) +\frac{1}{e} \partial_\mu \alpha (x)$; here $\cancel{D}$ abbreviates $\gamma^\mu D_\mu$ and 
\begin{equation}
   F_{\mu\nu} = \partial_\mu A_\nu -\partial_\nu A_\mu \label{fmunu}
\end{equation}
is the EM field strength. Because of the redundancy described above, we can add extra conditions by fixing the gauge: a common choice in this case is the Lorenz one, that reads $\partial_\mu A^\mu =0$ and makes that the momentum and polarization vector of $A_\mu$ orthogonal in that gauge. Still, another requisite is needed for massless $A_\mu$ fields; as an example, if a scalar function $\phi$ is introduced such that $\partial_\mu \partial^\mu \phi = 0$, then the theory does not change if $A_\mu \rightarrow A_\mu +\partial_\mu \phi$ is applied.

It is easy to see that a mass term $m_\gamma^2 A_\mu A^\mu$ for the EM field would not be invariant under such gauge transformations, so the requirement of gauge invariance in the Abelian case implies $m_\gamma$ to be null, as mentioned above. Because of the redundancies, though, it is possible to add gauge-fixing terms like $\left( \partial_\mu A^\mu \right)^2$ to $\mathcal{L}$ and still obtain consistent theories for massive photons.

\subsection{Non-Abelian theories}
A similar procedure can be followed in the non-Abelian case $SU(n)$, by imposing a local transformation $\phi(x) \rightarrow G(x) \psi(x)$, where $G(x) = e^{i \alpha(x) \cdot \tau /2}$ is now a $(n \times n)$-matrix that can be expressed in terms of the $SU(n)$ generators $\tau^i$. A covariant derivative matrix can be introduced,
\begin{equation}
   D_\mu = \mathbb{I}_n \partial_\mu +i g A_\mu,
\end{equation}
including new gauge fields $A_\mu(x)$ whose transformation is determined by imposing 
\begin{equation}
   D_\mu(x) \psi(x) \rightarrow G(x) \left[ D_\mu(x)\psi(x) \right].
\end{equation} 
When the new derivative is used, the Lagrangian density contains conserved currents that describe the interactions among the spinors and the gauge fields, analogously to the QED example. In the non-Abelian case, though, the vertex terms include matrices $(\tau^i)_{\alpha\beta} (\gamma^\mu)_{ab}$ with their own indices, that project the product of two spinors onto a vector in the internal space of the non-Abelian group, in addition to the Lorentz one.

A kinetic term for the propagation of the gauge field can be generalised from the Abelian one in Eq. \eqref{l_abelian}, but the field strength has to be rewritten as
\begin{equation}
   \mathbb{F}_{\mu\nu} = \partial_\mu A_\nu - \partial_\nu A_\mu -i g [A_\mu,A_\nu]. \label{fmunu_noab}
\end{equation}
From the commutator, that contains $\tau$ matrices, new terms appear that depend on the coupling constant $g$ and describe self-interactions of the gauge field.

The group that describes the weak interaction is $SU(2)$ and presents three generators that can be expressed in terms of the $(2\times 2)$ Pauli matrices as $\tau^I = \sigma^I /2$, with $I=1,2,3$. For what concerns the strong interaction, $SU(3)$ shows eight generators $T^a = \lambda^a /2$, where $a = 1, \ldots, 8$ and $\lambda^a$ are the $(3\times 3)$ Gell-Mann matrices. The generators of both groups have null traces. They satisfy
\begin{equation}
   [\tau^I,\tau^J] = i \varepsilon^{IJK} \tau^K, \hspace{5mm} [T^a,T^b] = i f^{abc} T^c, \label{struct_const}
\end{equation}
with $\varepsilon^{IJK}$ and $f^{abc}$ the structure constants of $SU(2)$ and $SU(3)$. The generator normalisations are fixed by the conditions $tr(T^a T^b) = \frac{1}{2}\delta_{ab}$ and $tr(\tau^I \tau^J) = \frac{1}{2}\delta_{IJ}$ on the product traces.

Since a fermion and an antifermion can interact in a vertex with a gluon, they are in the fundamental and antifundamental representations of $SU(3)_c$, respectively. Quarks have never been seen isolated, and all physical objects observed so far are invariant under the same group transformations. It can be shown that this is possible only if the difference between the number of quarks and antiquarks in the particle is zero in modulus three: the most simple cases for this condition are represented by the mesons and baryons introduced before. Since baryons have integer electric charges, quarks must possess multiples of $1/3$ as a charge.

\begin{figure}
   \centering
   \caption{\footnotesize{Notation for the colour indices $i,j,k,\ell$ in a photon ({\it left}) and a gluon ({\it right}) exchanges between a quark and an antiquark lines. $T^a$ with $a=1,\ldots,8$ are the $SU(3)$ generators}} \label{fig:colour_contract}
   \includegraphics[width=.7\textwidth]{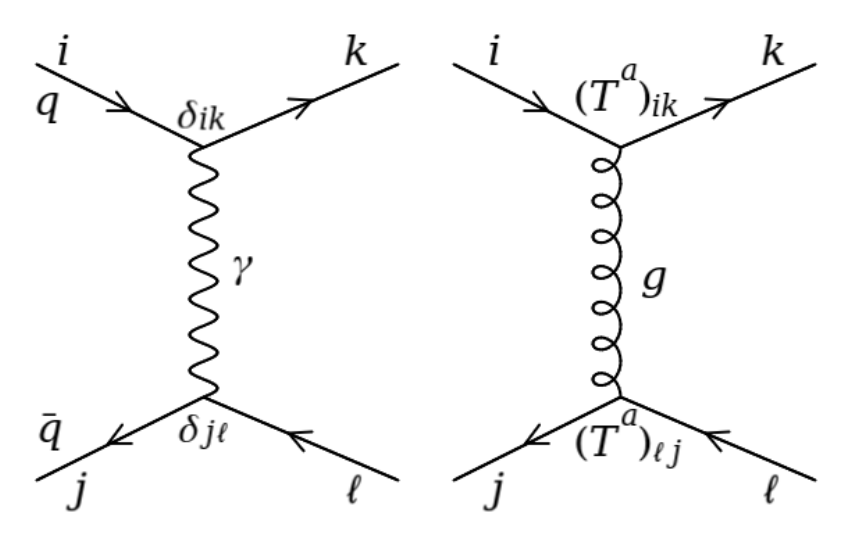}
\end{figure}

Let us consider the exchange of a photon or a gluon among two quark lines, as in Fig. \ref{fig:colour_contract}. In the QED case, the interaction between a quark and an antiquark is always attractive because of their opposite electric charges. In QCD, the nature of the interaction depends on the state of the system itself. Thanks to the Fierz identities, the product of $SU(3)$ matrices that comes from the two vertices with the gluon can be written as
\begin{equation}
   (T^a)_{ik}(T^a)_{\ell j} = \frac{1}{2} \left( \delta_{ij} \delta_{\ell k} -\frac{1}{n} \delta_{ik} \delta_{\ell j} \right), \label{fiertz_ids}
\end{equation}
with $n$ the number of colours. This means that the product can be projected either on a singlet or an octet state. In the first case, the quark and the antiquark, that might come from a $e^+ e^- \rightarrow \gamma^* \rightarrow q \bar q$ interaction, attract each other through the exchange of gluons and bind together; in the second one, like in $q \bar q \rightarrow g^* \rightarrow q \bar q$, they would repel if not emitting gluons all around them, and selection rules apply for a bound-state formation.

When asking for gauge invariance for non-Abelian theories in some gauge choices, non-physical d.o.f. appear in the Lagrangian and new special ones have to be introduced to cancel them off. This trick comes from the Fadeev-Popov quantisation and the new states are referred to as ``ghost fields'' $\chi$: they are anticommuting scalars that couple to the gluons and propagate. Their Lagrangian density has the form
\begin{equation}
   \mathcal{L}_\text{Ghosts} = -\bar \chi^a \left( \partial_\mu \partial^\mu \delta_{ac} +g \partial^\mu f^{abc} A_\mu^b \right) \chi^c. \label{l_ghosts}
\end{equation}
The amplitude for ghost-antighost production $q \bar q \rightarrow \chi \bar \chi$ cancels exactly the non-physical d.o.f. in the $q \bar q \rightarrow gg$ process.

\section{\label{sect:beh}Hidden symmetries and the BEH mechanism}
As mentioned in the previous section, the N\"{o}ther theorem relates continuous exact symmetries to conservation laws, making gauge invariance a dynamical principle to build interacting theories. Because of this, though, it is not possible to explicitly write mass terms for the gauge bosons in non-Abelian theories.

Symmetries, on the other hand, can also be non-exact: in a magnet, for example, it is only below a certain critical temperature $T_c$ that atoms can arrange in domains with the same spin, producing a non-zero magnetic field. Above $T_c$, instead, thermal disorder dominates. The Lagrangian density is the same in both cases, but the vacuum is symmetric under $O(3)$ above $T_c$ and only under $O(2)$ below $T_c$. The energy minima are different in the two cases: this phenomenon is called ``spontaneous symmetry breaking'' (SSB). % and can only happen if there is an infinite number of d.o.f., as in field theories.

If we consider two scalar fields $\phi_1$ and $\phi_2$, we can write for them a potential that only depends on the square of the length of the $(\phi_1, \phi_2)^T$ vector, as
\begin{equation}
   V(\phi_1^2+\phi_2^2) = \frac{\mu^2}{2}\left( \phi_1^2+\phi_2^2 \right) +\frac{\lambda}{4}\left( \phi_1^2+\phi_2^2 \right)^2. \label{potential_example}
\end{equation}
Here, $\lambda$ has to be positive to bound the system energy from below. The resulting Lagrangian density is invariant under $SO(2)$ rotations. The system shows a unique minimum for the energy at $\phi_1 = \phi_2 = 0$ if $\mu^2 >0$; if $\mu^2<0$ instead, the same point becomes a relative maximum and the minima are identified by the condition $\phi^2_1+\phi_2^2 = |\mu^2|/\lambda=v^2$. This represents a circle in the $\phi_1$ {\it vs} $\phi_2$ plane, so there is an infinite set of minima that the system can assume, all equivalent in energy. It is not possible to predict which one the system will choose, but we can set the axes so that the actual minimum is parametrised as $\binom{\phi_1}{\phi_2} = \binom{v}{0} = \vec{\phi}_0$. If we define $\vec{\phi}^\prime = \binom{\phi_1}{\phi_2}-\vec{\phi}_0 = \binom{h}{\xi}$ and expand the Lagrangian density around $\vec{\phi}_0$, we can see that $h$ gets a mass term with $m_h^2 = 2\mu^2$, while $\xi$ does not. Indeed, $h$ oscillates in radial direction, while $\xi$ moves from one vacuum to another along the angular direction.

This situation is formally explained by the Goldstone theorem, which affirms that when a symmetry is broken, massless states called ``Nambu-Goldstone bosons'' (NGB) are generated in the same number as that of the broken generators.

The Brout-Englert-Higgs (BEH) mechanism combines gauge symmetries and SSB to introduce longitudinal modes that give mass to the massive bosons while preserving unitarity.

In the Abelian case, a complex scalar $\phi$ is introduced with a potential $V(|\phi|) = \mu^2 |\phi|^2 +\lambda (|\phi|^2)^2$ analogous to the example above, and it has to couple with the gauge field $A_\mu$. If $\mu^2<0$, non-trivial minima $\left< \phi_0 \right> = \pm \frac{v}{\sqrt{2}}$ can be identified (again, the axes are set in a way to pick this configuration among all the possible phases), and small fluctuations around them can be parametrised via a real field $h$, that is the Higgs field and gets a mass, and an imaginary one $\xi$, the NGB. Indeed, $\xi$ only appears in the Lagrangian through its derivatives, and $\mathcal{L}$ is invariant under shifts $\xi \rightarrow \xi-v$ because $\partial_\mu v = 0$. Furthermore, it is possible to fix the gauge such that it disappears, as it is not physical: $\xi$ is absorbed by $A_\mu$, that gains a longitudinal d.o.f. and becomes massive. This gauge is called ``unitary'' and it is the one where the mass content is evident. The total number of d.o.f. is the same before and after SSB and gauge transformation.

The non-Abelian case is analogous. Let us consider the triplet representation of $SU(2)$, for example, where $\vec{\phi}$ is a vector of three fields; the effective potential depends on $|\vec{\phi}|^2$. In this representation, th generators take the form
\begin{equation}
   \tau^1 = i \left( \begin{array}{ccc} 0 & 0 & 0 \\ 0 & 0 & -1 \\ 0 & 1 & 0 \\ \end{array} \right), \hspace{5mm} 
   \tau^2 = i \left( \begin{array}{ccc} 0 & 0 & 1 \\ 0 & 0 & 0 \\ -1 & 0 & 0 \\ \end{array} \right), \hspace{5mm}
   \tau^3 = i \left( \begin{array}{ccc} 0 & -1 & 0 \\ 1 & 0 & 0 \\ 0 & 0 & 0 \\ \end{array} \right).
\end{equation}
When the potential has only one minimum for $\vec{\phi}=\vec{0}$, the theory reduces to the Yang-Mills one, with three massless gauge bosons. In case of SSB, the new minimum can be parametrised as $\left< \vec{\phi}_0 \right> = \left( 0,0,v \right)^T$, that is invariant under the action of $\tau^3$, but not of $\tau^1$ and $\tau^2$. Since two generators are now broken, two NGBs are introduced. In the unitary gauge, the oscillations around our choice for the vacuum can be written as $\left( 0,0,v+h \right)^T$, while $\xi$ does not appear in the Lagrangian density. As a result, the two gauge bosons related to $\tau^{1,2}$ absorb two components of $\xi$, gaining a longitudinal d.o.f. each and thus a mass; the third gauge boson remains massless. The Higgs boson $h$ obtains a mass term as well.

This is the same scenario that is requested in the SM, where the $W^\pm$ and $Z$ bosons need mass terms, while the photon has to remain massless. In order to achieve this, a gauge symmetry needs to be partially broken as
\begin{equation}
   SU(2)_L \otimes U(1)_Y \rightarrow U(1)_\text{EM},
\end{equation}
giving masses to the $W$ and $Z$ bosons, while the remaining symmetry after SSB ensures that $\gamma$ does not get any mass terms, as the gauge invariance of EM requires. Above, $U(1)_Y$ is the symmetry associated to the hypercharge $Y$, whose gauge field is indicated as $B_\mu$. Since it has to interact with the $SU(2)$ part of the gauge theory, the simplest representation for the Higgs field is a $SU(2)$ doublet.
Three of the four generators of the initial group are broken to become the longitudinal d.o.f. of the new massive bosons, so the scalar field $\phi$ must have at least three real scalar components that play the NGB role. After SSB, we are left with a massless gauge boson related to the unbroken generator, recognised as the photon, and a real scalar field for the radial mode $h$ that is the Higgs boson. Then, the potential for the field $\phi$ takes the form
\begin{equation}
   V(\phi) = \frac{1}{2} m_h^2 h^2 +\lambda v h^3 +\frac{\lambda}{4} h^4,
\end{equation}
with $m_h^2 = 2 \lambda v^2$ the Higgs mass. Its value, together with the vacuum $v$ one, fixes the cubic and quartic self-interaction couplings, at least in the SM. The EM generator can be written as $Q=\tau^3 +\frac{Y}{2}\mathbb{I}_2$ and transformations along it give rise to the unbroken symmetry.

The physical vector bosons of the EW sector are defined from the ones that are present before SSB through rotations
\begin{align}
   W^\pm_\mu &= \frac{1}{\sqrt{2}}\left( W_\mu^1 \mp i W_\mu^2 \right), \\
   Z_\mu &= c_W W_\mu^3-s_W B_\mu, \nonumber \\
   A_\mu &= s_W W_\mu^3 +c_W B_\mu, \nonumber
\end{align}
where $c_W$ and $s_W$ are respectively the cosine and sine of the rotation angle $\theta_W$ between the planes containing the original and the physical vector bosons. In the unitary gauge, the kinetic term for the Higgs field contains the $W^\pm$ and $Z$ mass terms and their interactions with the Higgs, as
\begin{align}
   \mathcal{L}_\text{Higgs} \supset \frac{1}{2} \partial_\mu h \partial^\mu h &+ M_W^2 W^+_\mu W^{-,\mu} +\frac{1}{2} M_Z^2 Z_\mu Z^\mu \label{l_higgs} \\
   &+\frac{1}{8}\left( 2vh+h^2 \right)\left(2 g^2_\text{w} W_\mu^+ W^{-,\mu}+g_Z^2 Z_\mu Z^\mu \right). \nonumber
\end{align}
As explained before, in this gauge the NGBs do not appear explicitly, as they are absorbed by the gauge fields as longitudinal d.o.f. The boson masses and $v$ fix the couplings among the Higgs and $W,Z$.

The masses for the fermions are also induced by their interaction with the Higgs field, as a canonical $m_\psi \bar \psi \psi$ term would spoil the $SU(2)_L$ symmetry: as seen before, fermions have components of different chiralities that transform differently. Instead, each of them couples to the Higgs field vacuum $v$ with different strengths, called ``Yukawa couplings'' $y$, and gain mass terms in form of matrices $M_\psi = v y_\psi /\sqrt{2}$ in the unitary gauge. If we name $\widetilde{\phi} = i\tau^2 \phi^*$, the correspondent part of the Lagrangian reads
\begin{equation}
   \mathcal{L}_\text{Yukawa} = -(y_u)_{pr}\ \bar q_p \widetilde{\phi} u_r -(y_d)_{pr}\ \bar q_p \phi d_r -(y_e)_{pr}\ \bar \ell_p \phi e_r +\text{h.c.}, \label{l_yukawa}
\end{equation}
where $p,r$ are generation indices, $q$ and $\ell$ are the $SU(2)$ doublets \eqref{su2_doublets} and $u,d,e$ are the right-handed singlets. The interaction between the Higgs and each fermion is proportional to the mass of the latter, with the top quark showing the largest Yukawa coupling.

The mass matrices $M_\psi$ can be diagonalised as $M_\psi^\text{diag} = L_\psi M_\psi R_\psi^\dagger$ through field rotations $u_L \rightarrow L_u u_L$ and $d_L \rightarrow L_d d_L$, and analogously for the right-handed components through $R_{u,d}$. The changes to the interaction terms among the fermions driven by this are described by the Cabibbo-Kobayashi-Maskawa (CKM) mixing matrix $V_\text{CKM} = L_u L_d^\dagger$, that can be parametrised through three angles and a complex phase for $CP$-violation.
A similar mechanism exists for the leptons, with the difference that no right-handed neutrinos are present in the SM, so such Yukawa interactions cannot be built for them and they remain massless at the Lagrangian level.

\section{\label{sect:renorm}Regularisation and renormalisation}
Given a QFT in (3+1) dimensions, described by a Lagrangian density $\mathcal{L}$ with its content of fields, derivatives, mass terms and couplings, there will always be loop diagrams that are not well-defined and diverge. A theory is renormalisable if these infinities can be reabsorbed inside a finite number of parameters through field and coupling redefinitions; this has to happen order by order in the perturbative expansion.

Every theory is an effective one, as it has limits in the infrared (IR) and ultraviolet (UV) regimes. Renormalisable theories do not provide any information on their upper cutoff $\Lambda$, which represents the energy scale at which they would break. This is the case for the SM. A fundamental theorem in QFT states that only theories that include operators with dimension lower or equal to four are renormalisable.

The renormalisation procedure of a theory happens in two steps. The first is called ``regularisation'' and consists in making the infinities explicit by quantifying the divergencies; then these have to be reabsorbed inside bare quantities, that can never be accessed in measurements. It is usually performed with the insertion of a cutoff $\Lambda$ in the diverging integrals, or by computing them in $4-\varepsilon$ dimensions. These two methods are equivalent, as logarithms of the cutoff scale have the same meaning as $1/\varepsilon$ poles when the $\Lambda \rightarrow \infty$ and $\varepsilon \rightarrow 0$ limits are taken.

\begin{figure}
   \centering
   \caption{\footnotesize{Examples of corrections to the fundamental Feynman diagrams that need to be computed to renormalise QED at one-loop: fermion ({\it a}) and photon ({\it b}) propagators, and fermion-photon interaction vertex ({\it c}). The second row shows examples of a real-emission diagram ({\it d}) and a virtual one ({\it e})}} \label{fig:qed_renorm}
   \includegraphics[width=.32\textwidth]{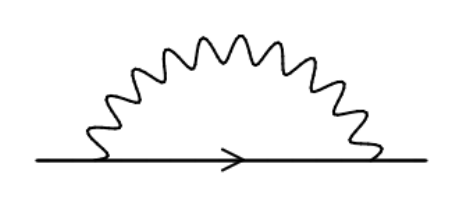}
   \includegraphics[width=.32\textwidth]{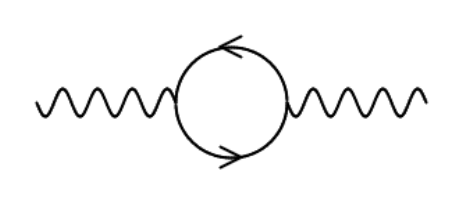}
   \includegraphics[width=.32\textwidth]{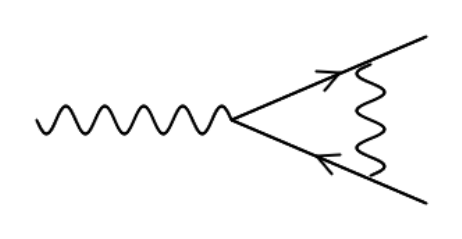}
   \begin{minipage}{.32\textwidth}
      \centering
      (a)
   \end{minipage}
   \begin{minipage}{.32\textwidth}
      \centering
      (b)
   \end{minipage}
   \begin{minipage}{.32\textwidth}
      \centering
      (c)
   \end{minipage}
   \begin{minipage}{.16\textwidth}
      \hspace{1mm}
   \end{minipage}
   \includegraphics[width=.32\textwidth]{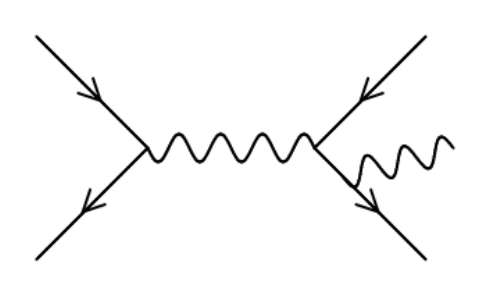}
   \includegraphics[width=.32\textwidth]{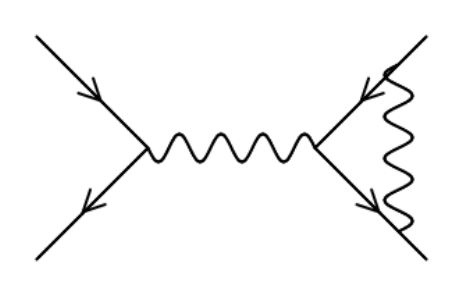}
   \begin{minipage}{.16\textwidth}
      \hspace{1mm}
   \end{minipage}
   \begin{minipage}{.32\textwidth}
      \centering
      (d)
   \end{minipage}
   \begin{minipage}{.32\textwidth}
      \centering
      (e)
   \end{minipage}
\end{figure}

In order to renormalise an Abelian theory like QED at one-loop, the corrections to the self-energies of fermions and photons and to their interaction vertex need to be computed: their representations are sketched in the diagrams (a,b,c) of Fig. \ref{fig:qed_renorm}, respectively. Real-emission diagrams, like the one depicted in Fig. \ref{fig:qed_renorm}d, interfere among themselves, while virtual ones can only interfere with the tree-level at this order. No UV divergencies can come from the real diagrams, because the energy of the real photon is bounded from above by the total energy of the system; neither they are problematic in the IR regime as the poles cancel off when the virtual contribution is included. Virtual diagrams, on the other side, contain loops where the momentum running inside is not bounded and can yield divergencies in the UV; an example is shown in Fig. \ref{fig:qed_renorm}e.

Since the UV behaviour is  the only one that matters here, one can assume that the momentum in the loop is much larger than the external-state ones. This implies that a cutoff $\mu$ has to be introduced in the IR as well, because the approximation above would not be valid there. This quantity is not physical and disappears when all the orders in the expansion are included, but it is needed in the calculations if a truncation is performed, as it usually is the case.

The renormalisation procedure in QED is based on the fact that the fermions contributing to EM processes posses a bare electric charge $e_B$ and a renormalised and physical one $e_R$, and we are only able to measure the latter. The relation among the two is expressed through multiplicative factors that contain all the diverging corrections at a certain order, and are thus infinite. In the QED case, it is possible to see that only the photon self-energy is responsible for corrections to the fermion charges. As a result of this mechanism, the physical charge is not fixed and its value varies with the energy $\mu$ as
\begin{equation}
   e_R \left( \mu^2, \Lambda^2 \right)-e_R \left( \mu_0^2, \Lambda^2 \right) = \frac{e_R^3}{24\pi^2} \log \frac{\mu^2}{\mu_0^2}.
\end{equation}
This expression does not explicitly depend on the bare charge, which cannot be measured, and on the UV cutoff $\Lambda$, as it is expected from a renormalisable theory. Only the value of the charge relative to a different energy scale can be computed, because its absolute value depends on $\Lambda$. In the QED case, the charge increases with the energy $\mu$.

By imposing that the bare charge has to remain constant with the energy, one obtains the $\beta$-function for the renormalised fine-structure constant $\alpha_R = e_R^2 /(4\pi)$ at LO in QED, namely
\begin{equation}
   \beta\left( \alpha_R(\mu^2) \right) = \frac{d\ \alpha_R}{d\ \log \mu^2} = \frac{\alpha_R^2}{3\pi}.
\end{equation}
This effect can be explained through vacuum polarisation: more energetic probes are able to get closer to the charge and measure a higher value for it, as the screening from the virtual particles around it is less intense.

In QCD, dimensional regularisation is usually employed. At Next-to-Leading Order (NLO), the corrections to the gluon and quark self-energies and to their vertex have to be computed, analogously to the QED case, but gluon self-interactions have to be included. The $\beta$-function for $\alpha_s(\mu^2)$ is negative this time, meaning that the coupling constant decreases when the scale increases. This is consistent with the facts that quarks are never observed in an asymptotic-freedom state and that, despite the gluons null mass, the strong interaction has such a short range. It also makes the gluon and quark masses challenging to measure.

In its running with $\mu$, $\alpha_s$ diverges at the scale $\Lambda_\text{QCD} \sim 250$ GeV. At this energy magnitude, the theory is no longer predictive because the interaction is not perturbative anymore and hadrons appear as bound states. The same divergence happens for the EM $\alpha$, but at much higher energies $\Lambda_\text{QED} \sim 10^{250}$ MeV: for this reason, QED processes can always be treated in a perturbative way.

The fact that the theory changes with $\mu$ means that scale invariance is broken. These variations are defined by the ``renormalisation group'': each operator $O$ is related to a renormalisation constant that allows to match its bare behaviour $O_B(g_B)$ with the renormalised one $O_R \big( g_R(\mu),\mu \big)$, with $g_R$ a coupling constant. The $\beta$-function is generalised as $\beta = \frac{d\ g_R}{d\ \log\mu}$ and the running of the operator is described by the Callan-Symanzik equation
\begin{equation}
   \left( \frac{\partial}{\partial\ \log\mu}+\beta \frac{\partial}{\partial g_R} -\gamma\right) O_R = 0. \label{ren_group}
\end{equation}
$\gamma$ is called ``anomalous dimension'' of $O$ and is a number; it is null for dimensionless operators. This result holds at any order of perturbation.

\section{The general structure of the SM, and some problems}
By adding together the colour and EW structures described in the previous sections, we obtain the gauge symmetry for the SM $SU(3)_c \otimes SU(2)_L \otimes U(1)_Y$, with total covariant derivative
\begin{equation}
   D_\mu = \partial_\mu +i g_S^{} G^a_\mu T^a +i g_W^{} W^I_\mu \tau^I +\frac{i}{2} g_Y^{} Y B_\mu. 
\end{equation}
Here, $G_\mu^a$ are the eight gauge fields of $SU(3)_c$, $W^I_\mu$ are the three ones of $SU(2)_L$ and $B_\mu$ is the gauge field of $U(1)_Y$; $g_S^{}$, $g_W^{}$ and $g_Y^{}$ are the gauge couplings for the respective interactions. The field strengths defined in Eqs. \eqref{fmunu} and \eqref{fmunu_noab} are
\begin{align}
   G^a_{\mu\nu} &= \partial_\mu G^a_\nu -\partial_\nu G^a_\mu +g_S^{} f^{abc} G^b_\mu G^c_\nu, \label{field_strengths} \\
   W^I_{\mu\nu} &= \partial_\mu W^I_\nu -\partial_\nu W^I_\mu +g_W^{} \varepsilon^{IJK} W_\mu^J W_\nu^K, \nonumber \\ 
   B_{\mu\nu} &= \partial_\mu B_\nu - \partial_\nu B_\mu. \nonumber 
\end{align}
The complete SM Lagrangian density is obtained by adding up the different pieces
\begin{equation}
   \mathcal{L}_\text{SM} = \mathcal{L}_\text{Gauge} + \mathcal{L}_\text{Fermions}+\mathcal{L}_\text{Higgs} +\mathcal{L}_\text{Yukawa}+\mathcal{L}_\text{Ghosts},
\end{equation}
where $\mathcal{L}_\text{Ghosts}$ is defined in Eq. \eqref{l_ghosts}, $\mathcal{L}_\text{Yukawa}$ in Eq. \eqref{l_yukawa}, $\mathcal{L}_\text{Higgs}$ includes the part in Eq. \eqref{l_higgs} plus $V(\phi)$, and
\begin{align}
   \mathcal{L}_\text{Gauge} &= -\frac{1}{4} G^a_{\mu\nu} G^{a,\mu\nu} -\frac{1}{4} W^I_{\mu\nu} W^{I,\mu\nu} -\frac{1}{4} B_{\mu\nu} B^{\mu\nu}, \\
   \mathcal{L}_\text{Fermions} &= \sum_p i\bar q_p \cancel{D} q_p + i\bar u_p \cancel{D} u_p +i\bar d_p \cancel{D} d_p +i\bar \ell_p \cancel{D} \ell_p +i \bar e_p \cancel{D} e_p. \nonumber
\end{align}

The SM presents eighteen free parameters: three gauge couplings, nine lepton and quark masses, four CKM parameters and two constants in the Higgs potential. The QCD vacuum angle $\theta$, associated to $CP$-violating effects, should also be added to the list, but it is much smaller than one. Among these quantities, only the Higgs triple and quartic self-interaction couplings are yet to be directly evaluated at experiments and compared against their predictions from the SM, which states that they are functions of the Higgs mass and vacuum.

The most precisely-measured quantities in the SM are the electron magnetic moment $g_e$, the muon lifetime $\tau_\mu$ and the $Z$-boson pole mass $M_Z$. The first one allows to determine the fine-structure constant via the one-loop relation $g_e-2 = \alpha(\mu=0) /\pi$. The second one, as it will be mentioned in Sect. \ref{sect:fermi_th}, is related to the Fermi constant $G_F$ and the value of the Higgs vacuum $v$. All the other parameters in the EW sector of the SM can be obtained as combinations of these: for example, $\theta_W$ is a function of $\alpha$, $G_F$ and $M_Z$, and it fixes $M_W$ as well. Nonetheless, other schemes exist that make use of different quantities as inputs to determine all the others.

Despite the vast agreement between SM predictions and measurements for some EW quantities, some experimental evidence seems to suggest that a more comprehensive theory is needed, and most of these unexplained phenomena have an astronomical origin. For example, the rotation curves for the galaxies that we measure do not match the predictions \cite{Zwicky:1933gu,Rubin:1970zza}, hinting that particles that do not interact under the SM forces might constitute a large fraction of the matter content of the universe. The observed anisotropies in the Cosmological Microwave Background \cite{planck15} and the Bullet Cluster \cite{Clowe_2004} further support this statement. These hypothetical new stable and SM-neutral states are referred to as ``dark matter'': only hypotheses can be made on their origin, mass range and couplings to the SM, if any \cite{dark_matter_1,dark_matter_2}.

Furthermore, we saw that the neutrinos do not receive a mass term in the SM, but the oscillations between their three flavours that we observe in solar, atmospheric and accelerator measurements seem to be explainable only if at least two of the generations have a non-zero mass. Unluckily, these oscillations only depend on the mass differences and do not provide any information about their absolute scales \cite{Fukuda_1998,Ahmad_2001,Ahmad_2002}.

Concerning the Higgs boson, it gains its mass $m_h$ from the vacuum $v$, but this receives corrections at loop level that diverge quadratically: normally, they are cancelled out by counter-terms in dimensional regularisation. However, because of this the Higgs mass becomes sensitive to UV parameters through loops when completions to the SM are considered, and the corrections can be much larger than the measured mass value. Only a fine-tuning of the bare parameters can keep the variations under control, and many new-physics models have been introduced where symmetries protect the $m_h$ value \cite{higgs_1,higgs_2,higgs_3}.
There is also no mechanism that explains the hierarchy of masses of the SM states, as they span over a large range: the Yukawa couplings remain indeed free parameters of the theory.

Recently, a $\sim 5\sigma$ discrepancy among the experimental value for the muon magnetic moment $g_\mu$ and its SM prediction was observed, but more complete calculations that make use of lattice-QCD techniques verified that the gap is in reality much smaller than that \cite{muongminus2}. 

This is only a short list of the puzzles that remain unsolved and are still debated by the scientific community. They suggest that the SM might be an approximation at low energy of a more complete theory; many proposals for it have been suggested over the years, but the lack of experimental evidence does not allow us to rule many of them out.

%!TEX root = main.tex

%%%%%%%%%%%%%%%%%%%%%%%%%%%%%%%%%%%%%%%%%%%%%%%%%%%%
%
%      Chapter II :
%
%
%%%%%%%%%%%%%%%%%%%%%%%%%%%%%%%%%%%%%%%%%%%%%%%%%%%

\chapter{The Standard Model as an Effective Field Theory}
\label{chap:smeft}
\pagestyle{fancy}

\hfill
\begin{minipage}{\textwidth}
   \centering
   \includegraphics[width=.7\textwidth]{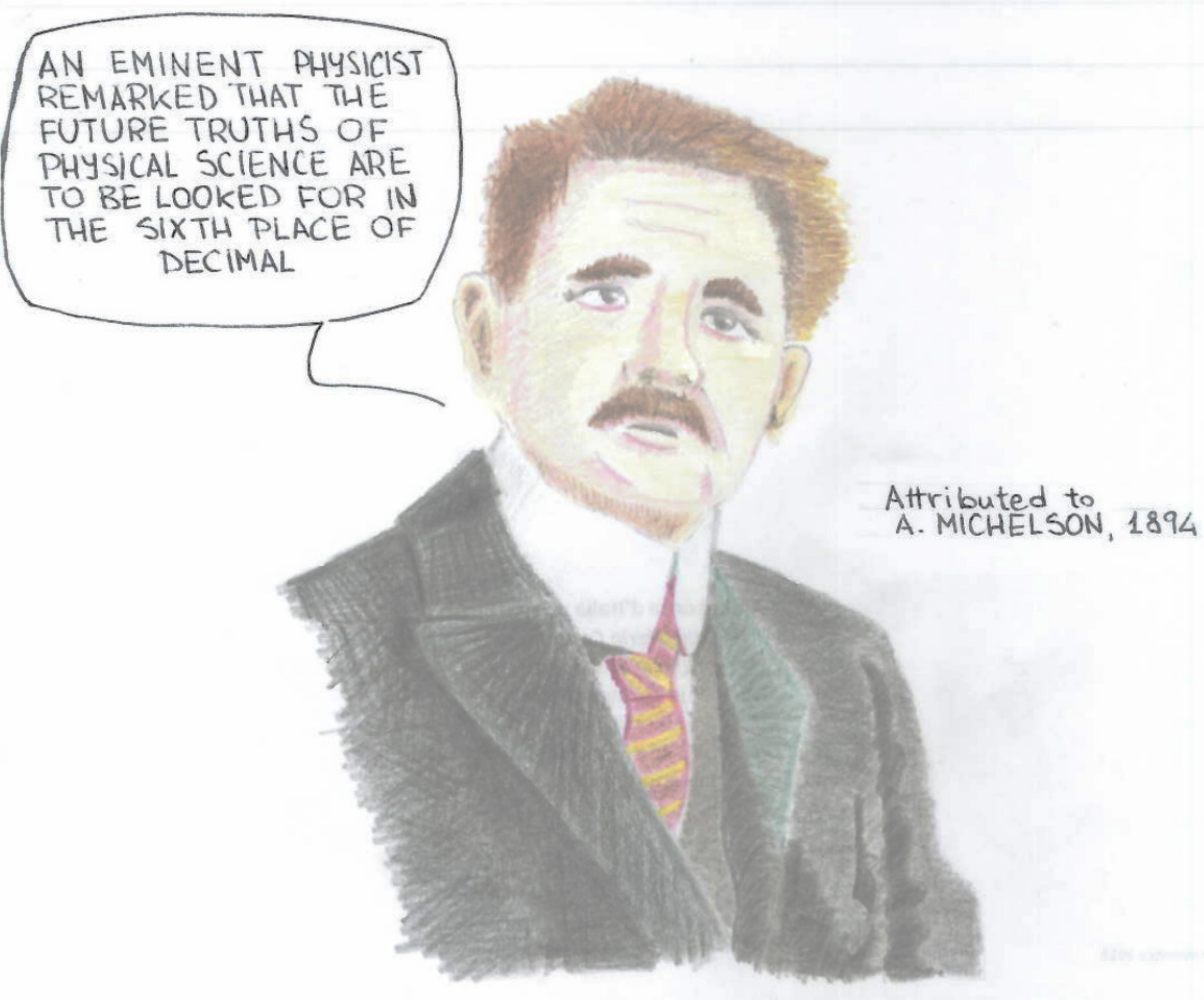}
\end{minipage}

\vspace{0.5cm}

As briefly introduced in Sect. \ref{sect:renorm} of the previous chapter, an Effective Field Theory (EFT) is a description of physics that is only valid at a certain energy scale. It is usually introduced as an expansion around a QFT to an arbitrarily high order, and the structure of the added terms can highlight new symmetries. It only captures the essential features of the phenomena it is conceived to describe, which can then be improved to any desired accuracy by including more and more terms in the expansion.

If the relevant scales of a QFT span over many orders of magnitude, EFTs offer a way to organise them in a description controlled by the ratios of such scales or their logarithm. Furthermore, an EFT working in a certain energy range does not need information about the dynamics at higher scales. As an example, let us consider a scattering amplitude between some external states with mass $m$ and momentum $p$, mediated by a boson with mass $M$ such that $M^2 \gg m^2 \sim p^2$: the terms divided by the heavy mass yield a negligible effect and only appear as higher-order corrections in the expansion in powers of $p/M \sim m/M$.

This is formalised by the ``decoupling theorem'' by Applequist and Carazzone: the effects of physics with mass $M$ decouple at small momentum $p^2 \ll M^2$ and only result in the shift of low-energy renormalisation constants, with effects of order $p^2/M^2+\ldots$ \cite{Appelquist:1974tg}.

This means that one can only consider the particles that are light enough to be produced at the energy scale described by the EFT, while the heavier ones can be integrated out:
\begin{equation}
   \mathcal{L}_\text{Full}\left( \phi_\text{light},\phi_\text{heavy} \right) \rightarrow \mathcal{L}_\text{Full} \left( \phi_\text{light} \right)+\mathcal{L}_\text{EFT} \left(\phi_\text{light}, M^2_\text{heavy} \right).
\end{equation}
This introduces changes in the couplings of the interactions among the light fields.

\section{\label{sect:fermi_th}Fermi theory and other examples}
\begin{figure}
   \centering
   \caption{\footnotesize{Feynman diagrams for the muon decay into an electron in the SM ({\it left}) and in Fermi theory ({\it right}). $p$ is the muon momentum}} \label{fig:muon_decay}
   \includegraphics[width=.8\textwidth]{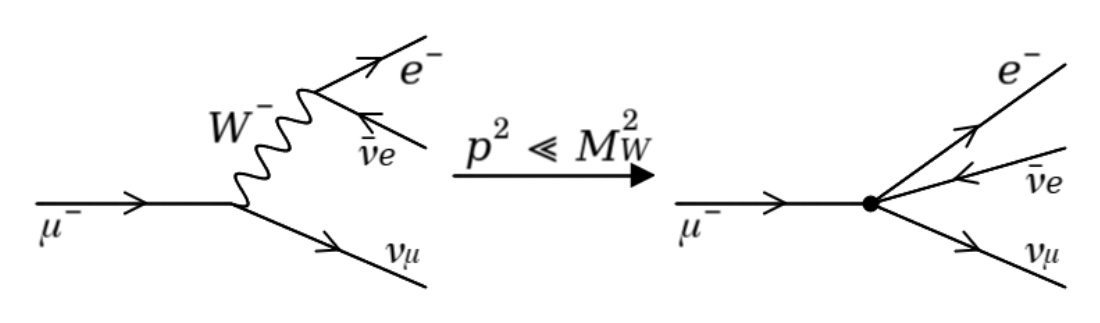}
\end{figure}

%The $\beta$-decay and the decay of the muon into an electron ($\mu^- \rightarrow e^- \bar \nu_e \nu_\mu$) remained open issues for years because of the missing energy in the final states. This led Pauli to postulate the existence of a particle that could escape the detectors, the neutrino. 
Nowadays, we know that the decay of a muon into an electron is mediated by the $W$ boson, but thirty years before the formulation of the EW theory, Fermi introduced a four-fermion contact term to explain it as
\begin{equation}
   \mathcal{L}_\text{Fermi} = -\frac{G_F}{\sqrt{2}}\left( \bar \nu_\mu \gamma^\alpha (1-\gamma_5) \mu \right) \left(\bar e \gamma_\alpha (1-\gamma_5) \nu_e \right). \label{l_fermi_muon}
\end{equation}
The $1-\gamma_5$ terms were added for phenomenological reasons, to select only the left-handed components of the Dirac spinors; $G_F$ plays the role of a coupling constant. The Feynman diagrams for the muon decay in the SM and in Fermi theory are depicted in Fig. \ref{fig:muon_decay}.

The Lagrangian density \eqref{l_fermi_muon} enables the computation of the amplitude for muon decay in Fermi theory and the width has the form
\begin{equation}
   \Gamma (\mu^- \rightarrow e^- \bar \nu_e \nu_\mu ) = \frac{G_F^2 m_\mu^5}{192 \pi^3}.
\end{equation}
The same computation in the SM returns an analogous result in terms of the EW quantities, and the comparison among the two gives
\begin{equation}
   G_F = \frac{ g_W^2}{4\sqrt{2} M_W^2} = \frac{1}{\sqrt{2} v^2}.
\end{equation}
Since the Fermi constant $G_F$ is known from experiments such as the measurement of the muon decay lifetime, this relation allows us to extract $v \sim 250$ GeV for the Higgs vacuum. The calculation is easier in Fermi theory, as no knowledge about the $W$-boson nature is needed. The coupling constant, though, remained a free parameter until the EW theory was formulated, allowing the matching above with the more fundamental parameters $g_W^{},M_W$: before that, its value could only be measured. 

When the muon energy approaches the pole mass of the $W$ boson, on-shell effects start to dominate and Fermi theory is not predictive anymore. This formalism is not renormalisable, as the Lagrangian density \eqref{l_fermi_muon} has a dimension larger than four: this means that it carries information about the scale at which it breaks. Indeed, the amplitude for the muon decay can be expanded in partial waves through functions $a_\ell (s)$ that, for the optical theorem, have to satisfy $|a_\ell(s)|\leq 1$ and $\text{Re}(a_0) < 1/2$. These requirements ensure the unitarity of the scattering matrix, and they lead to an upper bound on the breaking scale of $\sim 875$ GeV. Fermi theory returns meaningful predictions only for $\sqrt{s} < M_W$, that is much smaller than that: the introduction of the $W$ boson solves the breaking of the EFT, and a more complete theory must be employed for calculations already at lower energies than the ones at which unitarity is violated.

The features highlighted here are common to other EFTs. Fermi theory can be employed to perform calculations in the $p^2 \ll M_W^2$ regime in a simpler way, since most of the SM fields can be integrated out and only included in the values of low-energy constants like $G_F$. In this scenario, called ``top-down approach'', new operators like the four-fermion one in \eqref{l_fermi_muon} appear, as they are not present in the full theory. All the predicted quantities at colliders can be expanded in powers of $p/M$, and the series is infinite and in decreasing relevance: it can be truncated when the desired precision is achieved.

EFTs can also be employed the other way around, to investigate new physics when the underlying theory is not known. This approach is called ``bottom-up'' and is analogous to the Fermi-theory case when the SM was yet to be formulated. In this methodology, the Lagrangian at every order $(d)$ in the expansion is written including the most general sets of operators $O$ that are consistent with the known symmetries, namely
\begin{equation}
   \mathcal{L}_\text{EFT}^{(d)} = \sum_i \widetilde{C}_i(\Lambda) O_i(\phi_\text{light}). \label{l_eft_gen}
\end{equation}
Here, $\Lambda$ is a cutoff scale for the EFT and the $\widetilde{C}_i$ are coefficients. All the terms of a given dimension should be included, as each of them can be generated by multiple new-physics scenarios and the renormalisation group \eqref{ren_group} can mix them. Only symmetries can forbid some of them, but usually their coefficients can be set to zero only at particular scales. All the non-redundant operators at a certain dimension form a basis: the others are related to these through integration by parts, Fierz identities, equations of motion and field redefinitions. Redundant operators are not necessary to compute physical quantities, but are needed for the off-shell ones.

Most of the couplings are not known and have to be fit into the data. The final goal of the bottom-up approach is to obtain precise values of the $\widetilde{C}_i$ thanks to the measurements, so that any UV theory that cannot reproduce them would be automatically ruled out. The series is stopped at a given $(d)$ depending on the accuracy, but every new order introduces new operators and requires more data to be constrained. As for Fermi theory, it is in general possible to obtain an estimate of the energy at which a certain EFT breaks unitarity, and this also gives an upper bound for the resonances of any UV model that matches to it.

Other examples of this approach are the chiral perturbation theory $\chi$-PT that describes kaons and pions, and the Standard Model Effective Field Theory (SMEFT), that will be introduced in more detail.

\section{Power counting and operator dimension}
The operators in the Lagrangian of an EFT form an infinite tower, but they can be systematically organised, allowing to identify the leading contributions among them. A generic operator that is built according to the symmetries takes the form
\begin{equation}
   O = (D_\mu)^{d_D} (\bar \psi \psi)^{d_\psi} (F_{\mu\nu})^{d_F} (\phi)^{d_\phi},
\end{equation}
so that its mass dimension is $d = [O] = d_D+3d_\psi+2d_F +d_\phi$. In four dimensions and natural units, the Lagrangian density shows $[\mathcal{L}] = 4$. In order to obtain terms that can fit in it, any operator of dimension $d$ in \eqref{l_eft_gen} has to be divided by powers of $\Lambda$, and its coefficient becomes $\widetilde{C}_i = C_i /\Lambda^{d-4}$. The $C_i$ are usually called ``Wilson coefficients''. Normally in the bottom-up approach, $\Lambda$ is a fictitious scale introduced to make the Wilson coefficients dimensionless: it cannot be identified with the heavy new-physics mass. In general, though, the EFT can be considered predictive only if $\Lambda$ is much larger than the mass scales involved in the calculations.

From the powers of $\Lambda$ in the denominator, it follows that operators with $d>4$ are suppressed compared to the $d \leq 4$ ones. The former are then called ``irrelevant'', while the latter are ``relevant'' or ``marginal''. The number of irrelevant operators in a basis at a given $(d)$ grows with the dimension, unless symmetries affect them. Each order introduces new corrections to the amplitudes in powers of $\sqrt{s}/\Lambda$, with the leading ones coming from $d=5$ operators; $(\sqrt{s}/\Lambda)^2$ corrections involve $d=6$ operators and double insertions of $d=5$ ones, and so forth.

Since they involve irrelevant operators with $d > 4$, EFTs are not renormalisable in the traditional sense. Nonetheless, if only a certain order $(d)$ is considered, all the divergencies can be reabsorbed within a finite number of counterterms, which means that EFTs are order-by-order renormalisable in power-counting. These counterterms have to be computed at an order higher than $d$, even though the operators with those dimensions were not included in first place. Furthermore, the UV counterterms for a certain coefficient can depend on other ones: they are contained in the anomalous dimension matrix $\gamma$ for the theory, a generalisation of the quantity that appears in Eq. \eqref{ren_group}. The Wilson coefficients run and mix according to its entries as
\begin{equation}
   \frac{d\ C_i}{d\ \log\mu} = \gamma_{ij}C_j.
\end{equation}

\section{\label{sect:smeft}The SMEFT}
The SM is a great EFT candidate, as it is very well understood at the weak scale $v$ and experiments suggest a large separation in energy with hypothetical new physics. New fields and extra derivatives can be added, but they need to respect the SM structure: operators must be singlets under $SU(3)_c \otimes SU(2)_L \otimes U(1)_Y$ and the Higgs has to remain a doublet. Furthermore, the SM Lagrangian should be recovered in the $\Lambda \rightarrow \infty$ limit.

Since the Higgs nature is included in the premises, it is important to remind that the couplings that shape its potential have not been measured yet. It is possible to write EFTs where the SM does not completely explain the SSB and another massive singlet scalar than the one in Sect. \ref{sect:beh} accounts for the particle discovered at the Large Hadron Collider (LHC). The most notable result in this direction is the Higgs Effective Field Theory (HEFT). This more general approach is useful when the EFT is used to simplify the low-energy effects of UV theories where the SSB is non-linear; a larger number of free parameters is though needed, with respect to other alternatives \cite{heft_1,heft_2,heft_3}.

At the current stage, all measurements suggest that the scalar boson discovered at the LHC is indeed the SM Higgs boson. Under this assumption and the ones at the beginning of this section, the most used resulting theory is the SMEFT, which is the framework under which all the findings in this thesis will be presented \cite{Brivio_2019}.

The operators with odd dimension $d$ are not always included, as they violate some global symmetries of the SM and are thus assumed to be suppressed. As an example, the only operator that is present at $d=5$, the Weinberg one, violates lepton number $L$ by two units; since right-handed neutrinos are not included in the SMEFT, this is the only term that produces masses for neutrinos in the broken phase. The non-redundant operators at $d=6$ in the SMEFT form the ``Warsaw basis'', that had been developed during many years before its final version \cite{Buchmuller:1985jz,Grzadkowski:2010es}. All the operators with this dimension conserve $L$, but some of them violate baryon number $B$. If the fully general flavour structure and $B$ conservation are assumed, the Warsaw basis counts 2499 operators: we are interested in probing them at colliders. If we use the symbol $X$ for the field strengths \eqref{field_strengths}, they can be divided into the following groups, depending on their field and derivative content:
\begin{itemize}
   \item $X^3$: these operators feature products of three field-strength tensors. The $O_G$ and $O_W$ terms, that will be discussed in the next chapters, belong to this category. They modify the three- and four-boson vertices of the SM and introduce new five- and six-point interactions. They are constrained through diboson and multijet processes. Their $CP$-odd counterparts, $O_{\widetilde{G}}$ and $O_{\widetilde{W}}$, also belong here, with $\widetilde{X} = \varepsilon^{\mu\nu\rho\sigma} X_{\rho\mu}/2$.
   \item $\phi^6, \phi^4 D^2$: among them, $O_{\phi}$ affects the SM self-interactions of the Higgs, and $O_{\phi\square}, O_{\phi D}$ modify its couplings with all other particles. For this reason, they affect Higgs-production processes and decays.
   \item $\psi^2 \phi^3$: in these operators, the SM Yukawa couplings are multiplied by $\phi^\dagger \phi$. This yields a $v^2$ factor after SSB, that modifies the relation among fermion masses and Yukawa couplings and introduces a growth of the cross sections with the energy, that spoils unitarity at some scale.
   \item $X^2 \phi^2$: these operators modify the interactions between the Higgs and the gauge bosons through the field strengths. The most interesting example is $O_{\phi G}$, that contributes to $gg \rightarrow h$ production at tree-level, while the same process is loop-induced in the SM.
   \item $\psi^2 X \phi$: these operators yield dipole interactions in the broken phase, so they are generally very well constrained through measurements.
   \item $\psi^2 \phi^2 D$: they modify the interactions of fermions and gauge bosons and introduce new vertices that also feature the Higgs boson.
   \item $\psi^4$: these four-fermion operators are analogous to the Fermi one in Eq. \eqref{l_fermi_muon}. They can be further divided according to the chiralities of the fields they involve. The operators in this class that only feature light quarks will be discussed in Chapter \ref{chap:fourLQ}.
\end{itemize}
Complete bases for the operators from $d = 7$ to 9 are also available, and the techniques used to write all the possible Lorentz structures can be systematically applied at any dimension \cite{dim7_basis,dim8_basis,dim9_basis}.

Normally, since $\Lambda$ is assumed to be much larger than most of the SM masses, the Yukawa couplings for all the fermions can be set to zero in the SMEFT. The SM shows a global symmetry under flavour-field transformations in the massless limit: since there are five multiplets $\ell_p,q_p,e_p,u_p,d_p$, the associated group is $U(3)^5$, that can be written as the combination of five phase rotations $U(1)^5$ and flavour-space rotations under $SU(3)$. Imposing this symmetry allows neglecting all the operators that contract $SU(2)$-multiplets with different fermion types, and makes the Wilson coefficients diagonal in flavour space. The flavour assumptions in the SMEFT can then be tailored to the analysed scenarios: for example, it is sometimes useful to separate the third generation from the first lighter two.

\section{Fits of the SMEFT parameters}
As explained before, the $C_i/\Lambda^{d-4}$ coefficients in front of the SMEFT operators remain free parameters, so they need to be fit to the experimental data. This is not an easy task, as the operators are in large number already at $d=6$ and they contribute to multiple processes; furthermore, the renormalisation group allows setting some coefficients to zero only at certain scales.

This procedure requires large and broad amounts of data, with precise estimates of all the uncertainty sources. The likelihood $\mathcal{L}(C_i, \theta |X)$ is the probability to observe some data $X$, given the Wilson coefficient values and some nuisance parameter $\theta$ that can represent the uncertainties. Measurements are normally obtained from event detection distributed in histograms $x^\text{exp}$ of $N_\text{bins}$, with a large enough number of events in each bin. From this, a test statistic can be defined, and the best-fit points are the values of the Wilson coefficients and $\theta$ for which the likelihood is maximum. Close to this point, the test statistics can be approximated through a $\chi^2$ distribution, whose d.o.f. are determined by the number of Wilson coefficients in the fit and of data points. Its generic expression is
\begin{equation}
   \chi^2 (C_i,\theta) = \left[ x^\text{exp}-x^\text{th}(C_i,\theta)\right]^T_j V^{-1}_{jk} \left[ x^\text{exp}-x^\text{th}(C_i,\theta) \right]_k, \label{chisq_general}
\end{equation}
where $V$ is the covariance matrix, that includes the uncertainties, and $x^\text{th}$ are the SM and SMEFT predictions for the measured distributions. The constraints are placed for the $C_i/\Lambda^{d-4}$ ratios by comparing the $\chi^2$ function with a threshold value, that only depends on the number of d.o.f. and on the Confidence Level (CL), usually 95\%.

Rough estimates of the limit on a coefficient are called ``individual bounds'' and are obtained by setting all the other Wilson coefficients to zero. If instead all the $C_i$ included in the fit are allowed to vary at the same time, in a more realistic set up, the limits are called ``marginalised''. Only constraints on the ratios of the Wilson coefficients and $\Lambda$ are meaningful, so no information can be directly extracted on the cutoff scale alone. It is possible to make assumptions on the Wilson coefficient values, \eg\ that they should be $\mathcal{O}(1)$ because of naturalness, but no valid reasons back this type of statements.

In order to remain inside the validity range of the EFT, the strength of the interactions that are mediated by heavy new states has to be small, so that the SMEFT only slightly deforms the SM quantities at the experimental energies. For this reason, the measurements and theoretical predictions have to be obtained with high accuracy.

%!TEX root = main.tex

%%%%%%%%%%%%%%%%%%%%%%%%%%%%%%%%%%%%%%%%%%%%%%%%%%%%
%
%      Chapter III :
%
%
%%%%%%%%%%%%%%%%%%%%%%%%%%%%%%%%%%%%%%%%%%%%%%%%%%%

\chapter{Interference resurrection and the case of $O_G$}
\label{chap:interf_OG}
\pagestyle{fancy}

\hfill
\begin{minipage}{\textwidth}
   \centering
   \includegraphics[width=.7\textwidth]{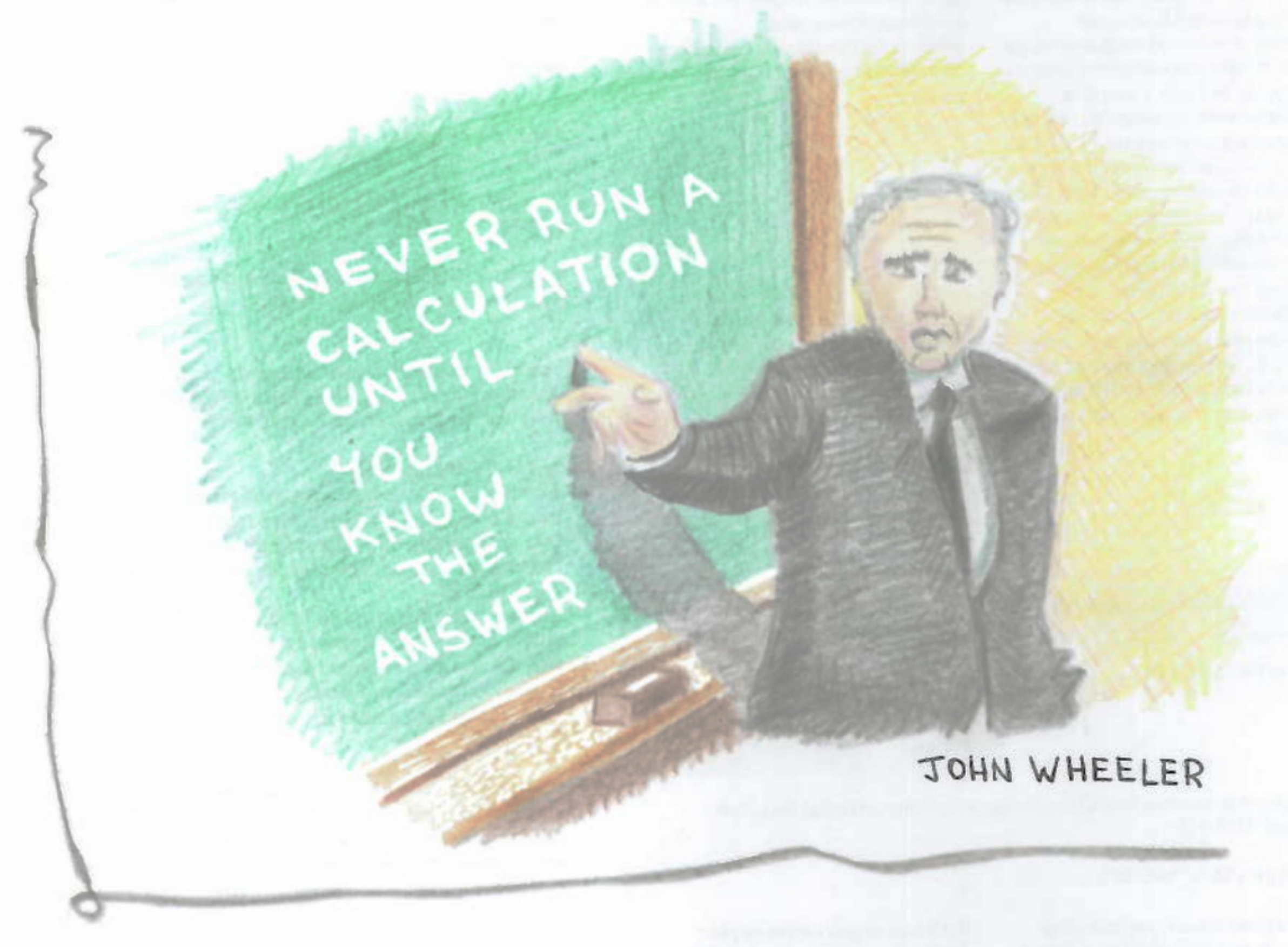}
\end{minipage}

\vspace{0.5cm}

In this chapter, I will introduce the issue of interference suppression and the strategies and quantities that we employed to look for observables that could restore it.
To showcase their potential, I will show how we applied them in the case of the $O_G$ SMEFT operator in three-jet production. This operator is a dimension-6 and CP-even element of the Warsaw basis. Throughout this work, it is defined as
\begin{equation}
   O_G = g_S^{} f^{ABC} \hspace{1mm} G_\mu^{A,\nu} G_\nu^{B,\rho} G_\rho^{C,\mu}, \label{oG_def}
\end{equation}
with $g_S^{}$ the strong coupling, $f$ the $SU(3)$ structure constant in Eq. \eqref{struct_const} and $G_{\mu\nu}$ the gluon field strength, defined in Eq. \eqref{field_strengths}. This operator modifies the three- and four-gluon vertices in the SM and also introduces five- and six-gluon additional interactions. At tree-level, it contributes to multijet and heavy-quark  production, and scattering processes with gluon self-interactions. The associated Wilson coefficient is indicated with $C_G$.

\section{Interference suppression}
When the SMEFT dimension-6 operators are added to the SM Lagrangian, the amplitudes for the processes they could contribute to can be written as
\begin{equation}
   \mathcal{M} = \mathcal{M}_{SM} + \mathcal{M}_{1/\Lambda^2} + \ldots, \label{smeft_ampl}
\end{equation}
where the $1/\Lambda^2$ term contains the diagrams that feature the insertion of one dimension-6 operator. The resulting cross section can then be expanded around the SM one in powers of $1/\Lambda$:
\begin{equation}
   \sigma = \sigma^{SM} + \sum\limits_i \frac{C_i}{\Lambda^2} \sigma^{1/\Lambda^2}_i + \sum\limits_i \sum\limits_j \frac{C_i C_j}{\Lambda^4} \sigma^{1/\Lambda^4}_{ij} + \ldots, \label{smeft_xsect}
\end{equation}
where the $i,j$ indices run over the dimension-6 operators. In principle, both the square of amplitudes including dimension-6 operators and the interference between the SM and dimension-8 operators can contribute at $\mathcal{O}(1/\Lambda^4)$ level, but the latter is not considered if the amplitude is truncated as in \eqref{smeft_ampl}.

Due to the $1/\Lambda$-powers suppression, the $\mathcal{O}(1/\Lambda^2)$ term, that is linear in the Wilson coefficients, is usually the leading correction to the SM at colliders, and all the other ones, starting from those quadratic in the coefficients, are subleading. The $\mathcal{O}(1/\Lambda^4)$ order eventually grows larger than the linear one at high energies because of the different energy dependences in their amplitudes, but the statement above normally holds within the validity region of the EFT, $\sqrt{s} \ll \Lambda$.

Being the integral over the phase space $\Phi$ of an interference between the SM and diagrams with the insertion of a dimension-6 operator, though, the linear term is not positive-definite:
\begin{equation}
   \sigma = \int d\Phi \hspace{1mm} \left[|\mathcal{M}_{SM}|^2 +2 \text{Re} \left(\mathcal{M}_{SM} \mathcal{M}^*_{1/\Lambda^2} \right) + |\mathcal{M}_{1/\Lambda^2}|^2 + \ldots \right].
\end{equation}
This means that differential cross sections may change their sign over different areas of the phase space, leading to a partial cancellation among them when more global observables, such as the total cross section, are measured. The same thing would happen with the interference between the SM and dimension-8 diagrams at $\mathcal{O}(1/\Lambda^4)$ order, if it had to be included.

In particular, if the positive- and negative-weighted contributions to the linear term are almost equal in magnitude, they can result in a very small total cross section at interference level, after the integration. In these cases, the most stringent bounds on the Wilson coefficients would come from the quadratic order, at least if the total cross section is employed as the observable to compute them.
This brings some disadvantages: as already stated, the quadratic term in the coefficients represents only part of the $\mathcal{O}(1/\Lambda^4)$ level, and the missing one is usually hard to estimate because of the large number of dimension-8 operators that could contribute to the considered processes \cite{dim8_basis}. Furthermore, the quadratic contribution is positive-definite, so it would not be sensitive to the sign of the coefficients, as the linear one would instead be. The interference is also less suppressed by the powers of $1/\Lambda$.

These reasons motivate an attempt to revive the interference term, if possible, in order to obtain competitive bounds on the Wilson coefficients from it. Of course, the linear term might also be small across the entire phase space, if the SM and the SMEFT operators give vanishing contributions to the same amplitudes: a more detailed discussion about the ``non-interference rule'' can be found in Sect. \ref{sect:rev_strategies}. Because of this, we need a way to determine if the linear term is suppressed by a cancellation between large positive- and negative-weighted contributions, or if it is really close to zero all over the phase space.

\section{Some useful quantities to investigate the suppression}
The total cross section at interference level is usually computed as the integral over the phase space of some differential distribution. In practice, when events are simulated for a certain process through a Monte Carlo (MC) generator, each of them is associated to a weight $w$, and the total linear cross section can be approximated by summing all the weights of the events that pass certain generation cuts, if they are in a large enough number $N$:
\begin{equation}
   \sigma^{1/\Lambda^2} = \int d\Phi \hspace{1mm} \frac{d \sigma^{1/\Lambda^2}}{d \Phi} = \lim_{N\rightarrow \infty} \sum\limits_{i=1}^{N} w_i. \label{xsect_tot}
\end{equation}
These $w_i$ are not positive-definite, so this quantity can be suppressed if the numbers of positive- and negative-weighted events are close to each other. When computing the following observables presented in this section, the events will always be assumed as unweighted: all the ones from the same sample show the same weight in absolute value. Also, we will focus on the generation at LO parton level, since NLO and parton-shower (PS) effects can generate additional negative weights that are not distinguishable from the interference ones \cite{nlo_events,powheg,neg_wgt_reduct}.
If the linear cross section is suppressed, then the ``cancellation level''
\begin{equation}
   R_{w\pm} = \frac{ N_{w+}-N_{w-} }{ N_{w+}+N_{w-} }, \label{rwpm}
\end{equation}
with $N_{w+}$ and $N_{w-}$ the numbers of positive and negative weights in the sample, would be $\sim 0$. It can be computed in each bin of any differential distribution simulated with MC, to check which phase-space portions present a larger cancellation. This variable is not always enough to quantify the suppression, though, because it does not carry any information on the bin contribution to the total cross section.

\begin{figure}
   \centering
   \caption{\footnotesize{The plot on the left shows examples of a SM and a $\mathcal{O}(1/\Lambda^2)$ amplitudes as functions of the CoM-scattering angle $\theta$ in a $2\rightarrow 2$ process. The plot on the right sketches the corresponding differential interference cross section, signed and in absolute value, as a function of the same angle}} \label{fig:sigma_int_cartoon}
   \includegraphics[width=.6\textwidth]{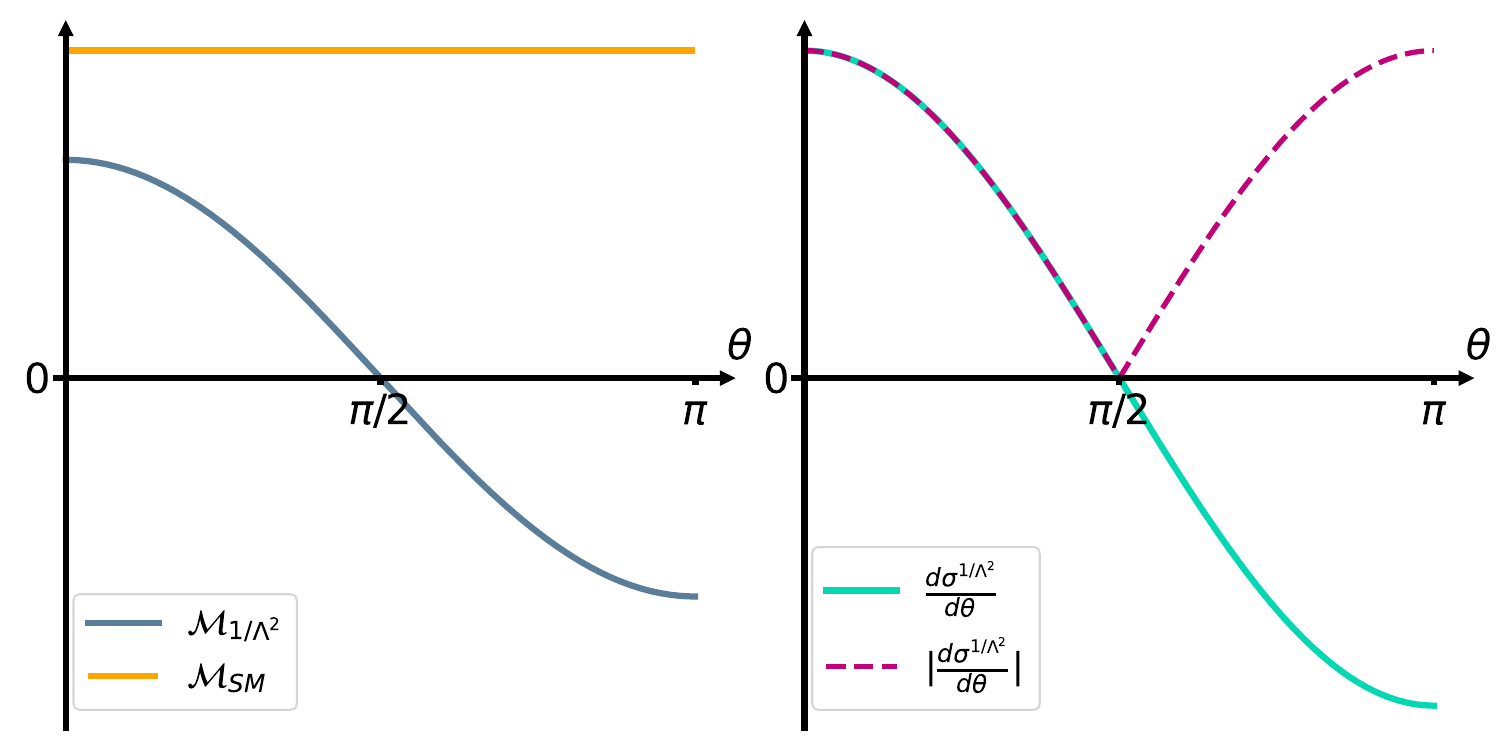}
\end{figure}

In addition to that, one can integrate a certain differential distribution taken in absolute value, or analogously sum over the moduli of the MC weights:
\begin{equation}
   \sigma^{|\text{int}|} = \int d \Phi \hspace{1mm} \left| \frac{d \sigma^{1/\Lambda^2}}{d \Phi} \right| = \lim_{N\rightarrow \infty} \sum\limits_{i=1}^{N} | w_i |. \label{xsect_int}
\end{equation}
We refer to this quantity as ``integrable cross section''. If its value, for the linear contribution to a certain process, is much larger than the total cross section $\sigma^{1/\Lambda^2}$, that would be a hint of a phase-space cancellation between two large contributions with opposite sign for the interference. An example in the case of a $2\rightarrow 2$ scattering is sketched in Fig. \ref{fig:sigma_int_cartoon}: if the SM amplitude is constant with respect to the scattering angle in the centre-of-mass (CoM) frame and the $\mathcal{O}(1/\Lambda^2)$ one oscillates with it, the differential cross section for the linear term oscillates around the horizontal axis as well, since it is a function of the product of the two. When the integration over the angle is performed to compute the total cross section for the interference, the areas below the positive portion of the curve and above the negative one cancel out, yielding a null result. In this case, when the differential cross section is taken in absolute value, its integral is much larger than the previous one, suggesting that a cancellation is occurring for the total interference cross section.

$\sigma^{|\text{int}|}$ can be easily and quickly computed for a set of simulated events, but it is not always accessible in experiments: its calculation requires complete knowledge over the four-momentum, spin and identity of each particle in the events. As an example, at a real proton-proton collider like the LHC, the nature of the initial-state particles in each interaction is not known, and the total longitudinal energy of the event cannot be determined if neutrinos are generated, since they usually leave the detector. Moreover, the final-state particles of the main interaction have to be reconstructed from the tracks that are deposited in the calorimeters by the showers that are radiated from them and their decay products, making the measurement of their four-momentum components only approximate and their identities difficult to guess.

It is possible, therefore, that even if some interference suppression occurs for an operator and process, it could not be accessible at a certain experiment, because of all the unknown information and measurement smearing that are intrinsic to the setup.

In order to establish this, we introduce the ``measurable integrable cross section'' $\sigma^{|\text{meas}|}$: it is defined as the integral over the experimentally-accessible phase space of the absolute value of a differential cross-section distribution. In practice, we compute it by taking the integrals (or sums) of the interference squared amplitude over the unmeasurable quantities $\{\text{um}\}$, like the neutrino longitudinal components or the identities and helicities of the initial and final states. Then, we compare its sign against the one of the MC weight associated to the event, as
\begin{align}
   \sigma^{|\text{meas}|} &= \int d \Phi_\text{meas} \hspace{1mm} \left| \sum\limits_{\{\text{um}\}} \frac{d \sigma^{1/\Lambda^2}}{d \Phi} \right|  \nonumber \\
   &= \lim_{N\rightarrow \infty} \sum\limits_{i=1}^N w_i \cdot \text{sign} \left( \sum\limits_{\{\text{um}\}} \mathcal{M}_{SM} \mathcal{M}_{1/\Lambda^2}^* (\vec{p}_i,\{\text{um}\}) \right). \label{xsect_meas}
\end{align}
In the above formula, the $\vec{p}_i$ label the final-states momenta. In some cases, the integrated matrix element is null or of opposite sign than the weight, so $\sigma^{|\text{meas}|} \leq \sigma^{|\text{int}|}$. The quantity \eqref{xsect_meas} represents an upper bound for any asymmetry that is built out of experimentally-available information and that aims to restore the interference for a particular process at a given collider. %If these two values are close, it is then possible to build a quantity to restore the interference for the given process, at the considered collider. 
Its computation is close to a matrix-element method (MEM) at parton level \cite{mem_1,mem_2,mem_3,mem_4,mem_5,mem_6,mem_7}.
By varying the set of unmeasurable quantities that are taken into account, it is possible to obtain an idea of which ones are more responsible for the suppression.

\section{\label{sect:rev_strategies}Strategies for interference revival}
There are cases where the interference is actually small over the whole phase space. In these situations, both $\sigma^{|\text{meas}|}$ and $\sigma^{|\text{int}|}$ are of the same order of magnitude as the total cross section, and small compared to the following-order term in the expansion \eqref{smeft_xsect}.

For instance, it was proven that the interference among the SM and the $O_G$ operator, defined in Eq. \eqref{oG_def}, is exactly null for dijet production. This happens because the helicity structure of $2 \rightarrow 2$ amplitudes involving three field-strength operators is usually orthogonal to the QCD one. In particular, the SMEFT members couple together only outgoing particles with the same helicity, while the SM tends to generate opposite-helicity ones. The difference is exact for massless particles only, but it still takes place approximately in the massive case and it becomes better at high energies, where the masses can be neglected \cite{Simmons:1990dh,Azatov:2017,Riva:2018}. A clear example for this is the $O_W$ interference in $WZ$ production, that will be discussed in Chapter \ref{chap:interf_OW}.

In this scenario, bounds can only come from terms of order $\mathcal{O}(1/\Lambda^4)$ or higher, unless some changes are made to the process.
\begin{itemize}
   \item Adding extra jets, for instance, would alter the topology from the $2 \rightarrow 2$ one and significantly increase the linear term. For dijet production, this would mean switching to processes like three- or four-jets, as we will see in Section \ref{sect:oG_case}. Heavy-quark production is also a candidate to consider: since in $gg \rightarrow q \bar q$ the interference is proportional to the quark mass, $t \bar t$ represents a case where the helicity mismatch is only approximate. Unluckily, as we will see, the colour-octet contribution to it is small in the SM, so it is not as competitive as multijets to set bounds on $C_G/\Lambda^2$ \cite{Zhang_2011,Hirschi_2018}.
   \item Another strategy, that we employed for $WZ$ production, is the inclusion in the calculations of the decay products from the EW bosons: this also changes the number of final states. The cleanest decay channels at a proton collider like the LHC are the leptonic ones, which means that a neutrino is generated in the $W$-boson disintegration. The impossibility to reconstruct its longitudinal momentum component with precision introduces smearing in the predictions for many quantities, but we will show that suitable cuts over specific variables, computed from the $W$ and $Z$ decay-products properties, can still help restore the linear term.
   \item In the same fashion, going to NLO can also help to lift part of the interference suppression. For some processes, this can open new channels, as in the $WZ$ case. Moreover, the inclusion of real-emission diagrams changes the $2 \rightarrow 2$ topology that characterises the process at LO.
   \item Investigating azimuthal differential distributions might be useful, too. We call $\phi_V$, for $V = W,Z$ or other massive intermediate states, the azimuthal angle between the plane containing the beam axis $\hat{z}_\text{lab}$ and $V$, and the plane where its decay products lie, in the lab frame. Namely, if we consider the generic decay $V \rightarrow \ell_+ \ell_-$, with the signs identifying the lepton helicities, we can define two unitary vectors $\hat{n}_\text{decay}$ and $\hat{n}_\text{scat}$, respectively along the directions of the vector products $\vec{p}_{\ell +} \times \vec{p}_{\ell -}$ and $\hat{z}_\text{lab} \times \vec{p}_V$. It is then possible to obtain the aforementioned azimuthal angle as
   \begin{equation}
      \phi_V = \text{sign} \left[ \left( \hat{n}_\text{scat} \times \hat{n}_\text{decay} \right) \cdot \vec{p}_V \right] \hspace{1mm} \arccos \left( \hat{n}_\text{scat} \cdot \hat{n}_\text{decay} \right), \label{phiV_def}
   \end{equation}
   as described in \cite{Barducci:2019}. The interference between the two helicity configurations of the SM and the three field-strength linear amplitudes introduces opposite-sign $\pm i \phi_V$ phases that add together in the linear cross-section term. This induces an azimuthal asymmetry for the decay products, around the $V$ flight direction.
   %\item $\ldots$
\end{itemize}

In the ``restorable'' interference suppression scenario, on the other hand, $\sigma^{|\text{meas}|}$ and $\sigma^{|\text{int}|}$ have similar values and both are much larger than the total cross section: in this case, bounds can be obtained at $\mathcal{O}(1/\Lambda^2)$ order from the process as it is, and the measurable cross section represents the most suitable observable to do it.

The problem with $\sigma^{|\text{meas}|}$, though,  is that it requires the computation of the interference part of the squared matrix element for each event in the MC sample, and the computational and time costs of this operation scale pretty quickly with the complexity of the process. %Moreover, the quantity is model-dependent and might vary with different assumptions on the new physics.

Because of this, our goal is to find observables that can be easily measured at colliders and that could approximate the $\sigma^{|\text{meas}|}$ value for a certain process, meaning that they could separate the positive- and negative-weighted regions of the phase space much more efficiently than the total cross section. These variables, built from kinematic quantities only, could never revive the interference more than $\sigma^{|\text{meas}|}$, but they would provide a more general tool that is applicable even outside the SMEFT, or where the EFT approximation breaks down.
%In order to find the most suitable observables, it is necessary to understand the source of the suppression that 

\section{\label{sect:oG_case}A proof-of-principle: the case of $O_G$}
The $O_G$ operator is defined in Eq. \eqref{oG_def} at the beginning of this chapter. The most simple class of processes to which it contributes at linear level is multijet production, but its interference in $pp \rightarrow jj$ is exactly zero because of the helicity argument introduced in the previous section.

The impact of $O_G$ over various jet observables has been investigated widely in the literature. In particular, \cite{Hirschi_2018} showed that the dimension-6 squared contribution to the $\mathcal{O}(1/\Lambda^4)$ order usually dominates over the interference one. They also estimated the impact of the SM interference with the dimension-8 operators, validating the EFT expansion. In the same study, it was checked that the linear term is non-zero for dijet production at NLO with up-to-one insertion of $O_G$, but it is still too small compared either to the SM or to the following one in the cross-section series, depending on the energy window.

Indeed, existing bounds on the coefficient were obtained assuming the $\mathcal{O}(1/\Lambda^4)$ order, without the inclusion of dimension-8 operators. In \cite{Krauss_2017}, the authors employ an observable built out of any transverse momenta of jets or missing particles with $p_T> 50$ GeV, namely
\begin{equation}
   S_T = \sum\limits_{i=1}^{N_\text{jets}} p_{T,i} + p_T^\text{miss}. \label{sT}
\end{equation}
This variable had been measured in a CMS search for black holes with Run II data \cite{cms_bh_run2}, that was recast for the $O_G$ study, and the final limit they obtained for $C_G/\Lambda^2$ is of order $4 \cdot 10^{-2}$ TeV${}^{-2}$  at 95\% CL.

Another analysis that aimed to constrain this coefficient is presented in \cite{Goldouzian_2020}, where the limit was further improved to $3.1 \cdot 10^{-2}$ TeV${}^{-2}$, thanks to predictions on the dimension-6 squared contribution to multijet production. They used public CMS data \cite{2j_2018}, that measured the exponential of the absolute value of the rapidity difference between the two leading jets in $p_T$,
\begin{equation}
   \chi_{jj} = e^{| y_{j1}-y_{j2} |}. \label{chi_jj}
\end{equation}
This quantity is related to the scattering angle in the CoM frame.

\subsection{Choice of the most suitable process}
In our study, we compared the $\mathcal{O}( 1/\Lambda^2 )$ cross sections and $R_{w\pm}$ values for different processes that could be impacted by $O_G$ at linear level. We simulated events at LO parton level via \amc\ \cite{mg5} v2.8.2, feeding it the TopEffTh\footnote{See \url{https://cp3.irmp.ucl.ac.be/projects/madgraph/attachment/wiki/Models/TopEffTh/note.pdf}.} Universal Feynrules Output (UFO) \cite{ufo,feynrules}. The value of $C_G$ in the model was set to 1, with $\Lambda = 5$ TeV. No more than one $O_G$ insertion was included in each diagram, and QED contributions were ignored. NNPDF2.3 was used as parton distribution function (PDF) set \cite{nnpdf23}. Only the top quark was considered to be massive, and we applied different cuts on the minimum transverse momenta of the other jets, namely at $p_T^j > 50$, 200 and 1000 GeV. Furthermore, we asked for a minimum $\Delta R_{jj}$ separation of 0.4 between the light jets, when applicable. The total transverse energy of the event divided by 2, $H_T /2$, was chosen as dynamical scale. The uncertainties that are shown are numerical and are due to the limited number of events generated. Results are presented for the LHC at 13 TeV.

Four different processes were compared, other than dijets: $t \bar t$, $t \bar t j$, three- and four-jet production. The decays of the top quarks were not included for the first two. The results are shown in Table \ref{tab:oG_xsects}: the largest cancellation seems to occur for three-jet production. Since this process presents also a cross section that is large enough to allow differential measurements, at least for the lower cuts on $p_T^j$, we focused on it to compute bounds on $C_G/\Lambda^2$ at linear level. On the other hand, top-antitop processes have both smaller interference cross sections and cancellations.

\begin{table}
\centering
\caption{\footnotesize{LO $\mathcal{O}(1/\Lambda^2)$ cross sections, in pb, and $R_{w\pm}$ values for different processes affected by the $O_G$ operator at linear level, for different cuts on the transverse momenta of the massless jets. $C_G$ is set to 1 and $\Lambda$ to 5 TeV. The numerical uncertainties are reported. $R_{w\pm}$ is defined in Eq. \eqref{rwpm}: the closer it is to zero, the largest might the interference suppression be}} \label{tab:oG_xsects}
%\resizebox{\textwidth}{!}{
\begin{tabular}{c|cc|cc|cc}
    & \multicolumn{2}{c|}{$p_T^j > 50$ GeV} & \multicolumn{2}{c|}{$p_T^j > 200$ GeV} & \multicolumn{2}{c}{$p_T^j > 1$ TeV} \\
    & $\sigma$ & $R_{w\pm}$ & $\sigma$ & $R_{w\pm}$ & $\sigma$ & $R_{w\pm}$ \\ \hline
   $t \bar t$ & 1.384$\pm 0.7\%$ & 0.70 & 1.384$\pm 0.7\%$ & 0.70 & 1.384$\pm 0.7\%$ & 0.70 \\
   $t \bar t j$ & 5.1$\cdot 10^{-1} \pm 2\%$ & 0.20 & 1.17$\cdot 10^{-1} \pm 2\%$ & 0.20 & 1.44$\cdot 10^{-3}\pm 2\%$ & 0.30 \\
   $jjj$ & 3.7$\cdot 10^1 \pm 4\%$ & 0.04 & 7.3$\cdot 10^{-1} \pm 2\%$ & 0.06 & 4.55$\cdot 10^{-4} \pm 0.7\%$ & 0.20 \\
   $jjjj$ & -3.1$\cdot 10^1 \pm 5\%$ & -0.09 & -1.53$\cdot 10^{-1} \pm 5\%$ & -0.10 & -3.8$\cdot 10^{-6} \pm 3\%$ & -0.19 \\
   \hline
\end{tabular}
%}
\end{table}

\subsection{Observable search in three-jet production}
We generated different three-jet samples at LO for the SM, the linear and the quadratic terms with different minimum cuts on $p_T^j$. Depending on such cuts, both the renormalisation and factorisation scales $\mu_R,\mu_F$ were fixed to certain values: 150 GeV for $p_T^j > 50$ GeV, 500 GeV for $p_T^j > 200$ GeV, 1 TeV for $p_T^j > 500$ GeV and 2 TeV for $p_T^j > 1$ TeV. Scale variation uncertainties were not included in the analysis. The other settings remained as described above.

The total cross section of each contribution is summarised in Table \ref{tab:3j_xsects}, with $C_G=1$ and $\Lambda = 5$ TeV. For the linear term, the values of $R_{w\pm}$, $\sigma^{|\text{meas}|}$ and $\sigma^{|\text{int}|}$ are also reported. 
To compute the measurable cross sections, for each event we calculated the interference squared amplitudes for all the possible permutations of the initial- and final-state quark flavours in all the three-jet subprocesses, and checked the sign of their sum. Each squared amplitude was weighted by its PDF. We assumed perfect momenta reconstruction, as no neutrinos are present and PS effects are ignored for simplicity.

The comparison of the total interference cross sections with the integrable ones highlights a large cancellation between opposite-sign contributions, resulting in the suppression of the total linear term. The values of the measurable cross section show that a large fraction of the interference total magnitude can be accessed at the LHC, at least in this simplified scenario. %that we are considering here.

\begin{table}
\centering
\caption{\footnotesize{LO total cross sections for the SM, linear and quadratic contributions to three-jet production, in pb, for different $p_T^j$ cuts in GeV. Visual representations through bar charts of the $p_T^j>200$ GeV and 1 TeV cases are shown at the bottom. $C_G$ is set to 1 and $\Lambda$ to 5 TeV. For the interference, the values of $R_{w\pm}$ and of the measurable and integrable cross sections are also shown. The numerical uncertainties are reported in the table}} \label{tab:3j_xsects}
\begin{minipage}{\textwidth}
\resizebox{\textwidth}{!}{
\begin{tabular}{c|c|cccc|c}
   %\hline
   \multicolumn{7}{c}{$pp \rightarrow jjj$ LO} \\ %\hline
    & SM & \multicolumn{4}{c|}{$\mathcal{O}(1/\Lambda^2)$} & $\mathcal{O}(1/\Lambda^4)$ \\ %\hline
   min $p_T^j$ & $\sigma$ & $\sigma$ & $R_{w\pm}$ & $\sigma^{|\text{meas}|}$ & $\sigma^{|\text{int}|}$ & $\sigma$ \\
   \hline
   50 & 9.73$\cdot 10^5 \pm 0.15\%$ & 1.5$\cdot 10^1 \pm 10\%$ & 0.01 & 7.81$\cdot 10^2 \pm 0.19\%$ & 1.051$\cdot 10^3 \pm 0.15\%$ & 3.922$\cdot 10^1 \pm 0.15\%$ \\
   200 & 8.96$\cdot 10^2 \pm 0.17\%$ & 4.6$\cdot 10^{-1} \pm 4\%$ & 0.04 & 8.77$\pm 0.2\%$ & 1.251$\cdot 10^1 \pm 0.16\%$ & 2.737$\pm 0.16\%$ \\
   500 & 3.11$\pm 0.16\%$ & 1.87$\cdot 10^{-2} \pm 1.9\%$ & 0.08 & 1.508$\cdot 10^{-1} \pm 0.2\%$ & 2.243$\cdot 10^{-1} \pm 0.16\%$ & 1.484$\cdot 10^{-1} \pm 0.16\%$ \\
   1000 & 9.08$\cdot 10^{-3} \pm 0.17\%$ & 4.58$\cdot 10^{-4} \pm 0.8\%$ & 0.20 & 1.470$\cdot 10^{-3} \pm 0.2\%$ & 2.297$\cdot 10^{-3} \pm 0.15\%$ & 3.062$\cdot 10^{-3} \pm 0.16\%$ \\
   \hline
\end{tabular}
}
\end{minipage}
\begin{minipage}{.7\textwidth}
   \includegraphics[width=\textwidth]{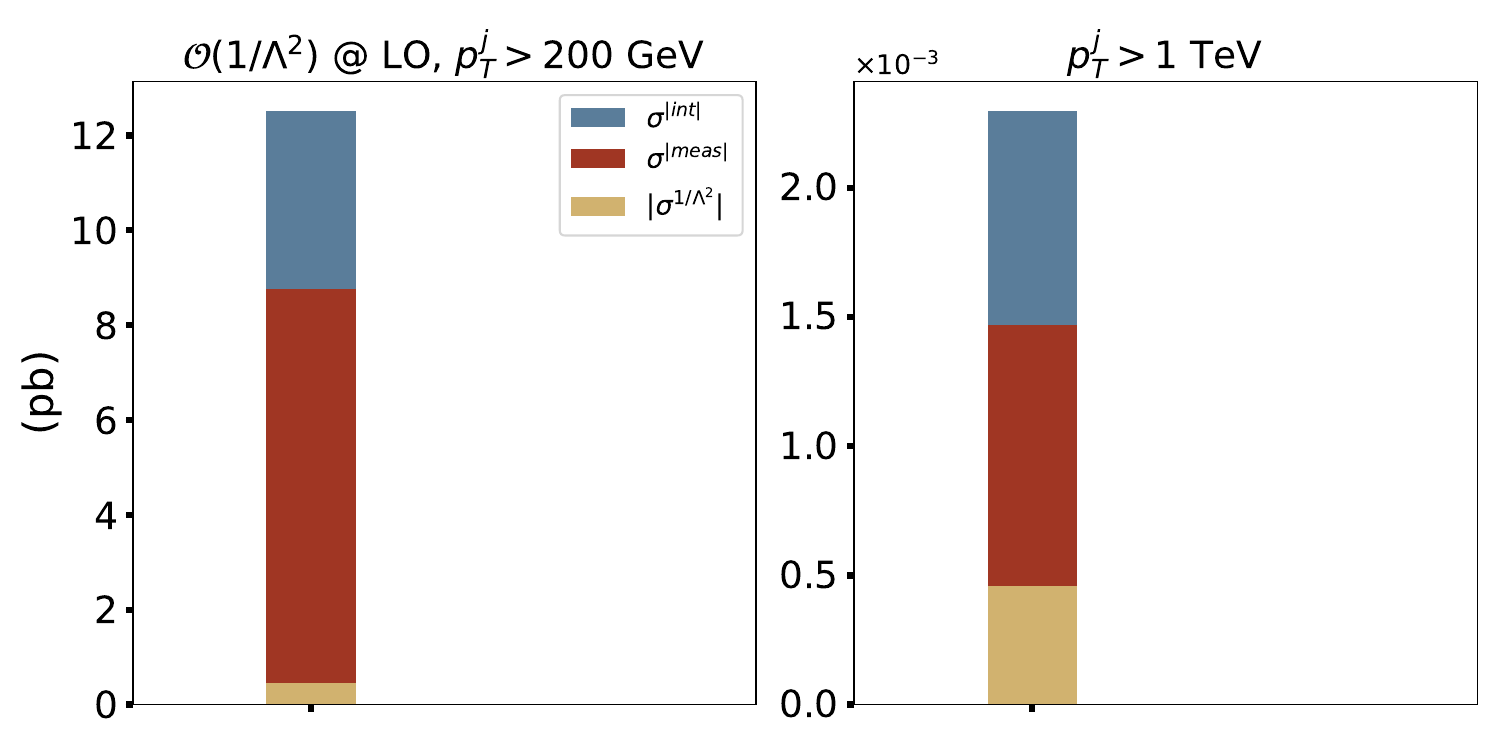}
\end{minipage}
\end{table}

All the observables that we defined seemed to suggest that a revival of the $O_G$ interference in three-jets is indeed possible, so we checked if any simple kinematic variables could separate the positive- and negative-weights in the linear term samples with a similar efficiency to the measurable cross section. We considered the $p_T$ of the jets, ordered decreasingly, their pseudorapidities $\eta$ and invariant masses $M_{jj}$, and the angular distances $\Delta R_{jj}$ among them. We also simulated the differential distributions for some event-shape observables: the normalised triple product $\frac{ (\vec{p}_1 \times \vec{p}_2) \cdot \vec{p}_3 }{ |\vec{p}_1 \times \vec{p}_2||\vec{p}_3| }$ of the jets three-momenta, the transverse sphericity $Sph_T$, the transverse thrust $Thr_T$ and the jet broadening \cite{sphT,evt_shape_pheno}.
The transverse sphericity is defined as
\begin{equation}
   Sph_T = \frac{ 2\lambda_2 }{ \lambda_1 + \lambda_2 }, \label{sphT}
\end{equation}
where $\lambda_1 > \lambda_2$ are the eigenvalues of the transverse-momentum tensor
\begin{equation}
   M_{xy} = \frac{1}{ \sum_i |\vec{p}_{T,i}| } \sum_{i=1}^{N_\text{jets}} \frac{1}{ |\vec{p}_{T,i}| } \left( 
              \begin{array}{cc} 
                 p_{x,i}^2 & p_{x,i}p_{y,i} \\ p_{y,i}p_{x,i} & p_{y,i}^2 
              \end{array} \right).
\end{equation}
It represents a useful quantity to differentiate pencil-like events ($Sph_T \sim 0$) from more isotropic ones ($Sph_T \sim 1$).
Analogously, we call ``transverse thrust'' the quantity
\begin{equation}
   Thr_T = 1-\frac{ \sum_{i=1}^{N_\text{jets}} |\vec{p}_{T,i}\cdot \hat{n}_T| }{ \sum_{i=1}^{N_\text{jets}} |\vec{p}_{T,i}| }, \label{thrT}
\end{equation}
with $\hat{n}_T$ the unit vector on the transverse plane that minimises the projection, on the same plane, of the jet momenta. This quantity is, as the one above, related to the event topology: it equals zero for balanced dijet events, while it tends to $1-2/\pi$ for isotropic multijet ones. $Thr_T$ is a linear function of the momenta and, thus, infrared- and collinear-safe (IRC-safe); the same cannot be stated about $Sph_T$, but the lower cuts on the $p_T$ of all the jets avoid any possible strong dependence on the hadronisation.

For each of these observables, we defined an asymmetry by summing the bin contents of their linear differential distributions in absolute value, and then we compared the results against $\sigma^{|\text{meas}|}$ to determine which ones are more efficient in separating the positive- and negative-weighted events in our samples. The results for some of those are shown in Table \ref{tab:3j_asymm}. Some variables, like the $p_T$ of the leading jet and $\Delta R_{j2 j3}$, can achieve an efficiency of about 40\%, as well as some double distributions of the jets $p_T$ and $\eta$. As a comparison, the total linear cross section can only revive 5\% of the measurable one.

The most effective observables, however, turned out to be $Sph_T$ and $Thr_T$, whose asymmetries reach above 80\% of the measurable cross-section value. This suggests that the events that are almost back-to-back and the more isotropic ones contribute with opposite signs to the linear term, and a sign flip occurs between these two extremal topologies. This can explain the decrease of the cancellation with higher $p_T^j$ cuts that we observe in Table \ref{tab:3j_xsects}, since a strong hierarchy between the jets becomes less probable at high energies.

\begin{table}
\centering
\caption{\footnotesize{ Comparison for some observables of their power to revive the interference for three-jet production at LO, with a minimum $p_T$ for all the jets of 200 GeV. For each kinematic variable, the asymmetry is built by summing the bin contents of its cross-section distribution in absolute value. The numerical uncertainties are shown }} \label{tab:3j_asymm}
\begin{tabular}{c|ccc}
   %\hline
   \multicolumn{4}{c}{ $pp \rightarrow jjj$ LO, $p_T^j > 200$ GeV } \\ %\hline
   Obs. & Asymm. (pb) & \% of $\sigma^{|\text{int}|}$ & \% of $\sigma^{|\text{meas}|}$ \\ \hline
   $\sigma^{|\text{int}|}$ & 1.251$\cdot 10^1 \pm 0.16\%$ & 100 & - \\
   $\sigma^{|\text{meas}|}$ & 8.77$\pm 0.2\%$ & 70 & 100 \\
   $Sph_T$ & 7.31$\pm 0.3\%$ & 58 & 83 \\
   $Thr_T$ & 7.27$\pm 0.3\%$ & 58 & 83 \\
   $\Delta R_{j2j3}$ & 4.31$\pm 0.5\%$ & 34 & 49 \\
   $p_T^{j1}$ & 3.73$\pm 0.5\%$ & 30 & 43 \\
   $S_T$ & 1.90$\pm 1.0\%$ & 15 & 22 \\
   $|\eta_{j3}|$ & 1.41$\pm 1.4\%$ & 11 & 16 \\
   $\frac{ (\vec{p}_1 \times \vec{p}_2)\cdot \vec{p}_3 }{ |\vec{p}_1 \times \vec{p}_2 ||\vec{p}_3| }$ & 1.28$\pm 1.6\%$ & 10 & 15 \\
   $|\eta_{j1}|$ & 4.7$\cdot 10^{-1} \pm 4\%$ & 4 & 5 \\
   $\sigma^{1/\Lambda^2}$ & 4.6$\cdot 10^{-1} \pm 4\%$ & 4 & 5 \\
   \hline
\end{tabular}
\end{table}

\begin{figure}
   \centering
   \caption{\footnotesize{ LO differential distributions of the leading-jet $p_T$ ({\it top}), the transverse sphericity ({\it centre}) and the leading-jet absolute rapidity ({\it bottom}) in three-jet production, for the SM divided by 100 ({\it black}), the linear ({\it orange}) and quadratic ({\it green}) orders. They all show the $p_T^j>200$ GeV case. The positive-and negative-weighted contributions to the linear term are shown separately through the shaded histograms. Numerical uncertainties are included. $C_G$ is set to 1 and $\Lambda$ to 5 TeV. In the top plot, the last bin contains the overflow; in the central one, the dotted line is the inverse of the negative part }} \label{fig:pTj200_diff}
   \includegraphics[width=.8\textwidth]{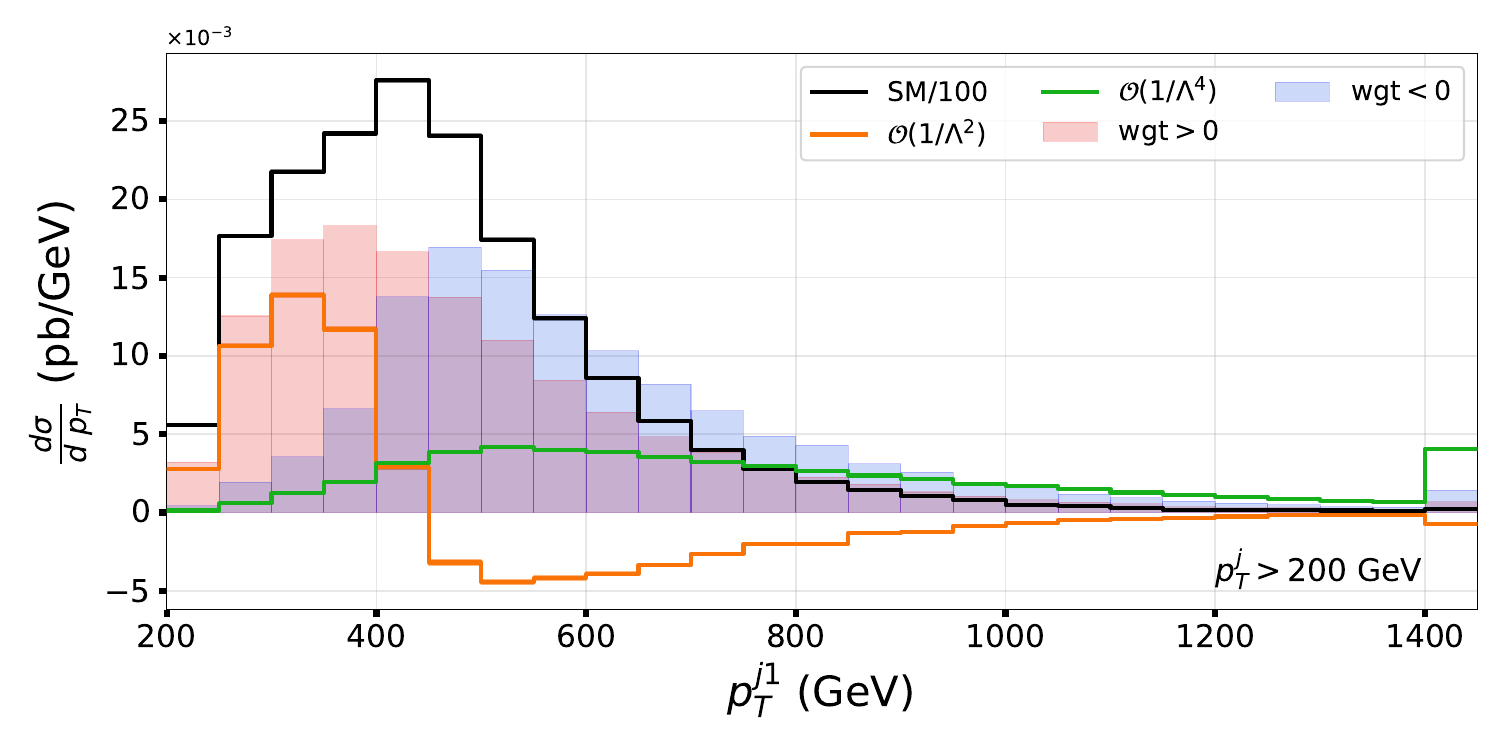}
   \includegraphics[width=.8\textwidth]{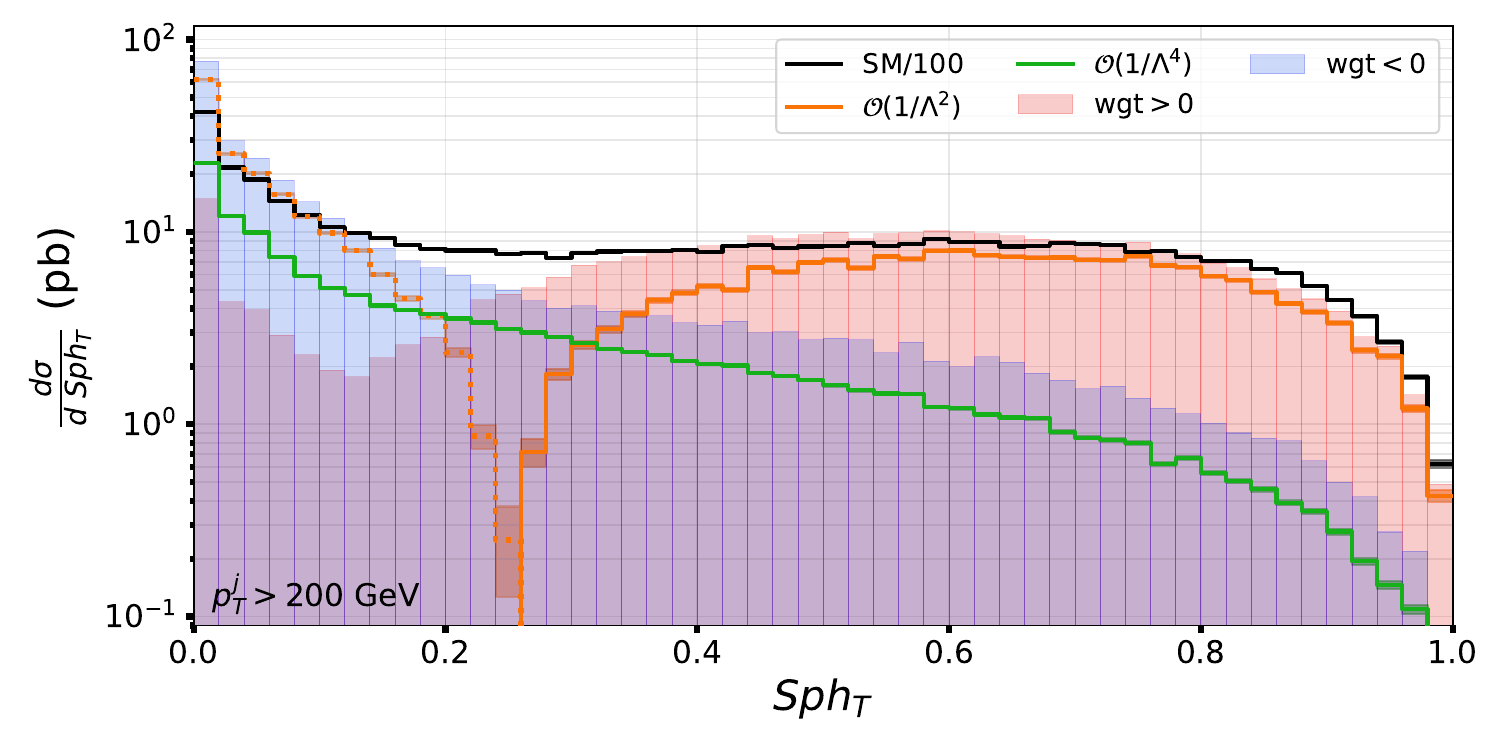}
   \includegraphics[width=.8\textwidth]{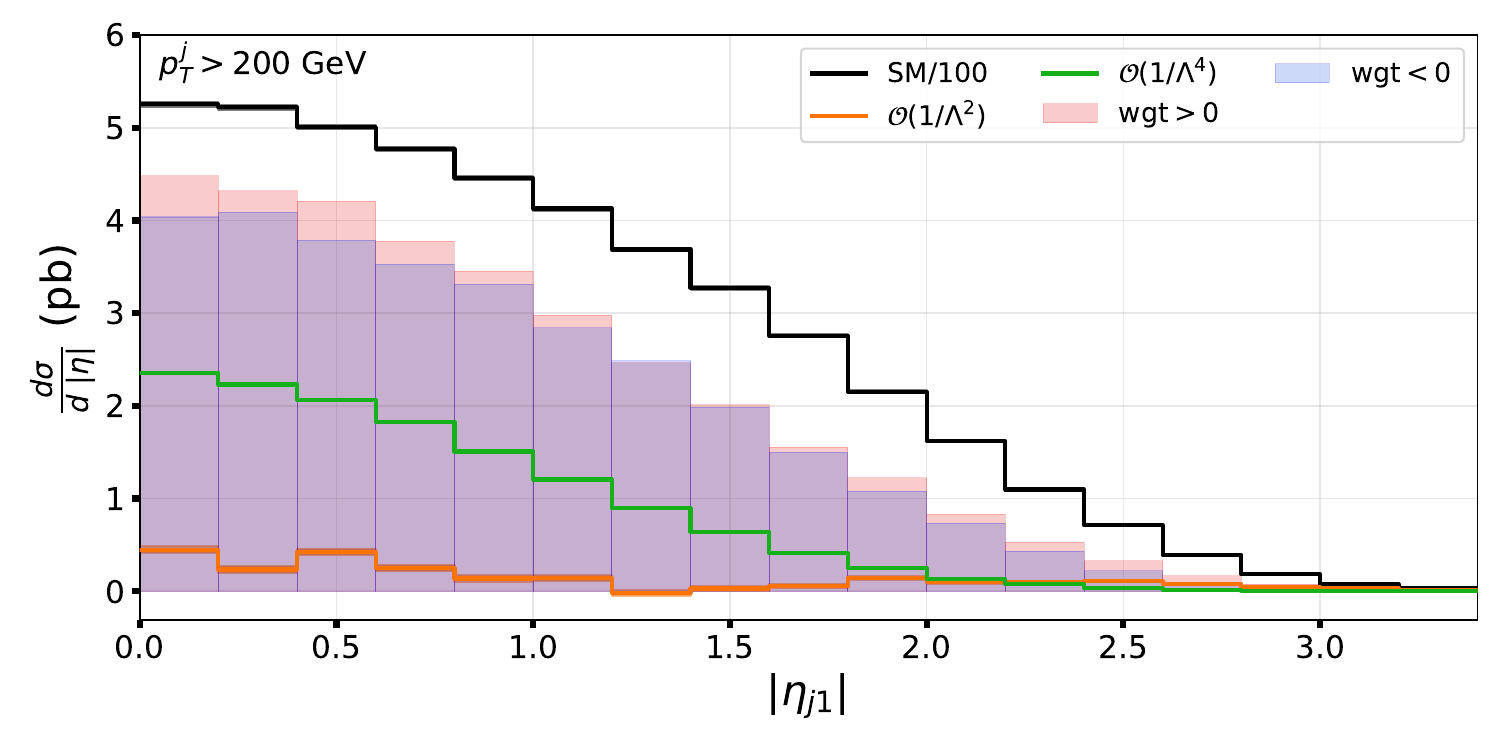}
\end{figure}

The differential distributions for $Sph_T$ and $p_T^{j1}$, in the $p_T^j > 200$ GeV case, are shown in Fig. \ref{fig:pTj200_diff}, for the SM, linear and quadratic contributions to three-jet production. It can be seen that the linear term changes sign over the phase space, unlike the other two. The absolute pseudorapidity of the leading jet is also shown: for this variable, the cancellation of the linear term is almost perfect all over its range. Indeed, its asymmetry in Table \ref{tab:3j_asymm} is close to the $\sigma^{1/\Lambda^2}$ one.

\subsection{Limits on $C_G/\Lambda^2$ from the linear term}
We combined the transverse sphericity with some observables that grow with the energy in double-differential distributions to obtain bounds over $C_G/\Lambda^2$, shown in Table \ref{tab:oG_bounds}. The energy-related variables are different with the $p_T^j$ cut. For $Sph_T$, we always employ two bins only, with their common border placed near the sign-flip value for that variable, reported in the table as well. The binning we used for $p_T^{j3}$ in the $p_T^j > 50$ GeV case is of 25 GeV bins from 50 to 200 GeV. For $S_T$ in the $p_T^j >200$ GeV case, we used bins of 100 GeV from 600 to 2100 GeV. In the $p_T^j> 500$ GeV, $M_{j2j3}$ was binned with 100 GeV bins from 200 to 2000 GeV, while in the $p_T^j > 1$ TeV we employed bins of 200 GeV from 400 to 3000 GeV. In all cases, the last bin includes the overflow.

\begin{table}
\caption{\footnotesize{ 95\% CL bounds on $C_G /\Lambda^2$, in TeV${}^{-2}$, obtained at the linear and quadratic level for three-jet production at LO. Different double-differential distributions are employed for each $p_T^j$ cut. The number of bins and the value at which $Sph_T$ changes sign are also shown }} \label{tab:oG_bounds}
\resizebox{\textwidth}{!}{
\begin{tabular}{c|ccc|cc}
   %\hline
   %\multicolumn{6}{c}{ $pp \rightarrow jjj$ } \\ \hline
   min $p_T^j$ (GeV) & Distribution & $Sph_T$ cut & $N_\text{bins}$ & $\mathcal{O}(1/\Lambda^2)$ bounds & $\mathcal{O}(1/\Lambda^4)$ bounds \\ \hline
   50 & $p_T^{j3}$ vs $Sph_T$ & 0.23 & 12 & [-1.5, 1.5] & [-4.1, 4.7]$\cdot 10^{-1}$ \\
   200 & $S_T$ vs $Sph_T$ & 0.25 & 32 & [-2.9, 2.9]$\cdot 10^{-1}$ & [-1.0, 1.0]$\cdot 10^{-1}$ \\
   500 & $M_{j2j3}$ vs $Sph_T$ & 0.31 & 32 & [-8.4, 8.4]$\cdot 10^{-2}$ & [-4.4, 6.1]$\cdot 10^{-2}$ \\
   1000 & $M_{j2j3}$ vs $Sph_T$ & 0.35 & 22 & [-3.1, 3.1]$\cdot 10^{-2}$ & [-1.7, 2.3]$\cdot 10^{-2}$ \\
   \hline
\end{tabular}
}
\end{table}

Since the measurements for these distributions are yet to be published, we assumed that the experimental data follows the SM one, resulting in the $\chi^2$ expression
\begin{equation}
   \chi^2 = \sum_{i=1}^{N_\text{bins}} \left( \frac{ x_i^\text{exp} - x_i^\text{th} }{ \Delta_i } \right)^2 = \sum_{i=1}^{N_\text{bins}} \left( \frac{C_G}{\Lambda^2} \frac{ x_i^{1/\Lambda^2} }{ \Delta_i } \right)^2,
\end{equation}
where $x_i^\text{exp}$ and $x_i^\text{th} = x_i^\text{SM} +\frac{C_G}{\Lambda^2} x_i^{1/\Lambda^2} \left( +\frac{C_G^2}{\Lambda^4} x_i^{1/\Lambda^4} \right)$ are the experimental and simulated bin contents. The uncertainty at the denominator, $\Delta$, was assigned to 10\% of the SM value in each bin \cite{atlas_unc}; we checked that the numerical errors in our predictions do not exceed that threshold for the chosen binning.

Our results show that, with suitable observable choices, it is possible to obtain bounds on $C_G/\Lambda^2$ from the linear term that are at the same order of magnitude as the ones from the quadratic level. Even though the latter still seems to provide better constraints, one should never forget that the $\mathcal{O}(1/\Lambda^4)$ term must also include the interference of the SM with dimension-8 operators to be complete. Furthermore, our linear limits from the highest-$p_T^j$ case are comparable to the ones already present in the literature. The inclusion of PS and detector effects would most likely worsen these results, nonetheless our simplified analysis suggests that dedicated measurements of $Sph_T$ in combination with observables that grow with the event energy can improve the limits over the $O_G$ operator.

\begin{figure}
\centering
\caption{\footnotesize{ 95\% CL limit on $\Lambda$ with respect to the CoM-energy upper cut, for $C_G =1$, in the $p_T^j > 200$ GeV and 1 TeV regions. The bounds were computed through the $S_T$ {\it vs} $Sph_T$ and $M_{j2j3}$ {\it vs} $Sph_T$ double-differential distributions, respectively. The limits from $S_T$ alone are also shown. The shaded areas cover the region where $\sqrt{s}$ is larger than the bound on $\Lambda$ }} \label{fig:bounds_sqrts}
   \includegraphics[width=.9\textwidth]{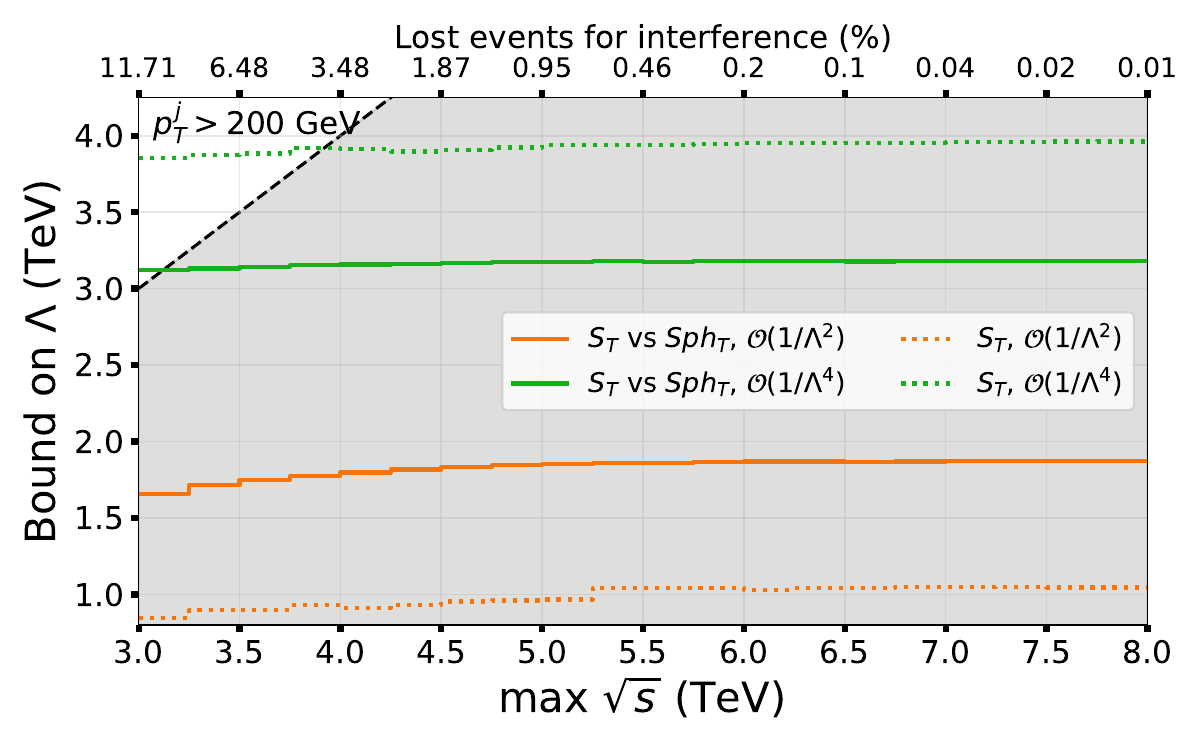}
   \includegraphics[width=.9\textwidth]{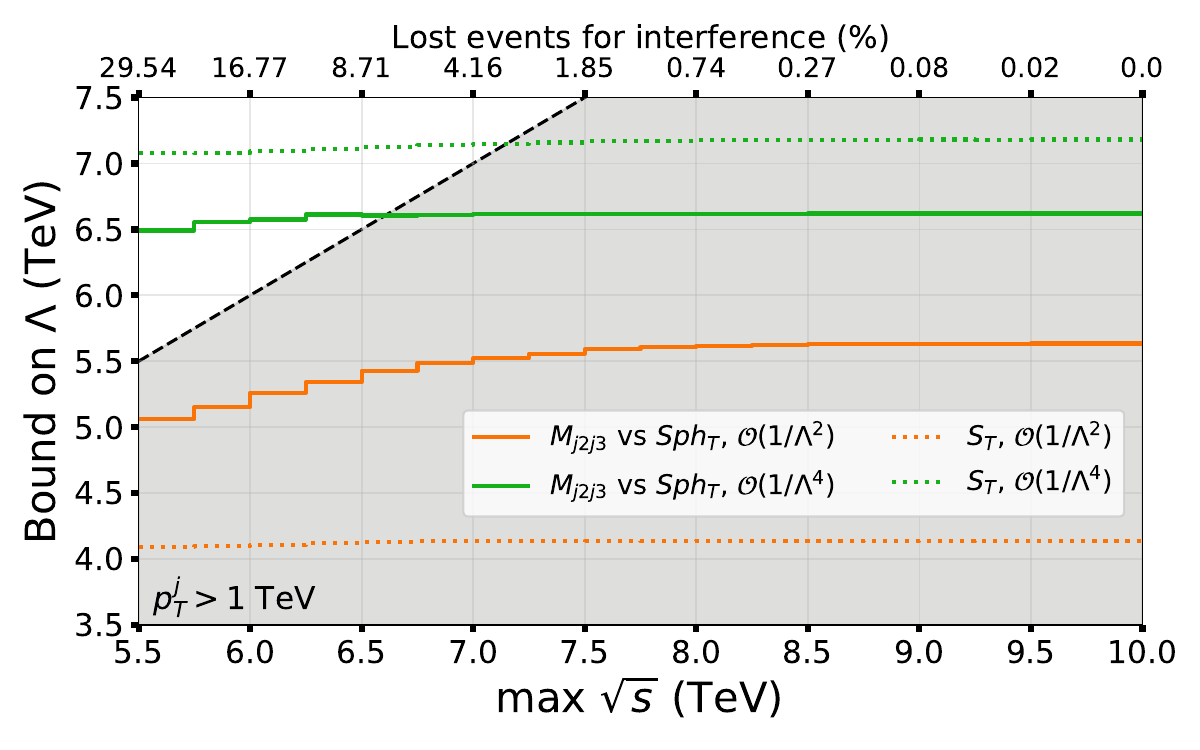}
\end{figure}

In order to verify the SMEFT expansion validity, we computed the limits on $\Lambda$ as a function of the upper cut on the CoM energy $\sqrt{s}$, assuming $C_G = 1$. The results are summarised in Fig. \ref{fig:bounds_sqrts} for the 200 and 1000 GeV $p_T^j$-cut cases: the bounds are computed through the best double-differential observables in those regions, and the ones from $S_T$ are also shown as a comparison. It can be observed that the linear limits grow faster than the quadratic ones, as one would expect from their different dependences on $\Lambda$, and that they barely change when the events with CoM energy above 5 (8) TeV are included in the $p_T^j > 200$ (1000) GeV case. 
As explained in Chapter \ref{chap:smeft}, the EFT framework can be trusted only if $\sqrt{s} < \Lambda$, as a rule of thumb. The limits we obtained seem to fall outside this region even for $\sqrt{s}$ cuts of few TeV, if $C_G$ is taken equal to 1. It is worth reminding that this assumption is not backed by any valid reason, and that only bounds over the $C_G/\Lambda^2$ ratio are meaningful.

\section{\label{sect:oG_conclusions}Conclusions and prospects}
This study targeted a process, three-jet production, for which the cross section of the $O_G$ interference with the SM is suppressed as the result of large cancellations between positive- and negative-weighted contributions over the phase space. It represents a different scenario than the dijet one, for which the linear term is exactly null due to helicity mismatch among the SM and new-physics amplitudes.

We introduced some quantities, like the integrable and measurable cross sections, that are able to estimate the total effect of the interference for given processes, and how much of it is accessible at a certain collider. These tools helped us identify an event-shape observable, the transverse sphericity, that is particularly sensitive to the linear-term suppression for three-jets, and therefore to the sign of $C_G$. When combined in double-differential distributions with other variables that grow with the energy, like $S_T$ or $M_{j2j3}$, it can provide bounds on $C_G/\Lambda^2$ that are of the same order of magnitude as the ones from the quadratic term, that does not suffer from the same cancellation, and that are comparable with the ones already published in the literature.

Future investigations are still needed to refine this simplistic analysis, as the inclusion of NLO and PS effects might alter the picture. As an example, the similar cross sections (in absolute value) for three- and four-jet production in Table \ref{tab:oG_xsects} suggest that a proper matching and merging algorithm should be applied to obtain reliable results, even at LO. Furthermore, the dimension-8 operators interference with the SM might strongly affect the quadratic limits at $\mathcal{O}(1/\Lambda^4)$ order. The measurable and integrable cross sections can only be computed at LO parton level, as NLO and PS generate additional negative weights that cannot be distinguished from the interference ones. As it will be illustrated in the next chapter, these quantities can be used at parton level to identify the observables that can restore the linear contribution of an operator to a certain process, and then the NLO, PS and detector effects over these variables need to be checked, to see how much they affect their distributions.

The strategy introduced in this chapter is quite generic, as it mainly relies on kinematic distributions, and in principle it can be employed in every scenario where large cancellations are present, even when the EFT approach is not feasible. $\sigma^{|\text{meas}|}$ can be used in combination with machine-learning techniques that aim at restoring the linear term \cite{CPasymmML:2022higgs,CPasymmML:2022ew}. As an example, the measurable cross section can help identify better inputs for any neural network trying to maximise the asymmetry, speeding the training and convergence. The employment of such techniques, though, presents a downside: it is not trivial, by just looking at its internal parameters after the training, to understand the analytical expression of the optimal observable that the network learned. More comments about the automation of the interference revival can be found in the conclusions of the next chapter, in Sect. \ref{sect:oW_conclusions}.

%!TEX root = main.tex

%%%%%%%%%%%%%%%%%%%%%%%%%%%%%%%%%%%%%%%%%%%%%%%%%%%%
%
%      Chapter IV :
%
%
%%%%%%%%%%%%%%%%%%%%%%%%%%%%%%%%%%%%%%%%%%%%%%%%%%%

\chapter{NLO-corrections stability through interference revival: the case of $O_W$}
\label{chap:interf_OW}
\pagestyle{fancy}

\hfill
\begin{minipage}{\textwidth}
   \centering
   \includegraphics[width=.7\textwidth]{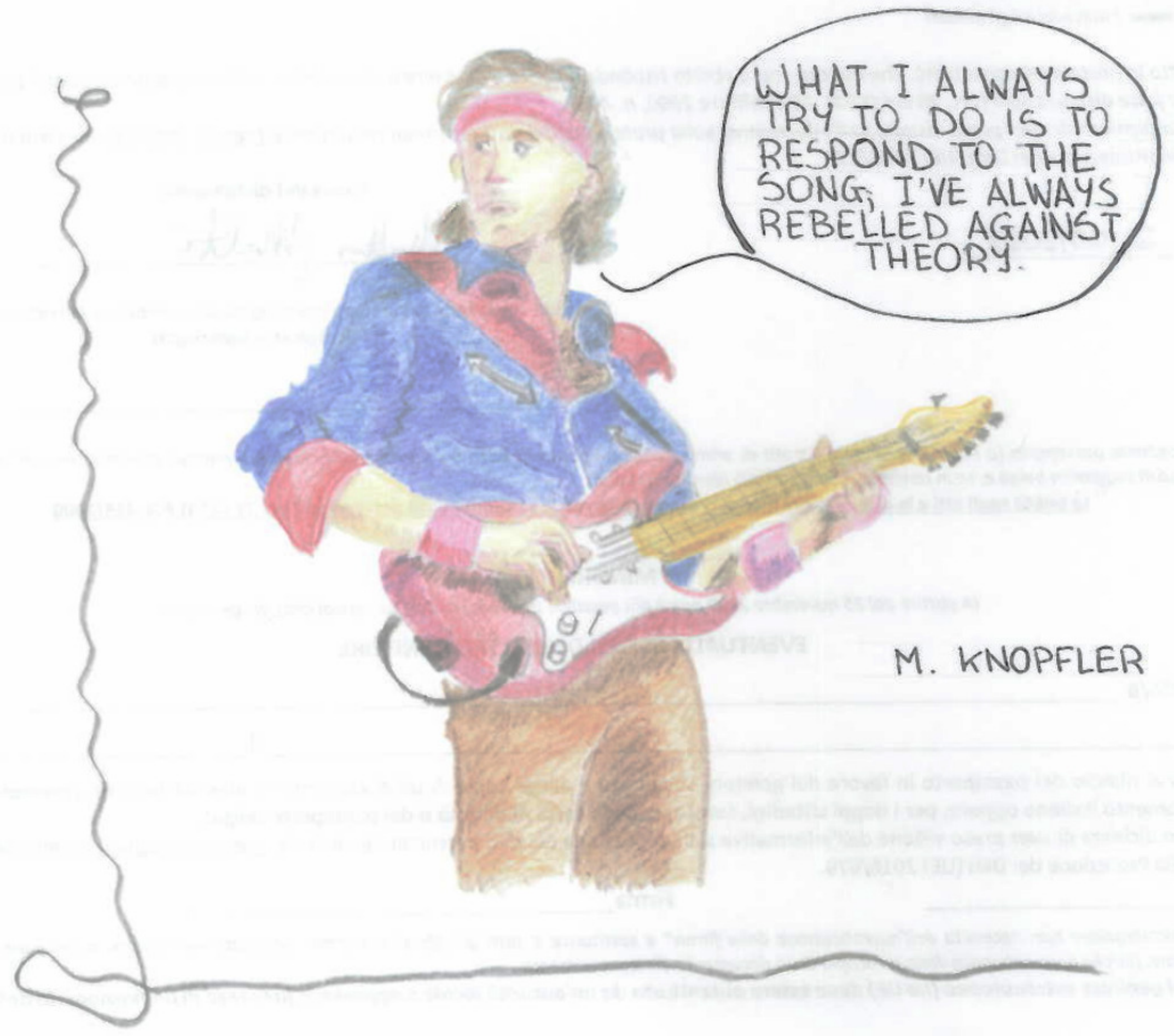}
\end{minipage}

\vspace{0.5cm}

Here below, I will describe how we employed the interference-reviving strategies from the previous chapter to obtain stable predictions at NLO for some processes that can be affected by the EW analogue of $O_G$, defined as
\begin{equation}
   O_W = \varepsilon^{IJK} \hspace{1mm} W_\mu^{I,\nu} W_\nu^{J,\rho} W_\rho^{K,\mu},
\end{equation}
with $W_{\mu\nu}$ the EW field strength, defined in Eq. \eqref{field_strengths}, and $\varepsilon$ the $SU(2)$ structure constant in Eq. \eqref{struct_const}. This is a dimension-6, CP-even term of the SMEFT that might contribute to diboson processes and triple-gauge couplings. Its Wilson coefficient is referred to as $C_W$.

As introduced in Sect. \ref{sect:rev_strategies}, the helicities induced by this operator in the $\mathcal{O}(1/\Lambda^2)$ amplitudes are orthogonal to the SM ones. Even though this statement is exact for massless final states in $2 \rightarrow 2$ processes, it is still valid approximately in the massive case, as for $W$- and $Z$-boson final states, especially at high energies where masses can be neglected. 

Previous studies \cite{smeft_at_nlo} computed the $K$-factors for its interference with the SM for different processes, and found large and/or negative results that seem to suggest a breaking of the EFT perturbativity. Throughout this work, we define the (differential) $K$-factor as the ratio of the (differential) cross sections at NLO and LO. With the findings from the previous chapter in mind, this might be explained as the result of a cancellation between two opposite-sign contributions to the linear terms at LO, that individually present more reasonable but different $K$-factors. If this is actually the case, suitable observables that are sensitive to the suppression may provide more stable predictions for the interference at NLO, and also better bounds on $C_W /\Lambda^2$ that could compete with the ones from the $\mathcal{O}(1/\Lambda^4)$ order.

\begin{figure}
   \centering
   \caption{\footnotesize{Representative diagrams for the three processes investigated in this chapter: $Zjj$ through VBF ({\it left}), fully-leptonic $WZ$ ({\it centre}) and leptonic $W\gamma$ production ({\it right})}} \label{fig:oW_diagrams}
   \includegraphics[width=.32\textwidth]{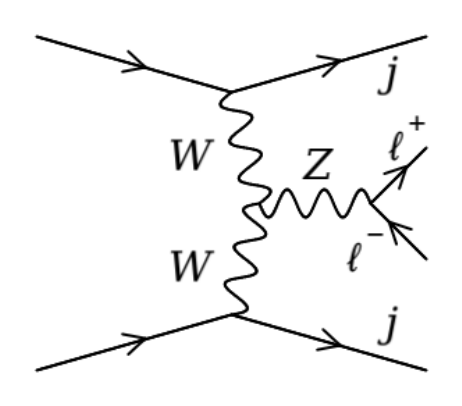}
   \includegraphics[width=.32\textwidth]{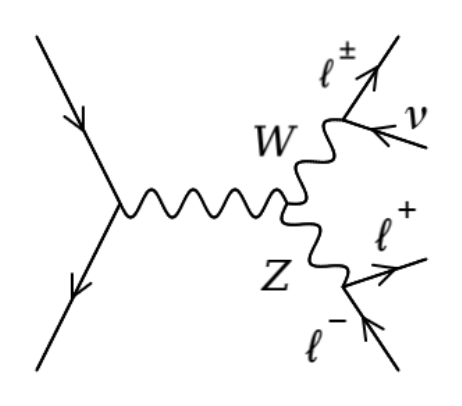}
   \includegraphics[width=.32\textwidth]{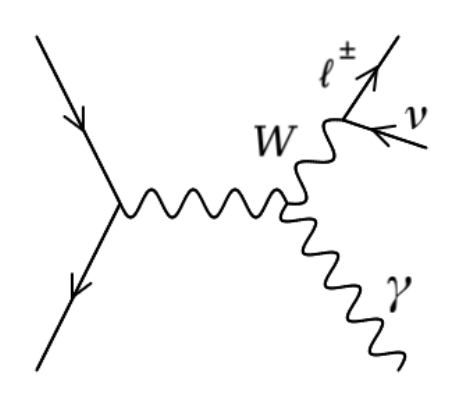}
   \begin{minipage}{.32\textwidth}
      \centering
      (a)
   \end{minipage}
   \begin{minipage}{.32\textwidth}
      \centering
      (b)
   \end{minipage}
   \begin{minipage}{.32\textwidth}
      \centering
      (c)
   \end{minipage}
\end{figure}

We focused on the fully-leptonic $Zjj$ production through Vector Boson Fusion (VBF), the fully-leptonic $W^\pm Z$ process and the leptonic $W^\pm \gamma$ one: some representative diagrams are shown in Fig. \ref{fig:oW_diagrams}. The leptonic decays were preferred over the hadronic ones because of the lower background they involve at the LHC; a study of the sensitivity to $O_W$ in case of hadronic and semileptonic decays of the $W$ bosons is reviewed in \cite{rafa_will_hadr}, where jet-substructure techniques are employed. For each process, we found observables that can at least partially lift the linear-term suppression, and we used them to obtain bounds on $C_W/\Lambda^2$.

%\section{General details on the calculation}
Our analysis was performed through \amc\ v3.4.2, which we fed the SMEFT@NLO UFO model \cite{smeft_at_nlo}. The rational and ultraviolet counterterms were obtained via \nloct\ \cite{nloct}. The leptons and all quarks but the top are taken as massless, and NNPDF3.0 with $\alpha_s (M_Z) = 0.118$ was employed as PDF set \cite{nnpdf30}. $C_W /\Lambda^2$ was set to 1 TeV${}^{-2}$.

Some NLO calculations were performed at fixed order (FO), with both $\mu_R$ and $\mu_F$ set to 1 TeV. When instead events were generated \cite{nlo_events}, we used \pyth\ or \hw\ to shower them \cite{pythia,hw1,hw2} and the dynamical scale was set to $H_T /2$ for each event. First, the dressed leptons were reconstructed through the $k_t$ algorithm with a radius parameter of 0.1 \cite{fastjet,kt_1,kt_2}. Then the anti-$k_t$ algorithm was employed to obtain the jets, with a radius parameter of 0.4 and excluding the dressed-leptons components \cite{antikt}.

All results were presented with numerical and scale-variation uncertainties, with the latter computed as the envelope of nine scale combinations of $\mu_R, \mu_F$ divided or multiplied by 2. All results are presented for the LHC at 13 TeV.

%%%%%%%%%%%%%%%%%%%%%%%%% Zjj %%%%%%%%%%%%%%%%%%%%%%%%%

\section{\label{sec:Zjj_oW}$Z$ + two jets production through VBF}

The first process we studied is the EW production of $Zjj$, characterised by an exchange of EW bosons in the $t$-channel and subject to the VBF mechanism. A representative diagram is shown in Fig. \ref{fig:oW_diagrams}a. We considered the $Z$-boson leptonic decay $Z \rightarrow \ell^+ \ell^-$. This is normally investigated as a background to VBF Higgs production \cite{higgsVBF_1,higgsVBF_2,higgsVBF_3}, but it can also be sensitive to new physics due to the high energies requested for the jets.

As explained in Section \ref{sect:rev_strategies}, the linear $O_W$ amplitude and the SM one tend to produce vector bosons in the final state with different helicity configurations. This argument motivates the employment of azimuthal observables for triple field-strength operators. Indeed, previous predictions \cite{atlas:2021Zjj,atlas:2021Comb} point out that the signed azimuthal distance between the two leading jets in $p_T$ shows a large sensitivity to the $O_W$ contribution to $Z+$jets production. This quantity is defined as $\Delta \phi_{jj} = \phi_{ja}-\phi_{jb}$, with $y_{ja} > y_{jb}$.

\begin{figure}
   \centering
   \caption{\footnotesize{Different orders in the strong and EW coupling constants contributing to the $pp \rightarrow Zjj$ cross section at LO and NLO. The $Z$-boson decay is not considered. The arrows specify the directions and natures of the NLO expansions. Some representative diagrams are shown}} \label{fig:zjj_orders}
   \includegraphics[width=.7\textwidth]{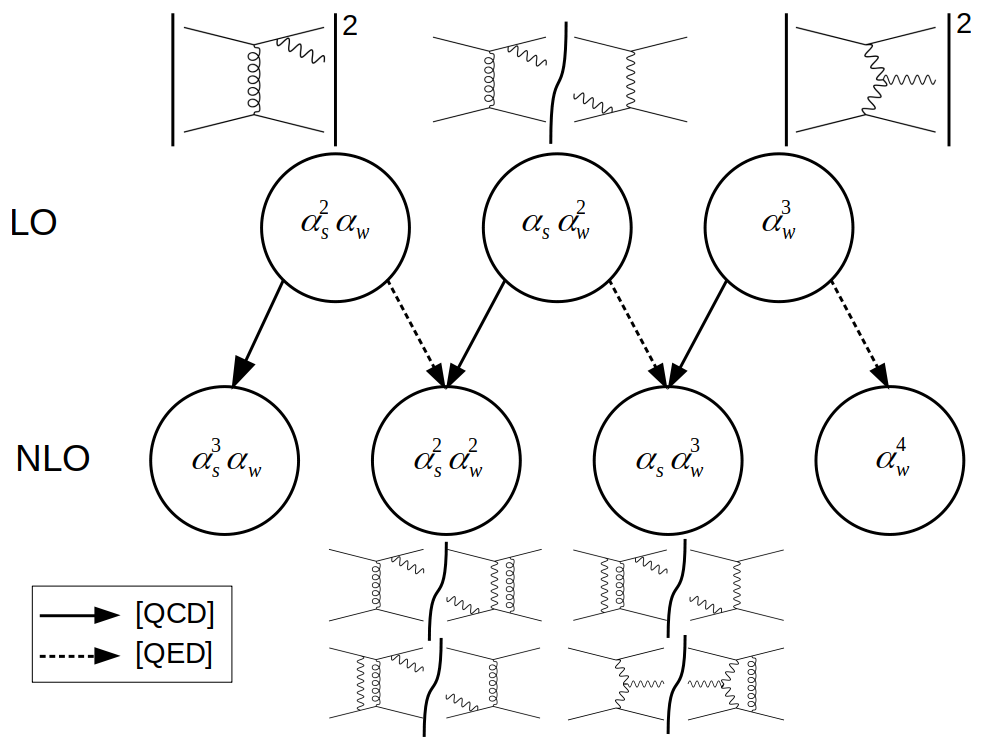}
\end{figure}

The simulation of this process presents multiple challenges at NLO. Different orders in $\alpha_S^{}$ and $\alpha_W^{}$ contribute at LO to the SM cross section: a fully-EW one ($\alpha_W^3$), one involving QCD interactions ($\alpha_S^2 \alpha_W^{}$), and an interference among the two ($\alpha_S^{} \alpha_W^2$). They are summarised in Fig. \ref{fig:zjj_orders}, where the leptonic decay of the $Z$ boson is not included. The NLO cross section can also be split into different orders in the same way, and some of them can be seen both as a QCD or a QED correction to two different Born contributions. For this reason, all the orders should be generated at the same time, or poles would not cancel and results could not be trusted; the computational and time costs, though, are too large, especially when looking at distributions.

In particular, we are interested in the pure-EW term, of order $\alpha_W^3$ at LO: the NLO component of the cross section that is obtained through QCD corrections to it ($\alpha_S^{} \alpha_W^3$) cannot be distinguished from the one that results from NLO corrections in QED to the LO interference of the fully-EW and the QCD terms. In this case the calculation can be simplified by including, in the pure-EW contribution, just the diagrams where only $W$ bosons are exchanged along the $t$-channel: the two quark lines could not feature the same flavour, the crossing would not interfere with the QCD diagrams and the $\alpha_S^{} \alpha_W^2$ order at LO would be null. To implement this, we modified the UFO model to add a new coupling for the $Wq \bar q$ interaction and used it to neglect diagrams with $Z$ bosons or photons along the $t$-channel. We checked that this changes the LO SM cross section by less than 1\%, for the phase-space cuts listed below.

\subsection{Description of the phase-space cuts}
We recast an analysis from the ATLAS Collaboration \cite{atlas:2021Zjj}, which requires for each event exactly two leptons ($\ell = e, \mu$) with $p_T > 25$ GeV and $|\eta| < 2.4$; their total invariant mass and total transverse momentum have to satisfy $81.2 < M_{\ell\ell} < 101.2$ and $p_T^{\ell \ell} > 20$ GeV.

At least two jets are needed, with minimum $p_T$ of 85 (80) GeV for the leading (subleading) one. Their rapidities have to satisfy $|y| < 4.4$ and they have to be separated from the leptons with a $\Delta R_{\ell j}$ of at least 0.4.

In addition to this, futher cuts are introduced to isolate the EW contribution to this process. The analysis requires a large invariant mass for the two leading jets, namely $M_{jj} > 1$ TeV, and also a large rapidity gap $|\Delta y_{jj}| > 2$ among them. No other jets with $p_T > 25$ GeV can be present in this rapidity gap. Moreover, the $Z$ boson needs to be produced centrally with respect to the dijet system and this is achieved by imposing $\xi_Z < 0.5$, where
\begin{equation}
   \xi_Z = \frac{ |y_{\ell\ell}-\frac{1}{2}(y_{j1}-y_{j2})| }{ |\Delta y_{jj}| }
\end{equation}
is the ``Zeppenfeld variable''. In combination with the previous ones, this last cut suppresses the QCD-induced background, where the hadronic and leptonic activities are more spread, compared to the EW VBF production, where the two jets tend to be back-to-back and the leptons to lie in the rapidity interval between the two leading jets \cite{zeppenfeld_var,zeppenfeld_var_2}.

\begin{figure}
   \centering
   \caption{\footnotesize{ NLO SM differential distributions for $\Delta\phi_{jj}$ in $Zjj$ production, at FO and matched to \pyth\ and \hw\ for the shower. Numerical uncertainties and scale variations are also shown. The dots mark the experimental measurements }} \label{fig:dphijj_ps_comparison}
   \includegraphics[width=.8\textwidth]{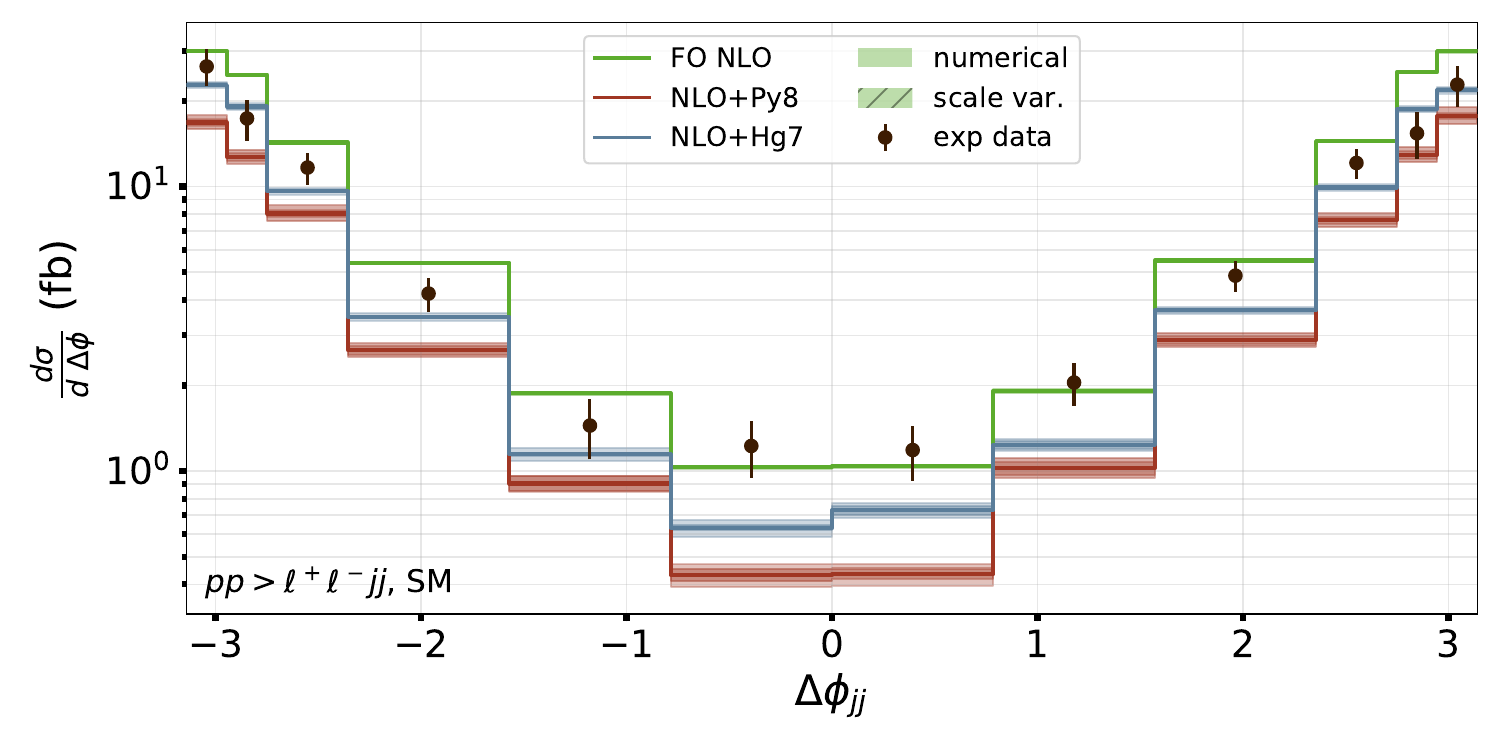}
\end{figure}

\subsection{Parton-shower effects}
It is known in the literature that different MC generators do not agree on their predictions for VBF processes and that the results are strongly influenced by the PS algorithm choice \cite{WWjj_2019,WWjj_1_2019,Hoche:2021,Jager:2020}. Indeed, the largest uncertainty source of the experimental measurement in \cite{atlas:2021Zjj} comes from the comparison of different MC codes and spans from $\sim 10\%$ in the central $\Delta\phi_{jj}$ bins to 30\% in the external ones. In order to assess this, we compared the NLO calculations for the SM at FO and matched to \pyth\ and \hw . In more detail, the global recoil scheme that is usually employed by \pyth\ does not suit this kind of processes, as it predicts too much QCD radiation between the two leading jets, that has to be discarded because of the imposed cuts. Another scheme, called ``dipole recoil'', should provide better results, but it is available in \amc\ only at LO and the shower counterterms at NLO can only be derived using the global scheme \cite{Jager:2020}.

At the total fiducial cross-section level, the best agreement between the NLO SM result and the ATLAS measurement is achieved when \hw\ is used for the shower. The comparison at differential level is shown in Fig. \ref{fig:dphijj_ps_comparison} for $\Delta \phi_{jj}$: it can be seen that no option is able to reproduce the experimental data all over the range of the variable. FO calculations, though, seem to be compatible with measurements in the central bins, so we picked this choice for our analysis as this differential distribution is quite important to study $O_W$. However, the other two NLO+PS possibilities present similar $K$-factors to the FO ones. It is important to point out that this comparison was carried out among the experimental data and the SM only, similarly to what was done in the ATLAS analysis \cite{atlas:2021Zjj} for other MC generators. In general, choosing the tools that are closer to the measurements with their SM predictions might restrict the sensitivity to new physics.

%\begin{adjustbox}{angle=90} \begin{sidewaystable}
\begin{table}
   \centering
   \caption{\footnotesize{Cross-section results in fb for the SM, linear and quadratic contributions, for {\it Zjj}, $WZ$ and $W \gamma$ production; the total \textit{K}-factors are also shown. $C_W/\Lambda^2$ is set to 1 TeV${}^{-2}$. For each result, the first uncertainty source is numerical, while the second ones come from scale variation. For the \textit{K}-factors, the numerical uncertainty is propagated in quadrature from the cross-section ones, while for the scale variations the total envelope is considered. FO computations were employed for the first two processes, while the last one is matched to PS. The $WZ$ results are averaged over four decay channels}} \label{tab:oW_xsects}
\begin{tabular}{c|ccc}
%\hline
 & SM & $\mathcal{O}(1/\Lambda^2)$ & $\mathcal{O}(1/\Lambda^4)$ \\ \hline \hline
\multicolumn{4}{c}{$pp \rightarrow \ell^+ \ell^- jj$ EW, $\ell=(e,\mu)$} \\
\hline
$\sigma_{LO}$ (fb) & 49$\pm 0.06\%^{+8\%}_{-6\%}$ & -1.67$\pm 0.4\%^{+6\%}_{-7\%}$ & 9.4$\pm 0.07\%^{+11\%}_{-10\%}$ \\
$\sigma_{NLO}$ (fb) & 52.2$\pm 0.19\%^{+0.8\%}_{-1.1\%}$ & -1.66$\pm 1.2\%^{+0.4\%}_{-0.8\%}$ & 11.1$\pm 0.18\%^{+3\%}_{-4\%}$ \\
\textit{K}-factor & 1.07$\pm 0.19\%^{+9\%}_{-7\%}$ & 0.99$\pm 1.2\%^{+6\%}_{-8\%}$ & 1.18$\pm 0.17\%^{+14\%}_{-14\%}$\\
\hline
\multicolumn{4}{c}{$pp \rightarrow \ell^\pm \overset{\scriptscriptstyle(-)}{\nu} \ell^+ \ell^-$, $\ell=(e,\mu)$} \\
\hline
$\sigma_{LO}$ (fb) & 34.6$\pm 0.012\%^{+1.2\%}_{-1.4\%}$ & 0.169$\pm 0.3\%^{+1.8\%}_{-2\%}$ & 6.2$\pm 0.06\%^{+2\%}_{-1.6\%}$\\
$\sigma_{NLO}$ (fb) & 50.5$\pm 0.02\%^{+1.6\%}_{-1.4\%}$ & -0.91$\pm 0.5\%^{+5\%}_{-7\%}$ & 7.34$\pm 0.07\%^{+0.8\%}_{-0.7\%}$\\
$\sigma_{N^2LO}$ (fb) & 62.8$\pm 0.3\%^{+1.4\%}_{-1.3\%}$ & - & - \\
\textit{K}-factor & 1.46$\pm 0.03\%^{+3\%}_{-3\%}$ & -5.4$\pm 0.6\%^{+7\%}_{-9\%}$ & 1.18$\pm 0.09\%^{+3\%}_{-3\%}$ \\
N$^2$LO / LO & 1.82$\pm 0.3\%^{+3\%}_{-3\%}$ & - & - \\
\hline
\multicolumn{4}{c}{$pp \rightarrow \ell^\pm \overset{\scriptscriptstyle(-)}{\nu} \gamma$, $\ell=(e,\mu,\tau)$} \\
\hline
$\sigma_{LO}$ (fb) & 20.7$\pm 0.4\%^{+1.4\%}_{-1.4\%}$ & -0.67$\pm 9\%^{+21\%}_{-9\%}$ & 110$\pm 0.5\%^{+5\%}_{-4\%}$ \\
$\sigma_{NLO}$ (fb) & 29.8$\pm 0.6\%^{+3\%}_{-2\%}$ & -3.4$\pm 9\%^{+9\%}_{-11\%}$ & 121$\pm 0.7\%^{+1.2\%}_{-1.2\%}$ \\
\textit{K}-factor & 1.44$\pm 0.5\%^{+4\%}_{-4\%}$ & 5.1$\pm 12\%^{+29\%}_{-22\%}$ & 1.10$\pm 0.7\%^{+6\%}_{-5\%}$ \\
\hline
\end{tabular}
%\end{adjustbox} \end{sidewaystable}
\vspace{1cm}
\caption{\footnotesize{Values of the LO $\mathcal{O}(1/\Lambda^2)$ integral and measurable cross sections for {\it Zjj}, in fb ({\it left}) and their graphical representation (bar chart on the {\it right}). The absolute asymmetry for $\Delta\phi_{jj}$ is reported. These results come from event generation at LO, without PS. The numerical uncertainties are shown in the table: they are computed separately on the positive- and negative-weighted events, then propagated in quadrature}} \label{tab:oW_zjj_meas}
\begin{minipage}{.55\textwidth}
\begin{tabular}{c|ccc}
%\hline
\multicolumn{4}{c}{$p p \rightarrow \ell^+ \ell^- j j$ EW, $\ell=(e,\mu)$} \\ %\hline
 & (fb) & \% of $\sigma^{|\text{int}|}$ & \% of $\sigma^{|\text{meas}|}$ \\ \hline
$\sigma^{|\text{int}|}$ & 13.27$\pm$0.3\% & 100 & -\\
$\sigma^{|\text{meas}|}$ & 12.81$\pm$0.3\% & 97 & 100 \\
$\Delta \phi_{jj}$ & 11.42$\pm$0.4\% & 86 & 89 \\
$\sigma^{1/\Lambda^2}_{LO}$ & -1.71$\pm$2\% & 13 & 13 \\
\hline
\end{tabular}
\end{minipage}
\begin{minipage}{.44\textwidth}
   \includegraphics[width=\textwidth]{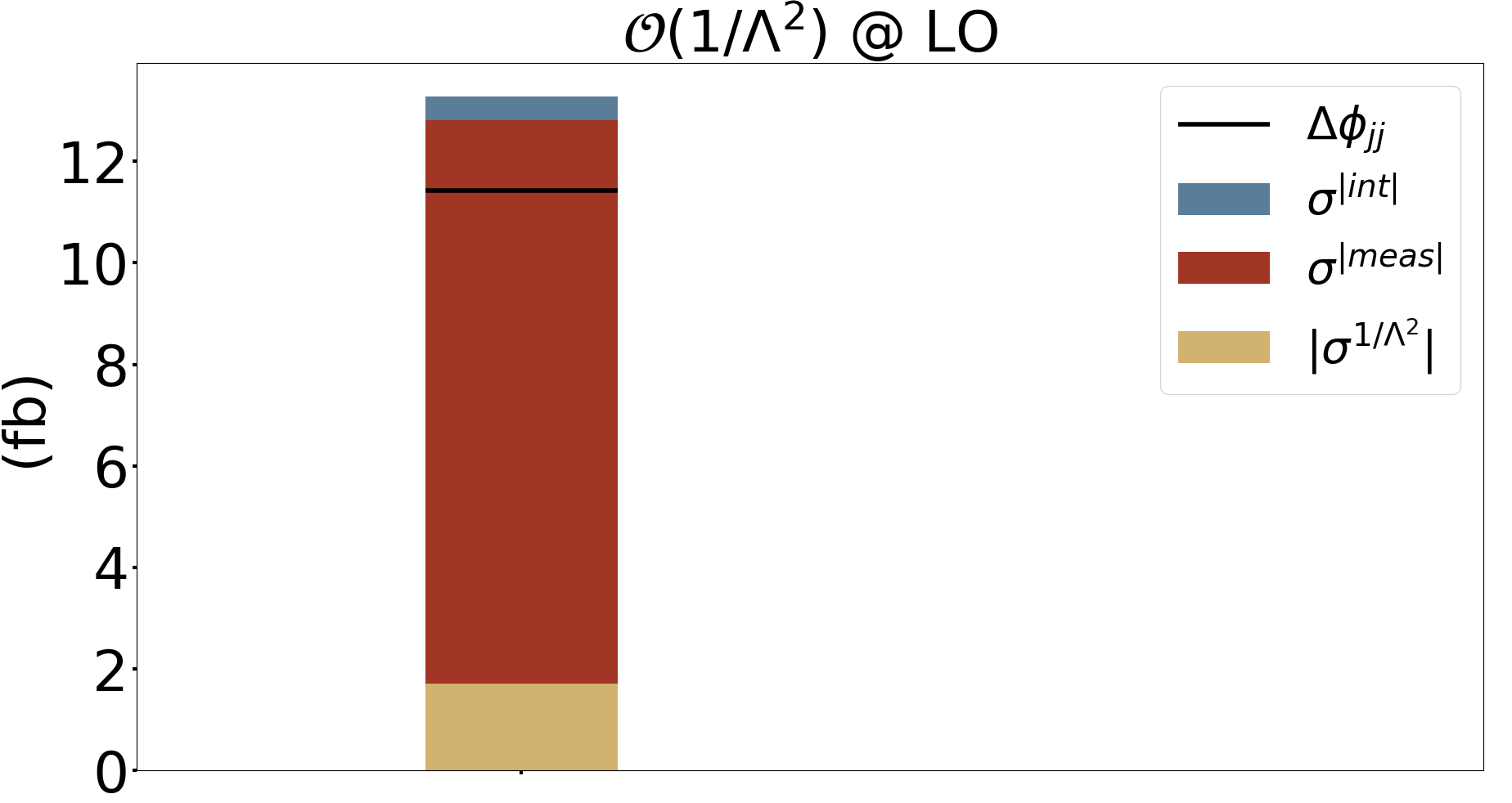}
\end{minipage}
\end{table}

\begin{figure}
   \centering
   \caption{\footnotesize{ LO differential cross section for $\Delta\phi_{jj}$ in $Zjj$ production, without PS. The SM is reproduced in black, while the linear term is in orange. The positive and negative contributions to the latter are shown separately in the red and blue shaded histograms, respectively. No uncertainties are reported }} \label{fig:oW_zjj_dphijj_lo}
   \includegraphics[width=.9\textwidth]{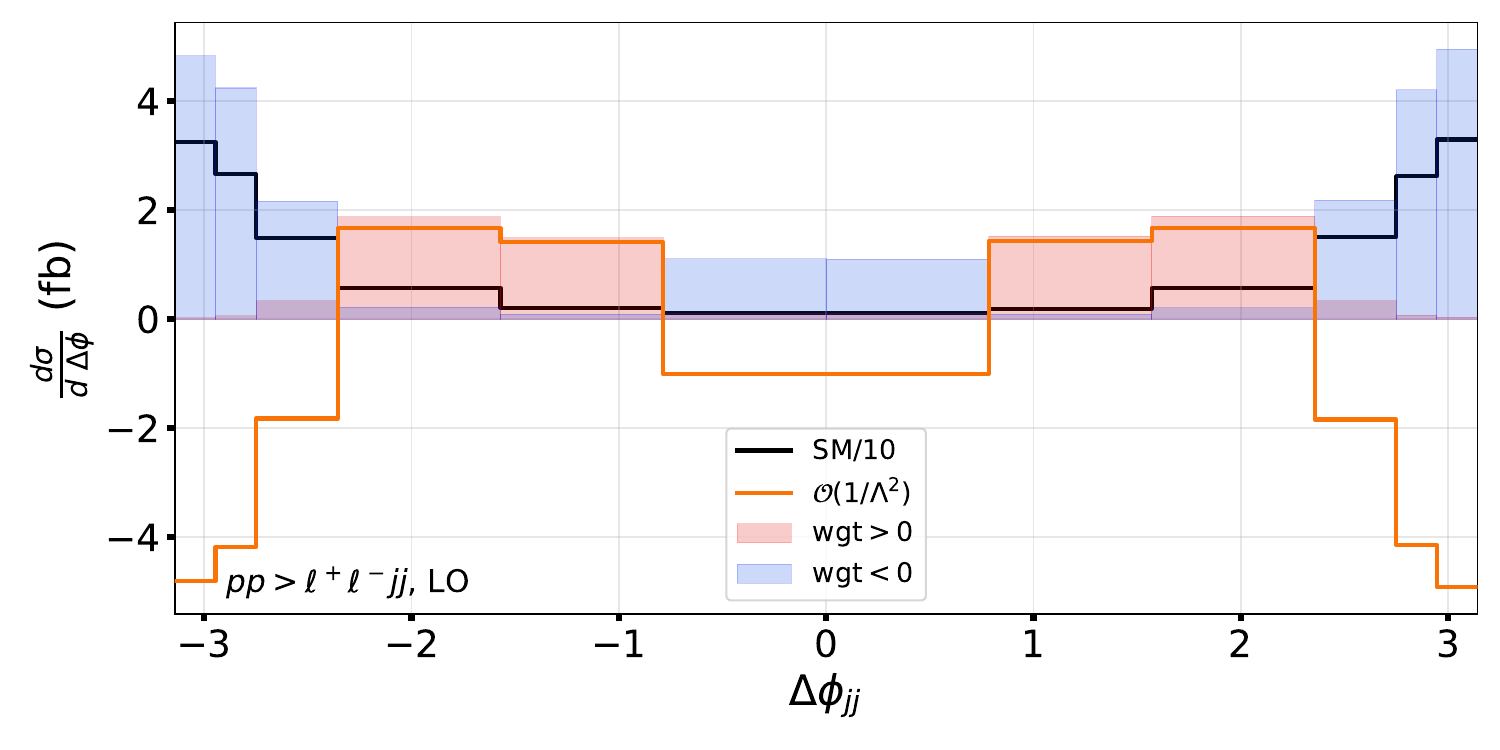}
   \caption{\footnotesize{FO differential distributions for $\Delta\phi_{jj}$ at NLO ({\it continuous}) and LO ({\it dotted}), together with the experimental measurements. The SM line is in black, the linear (quadratic) one is in orange (green). The differential $K$-factors and cancellation level \eqref{rwpm} are also shown in the second and third panels. The numerical and scale-variation uncertainties are also reported}} \label{fig:oW_zjj_dphijj_nlo}
   \includegraphics[width=.9\textwidth]{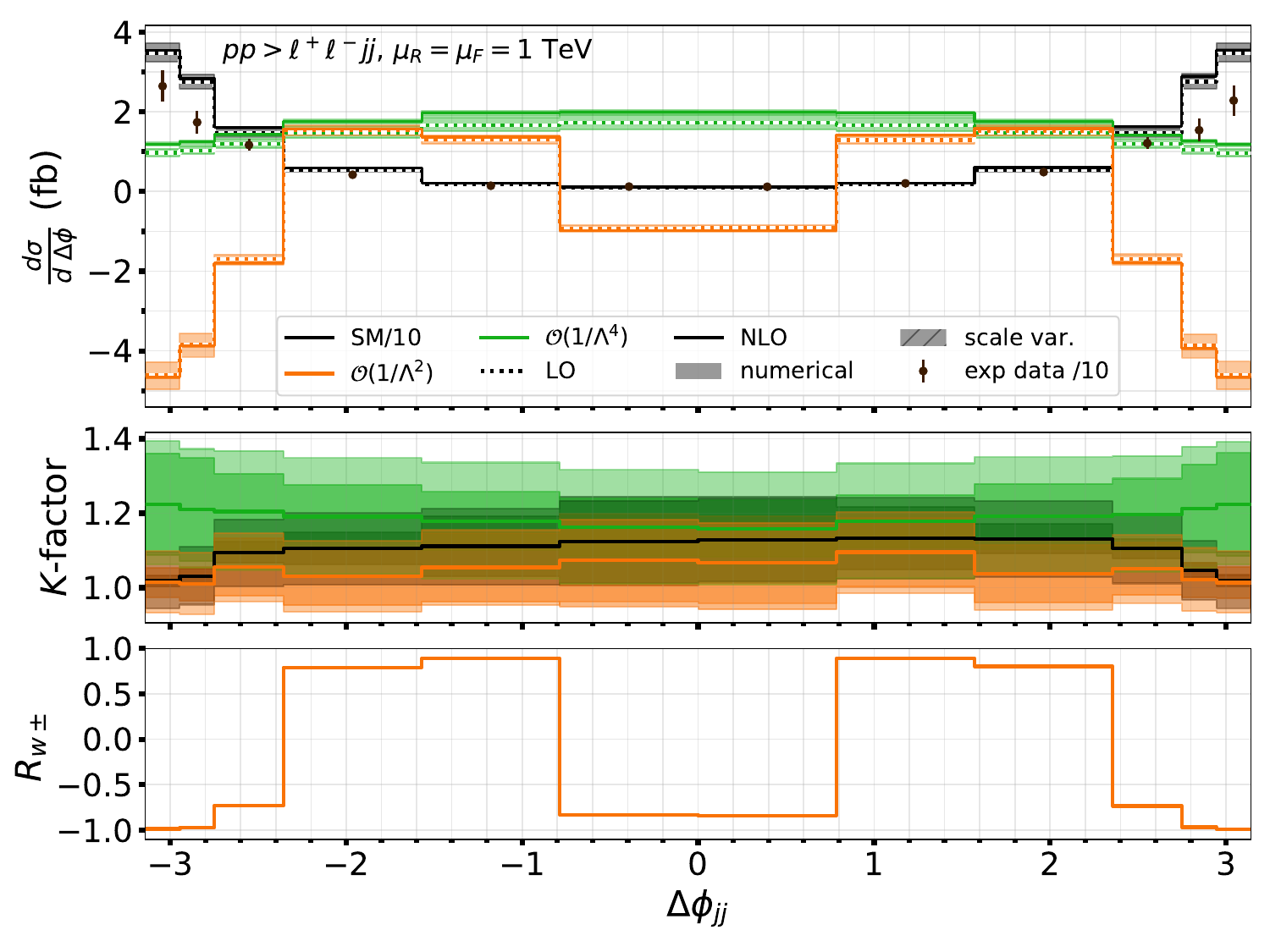}
\end{figure}

\subsection{Results and distributions}
The total FO cross sections at LO and NLO for the SM, linear and quadratic terms are summarised in Table \ref{tab:oW_xsects}. Our NLO SM result is larger than the one from the ATLAS collaboration in \cite{atlas:2021Zjj} by a factor $\sim 1.5$, as discussed above. The global $K$-factors present values close to one for all three contributions.

The values for the integrable and measurable cross sections, introduced in the previous chapter, are shown in Table \ref{tab:oW_zjj_meas}. As mentioned before, they were computed at LO parton level, as PS and NLO corrections can introduce negative weights in the samples that would be impossible to distinguish from the interference ones. $\sigma^{|\text{meas}|}$ was computed by summing the interference squared amplitude of each event over all the permutations of initial- and final-state momenta and over all the possible helicity configurations; each squared amplitude was weighted by the PDF. The asymmetry related to $\Delta\phi_{jj}$ is also reported: it was obtained by summing the bin contents of the cross-section distribution for that observable, in absolute value. The comparison of the linear cross section with $\sigma^{|\text{int}|}$ highlights a large suppression, but most of the total interference effect can be accessed at the LHC, as the $\sigma^{|\text{meas}|}$ value suggests. Furthermore, $\Delta \phi_{jj}$ seems to be very useful for lifting the cancellation, reaching an efficiency of almost 90\% compared to the measurable maximum.

The LO differential distributions for this observable are shown in Fig.~\ref{fig:oW_zjj_dphijj_lo}, with the positive- and negative-weighted contributions to the linear term separately. It can be seen that these two have different trends, explaining the ability of this observable in reviving the interference. The NLO predictions are reported in Fig.~\ref{fig:oW_zjj_dphijj_nlo}, for the SM, linear and quadratic terms. The differential $K$-factors are stable and close to one for all these contributions, as it is suggested from the fact that the cancellation level \eqref{rwpm} is far from zero in every bin. In both plots, the binning is $[0,\pi/4,\pi/2,3\pi/4,7\pi/8,15\pi/16,\pi]$ and its symmetrical around zero.

%%%%%%%%%%%%%%%%%%%%%%%%% WZ %%%%%%%%%%%%%%%%%%%%%%%%%

\section{\label{sec:WZ_oW}Fully-leptonic $WZ$ production}
We also generated predictions for the $W^\pm Z$ production, with both EW bosons decaying leptonically: $W^\pm \rightarrow \ell^\pm \overset{\scriptscriptstyle(-)}{\nu_\ell}$ and $Z \rightarrow \ell^+ \ell^-$, where $\ell = e, \mu$ and the sign stands for the electric charge. A representative diagram is shown in Fig. \ref{fig:oW_diagrams}b. Our results for the SM, linear and quadratic terms involving $O_W$ were obtained at FO.

Building on the azimuthal-observables argument in Section \ref{sect:rev_strategies}, the authors of \cite{Barducci:2019,NovelTGC:2017} prove that the interference of the $O_W$ with the SM for this process is proportional to
\begin{equation}
   \phi_{WZ} = \cos (2\phi_W) +\cos (2\phi_Z) \label{phiWZ}
\end{equation}
at high energy $\sqrt{s} \gg M_V$. As a reminder, $\phi_V$ for $V = W,Z$ is introduced in Eq. \eqref{phiV_def} and represents the azimuthal angle between the plane containing the beam axis and the EW boson, and the plane where its decay products lie, in the lab frame. The direction of the second plane is defined as the vectorial product of the positive- and negative-helicity lepton momenta: as the $Z$ boson couples to left- and right-handed leptons with similar strengths, this introduces an ambivalence $\phi_Z \leftrightarrow \phi_Z -\pi$, that anyway does not affect the $\phi_{WZ}$ value, being that a function of $\cos (2\phi_Z)$. In the $W$-boson case, the lepton helicities are fixed by the left-handed nature of the interaction, but another ambiguity is raised by the impossibility to exactly reconstruct the longitudinal component of the neutrino momentum, as it will be described below. This partially affects the $2\phi_W$ modulations in $\phi_{WZ}$.

A previous experimental analysis \cite{Atlas:2019wz} suggests the transverse mass of the $WZ$ system as a variable to study the SMEFT effects for this process, as it increases with the CoM energy. This quantity is defined as
\begin{equation}
   M_T^{WZ} = \sqrt{ \left( \sum_\ell p_T^\ell + p_T^\text{miss} \right)^2 - \left( \sum_\ell \vec{p}_T^{\hspace{0.7mm}\ell} + \vec{p}_T^{\hspace{0.7mm}\text{miss}} \right)^2 }, \label{mTWZ}
\end{equation}
where the two sums run over the charged leptons. It is not affected by the ignorance over the $z$-component of the neutrino momentum.

Another SMEFT analysis \cite{Dawson:2019} tackles the new-physics effects on other angular observables. Among those, we focused on the cosine of the angle between the negatively-charged $Z$-lepton in the $Z$-boson rest frame, and the direction of flight of the boson seen in the $WZ$ CoM frame. This variable is referred to as $\cos \theta^*_{\ell^-_Z Z}$ and the coordinate system is explained in more detail in \cite{Dixon:2011}.

\subsection{\label{sec:WZ_oW_cuts}Phase-space cuts and reconstruction procedures}
The phase space in our analysis was defined as in the experimental analysis \cite{Atlas:2019wz}. Events had to contain exactly three leptons, that needed to be assigned to their $W$ or $Z$ mothers for their reconstruction. 

The matching is trivial if two leptons have the same flavour, different from the one of the third lepton. In case they are all from the same family, two of them would present the same electric charge and one of those would be the $W$-boson lepton $\ell_W$. For each of these two candidates, we estimated the longitudinal component of the neutrino momentum $p_\nu^z$ from $p_T^\text{miss}$ and by assuming the $W$ boson to be on-shell; up to two solutions can come from this requirement and the smaller one in absolute value was chosen \cite{Riva:2018, Barducci:2019, Rahaman:2020}. If no real solution could be found, we obtained one by discarding the imaginary part.
For every combination of three leptons and reconstructed neutrino, we computed
\begin{equation}
   P = \left| \frac{1}{ M_{\ell + \ell -}^2 -M_Z^2 +i \Gamma_Z M_Z } \right|^2 \cdot \left| \frac{1}{ M_{\ell ' \nu}^2 -M_W^2 +i \Gamma_W M_W }\right|^2
\end{equation}
and considered the case that maximised it. $M_V$ and $\Gamma_V$, for $V=W,Z$, are the masses and widths of the EW bosons as set in the UFO model\footnote{The standard values in the SMEFT@NLO UFO model are $M_W=79.8244$, $M_Z=91.1876$, $\Gamma_W=2.00295$ and $\Gamma_Z =2.416023$, all in GeV}.

For the two $\ell_Z$, we asked a minimum $p_T$ of 15 GeV, while for $\ell_W$ the minimum cut was at 20 GeV. The maximum $|\eta|$ required for all leptons was 2.5. The angular distance had to be $\Delta R > 0.2$ among the two $\ell_Z$ and $\Delta R > 0.3$ between the $\ell_W$ and $\ell_Z$.
The mass of the $Z$-boson decay products had to satisfy $81.2 < M_{\ell^+_Z \ell^-_Z} < 101.2$ GeV, while the $W$-boson transverse mass needed to be above 30 GeV. This last quantity is defined as
\begin{equation}
   M_T^W = \sqrt{ 2 p_T^\text{miss} p_T^{\ell_W} \hspace{1mm} (1-\cos \Delta \phi_{\ell_W, \text{miss}}) }, \label{mTw}
\end{equation}
with $\Delta \phi_{\ell_W, \text{miss}}$ the azimuthal distance between the lepton and the missing transverse energy.

To be able to compare against the $M_T^{WZ}$ experimental measurements, all the results for this process were averaged over four decay channels: $e^\pm \nu_e e^+ e^-$, $e^\pm \nu_e \mu^+ \mu^-$, $\mu^\pm \nu_\mu \mu^+ \mu^-$, $\mu^\pm \nu_\mu e^+ e^-$. Complete correlation was assumed among them when propagating the uncertainties.

\begin{table}
\centering
\caption{\footnotesize{As in Table \ref{tab:oW_zjj_meas}, but for {\it WZ} in fb. The cases in which the $Z$-leptons helicities are separated are also shown, together with the regions in \eqref{oW_wz_cuts}. The graphical representation through bar charts is shown at the bottom}} \label{tab:oW_wz_meas}
\begin{minipage}{.55\textwidth}
\begin{tabular}{c|ccc}
\multicolumn{4}{c}{$p p \rightarrow \ell^\pm \overset{\scriptscriptstyle(-)}{\nu} \ell^+ \ell^-$, $\ell=(e,\mu)$} \\ %\hline
 & (fb) & \% of $\sigma^{|\text{int}|}$ & \% of $\sigma^{|\text{meas}|}$ \\ \hline
$\sigma^{|\text{int}|}$ & 4.93$\pm$0.4\% & 100 & - \\
$\sigma^{|\text{meas}|}$ & 2.04$\pm$1.0\% & 41 & 100\\
$p_T^Z$ \textit{vs} $\phi_{WZ}$ & 1.31$\pm$1.5\% & 27 & 64 \\
$\phi_{WZ}$ & 0.79$\pm$3\% & 16 & 39 \\
$M_T^{WZ}$ & 0.66$\pm$3\% & 13 & 32 \\
$\cos \theta^*_{\ell_Z^- Z}$ & 0.20$\pm$10\% & 4 & 10 \\
$\sigma^{1/\Lambda^2}_{LO}$ & 0.20$\pm$10\% & 4 & 10 \\
\hline\hline
\multicolumn{4}{c}{$h (\ell_Z^-)=-1, h (\ell_Z^+) =+1$} \\ \hline
$\sigma^{|\text{int}|}$ & 2.773$\pm$0.5\% & 100 & - \\
$\sigma^{|\text{meas}|}$ & 1.738$\pm$0.9\% & 63 & 100 \\
$M_T^{WZ}$ & 0.38$\pm$4\% & 14 & 21 \\
$\sigma^{1/\Lambda^2}_{LO}$ & 0.108$\pm$14\% & 4 & 6 \\
\hline
\multicolumn{4}{c}{$h (\ell_Z^-)=+1, h (\ell_Z^+) =-1$} \\ \hline
$\sigma^{|\text{int}|}$ & 2.135$\pm$0.6\% & 100 & - \\
$\sigma^{|\text{meas}|}$ & 1.067$\pm$1.1\% & 50 & 100 \\
$M_T^{WZ}$ & 0.289$\pm$4\% & 14 & 27 \\
$\sigma^{1/\Lambda^2}_{LO}$ & 0.087$\pm$14\% & 4 & 8 \\
\hline\hline
\multicolumn{4}{c}{$p_T^Z> 50$ GeV AND $\phi_{WZ}> -0.5$} \\
\hline
$\sigma^{|\text{int}|}$ & 2.260$\pm$0.7\% & 100 & - \\
$\sigma^{|\text{meas}|}$ & 0.873$\pm$1.7\% & 39 & 100 \\
$M_T^{WZ}$ & 0.660$\pm$2\% & 29 & 76 \\
$\sigma^{1/\Lambda^2}_{LO}$ & 0.660$\pm$2\% & 29 & 76 \\
\hline
\multicolumn{4}{c}{$p_T^Z< 40$ GeV OR $\phi_{WZ}< -1$} \\
\hline
$\sigma^{|\text{int}|}$ & 1.810$\pm$0.5\% & 100 & - \\
$\sigma^{|\text{meas}|}$ & 0.870$\pm$1.1\% & 48 & 100 \\
$M_T^{WZ}$ & 0.480$\pm$2\% & 27 & 55 \\
$\sigma^{1/\Lambda^2}_{LO}$ & -0.480$\pm$2\% & 27 & 55 \\
\hline
\end{tabular}
\end{minipage}
\begin{minipage}{.7\textwidth}
   \vspace{.1cm}
   \includegraphics[width=\textwidth]{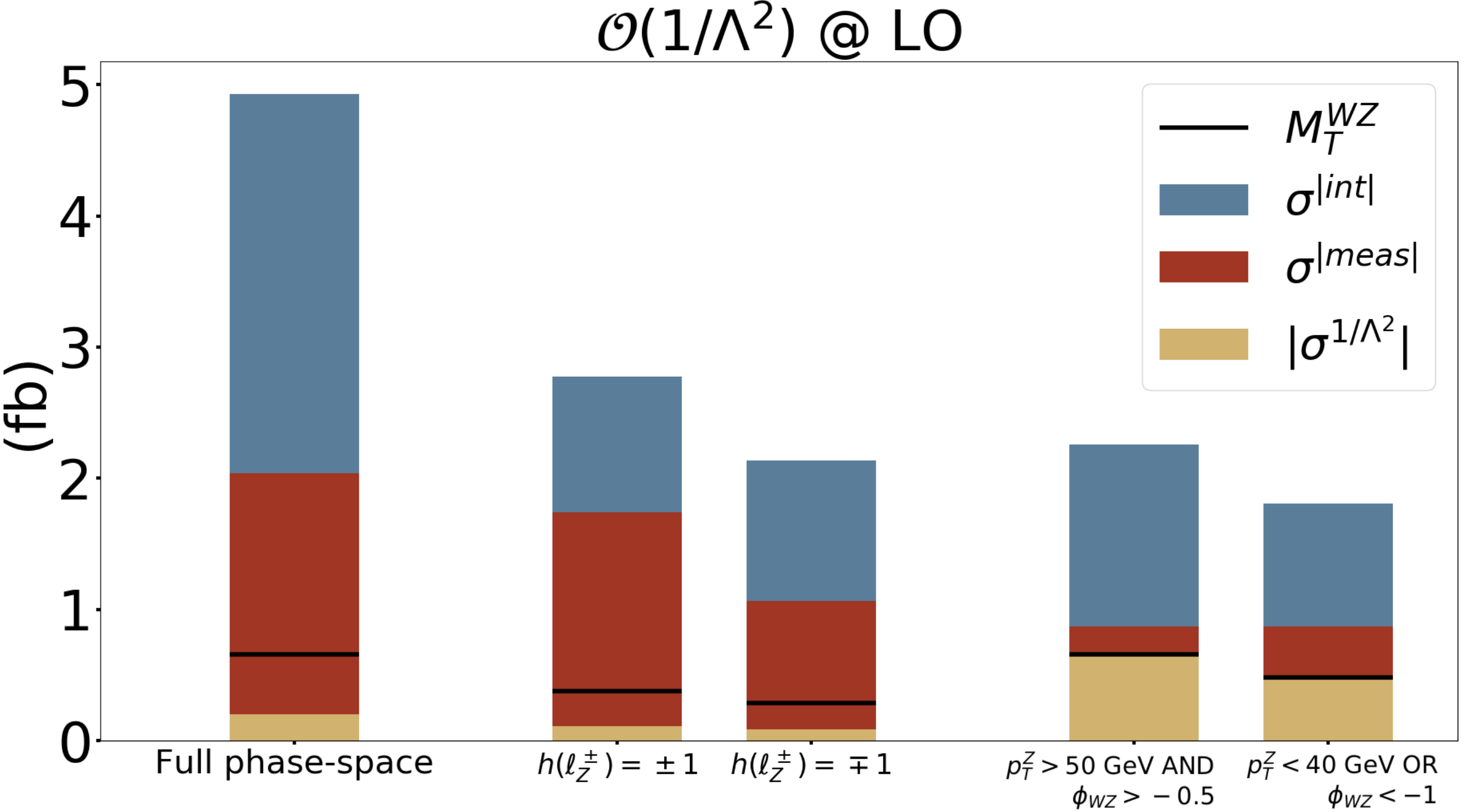}
\end{minipage}
\end{table}

\subsection{Results and distributions}
The total LO and NLO cross sections for the SM, interference and quadratic contributions are summarised in Table \ref{tab:oW_xsects}, with the relative numerical and scale-variation uncertainties and the global $K$-factors. The latter is of order 1.5 for the SM, as a new production channel with a gluon in the initial state opens at NLO. The linear $K$-factor is large and negative, with relative uncertainties that are twice as big as the SM and quadratic ones.
The N${}^2$LO SM cross section is also shown: it was obtained at FO through {\sc Matrix} \cite{Grazzini:2017mhc,Gehrmann:2015ora,Denner:2016kdg,Cascioli:2011va,Buccioni:2019sur,Buccioni:2017yxi,Catani:2012qa,Catani:2007vq}. The predictions are compatible with \cite{Grazzini:2017ckn,Grazzini:2016swo} and the results from ATLAS in the experimental analysis.

The integrable and measurable cross sections are reported in Table \ref{tab:oW_wz_meas}, and their comparison with the total linear cross section highlights a large cancellation for it. $\sigma^{|\text{meas}|}$ was computed by summing the interference squared amplitude of each event over all the permutations of initial- and final-state momenta, over all the possible helicity configurations and integrating over the longitudinal component of the neutrino momentum; each term was multiplied by its PDF factor. If the same calculation is performed without integration, but using the $p_\nu^z$ value obtained from the reconstruction procedure described in the previous section, the measurable-cross section result is almost 90\% of the one reported in the table: indeed, the on-shell assumption for the $W$ boson yields solutions that usually maximise the squared amplitude and, thus, dominate the integration.
The $\sigma^{|\text{meas}|}$ value totals to less than half of the $\sigma^{|\text{int}|}$ one: this is mostly due to the neutrino reconstruction, that is characteristic of  colliders like the LHC. It is not, though, the only source of suppression: if the measurable cross section is computed assuming that the $Z$-lepton helicities can be distinguished, its ratio with respect to the integrable one increases. In particular, the events where the positively- and negatively-charged leptons $\ell_Z^\pm$ have helicities $\pm 1$ and $\mp 1$ contribute with opposite sign to the linear cross section. The values for the integrable and measurable cross sections can be found in Table \ref{tab:oW_wz_meas} for these two cases.

\begin{figure}
   \centering
   \caption{\footnotesize{$WZ$ interference cross section per bin at LO without PS, as a function of $p_T^Z$ and $\phi_{WZ}$, in the two cases in which the {\it Z}-leptons $\ell_Z^\pm$ have helicities $\pm 1$ and $\mp 1$. Red (blue) areas mark where the cross section is positive (negative), as the positive- (negative-) weighted contribution dominates there. The black dashed lines separate the phase-space areas in \eqref{oW_wz_cuts}}} \label{fig:oW_wz_pTz_phiWZ_hel}
   \includegraphics[width=.9\textwidth]{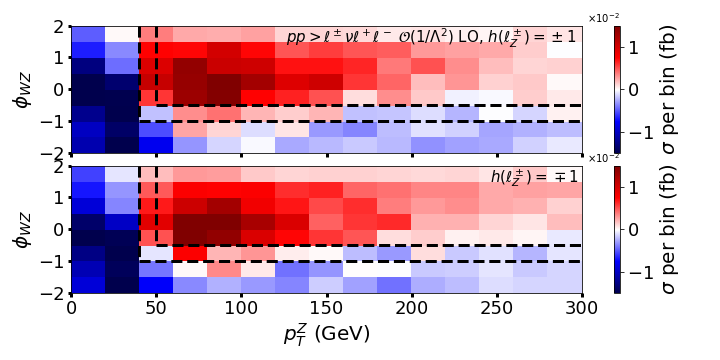}
\end{figure}

The lepton helicities cannot be easily measured at experiments, but we noticed that the double-differential distribution of the reconstructed $Z$-boson $p_T$ and $\phi_{WZ}$ show similar behaviours when the two helicity configurations are separated. Not many other observables we investigated presented the same property, and this one is able to restore more than 60\% of the measurable linear effect, as it can be seen in Table \ref{tab:oW_wz_meas}. In particular, the regions where the interference is mostly positive or negative for this variable are roughly the same in the two cases and are respectively delimited by the cuts
\begin{subequations} \label{oW_wz_cuts}
   \begin{gather}
      p_T^Z> 50 \text{ GeV AND } \phi_{WZ}> -0.5, \label{oW_wz_cuts_a} \\
      p_T^Z< 40 \text{ GeV OR } \phi_{WZ}< -1. \label{oW_wz_cuts_b}
   \end{gather}
\end{subequations}
In the phase-space strip between these two areas, the linear term changes sign, meaning that its cross section is small and unstable. The double distributions for the two helicity cases at parton level are shown in Fig. \ref{fig:oW_wz_pTz_phiWZ_hel}. The integrable and measurable cross sections computed over the two regions are shown in Table \ref{tab:oW_wz_meas}: even if more than half of the full interference effect cannot yet be accessed because of the neutrino longitudinal component, the $\ell_Z^\pm$ helicities and the initial-state flavours, the two total interference cross sections are now restored to larger fractions of their respective $\sigma^{|\text{meas}|}$ compared to the inclusive case. Indeed they have opposite signs, as most of the positive and negative weights are separated by the cuts on $\phi_{WZ}$ and $p_T^Z$. Similarly, the asymmetries for variables like $M_T^{WZ}$ are higher when the cuts are applied.

As it can be seen in Fig. \ref{fig:oW_wz_pTz_phiWZ_hel}, the way we delimited the two regions in \eqref{oW_wz_cuts} could be improved with more sophisticated shapes that better follow the separation between them. We checked more complicated options, like the Morse potential for real diatomic molecules, but we then had to deal with a larger number of free parameters for the curves, whose values can change quickly with the final-state cuts. For this reason, even if more refined choices could help revive the total interference cross section by a few percent more with respect to the measurable one, we decided to keep the analysis simple and adopt the squared shapes \eqref{oW_wz_cuts}.

\begin{figure}
   \centering
   \caption{\footnotesize{LO and NLO differential cross-section distributions for $\phi_{WZ}$ in $WZ$ production, over all the phase space (\textit{top}) and when cuts on $p_T^Z$ and $\phi_{WZ}$ are applied (\textit{centre} and \textit{bottom}). The black (orange, green) line represents the SM divided by 50 (interference, quadratic correction divided by 4). The {\it K}-factors are also shown, together with their numerical and scale uncertainties. For each case, the relative cancellation for LO interference is plotted. Note the different variable range in the central plot, due to the cuts}} \label{fig:oW_wz_phiWZ}
      \includegraphics[width=.6\textwidth]{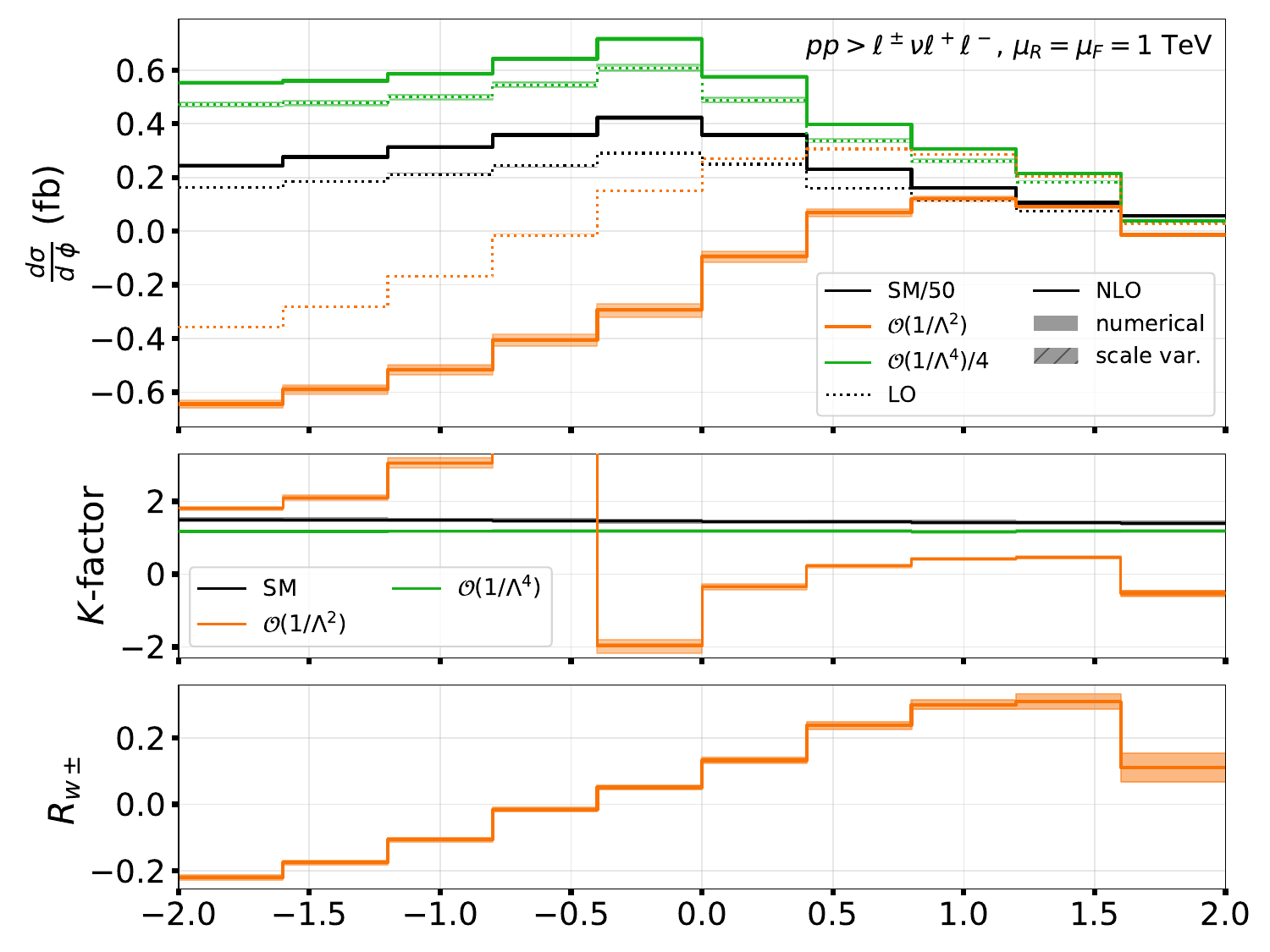}
      \includegraphics[width=.6\textwidth]{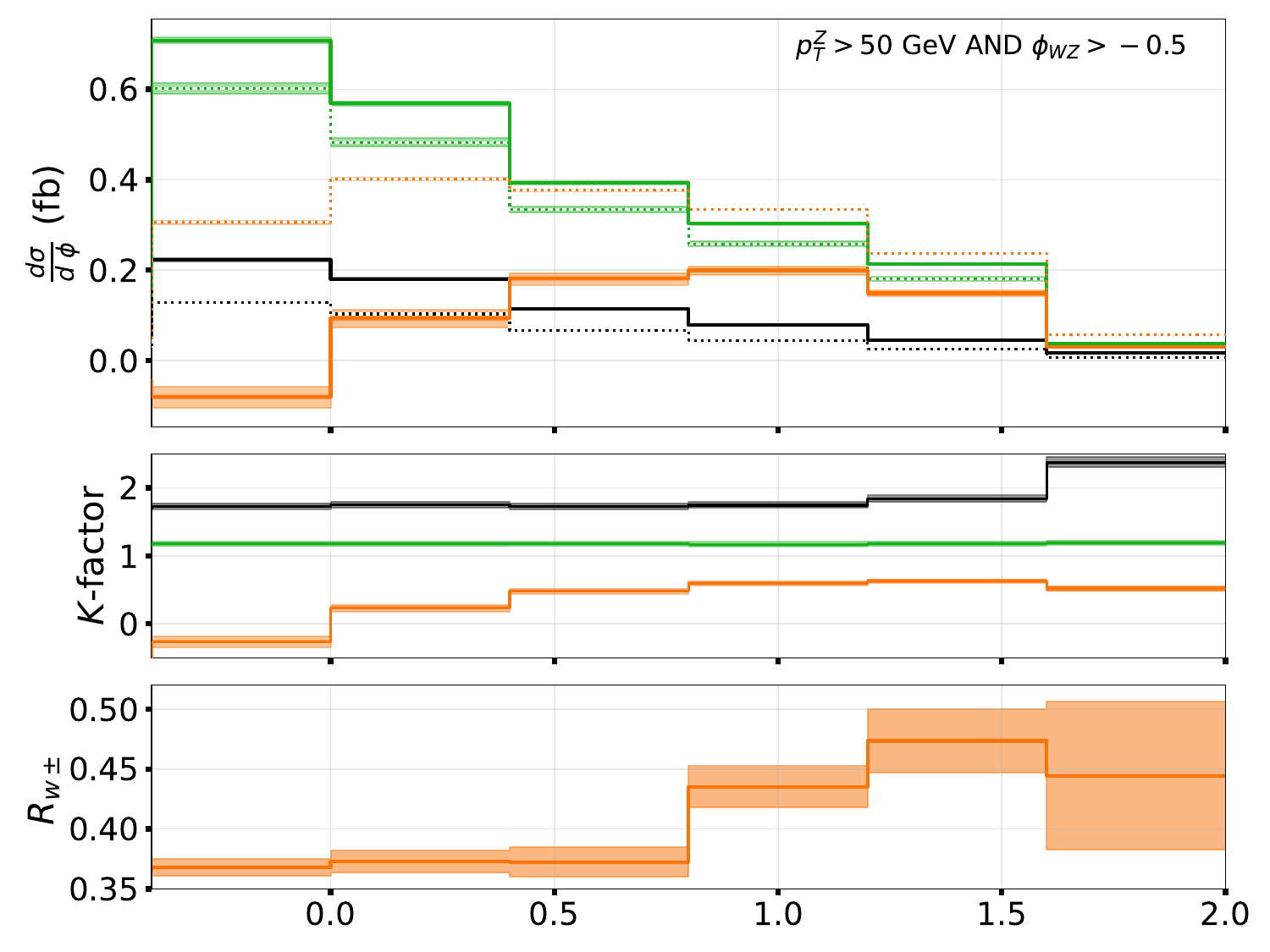}
      \includegraphics[width=.6\textwidth]{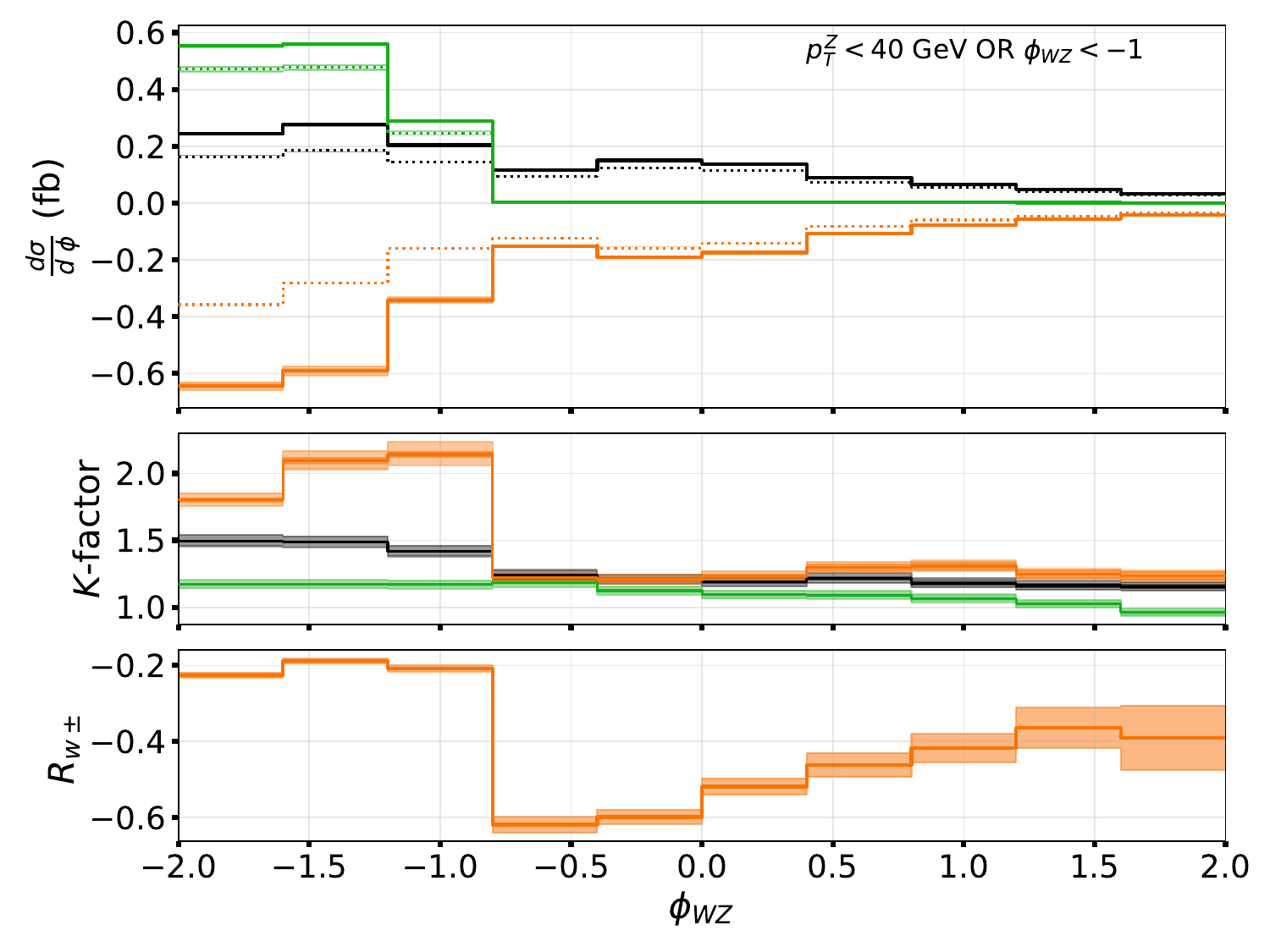}
\end{figure}

\begin{figure}
   \centering
   \caption{\footnotesize{Same as Fig. \ref{fig:oW_wz_phiWZ} but for the transverse mass of the $WZ$ system. In the top plot, the N${}^2$LO results at FO for the SM and the experimental data are also shown. The last bin contains the overflow}} \label{fig:oW_wz_mTwz}
   	  \includegraphics[width=.6\textwidth]{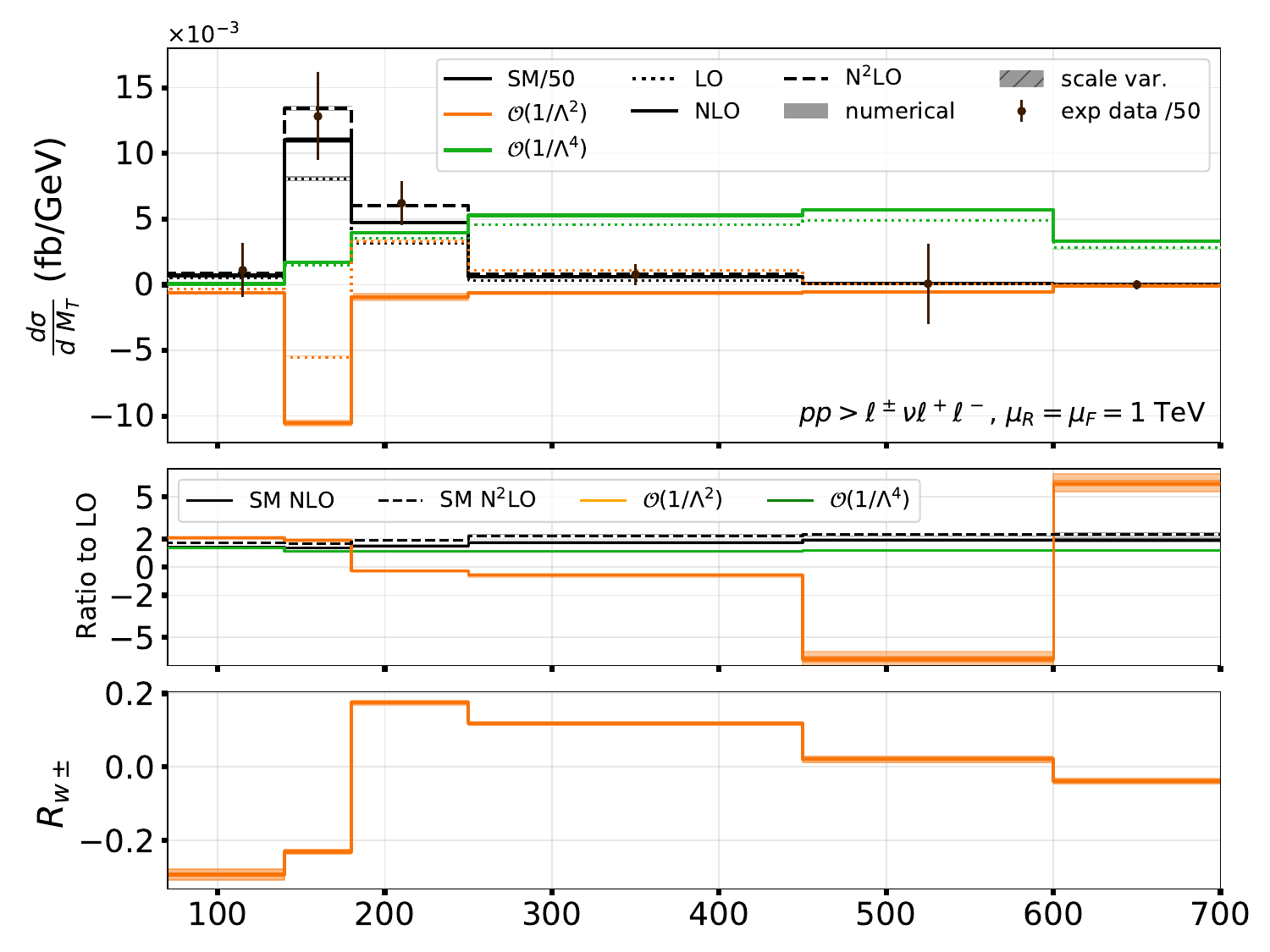}
   	  \includegraphics[width=.6\textwidth]{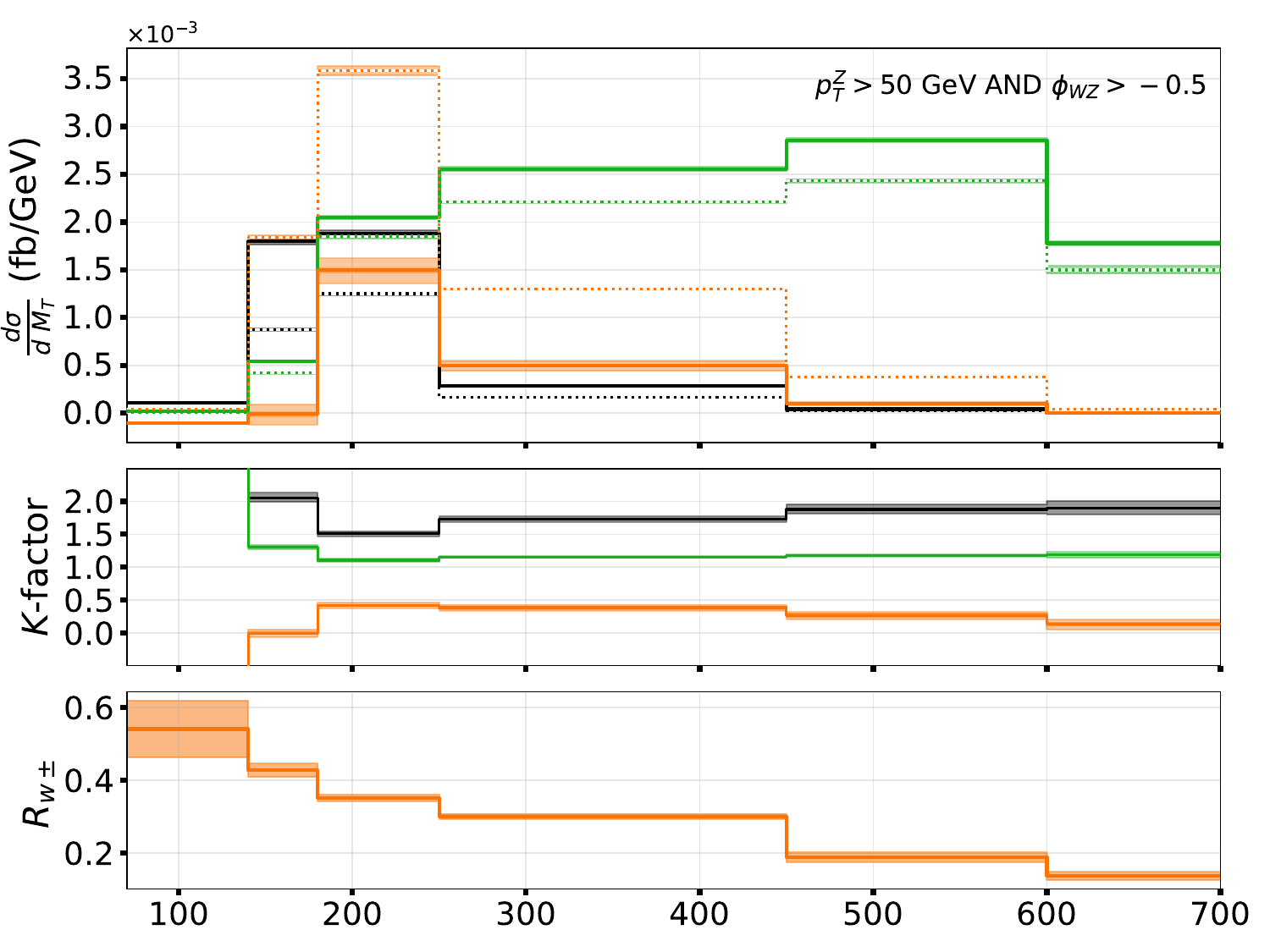}
      \includegraphics[width=.6\textwidth]{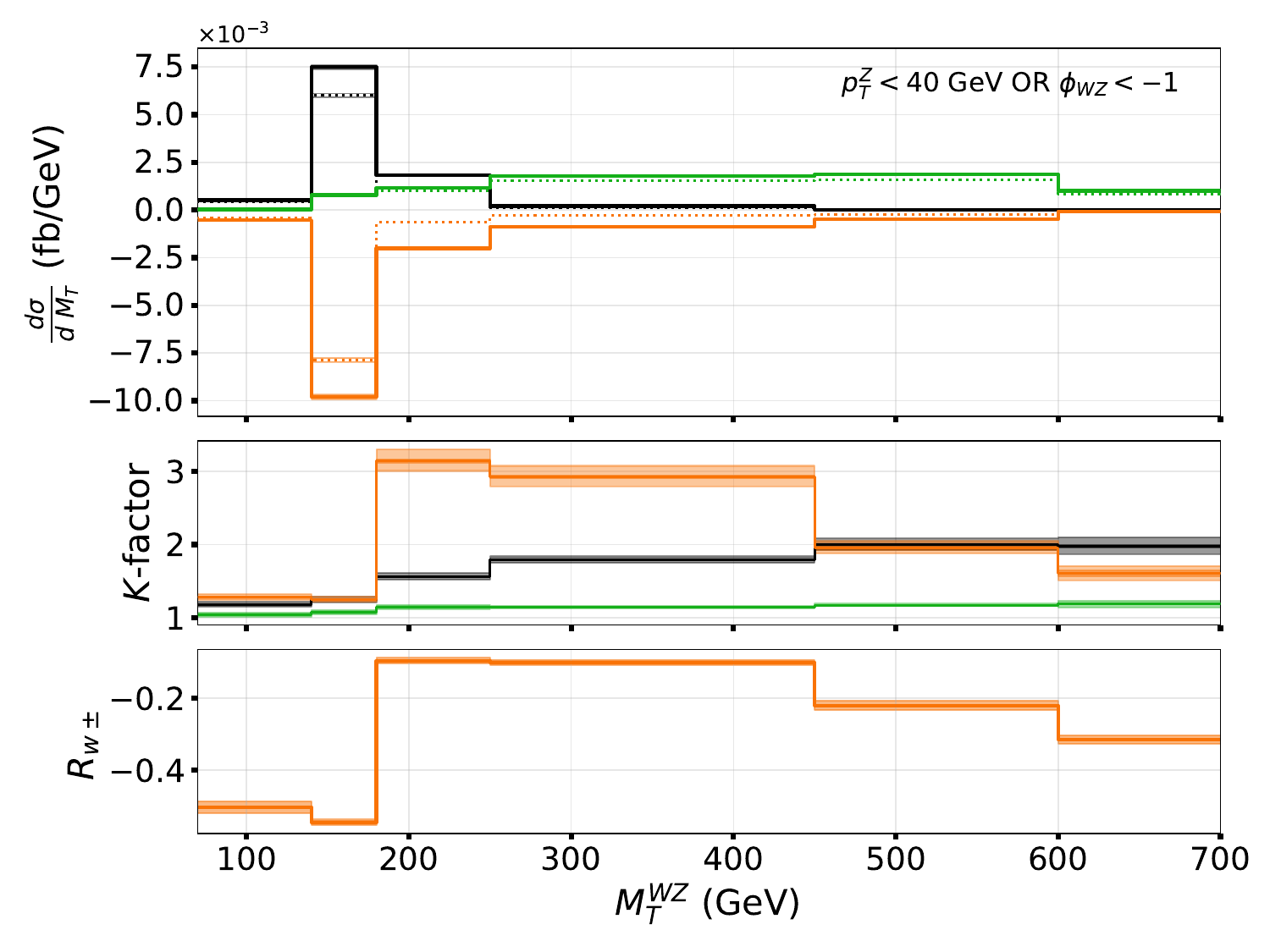}
\end{figure}

\begin{figure}
   \centering
   \caption{\footnotesize{Same as Fig. \ref{fig:oW_wz_phiWZ} but for the the angle between the $\ell_Z^-$ momentum in the $Z$-boson rest frame, and the direction of flight of the boson seen in the $WZ$ CoM frame}} \label{fig:oW_wz_costhZ}
      \includegraphics[width=.6\textwidth]{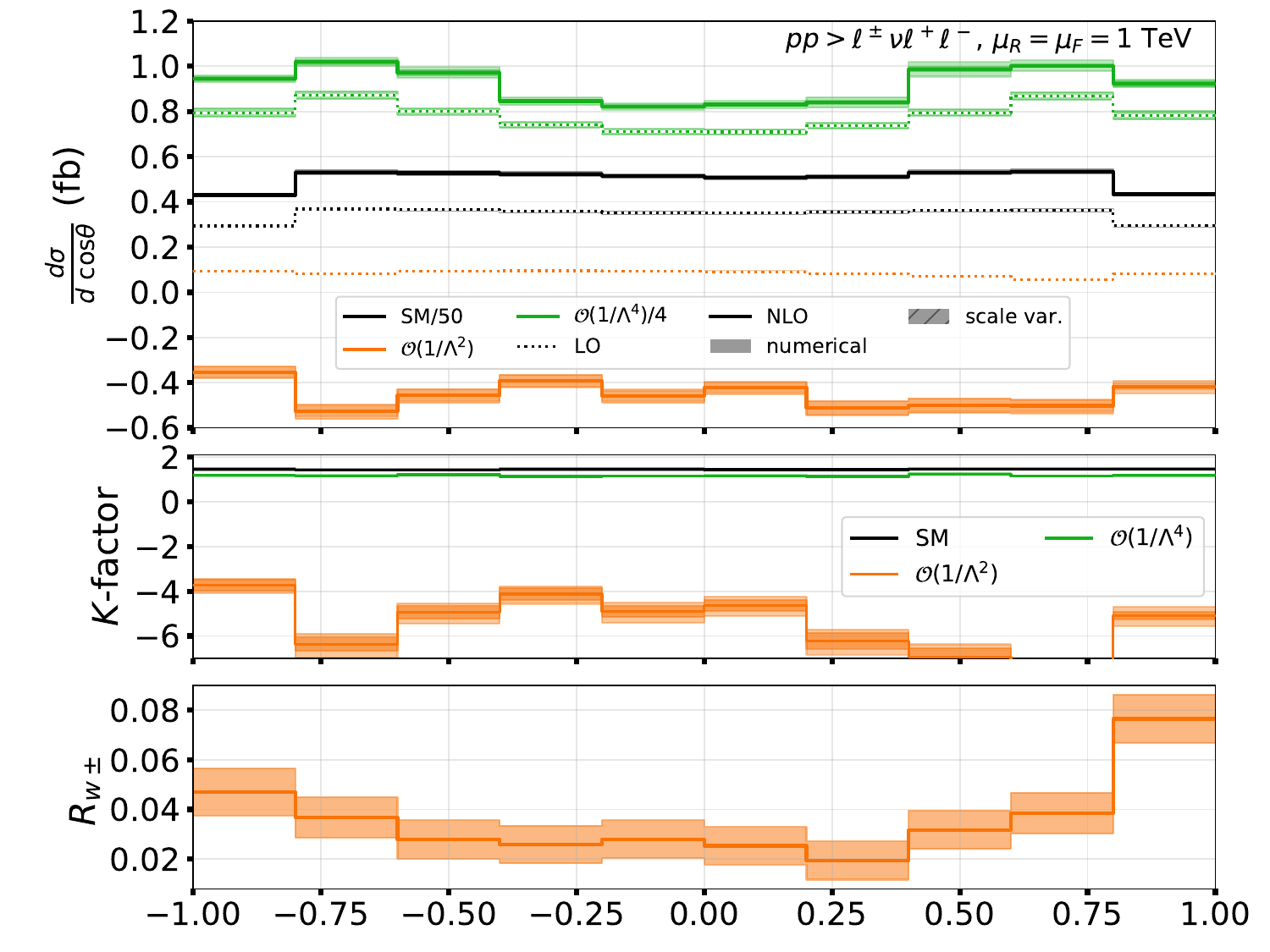}
      \includegraphics[width=.6\textwidth]{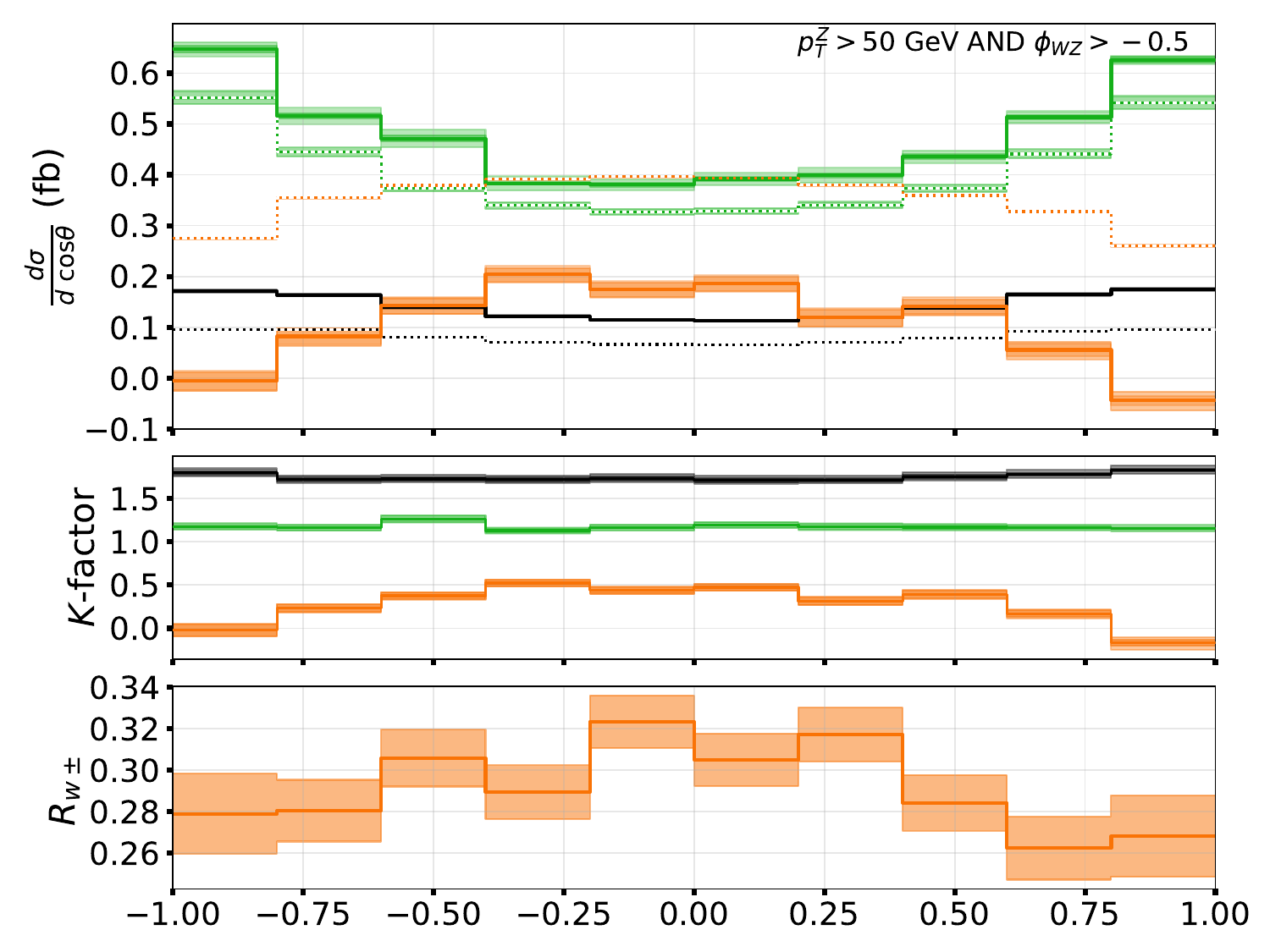}
      \includegraphics[width=.6\textwidth]{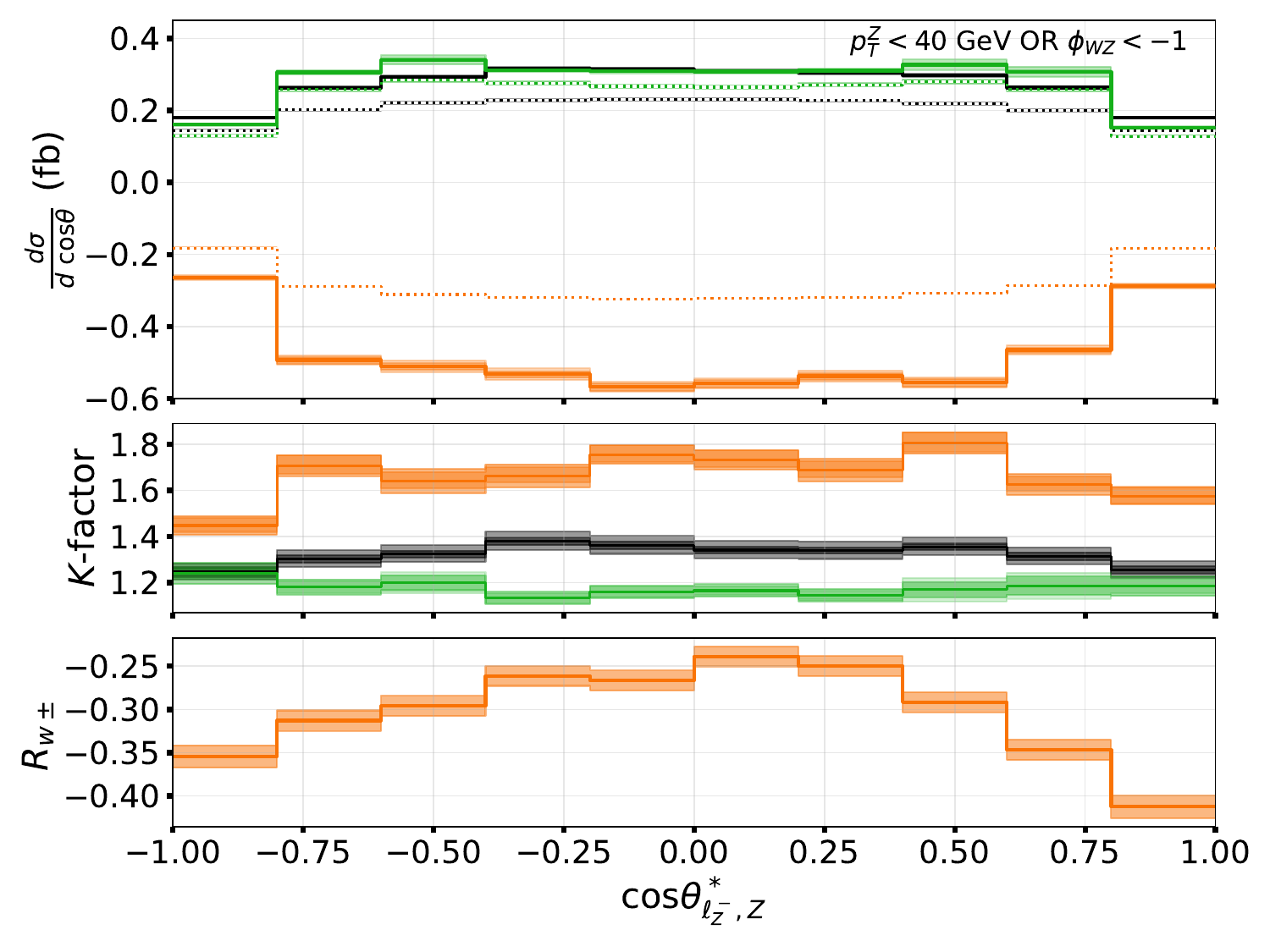}
\end{figure}

We plotted the predictions for the SM, linear and quadratic contributions to the relevant variables in this analysis, at LO and NLO, over the total phase space and in the two regions delimited by the cuts \eqref{oW_wz_cuts}: they are shown in Fig. \ref{fig:oW_wz_phiWZ} for $\phi_{WZ}$, in Fig. \ref{fig:oW_wz_mTwz} for $M_T^{WZ}$ and in Fig. \ref{fig:oW_wz_costhZ} for $\cos \theta^*_{\ell^-_Z Z}$. For all three, the differential $K$-factors are reported, together with the cancellation level $R_{w \pm}$ for the LO interference. Even if the SM and quadratic distributions present stable and reasonable $K$-factors in the inclusive case, the linear ones jump from positive to negative values or become large when the cancellation in a certain bin approaches or crosses zero. If the distributions are instead obtained separately over the two phase-space regions \eqref{oW_wz_cuts}, $R_{w \pm}$ is farther from zero (positive for the first cut and negative for the second) and the $K$-factor values are usually positive and more flat. They are still negative or large in some bins and, as discussed above, more tailored separations in $p_T^Z$ and $\phi_{WZ}$ could help improve the predictions in those as well, but these results already show how searches based on $\sigma^{|\text{meas}|}$ can lead to regions where the suppression is reduced. The first bin of the $M_T^{WZ}$ distribution in region \eqref{oW_wz_cuts_a} confirms that the cancellation level is not always enough to identify the bins with more suppression: both $R_{w\pm}$ and the $K$-factors are large because the cross section in the bin is small, even if no big cancellation occurs.

%%%%%%%%%%%%%%%%%%%%%%%%% Wa %%%%%%%%%%%%%%%%%%%%%%%%%

\section{\label{sec:Wa_oW}Leptonic $W\gamma$ production}
The last process we studied in this analysis is $W^\pm \gamma$ production, with the leptonic decay $W^\pm \rightarrow \ell^\pm \overset{\scriptscriptstyle(-)}{\nu_\ell}$. A representative diagram is shown in Fig. \ref{fig:oW_diagrams}c. To be able to compare against the experimental predictions, we considered all the three lepton families ($\ell = e, \mu, \tau$), with the $\tau$ decayed through \pyth .

Driven again by the azimuthal-observables argument in Section \ref{sect:rev_strategies}, we focused on the $\phi_W$ angle between the plane with the $W$ boson and the beam axis, and the plane where its decay products lie, in the lab frame. Its definition can again be found in Eq. \eqref{phiV_def}. As in $WZ$ production, the approximate reconstruction of the neutrino longitudinal component partially washes away the interference effects for this variable.

In the CMS analysis \cite{cms:2021wa}, they introduce a reference frame with the $\hat{z}$-axis along the $W^\pm$ flight direction in the $W\gamma$ CoM frame and $\hat{y} = \hat{r} \times \hat{z}$, where $\hat{r}$ denotes the Lorentz-boost direction from the lab frame. In this coordinate system, the angle $\phi$ is defined as the azimuthal angle of the lepton with its momentum in the $W$-rest frame. Due to the ambiguity raised by the neutrino reconstruction, the angle $\phi_f$ is used instead in the experimental analysis, defined as
\begin{equation}
   \phi_f = 
   \begin{cases}
      -(\pi+\phi), & \text{if}\ \phi<-\frac{\pi}{2} \\
      \phi, & \text{if}\ |\phi|<\frac{\pi}{2} \\
      \pi-\phi, &\text{if}\ \phi>\frac{\pi}{2}.
    \end{cases}
\end{equation}
In the reference, this variable is combined in a double-differential distribution with the photon $p_T$ to obtain bounds on the Wilson coefficient.

\subsection{Phase-space cuts}
We followed the prescriptions of the EFT section in the experimental study \cite{cms:2021wa}, that requires exactly one lepton with $p_T^\ell > 80$ GeV and $|\eta^\ell|< 2.5$. The missing transverse energy has to satisfy $p_T^\text{miss}> 40$ GeV. At least one photon needs to present $p_T^\gamma$ above 150 GeV and $|\eta^\gamma| < 2.5$. The angular distance $\Delta R_{\ell \gamma}$ between the lepton and the photon has to exceed 0.7. No jets with $p_T^j > 30$ GeV and $|\eta^j|< 2.5$ can be present in the final state. 

Because of this last cut and the $\tau$ decay, we generated our predictions with the matching to PS via \pyth . The neutrino is reconstructed as in Sect. \ref{sec:WZ_oW_cuts} for $WZ$ production.

\begin{table}
   \centering
   \caption{\footnotesize{As in Table \ref{tab:oW_zjj_meas}, but for $W\gamma$ in fb. The graphical representation with bar charts is shown on the right}} \label{tab:oW_wa_meas}
   \begin{minipage}{.55\textwidth}
   \begin{tabular}{c|ccc}
   \multicolumn{4}{c}{$p p \rightarrow \ell^\pm \overset{\scriptscriptstyle(-)}{\nu} \gamma$, $\ell=(e,\mu,\tau)$} \\
   %\hline
    & (fb) & \% of $\sigma^{|\text{int}|}$ & \% of $\sigma^{|\text{meas}|}$ \\ \hline
   $\sigma^{|\text{int}|}$ & 31.44$\pm$0.3\% & 100 & - \\
   $\sigma^{|\text{meas}|}$ & 12.50$\pm$0.9\% & 40 & 100 \\
   $\phi_W$ & 9.90$\pm$1.1\% & 31 & 79 \\
   \scriptsize{$p_T^\gamma \times |\phi_W|$} & 9.90$\pm$1.1\% & 31 & 79 \\
   \scriptsize{$p_T^\gamma \times |\phi_f|$} & 1.44$\pm$7\% & 5 & 12 \\
   $\sigma^{1/\Lambda^2}_{LO}$ & -1.44$\pm$7\% & 5 & 12 \\
   \hline
   \end{tabular}
   \end{minipage}
   \begin{minipage}{.44\textwidth}
      \includegraphics[width=\textwidth]{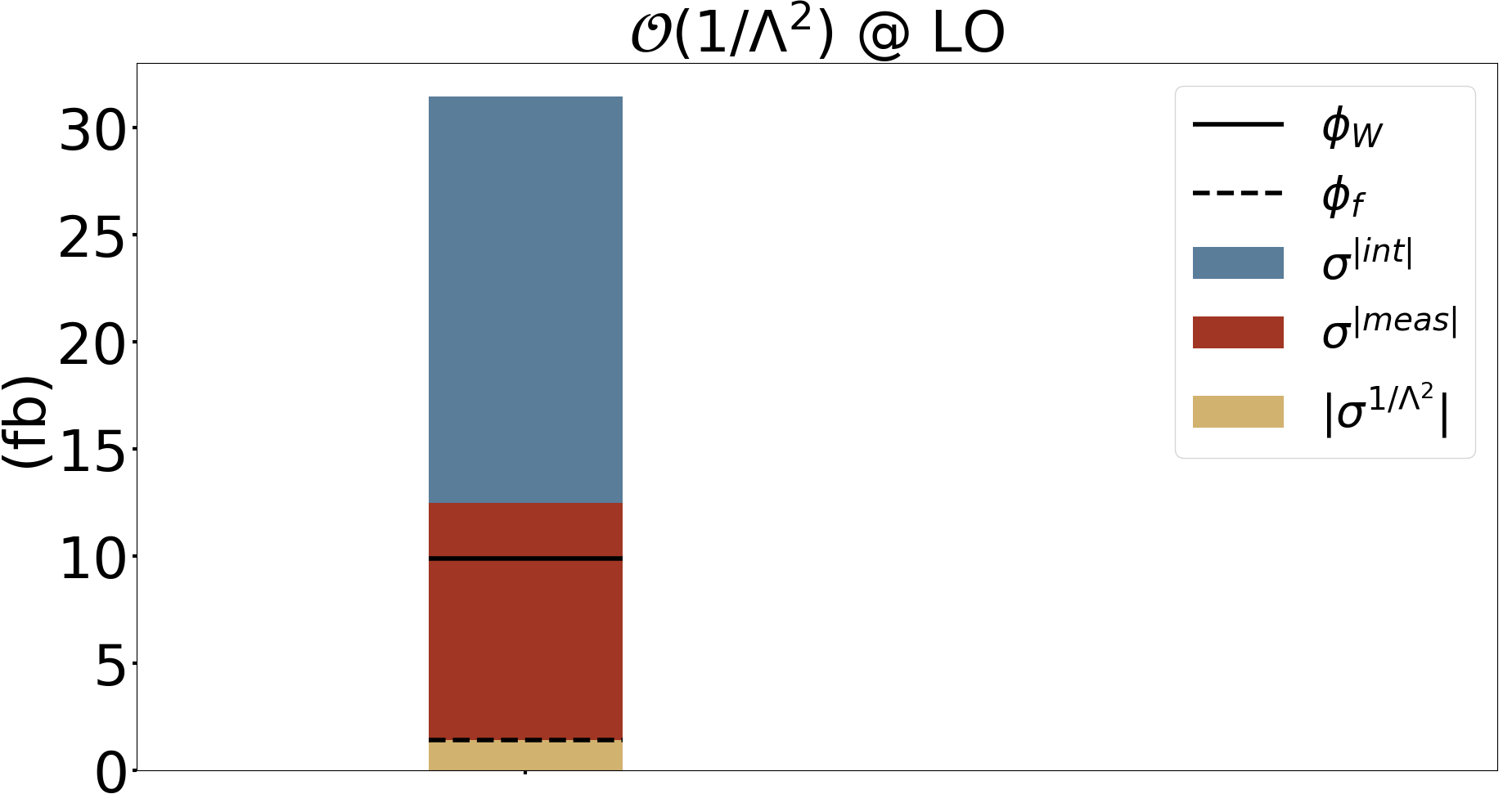}
   \end{minipage}
\end{table}

\subsection{Results and distributions}
The total LO and NLO cross sections for the SM and the linear and quadratic terms are shown in Table \ref{tab:oW_xsects}. As in the $WZ$ case, the global $K$-factor for the SM is larger than one due to the opening of a new channel at NLO, while the linear one suggests that the NLO corrections to the $O_W$ interference are not under control.

The integrable and measurable cross sections are summarised in Table \ref{tab:oW_wa_meas}: their comparison against the total interference cross section highlights a large suppression for the latter. $\sigma^{|\text{meas}|}$ was computed as in $WZ$ production, by summing the interference squared amplitude of each event over all the permutations of initial- and final-state momenta, over all the possible helicity configurations and integrating over the longitudinal component of the neutrino momentum; all the amplitudes were multiplied by the PDF factor. The large discrepancy between the linear cross sections before and after PS, respectively in Tables \ref{tab:oW_wa_meas} and \ref{tab:oW_xsects}, is due to the requirement of no jets in the final state. As in $WZ$ production, $\sigma^{|\text{meas}|}$ is less than half than $\sigma^{|\text{int}|}$ because of the neutrino presence. $\phi_W$ can nonetheless restore almost 80\% of the measurable cross section, analogously to $\Delta \phi_{jj}$ in $Zjj$ production.

\begin{figure}
   \centering
   \caption{\footnotesize{Same as Fig. \ref{fig:oW_zjj_dphijj_nlo}, but for $\phi_W$ in $W\gamma$ production}} \label{fig:oW_wa_phiW}
   	  \includegraphics[width=.7\textwidth]{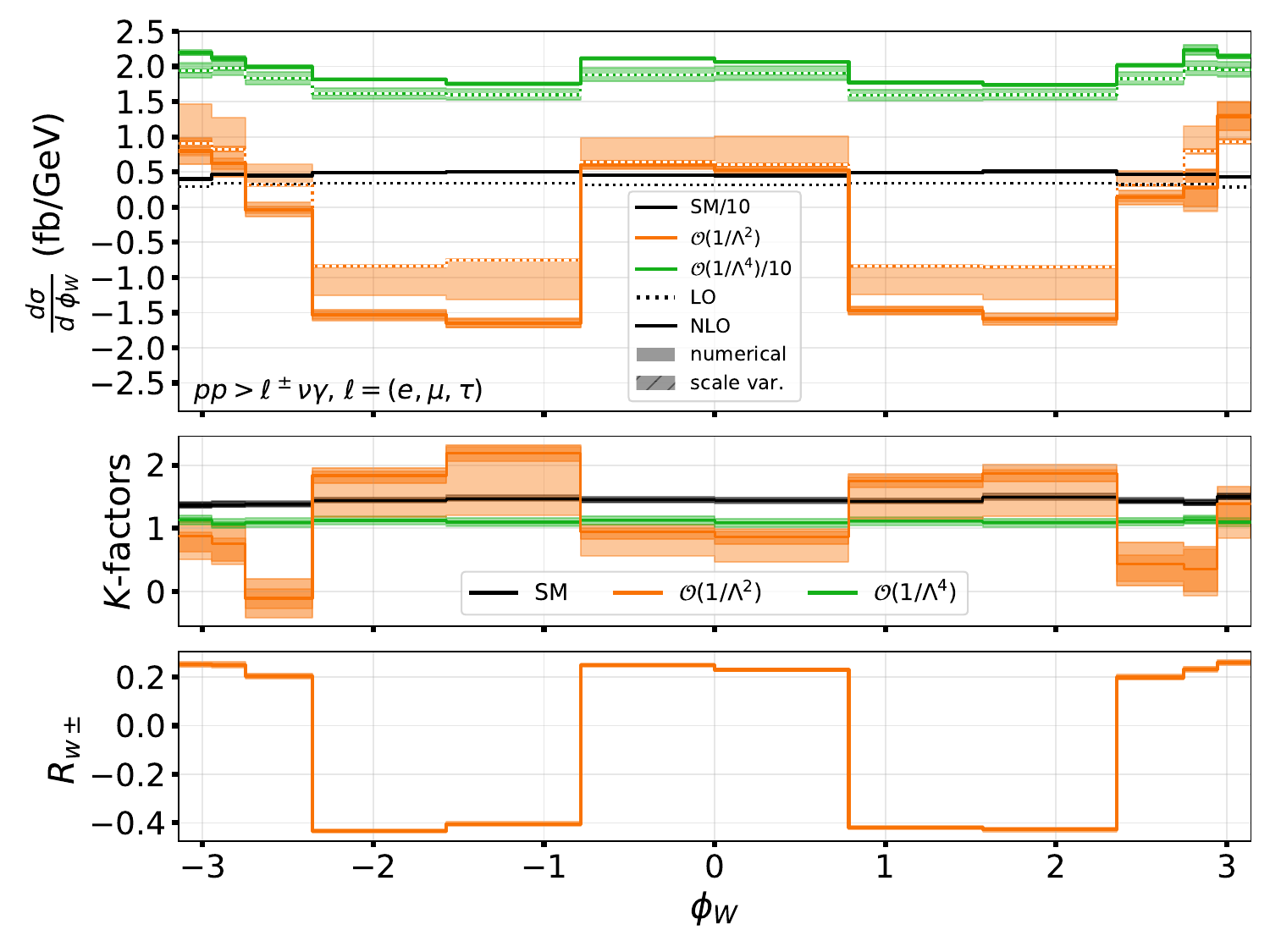}
\end{figure}

The LO and NLO differential distributions for this variable are shown in Fig. \ref{fig:oW_wa_phiW} for the SM, interference and quadratic terms. The bins are [0, $\pi/4$, $\pi/2$, $3\pi/4$, $7\pi/8$, $15\pi/16$, $\pi$] and their symmetrical around zero. Even if the differential cancellation level crosses zero in its trend, it is far enough from it to ensure reasonable $K$-factors for the interference in most of the bins (values of order $\sim 2$ can be considered normal for this process, due to the new channel opening at NLO). Some external bins present small and even negative $K$-factors, despite the $R_{w\pm}$ values: this quantity does not, indeed, contain any information about the cross-section magnitude, that is quite low in those bins.

\begin{figure}
   \centering
   \caption{\footnotesize{LO and NLO differential cross-section distributions for the photon transverse momentum in $W\gamma$ production, over different $|\phi_f|$ ranges ({\it left}) and $|\phi_W|$ ranges ({\it right}). The black (orange, green) line represents the SM, divided by 10 (interference, quadratic correction divided by 10). The experimental data is also shown on the left. The {\it K}-factors are reported, together with their statistical and scale uncertainties. For each case, the relative cancellation for the LO interference is plotted}} \label{fig:oW_wa_pTaPhi}
      \includegraphics[width=.49\textwidth]{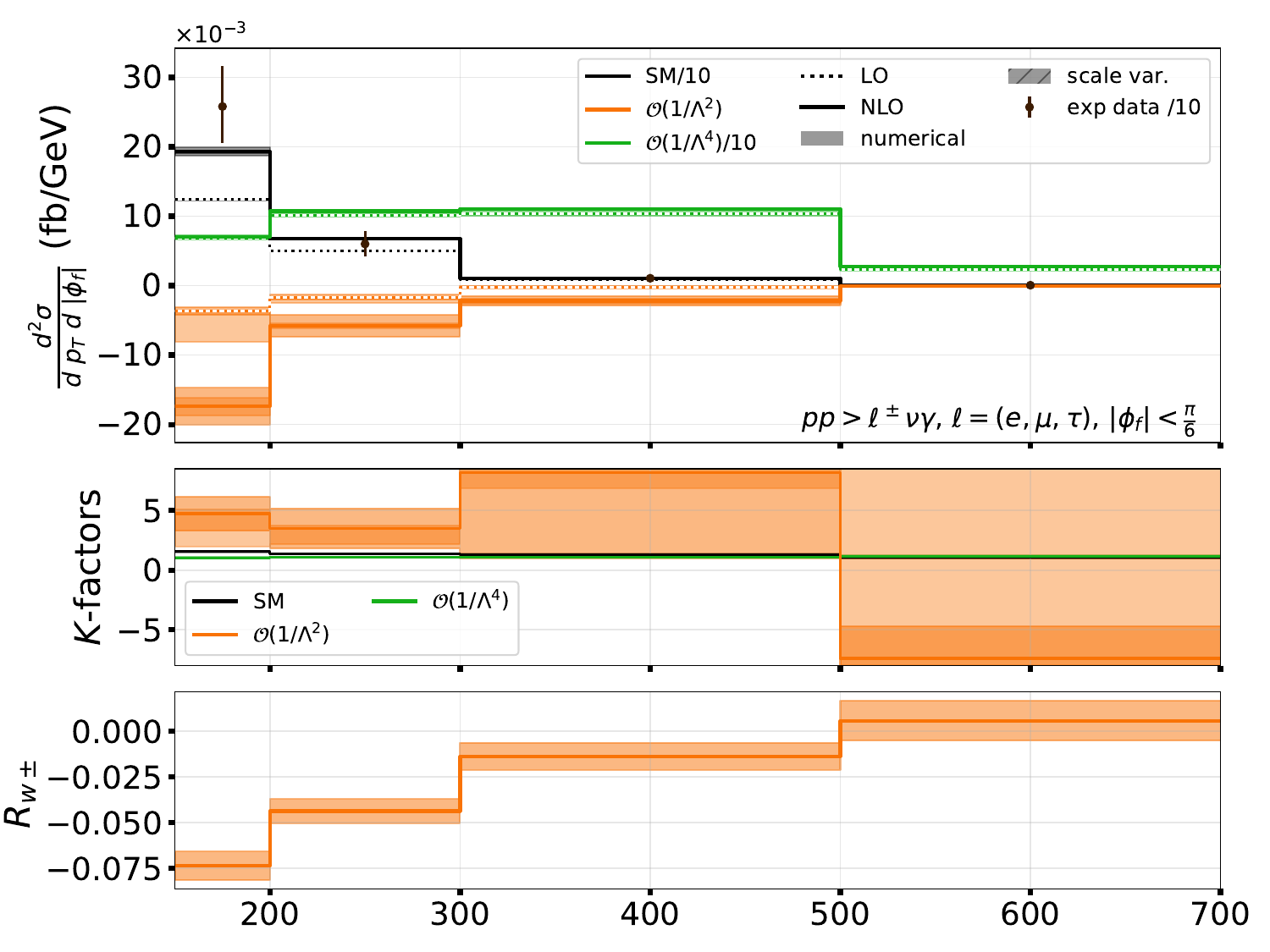}
      \includegraphics[width=.49\textwidth]{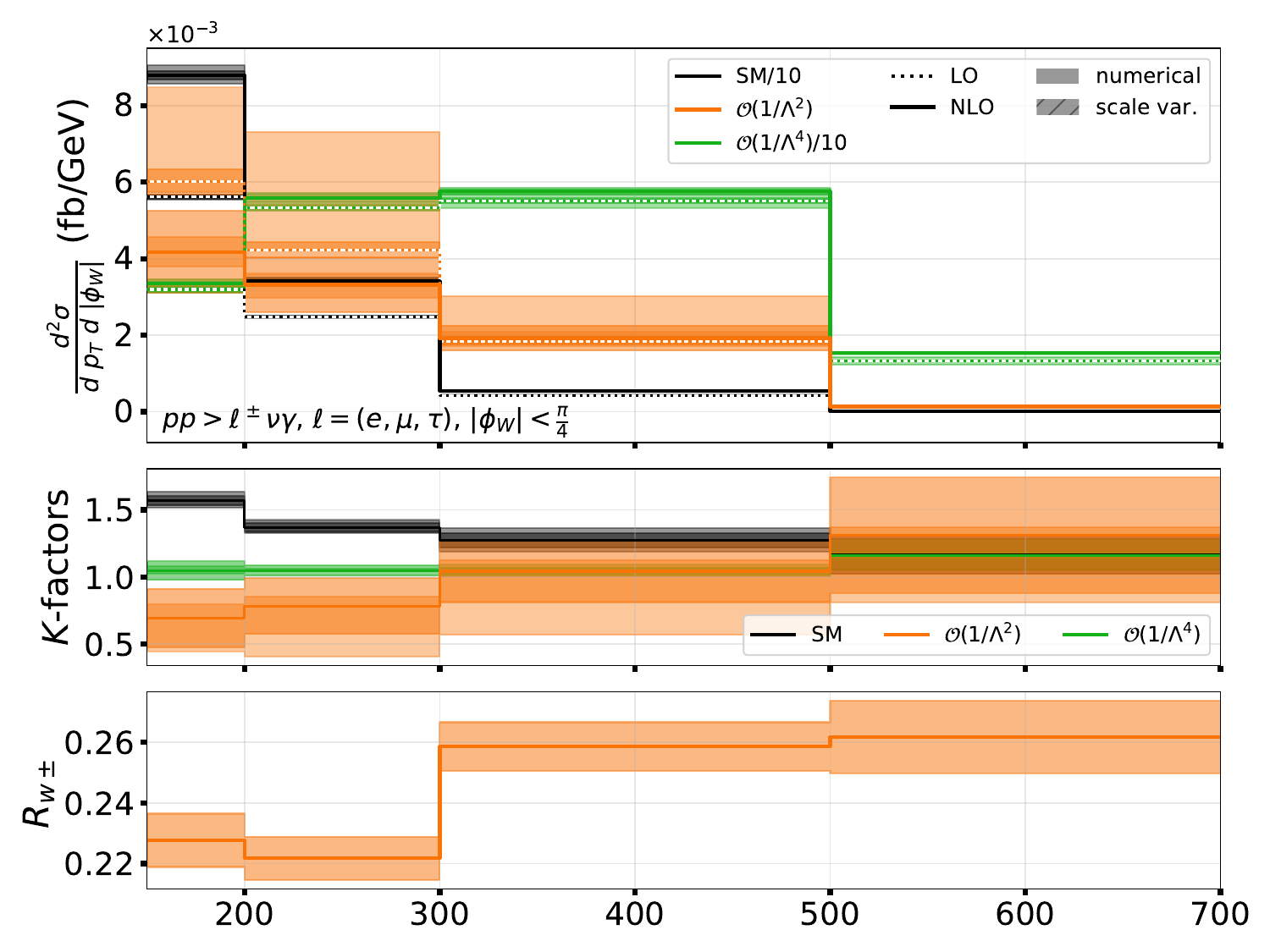}
      \includegraphics[width=.49\textwidth]{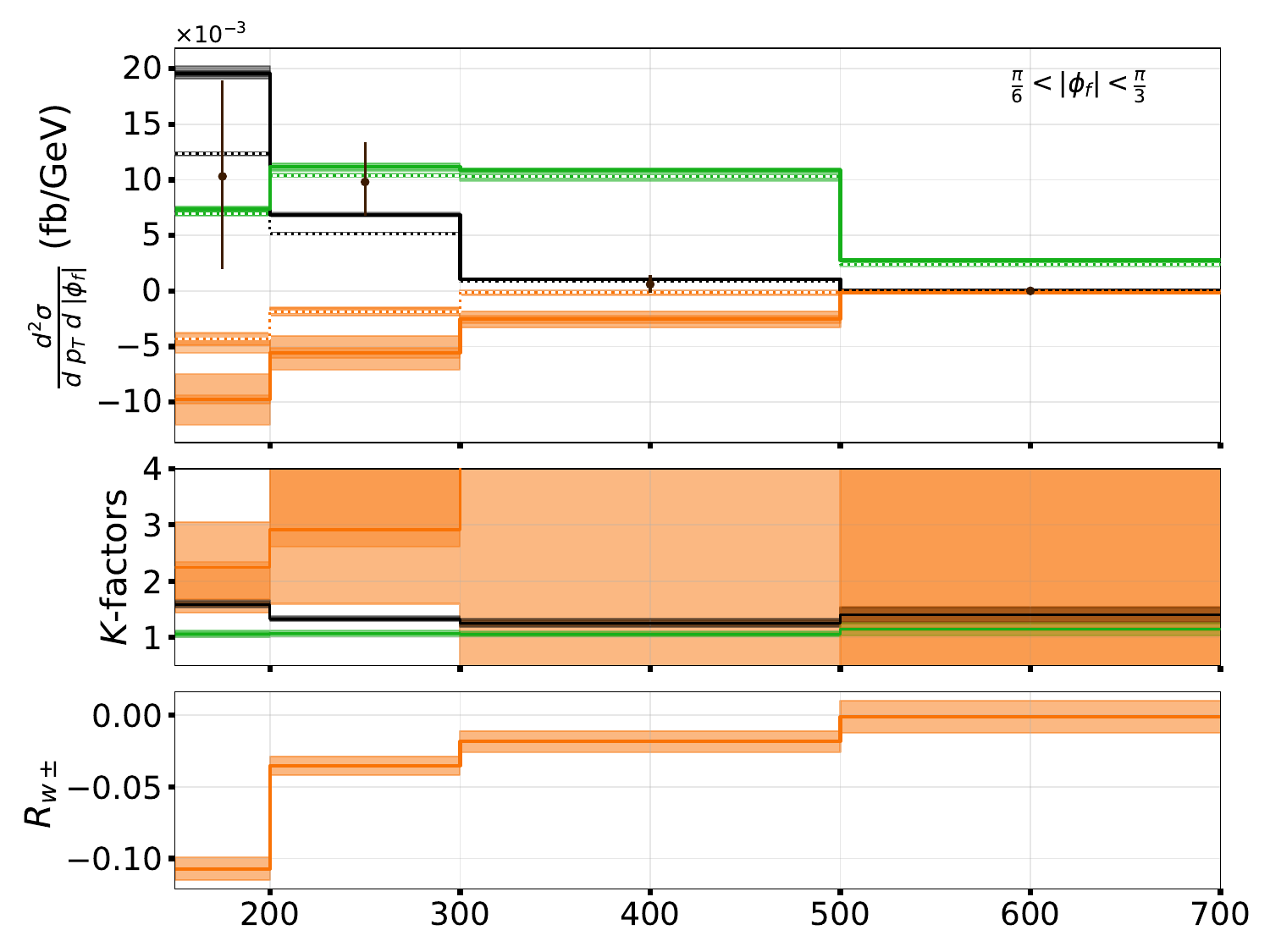}
      \includegraphics[width=.49\textwidth]{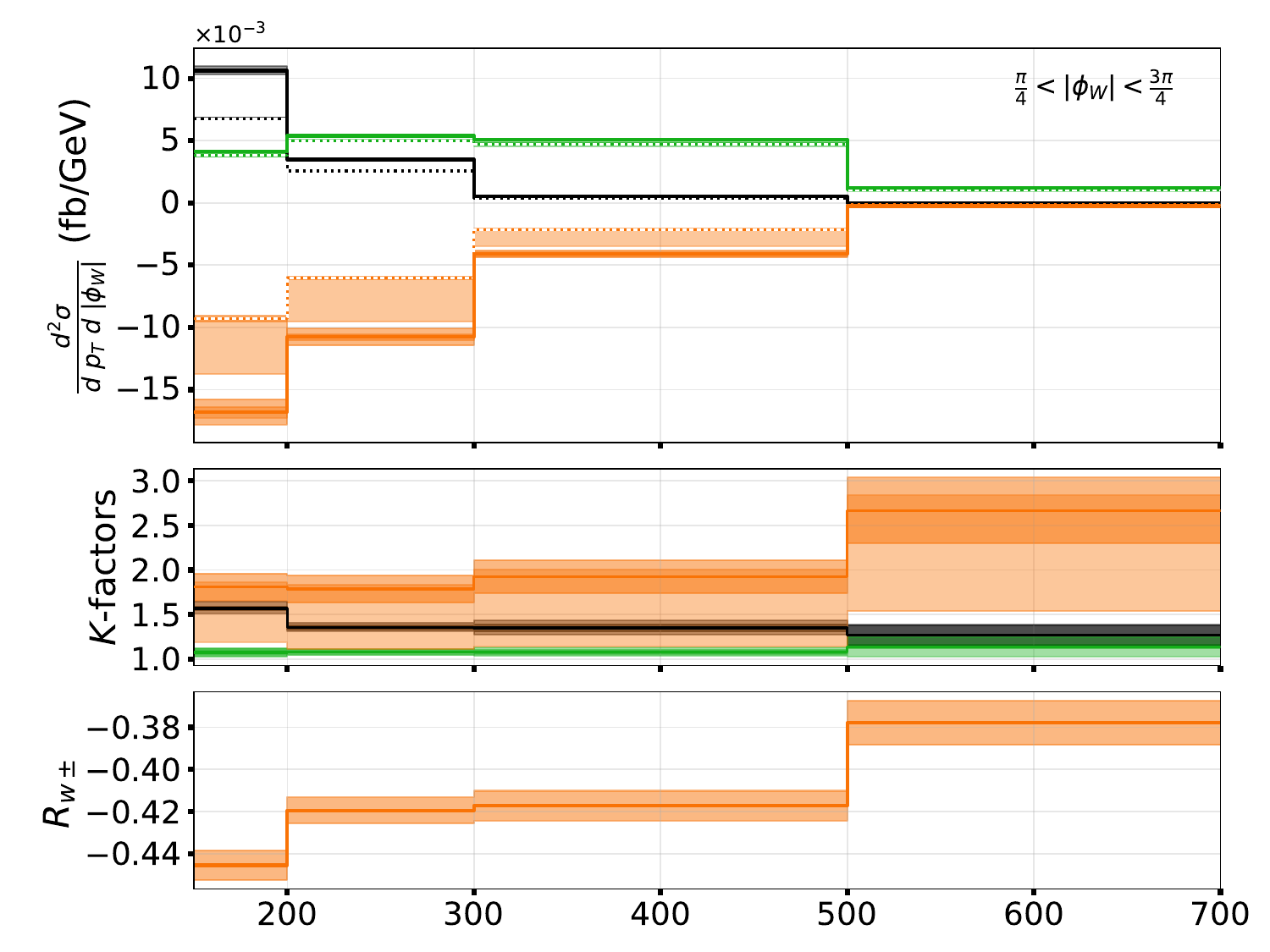}
      \includegraphics[width=.49\textwidth]{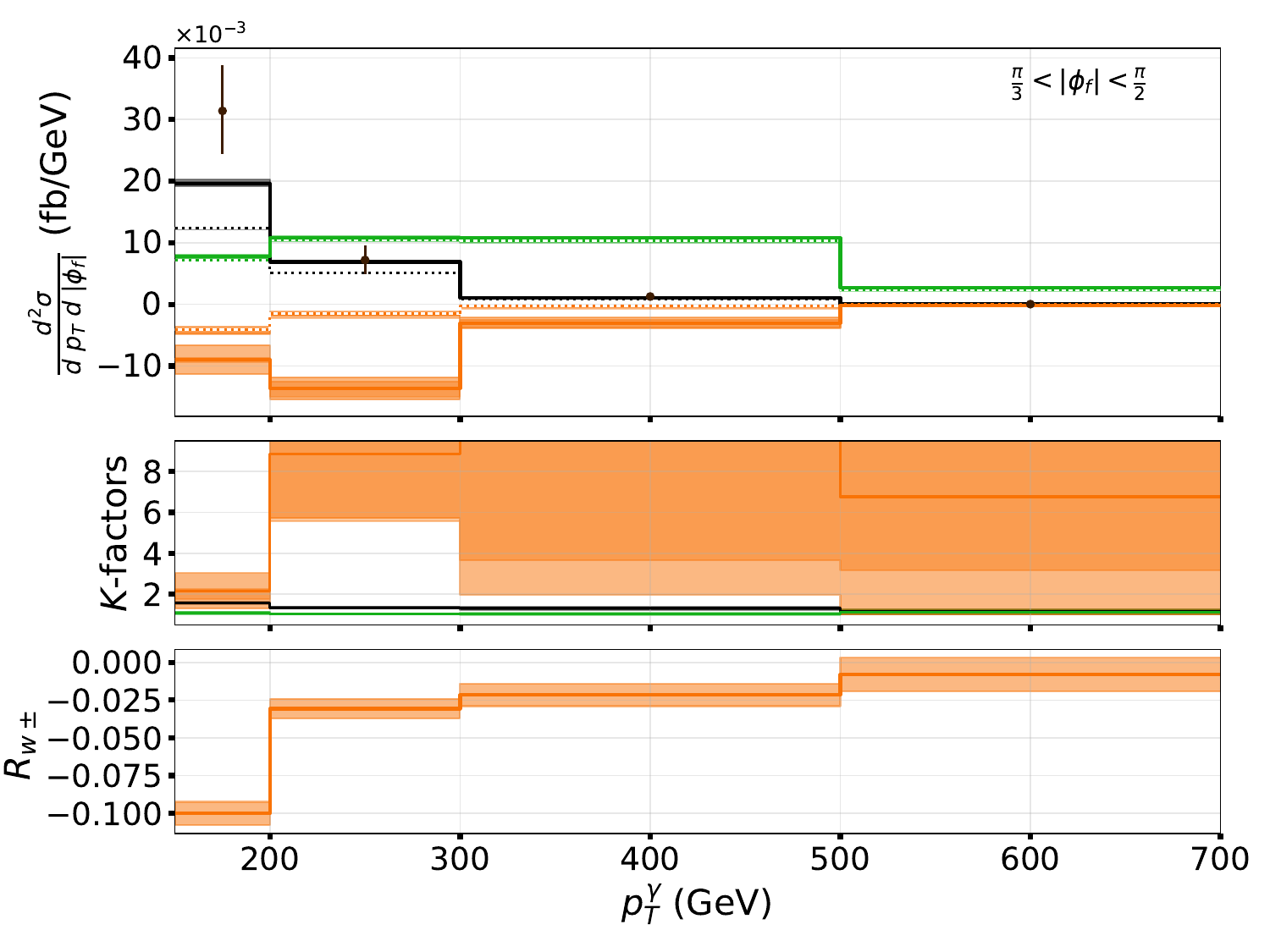}
      \includegraphics[width=.49\textwidth]{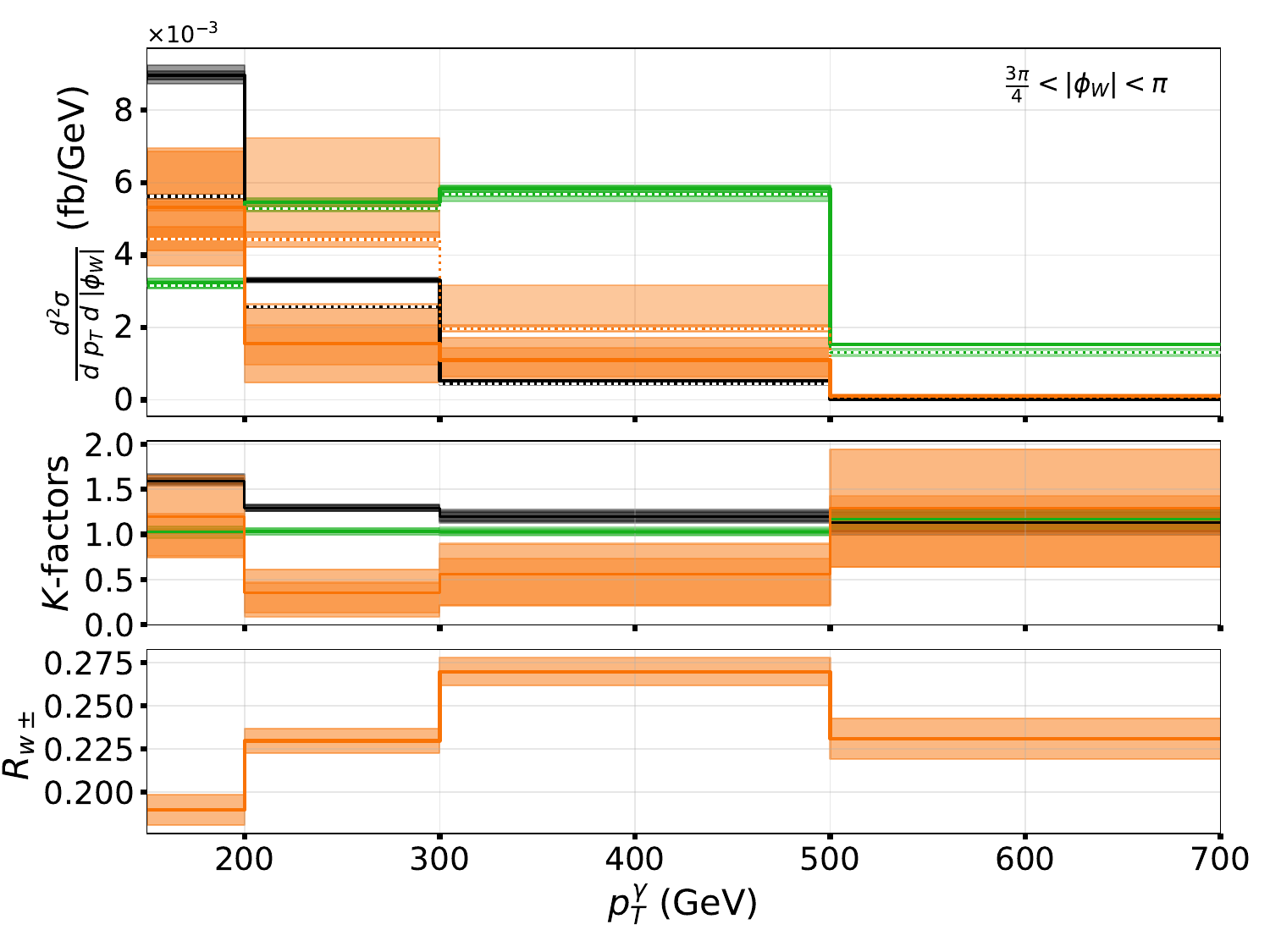}
\end{figure}

Our predictions for the double-differential distribution used in the experimental analysis, $p_T^\gamma$ {\it vs} $|\phi_f|$, are shown in the left column of Fig. \ref{fig:oW_wa_pTaPhi}. The binning is [150, 200, 300, 500, 800, 1500] GeV for the transverse momentum and [0, $\pi /6$, $\pi /3$, $\pi /2$] for the angle. While the SM and the $\mathcal{O}(1/\Lambda^4)$ correction present stable values for the differential $K$-factor, the linear ones are large and with big uncertainties almost everywhere. Indeed, the cancellation level is close to zero over the entire range of the variable.

As it can be seen in Table \ref{tab:oW_wa_meas}, this double distribution showcases the same reviving power than the total linear cross section, which is just above 10\% of the measurable effect for the interference. By combining $p_T^\gamma$ with another azimuthal angle like $\phi_W$, instead, the resurrecting efficiency achieves almost 80\% of the $\sigma^{|\text{meas}|}$ value. This can be understood by remembering that $\phi_f$ is measured in the CoM frame and, thus, more affected by the reconstruction procedure of the neutrino than $\phi_W$, that is computed in the lab frame. The LO and NLO distributions for this second double-differential observable are shown in the right column of Fig. \ref{fig:oW_wa_pTaPhi}, where we used the same binning as before for the transverse momentum and [0, $\pi/4$, $3\pi/4$, $\pi$] for the angle. The cancellation level for this observable is farther from zero, and in particular positive in the two external angle bins and negative in the central one, yielding more reasonable $K$-factors for the interference and with thinner error bars.

%\begin{figure}
%   \caption{\footnotesize{LO and NLO differential cross-section distributions for the photon transverse momentum in $W\gamma$ production, over different $|\phi_f|$ ranges. The black (orange, green) line represents the SM, divided by 10 (interference, quadratic correction divided by 10). The experimental data is also shown. The {\it K}-factors are also shown, together with their statistical and scale uncertainties. For each case, the relative cancellation for LO interference is plotted}} \label{fig:oW_wa_pTaPhif}
%   \begin{center}
%      \includegraphics[width=.6\textwidth]{figures/pTa_phifPi6.pdf}
%      \includegraphics[width=.6\textwidth]{figures/pTa_phifPi3.pdf}
%      \includegraphics[width=.6\textwidth]{figures/pTa_phifPi2.pdf}
%   \end{center}
%\end{figure}

%\begin{figure}
%   \caption{\footnotesize{Same as Fig. \ref{fig:oW_wa_pTaPhif}, but for the photon transverse momentum over different $\phi_W$ intervals}} \label{fig:oW_wa_pTaPhiW}
%   \begin{center}
%      \includegraphics[width=.6\textwidth]{figures/pTa_phiWPi4.pdf}
%      \includegraphics[width=.6\textwidth]{figures/pTa_phiW3Pi4.pdf}
%      \includegraphics[width=.6\textwidth]{figures/pTa_phiWPi.pdf}
%   \end{center}
%\end{figure}

%%%%%%%%%%%%%%%%%%%%%%%%% Bounds %%%%%%%%%%%%%%%%%%%%%%%%%

\section{Bounds on $C_W/\Lambda^2$}
We used the distributions described in the previous sections to obtain bounds on the $O_W$ Wilson coefficient. To do so, for a generic variable $X$ we employed the $\chi^2$ expression
\begin{equation}
   \chi^2 = \sum_{i=1}^{N_\text{bins}} \frac{1}{\Delta_i^2} \left[ \frac{d\sigma_i^\text{exp}}{d X} -\frac{d\sigma_{\text{best},i}^\text{SM}}{d X} -\frac{C_W}{\Lambda^2} \left( \frac{d\sigma_i^{1/\Lambda^2}}{d X}\pm \Delta_i^{1/\Lambda^2}\right) +\ldots \right]^2, \label{chisq_uncNum}
\end{equation}
%-\frac{C_W^2}{\Lambda^4} \left( \frac{d\sigma_i^{1/\Lambda^4}}{d X}\pm \Delta_i^{1/\Lambda^4} \right)
where the SM prediction was always the best available, while the linear and $\mathcal{O}(1/\Lambda^4)$ ones could be at LO or NLO. $\Delta_i$ at the denominator only contained the SM and experimental uncertainties; the others were included at the numerator to keep the expression above a quadratic or quartic function of $C_W/\Lambda^2$. Among the different options that could be obtained by summing or subtracting these sources of error, we picked the one that returned the widest bounds.

As it can be seen for the $M_T^{WZ}$ distributions in $WZ$ production in the top panel of Fig. \ref{fig:oW_wz_mTwz}, the NLO uncertainties from \amc\ are not large enough to include the N${}^2$LO order: this suggests that all the NLO predictions presented in this study might underestimate the real errors. Yet, these are the values we used, being the only ones available. For the linear and quadratic terms at LO, we associated to each bin a relative uncertainty equal to $|k_i -1|$, with $k_i$ the $K$-factor in the bin; if the latter is negative or above two, a 100\% was considered, since the scale variations cannot then be trusted as estimates of the missing corrections from higher orders.

In case of lack of real measurements, the experimental term was assumed to follow the best SM prediction, with a 10\% systematic uncertainty and a numerical one equal to $\sqrt{\sigma_{\text{best},i}^\text{SM} / \mathcal{L}_{LHC}}$, where $\mathcal{L}_{LHC} = 137$ fb${}^{-1}$ is the LHC luminosity at Run II. No correlation between the bins was assumed in this case. A list of the experimental datasets used in this calculation is reported in Table \ref{tab:oW_dataset}, for all the processes, signal regions and observables investigated. The observables we employed to obtain the limits and the integrated luminosities are also shown.

It is important to keep in mind that the $\mathcal{O}(1/\Lambda^4)$ term receives contributions from the SM interference with the dimension-8 operators, that are not included here but can induce comparable effects to the quadratic ones. Because of this, the bounds that will be shown at that order are not complete and should be taken only as a comparison for the linear ones and a test of the EFT validity.

\begin{figure}
   \centering
   \caption{\footnotesize{68\% and 95\% CL bounds on $C_W/\Lambda^2$ at LO ({\it dotted}) and NLO ({\it continuous}), with the inclusion of the quadratic term or without, for the processes presented in this chapter. The variables we used are noted next to the process names: there are two different ones for $W\gamma$. The limits in the inclusive case and over the regions \eqref{oW_wz_cuts} are shown for $WZ$. The limits with gray background come from comparison with the best SM distributions, the others with real data; a summary of the experimental measurements can be found in Table \ref{tab:oW_dataset}. The numerical values are on the right}} \label{fig:oW_bounds}
   \includegraphics[width=\textwidth]{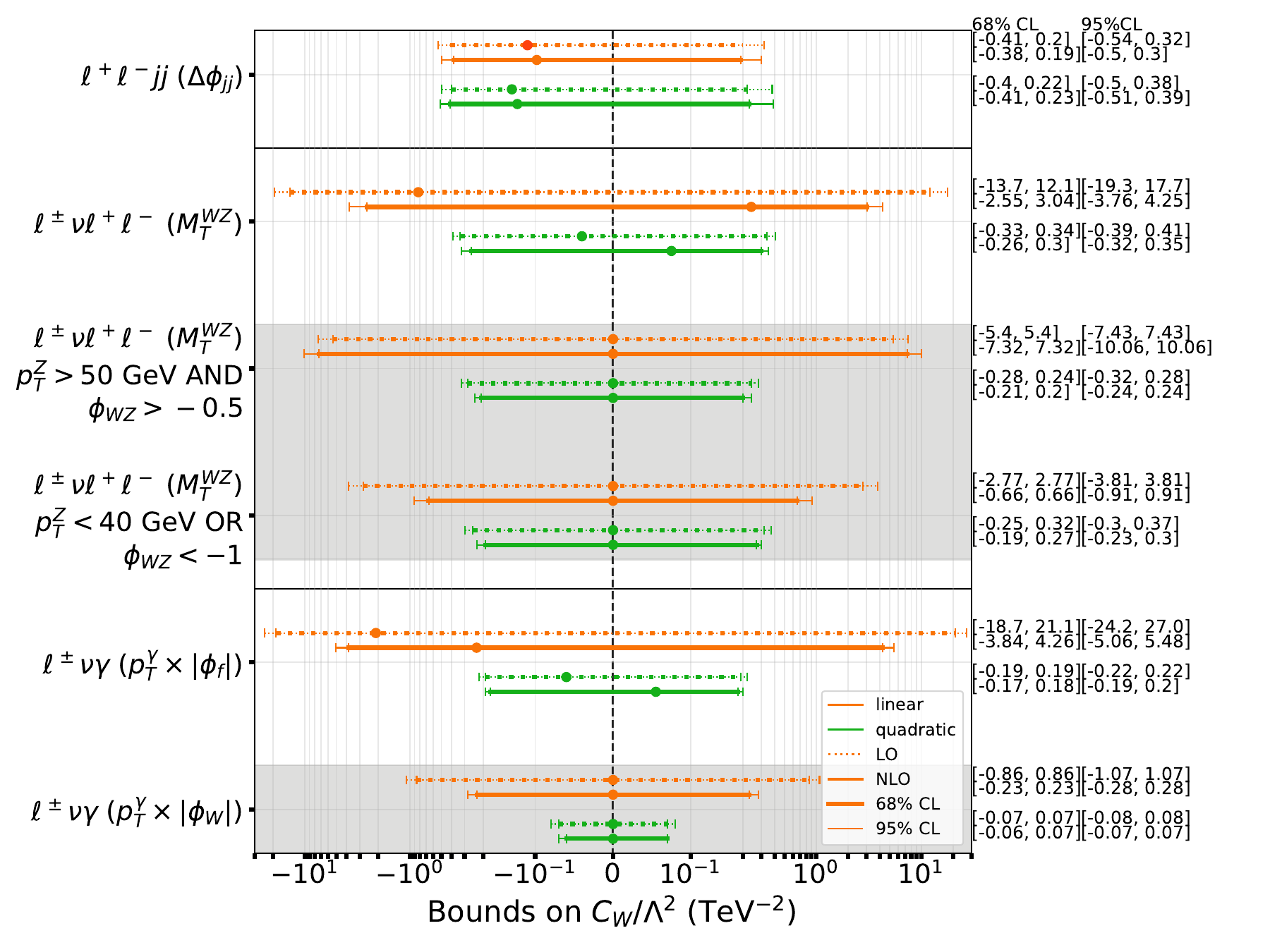}
\end{figure}

The limits we obtained on $C_W /\Lambda^2$ are summarised in Fig. \ref{fig:oW_bounds}. The best ones at linear level come from $\Delta \phi_{jj}$ in $Zjj$ production and they are comparable to the $\mathcal{O}(1/\Lambda^4)$ for the same observable. Because of the mismatch between the SM predictions from {\sc MadGraph} and the data, we used the NLO SM distribution generated with \hw +{\sc Vbfnlo} in the experimental analysis and added the FO linear and quadratic terms to that.

For the other processes, the $\mathcal{O}(1/\Lambda^2)$ bounds are at least three times larger than the quadratic ones. In the $WZ$ case, though, they sensibly improve when the cuts \eqref{oW_wz_cuts} are applied, especially for the second one, while the quadratic ones remain of the same order as in the inclusive case. For the first phase-space cut, the limits are better at LO than at NLO, since the differential $K$-factors for $M_T^{WZ}$ are smaller than one.

For $W\gamma$, the bounds from both the discussed double-differential distributions are presented, showing that observables with larger reviving efficiencies like $\phi_W$ can yield big improvements in the constraints both at linear and quadratic orders.

\begin{table}
\caption{\footnotesize{Summary of the experimental measurements used to obtain bounds on $C_W/\Lambda^2$, for each process investigated in this chapter. In the signal regions \eqref{oW_wz_cuts} and for the second $W\gamma$ observable, that have not been measured yet, the experimental distributions are considered to follow the NLO SM ones}} \label{tab:oW_dataset}
\resizebox{\textwidth}{!}{
\begin{tabular}{cc|cccc}
   Process & Observable & $\sqrt{s}$, $\mathcal{L}$ & Final state & $N_\text{data}$ & Ref. \\ \hline \hline
   $Zjj$ & $d\sigma/d \Delta \phi_{jj}$ & 13 TeV, 139 fb${}^{-1}$ & $\ell^+ \ell^- +$jets, $\ell=e,\mu$ & 12 & \cite{atlas:2021Zjj} \\ \hline
   $W^\pm Z$, full phase space & $d\sigma/d M_T^{WZ}$ & 13 TeV, 36.1 fb${}^{-1}$ & $\ell^\pm \nu \ell^+ \ell^-$, $\ell=e,\mu$ & 6 & \cite{Atlas:2019wz} \\
   \makecell{$W^\pm Z$, $p_T^Z > 50$ GeV \\ AND $\phi_{WZ}>-0.5$} & $d\sigma/d M_T^{WZ}$ & \multicolumn{4}{c}{Exp. data taken as NLO SM} \\
   \makecell{$W^\pm Z$, $p_T^Z < 40$ GeV \\ OR $\phi_{WZ}<-1$} & $d\sigma/d M_T^{WZ}$ & \multicolumn{4}{c}{Exp. data taken as NLO SM} \\ \hline
   $W^\pm \gamma$ & $d^2 \sigma /(d p_T^\gamma \hspace{1mm} d|\phi_f|)$ & 13 TeV, 138 fb${}^{-1}$ & $\ell^\pm \nu \gamma$, $\ell=e,\mu,\tau$ & 12 & \cite{cms:2021wa}\\
   $W^\pm \gamma$ & $d^2 \sigma /(d p_T^\gamma \hspace{1mm} d|\phi_W|)$ & \multicolumn{4}{c}{Exp. data taken as NLO SM} \\ \hline
\end{tabular}
}
\end{table}

\section{\label{sect:oW_conclusions}Conclusions and prospects}
The results illustrated in this chapter highlight how lifting the suppression of the interference term can be necessary to obtain meaningful predictions at NLO for some processes. With the aid of quantities like $\sigma^{|\text{meas}|}$ and $R_{w\pm}$, we found observables that present more reasonable differential $K$-factors in cases where the global ones are extremely large or even negative. We also showed that the same quantities can be used to better understand the nature of the interference cancellation and find phase-space regions where most of the positive- and negative-weighted contributions are concentrated, yielding different but more stable $K$-factors.

We also employed the same variables to infer limits on the $O_W$ Wilson coefficient, and obtained results from the $Zjj$ linear term that are comparable to the quadratic ones, which do not suffer from the same suppression as the interference between the SM and dimension-8 operators was not included. For the diboson processes, the bounds from the interference are still worse than the $\mathcal{O}(1/\Lambda^4)$ ones: a possible reason for that is the presence of a neutrino in the final state, which means that a lower portion of the total interference effect can be accessed at the LHC. The same processes could be studied with the $W$ boson decaying hadronically, but a larger background would be observed. On the other hand, a better final-state reconstruction is possible for $Zjj$.

As stated in the previous chapter, the variables proposed here only depend on kinematics and are fully generic; moreover, the predictions can be generalised to any order. For example, even if we only focused on NLO corrections in QCD, we can suppose from the good revival obtained via $\Delta \phi_{jj}$ in $Zjj$ that EW corrections should not move the $K$-factors too far away from unity, despite their magnitude. The same cannot be affirmed for diboson processes because of the reasons summarised above; nonetheless, $W\gamma$ showcases a similar constraining power when interference-reviving observables are employed.

The various sources of suppression for the processes presented in this chapter suggest that a dedicated study should be performed from scratch for any other scenario that shows room for interference revival: even if some extrapolations might be made in case of the presence of a neutrino, of a leptonically-decaying $Z$ boson or a similar event topology, the strategies can be different from one operator to another. For these reasons, regarding the possibility to automate the search for optimal observables to restore the interference, the best chance might come from machine-learning techniques, as briefly introduced in the conclusions of the previous chapter in Sect. \ref{sect:oG_conclusions}. Despite this, even if $\sigma^{|\text{meas}|}$ is used alongside neural networks to obtain even more optimal observables, with asymmetries that are a few percent larger than the best ones described here, the uncertainties over our predictions should be lowered in order to appreciate a difference in the $K$-factors and bounds.

%!TEX root = main.tex

%%%%%%%%%%%%%%%%%%%%%%%%%%%%%%%%%%%%%%%%%%%%%%%%%%%%
%
%      Chapter V :
%
%
%%%%%%%%%%%%%%%%%%%%%%%%%%%%%%%%%%%%%%%%%%%%%%%%%%%

\chapter{Constraints on the four-light quark operators at linear level}
\label{chap:fourLQ}
\pagestyle{fancy}

\hfill
\begin{minipage}{\textwidth}
   \centering
   \includegraphics[width=.7\textwidth]{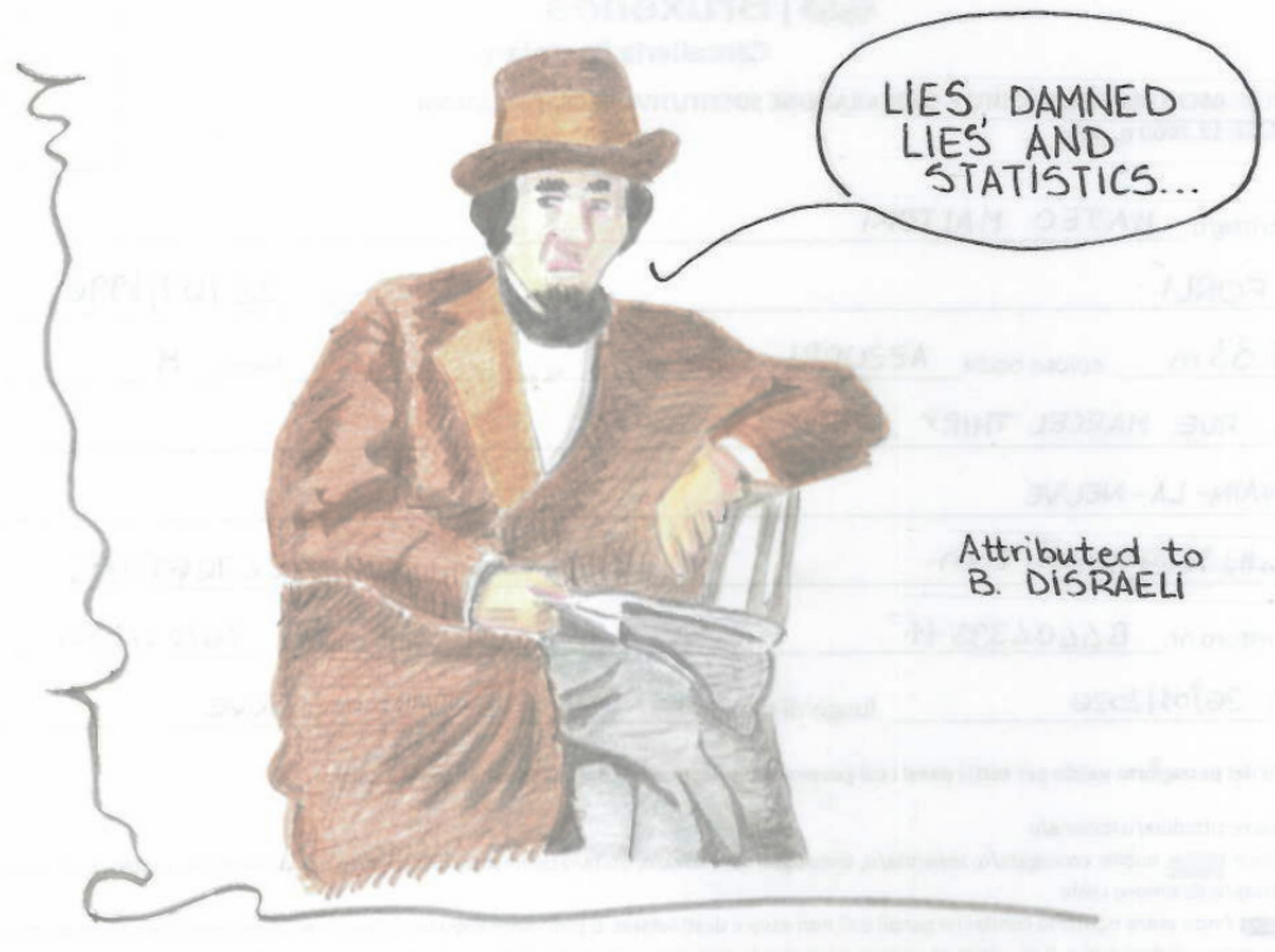}
\end{minipage}

\vspace{0.5cm}

This chapter deviates from the interference-resurrection topic of the previous two, as the main operators discussed here do not experience a large cancellation in their linear contributions to the considered processes. I will explain our study of ten four-light quarks (4LQ) operators in the SMEFT, and how we obtained bounds on their coefficients thanks to suitable observables at linear level.

The operators are defined in Table \ref{tab:4lq_def}. They are dimension-6 members of the Warsaw basis and introduce new four-fermion interactions among the $u,d,c,s,b$ quarks. They do not contribute to the main Higgs and top processes at LO, so they had been excluded from global fits until recently, as new physics is expected to couple preferentially to the heaviest SM states \cite{globalfits_1, globalfits_2, globalfits_3, globalfits_4, globalfits_5, smeft_global_mfv, globalfits_6}. Indeed, their effects are known to be strongly suppressed compared to the $O_G$ ones in multijet production \cite{Krauss_2017}. However, they could induce corrections to any process at NLO, if two of the quark lines in the diagrams they affect are closed into a loop.

\begin{table}
\centering
\caption{\footnotesize{List of 4LQ operators considered in this study. $q$ denotes the left-handed quark doublets \eqref{su2_doublets} of all three generations, while $u,d$ are the right-handed up-type and down-type quark fields, respectively. $\sigma^I$, with $I=\{1,2,3\}$ are the Pauli matrices, while $T^a$, $a=\{1,\ldots,8\}$ are the $SU(3)_c$ generators. $p,r,s,t$ are flavour indices; the spin and colour ones are omitted. For each process, $\checkmark$ and $\times$ specify if the interference of an operator with the SM QCD contributes to it or not}} \label{tab:4lq_def}
\begin{tabular}{c|cc|cccc}
   %\hline
   Operator & Coeff. & Definition & $jj$ & $Zjj$ & $Wjj$ & $\gamma jj$ \\
   \hline
   $O_{qq}^{(1)}$ & $C_{qq}^{(1)}$ & $(\bar{q}_p \gamma^\mu q_r)(\bar{q}_s \gamma_{\mu} q_t)$ & $\checkmark$ & $\checkmark$ & $\checkmark$ & $\checkmark$ \\
   $O_{qq}^{(3)}$ & $C_{qq}^{(3)}$ & $(\bar{q}_p \gamma^\mu \sigma^I q_r)(\bar{q}_s \gamma_{\mu} \sigma^I q_t)$ & $\checkmark$ & $\checkmark$ & $\checkmark$ & $\checkmark$ \\
   $O_{uu}$ & $C_{uu}$ & $(\bar{u}_p \gamma^\mu u_r)(\bar{u}_s \gamma_{\mu} u_t)$ & $\checkmark$ & $\checkmark$ & $\times$ & $\checkmark$ \\
   $O_{dd}$ & $C_{dd}$ & $(\bar{d}_p \gamma^\mu d_r)(\bar{d}_s \gamma_{\mu} d_t)$ & $\checkmark$ & $\checkmark$ & $\times$ & $\checkmark$ \\
   $O_{ud}^{(1)}$ & $C_{ud}^{(1)}$ & $(\bar{u}_p \gamma^\mu u_p)(\bar{d}_s \gamma_{\mu} d_s)$ & $\times$ & $\checkmark$ & $\times$ & $\checkmark$ \\
   $O_{ud}^{(8)}$ & $C_{ud}^{(8)}$ & $(\bar{u}_p \gamma^\mu T^A u_p)(\bar{d}_s \gamma_{\mu} T^A d_s)$ & $\checkmark$ & $\checkmark$ & $\times$ & $\checkmark$ \\
   $O_{qu}^{(1)}$ & $C_{qu}^{(1)}$ & $(\bar{q}_p \gamma^\mu q_p)(\bar{u}_s \gamma_{\mu} u_s)$ & $\times$ & $\checkmark$ & $\checkmark$ & $\checkmark$ \\
   $O_{qu}^{(8)}$ & $C_{qu}^{(8)}$ & $(\bar{q}_p \gamma^\mu T^A q_p)(\bar{u}_s \gamma_{\mu} T^A u_s)$ & $\checkmark$ & $\checkmark$ & $\checkmark$ & $\checkmark$ \\
   $O_{qd}^{(1)}$ & $C_{qd}^{(1)}$ & $(\bar{q}_p \gamma^\mu q_p)(\bar{d}_s \gamma_{\mu} d_s)$ & $\times$ & $\checkmark$ & $\checkmark$ & $\checkmark$ \\
   $O_{qd}^{(8)}$ & $C_{qd}^{(8)}$ & $(\bar{q}_p \gamma^\mu T^A q_p)(\bar{d}_s \gamma_{\mu} T^A d_s)$ & $\checkmark$ & $\checkmark$ & $\checkmark$ & $\checkmark$ \\
   \hline
\end{tabular}
\end{table}

In order to study them, we generated a UFO model that includes them on top of the SM. The top quark is included as well, but we excluded it when generating the processes. We employed \nloct\ to extract the rational and ultraviolet counterterms: the model can be used at NLO, even if all the results in this chapter are LO ones. The leptons and all the quarks but the top are taken as massless and the CKM matrix is assumed to be diagonal. We assumed $U(3)^5$ as flavour group, introduced at the end of Sect. \ref{sect:smeft}.

These operators feature different colour and flavour structures. $O_{qq}^{(1)}$, $O_{qq}^{(3)}$, $O_{uu}$ and $O_{dd}$ present two fermionic currents with the same chirality, left-handed for the first two and right-handed for the others. For this reason, there are two possible ways to contract their flavour indices: inside the fermion bilinears or between them. If we name the indices as in Table \ref{tab:4lq_def}, \eg\ $O_{qq}^{(1)} = \left( \bar q_p \gamma^\mu q_r \right) \left( \bar q_s \gamma_\mu q_t \right)$, the two options are $\delta_{pr}\delta_{st}$ and $\delta_{pt}\delta_{rs}$ \cite{smeft_global_mfv}. These two combinations are related by the $SU(3)$ Fierz identities \eqref{fiertz_ids} and carry respectively a colour-singlet and octet structure, so they can be constrained separately. In our UFO model, we instead summed the two contractions together and associated them to the same Wilson coefficient, for each operator. The contribution with four identical flavour indices, being identical in the two contractions, was however added only once. This choice was made because we wanted to keep the analysis simple and check what are the best bounds that can be placed upon these objects, leaving eventual studies of the colour-induced differences for the future. All the other operators, that show different chiralities in the two bilinears, only allow for a single contraction $\delta_{pr}\delta_{st}$.
%This can be justified by considering that the most impactful subprocesses to multijet production are the ones featuring first generation quarks because of PDF effects. Furthermore, a colour-octet contribution like $uu \rightarrow cc$ would produce distributions with the same shape as $uu \rightarrow uu$, so they would not be distinguishable if the differential results were normalised to the same value. For these reasons and for the sake of simplicity, we decided to tie the two contractions together and leave the study of the colour-induced differences for the future.

It is known that dijet production can only probe two directions in the Wilson-coefficient space \cite{2j_SMEFT}, so we focused on multijets but also processes where EW bosons are produced together with jets, like $Z,W,\gamma +$jets. The main diagrams that interfere with the SM are the ones where a quark line emits the boson, with the 4LQ operators influencing the contact interaction between the fermions. Since the EW bosons are sensitive to different quantum numbers of the quark fields, the idea is that their presence can rule some operators out or enhance some subprocesses, allowing to probing new directions in the coefficient space. For example, the left-handed nature of the $W$-boson interaction ensures that operators only affecting right-handed particles do not contribute in $W+$jets. Analogously, the different electric charge of quarks favours the interaction of a photon with up-like ones over the down-like. The operators that contribute to each of the processes we studied at linear level are summarised in Table \ref{tab:4lq_def}. Furthermore, we simulated a $b$- and $c$-jet tagging algorithm and applied it to the multijet case, in order to increase the sensitivity to subprocesses featuring those flavours.

Our analysis was performed via \amc\ v3.5.4 and the UFO model described above, with all the Wilson coefficients set to 1 TeV${}^{-2}$. In all cases, we included the diagrams featuring both QED and QCD vertices and generated up to three jets at LO parton level, that were then merged with {\sc MLM} and showered with \pyth . $H_T /2$ was chosen as dynamical scale for the events. As in the previous chapter, we reconstructed eventual dressed leptons first through the $k_t$ algorithm with a radius parameter $R$ of 0.1, and then we ran \fj\ again on the remaining final states with the anti-$k_t$ algorithm and $R=0.4$ to extract the jets.

Numerical and scale uncertainties are reported for our results. The latter were obtained as the envelope of nine scale combinations, with $\mu_R,\mu_F$ modified by factors 0.5 and 2. The distributions that will be shown are the ones that yield the most stringent limits, for each process. All the results are presented for the LHC at 13 TeV.

%%%%%%%%%%%%%%%%%%%%%%%%% jj %%%%%%%%%%%%%%%%%%%%%%%%%

\section{Multijet production}
The generation of jets is one of the most common processes at the LHC, with the final states that easily reach the highest energies accessible by that collider and large cross sections that allow for multidifferential distributions \cite{jets_1,jets_2,jets_3}. This class of processes represents the simplest one that could be affected by the 4LQ operators included in our study.

Not all the ten of them contribute at linear order, though, as it is not possible for a gluon to couple to quarks with different weak charges and same colour \cite{2j_SMEFT}. For this reason, $O_{ud}^{(1)}$, $O_{qu}^{(1)}$ and $O_{qd}^{(1)}$ do not interfere with the SM QCD in dijet production and they are not considered in this section, as the QED contribution in multijets is generally subleading. Similarly, $O_{qq}^{(1)}$ does not contribute at linear level to subprocesses that feature both up- and down-like quarks.

\begin{figure}
   \caption{\footnotesize{Shapes of the differential cross section with respect to $\chi_{jj}$ for the SM QCD and its interference with the 4LQ operators, for some subprocesses to dijet production at LO. PDF and PS effects are not included}} \label{fig:4lq_2j_shapes}
   \includegraphics[width=0.49\textwidth]{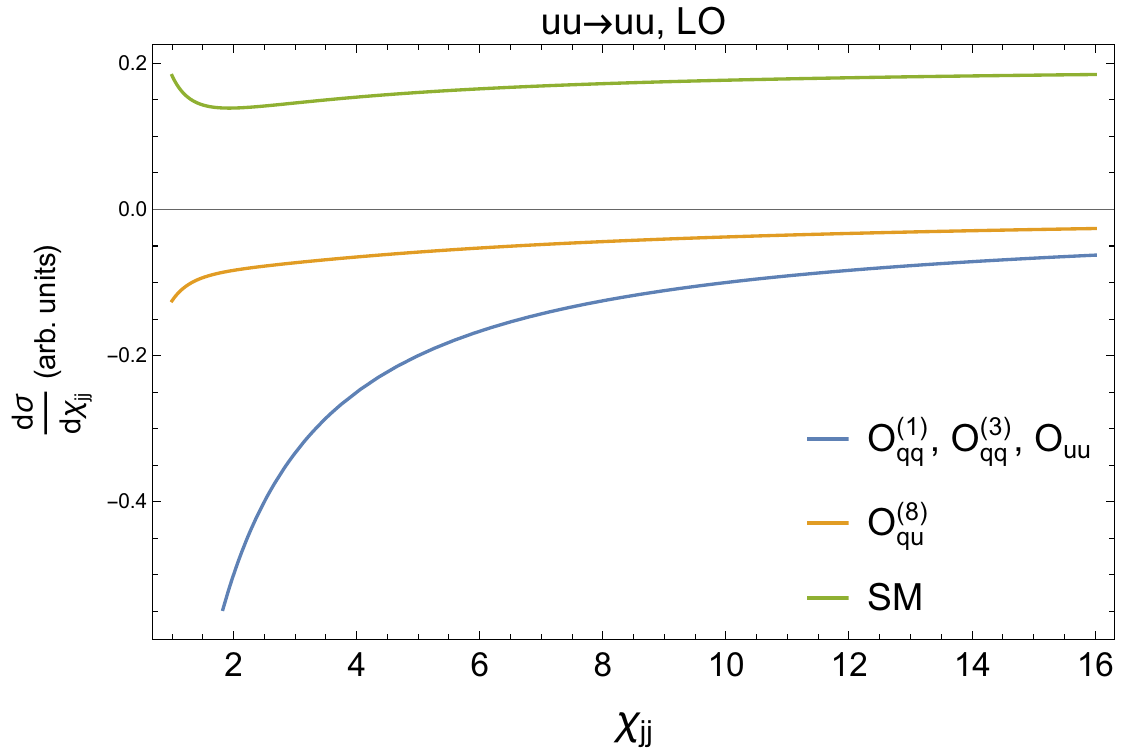}
   \includegraphics[width=0.49\textwidth]{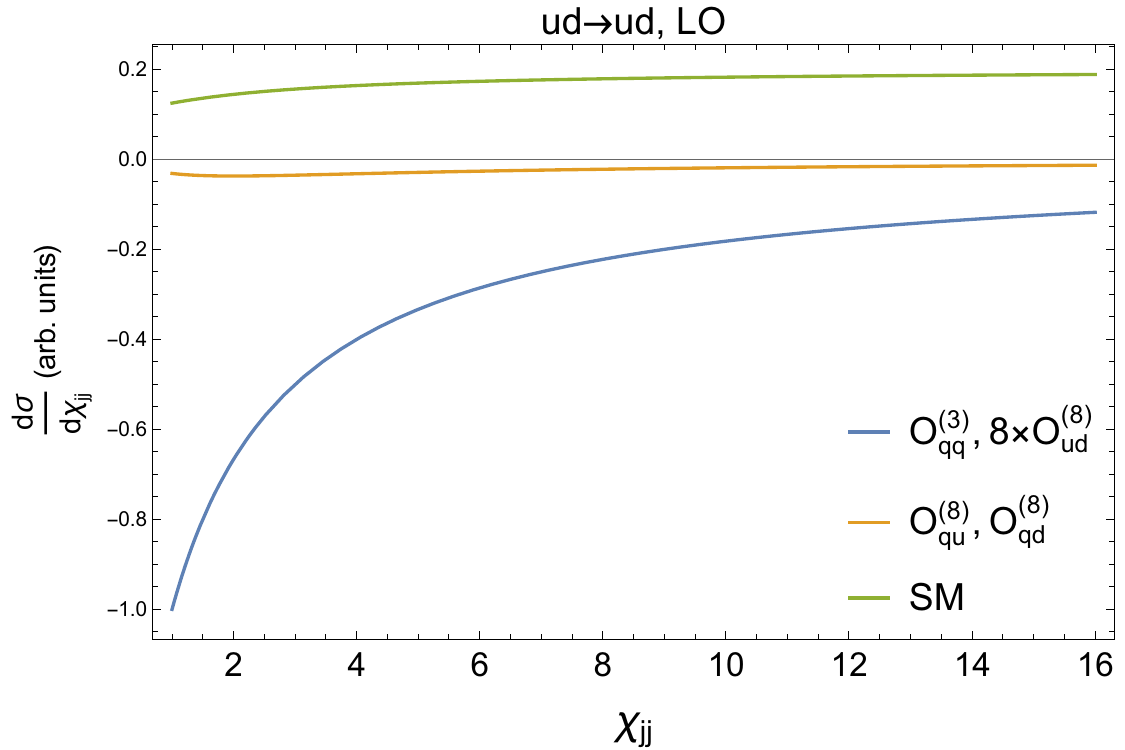}
   \includegraphics[width=0.49\textwidth]{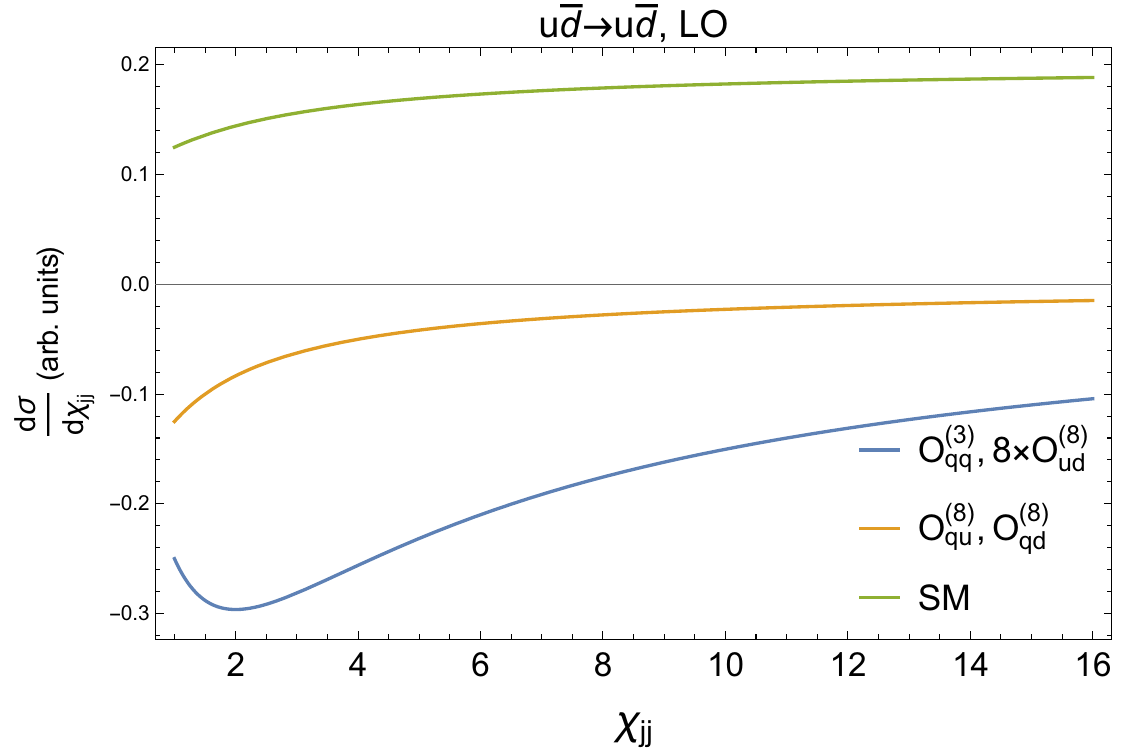}
   \includegraphics[width=0.49\textwidth]{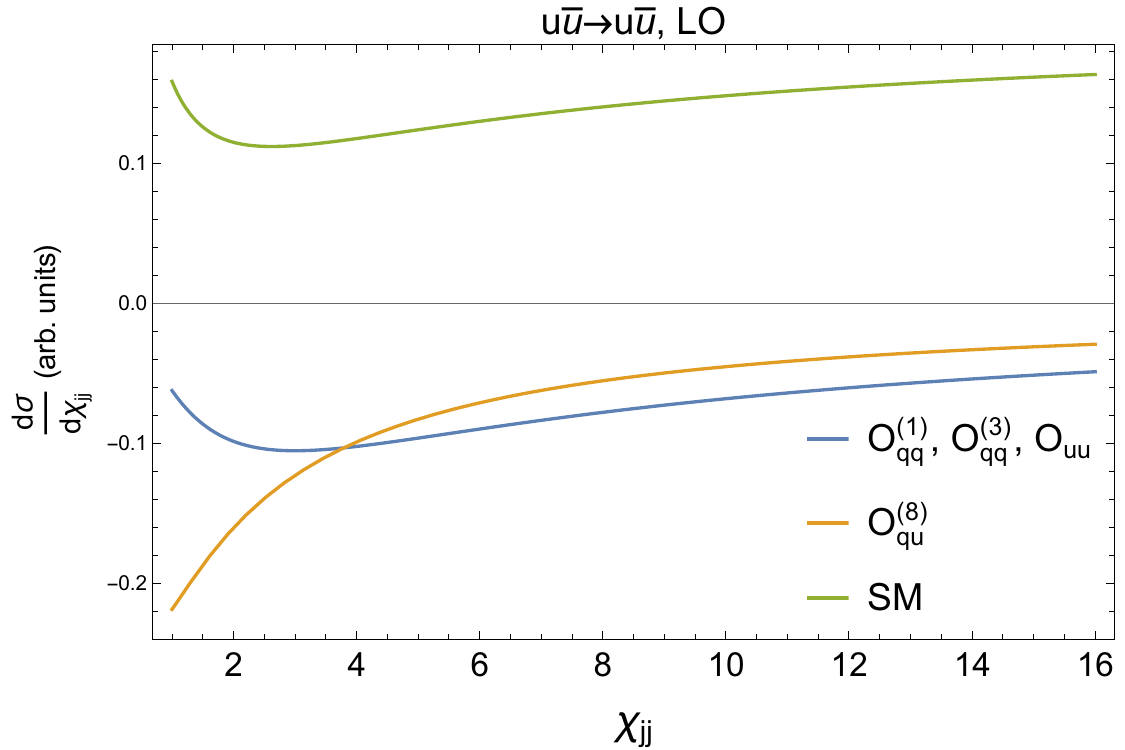}
\end{figure}

The exponential of the absolute difference among the two leading jets in transverse momentum, $\chi_{jj}$, is employed in multiple analysis to investigate these type of processes, as it is related to the scattering angle in the CoM frame. It is the same observable defined in Eq. \eqref{chi_jj}, also used to obtain bounds on $O_G$ from the quadratic order. Its differential cross-section distribution presents a trend that is almost flat for the SM, since for the dominant contribution, the gluon exchange along the $t$-channel, it is independent of the rapidity difference between the final-state jets. On the other side, the interference of the 4LQ operators shows a peak at $\chi_{jj} \sim 1$. The authors of \cite{2j_SMEFT} suggest that, for every subprocess to dijet production at parton level, only two shapes are possible for the linear $\frac{d \sigma^{1/\Lambda^2}}{d \chi_{jj}}$ differential cross section, as all the operators in the following two groups generate the same up to normalisations:
\begin{subequations} \label{4lq_shape}
   \begin{gather}
      \{ O_{qq}^{(1)}, O_{qq}^{(3)}, O_{uu}, O_{dd}, O_{ud}^{(8)} \}, \label{4lq_shape_a} \\
      \{ O_{qu}^{(8)}, O_{qd}^{(8)} \}. \label{4lq_shape_b}
   \end{gather}
\end{subequations}
The shapes for different subprocesses, obtained via \fa\ and \fc\ \cite{feynarts,formcalc}, are shown in Fig. \ref{fig:4lq_2j_shapes} for the SM and the 4LQ operators that contribute, and their analytical expressions can be found in Table \ref{tab:4lq_2j_shapes}.

\begin{table} 
\caption{\footnotesize{Analytical expressions of the differential cross sections for some dijet subprocesses, in the SM QCD and its interferences with the 4LQ operators, as functions of the $\theta^*$ scattering angle in the CoM frame and $\chi_{jj}$. $\theta^*$ is chosen so that $\cos\theta^*> 0$. PDF and PS effects are not included. Each interference term has a prefactor $4 C_i \alpha_S^{}/(9\Lambda^2)$, while for the SM ones it is $2\alpha_S^2 \pi /(9s)$, with $s$ the total energy squared of the process}} \label{tab:4lq_2j_shapes}
%\begin{tabularx}{\textwidth}{c|*{1}{>{\centering\arraybackslash}X}c}
\resizebox{\textwidth}{!}{
\begin{tabular}{c|cc}
   %\hline
   \multicolumn{3}{c}{$pp\rightarrow jj$, LO} \\
   %\hline\hline
   Operator & $\frac{d\sigma^{1/\Lambda^2}}{d\cos\theta^*}$ & $\frac{d\sigma^{1/\Lambda^2}}{d\chi_{jj}}$ \\
   \hline
   \multicolumn{3}{c}{$u u \rightarrow u u$, $d d \rightarrow d d$, $\bar{u} \bar{u} \rightarrow \bar{u} \bar{u}$, $\bar{d} \bar{d} \rightarrow \bar{d} \bar{d}$} \\
   \hline
   SM QCD & $\frac{11+34\cos^2 \theta^* +3\cos^4 \theta^*}{3(1-\cos^2 \theta^*)^4}$ & $\frac{2(3+2\chi+\chi^2+2\chi^3+3\chi^4)}{3\chi^2 (1+\chi)^2}$ \\
   $O_{qq}^{(1)}$, $O_{qq}^{(3)}$, $O_{uu}$, $O_{dd}$ & -$\frac{2}{1-\cos^2 \theta^*}$ & -$\frac{1}{\chi}$ \\
   $O_{ud}^{(8)}$ & 0 & 0 \\
   $O_{qu}^{(8)}$, $O_{qd}^{(8)}$ & -$\frac{1+3\cos^2 \theta^*}{4(1-\cos^2 \theta^*)}$ & -$\frac{1-\chi+\chi^2}{2 \chi (1+\chi)^2}$ \\
   \hline
   \multicolumn{3}{c}{$u d \rightarrow u d$, $\bar{u} \bar{d} \rightarrow \bar{u} \bar{d}$} \\
   \hline
   SM QCD & $\frac{5+2\cos\theta^* +\cos^2 \theta^*}{2(1-\cos\theta^*)^2}$ & $\frac{1+2\chi+2\chi^2}{(1+\chi)^2}$ \\
   $O_{qq}^{(1)}$, $O_{uu}$, $O_{dd}$ & 0 & 0 \\
   $O_{qq}^{(3)}$ & -$2\frac{1}{1-\cos\theta^*}$ & -$2\frac{1}{1+\chi}$ \\
   $O_{ud}^{(8)}$ & -$\frac{1}{4}\frac{1}{1-\cos\theta^*}$ & -$\frac{1}{4}\frac{1}{1+\chi}$ \\
   $O_{qu}^{(8)}$, $O_{qd}^{(8)}$ & -$\frac{(1+\cos\theta^*)^2}{16(1-\cos\theta^*)}$ & -$\frac{\chi^2}{4(1+\chi)^3}$ \\
   \hline
   \multicolumn{3}{c}{$u \bar{d} \rightarrow u \bar{d}$, $d \bar{u} \rightarrow d \bar{u}$} \\
   \hline
   SM QCD & $\frac{5+2\cos\theta^* +\cos^2 \theta^*}{2(1-\cos\theta^*)^2}$ & $\frac{1+2\chi+2\chi^2}{(1+\chi)^2}$ \\
   $O_{qq}^{(1)}$, $O_{uu}$, $O_{dd}$ & 0 & 0 \\
   $O_{qq}^{(3)}$ & -$\frac{1}{2}\frac{(1+\cos\theta^*)^2}{1-\cos\theta^*}$ & -$2\frac{\chi^2}{(1+\chi)^3}$ \\
   $O_{ud}^{(8)}$ & -$\frac{1}{16}\frac{(1+\cos\theta^*)^2}{1-\cos\theta^*}$ & -$\frac{1}{4}\frac{\chi^2}{(1+\chi)^3}$ \\
   $O_{qu}^{(8)}$, $O_{qd}^{(8)}$ & -$\frac{1}{4(1-\cos\theta^*)}$ & -$\frac{1}{4(1+\chi)}$ \\
   \hline
   \multicolumn{3}{c}{$u \bar{u} \rightarrow u \bar{u}$, $d \bar{d} \rightarrow d \bar{d}$} \\
   \hline
   SM QCD & $\frac{82-76\cos\theta^*+47\cos 2\theta^*-8\cos 3\theta^* +3\cos 4\theta^*}{12(1-\cos\theta^*)^2}$ & $\frac{27+12\chi+29\chi^2+2\chi^3+6\chi^4}{3(1+\chi)^4}$\\
   $O_{qq}^{(1)}$, $O_{qq}^{(3)}$, $O_{uu}$, $O_{dd}$ & -$\frac{(1+\cos\theta^*)^3}{8(1-\cos\theta^*)}$ & -$\frac{\chi^3}{(1+\chi)^4}$ \\
   $O_{ud}^{(8)}$ & 0 & 0 \\
   $O_{qu}^{(8)}$, $O_{qd}^{(8)}$ & -$\frac{8-(1-\cos\theta^*)^3}{16 (1-\cos\theta^*)}$ & -$ \frac{\chi (3+3\chi+\chi^2)}{2(1+\chi)^4}$ \\
   \hline
\end{tabular}
}
%\end{tabularx}
\end{table}

\subsection{Phase-space cuts and PS effects}
We recast the CMS analysis \cite{2j_2018}, where two upper cuts $\chi_{jj} < 16$ and $|y_\text{boost}| < 1.11$ are imposed; the last quantity is defined from the rapidities of the two leading jets as $y_\text{boost} = |y_{j1}+y_{j2}| /2$. They investigate different bins in the invariant mass of these two jets: we focused on the least and most energetic ones, respectively $2.4 < M_{jj} < 3$ TeV and $M_{jj} > 6$ TeV, as the behaviour of the intermediate ones can be extrapolated from those two. This is the same experimental analysis that is reproduced in \cite{Goldouzian_2020} to obtain bounds on $C_G/\Lambda^2$ through $\chi_{jj}$.

The {\sc CT18} PDF set was employed \cite{ct18pdf}. PS algorithms generally have troubles reproducing multijet measurements because scales drop quickly for these processes and colour reconnection spreads the simulated radiation at large angles, especially for the less-energetic jets \cite{2j_NLO}. A consistent fraction of events is thus lost when the above requirements are applied after PS and the generation cuts at parton level need to be strongly relaxed to account for this. Even if the cross section before matching and merging is not physical \cite{nlo_events}, we checked that its ratio with respect to the one after PS is roughly constant for all the 4LQ interference terms, meaning that the events that are lost do not contain specific flavours in the final states and that the PDF-induced kinematic discrepancies do not affect the rejection.

The total cross sections that the CMS Collaboration predicted for the SM and measured with their detector in \cite{2j_2018} are not public, so we assumed for both the LO SM value we obtained through \amc\ and we used that result to rescale the distributions presented in the same analysis, that are normalised to one. These numbers are shown in Table \ref{tab:4lq_xsects} for the two $M_{jj}$ regions we considered.

%\begin{sidewaystable} 
\begin{table}
\caption{\footnotesize{Cross sections for the contributions of the SM and its interference with the 4LQ operators, whose coefficients are set to 1 TeV${}^{-2}$, to the processes investigated in this analysis. The dim.-6 squared contributions to the $\mathcal{O}(1/\Lambda^4)$ term for $O_{qq}^{(3)}$ are also shown. The units are not the same in each column and can be found at the top. The 4LQ cross sections are computed at LO with {\sc MLM} and PS, and the first relative uncertainty is numerical, while the following numbers are scale variations. For the SM, we report the values from the respective experimental analyses for $Z+$jets and $\gamma +$jets, and the ones we obtained in our LO generations for $W+$jets and multijets. The experimentally-measured cross sections are also shown, when available, with their cumulative uncertainties}} \label{tab:4lq_xsects}
\resizebox{\textwidth}{!}{
\begin{tabular}{c|ccccc}
   %\hline
    & \multicolumn{2}{c}{Multijets} & $\ell^+ \ell^- +$jets & $\ell^\pm \nu +$jets & $\gamma +$jets \\
    & $2.4<M_{jj}<3$ TeV (pb) & $M_{jj}>6$ TeV {\bf (fb)} & {\bf (fb)} & (pb) & {\bf (fb)} \\ \hline \hline
   Exp. & Not avail. & Not avail. & 3.7$\cdot 10^1 {}^{+11\%}_{-11\%}$ & Not avail. & 2.3$\cdot 10^4 {}^{+5\%}_{-5\%}$ \\
   SM & 6.6$\pm 0.6\%^{+15\%}_{-29\%}$ & 2.7$\pm 0.5\%^{+26\%}_{-30\%}$ & 3.95$\cdot 10^1 {}^{+9\%}_{-9\%}$ & 4.2$\cdot 10^1\pm 0.5\%^{+10\%}_{-17\%}$ & 2.6$\cdot 10^4 {}^{+31\%}_{-31\%}$ \\ \hline
   & \multicolumn{5}{c}{$\mathcal{O}(1/\Lambda^2)$} \\ %\hline
   $O_{qq}^{(1)}$ & -5.9$\pm 0.7\%^{+27\%}_{-17\%}$ & -3.7$\cdot 10^1 \pm 0.5\%^{+35\%}_{-27\%}$ & -5.12$\pm 0.8\%^{+21\%}_{-14\%}$ & -1.3$\pm 0.8\%^{+31\%}_{-23\%}$ & -4.0$\cdot 10^2 \pm 0.3\%^{+13\%}_{-18\%}$ \\
   $O_{qq}^{(3)}$ & -1.8$\cdot 10^1 \pm 0.6\%^{+22\%}_{-11\%}$ & -7$\cdot 10^1 \pm 0.4\%^{+43\%}_{-23\%}$ & -3.7$\cdot 10^1\pm 0.5\%^{+16\%}_{-11\%}$ & -6.1$\pm 0.5\%^{+18\%}_{-11\%}$ & -1.9$\cdot 10^3\pm 0.15\%^{+9\%}_{-11\%}$ \\
   $O_{uu}$ & -4.4$\pm 0.5\%^{+16\%}_{-14\%}$ & -2.7$\cdot 10^1 \pm0.4\%^{+37\%}_{-26\%}$ & -3.8$\cdot 10^{-1}\pm 0.5\%^{+18\%}_{-8\%}$ & $\times$ & -3.7$\cdot 10^2 \pm 0.16\%^{+15\%}_{-14\%}$ \\
   $O_{dd}$ & -9.4$\cdot 10^{-1} \pm 0.5\%^{+16\%}_{-14\%}$ & -1.7$\pm 0.5\%^{+41\%}_{-24\%}$ & -3.3$\cdot 10^{-2}\pm 0.5\%^{+18\%}_{-9\%}$ & $\times$ & -3.9$\cdot 10^1 \pm 0.18\%^{+10\%}_{-13\%}$ \\
   $O_{ud}^{(1)}$ & $\times$ & $\times$ & 1.3$\cdot 10^{-2}\pm 0.6\%^{+15\%}_{-31\%}$ & $\times$ & 1.5$\cdot 10^1 \pm 0.2\%^{+27\%}_{-13\%}$ \\
   $O_{ud}^{(8)}$ & -1.03$\pm 0.6\%^{+16\%}_{-14\%}$ & -3.2$\pm 0.4\%^{+37\%}_{-25\%}$ & -1.02$\cdot 10^{-1}\pm 0.6\%^{+22\%}_{-14\%}$ & $\times$ & -1.13$\cdot 10^2 \pm0.18\%^{+8\%}_{-12\%}$ \\
   $O_{qu}^{(1)}$ & $\times$ & $\times$ & -5.6$\cdot 10^{-2}\pm 0.7\%^{+18\%}_{-13\%}$ & -1.1$\cdot 10^{-2}\pm 1.2\%^{+63\%}_{-46\%}$ & -1.3$\cdot 10^1 \pm 0.3\%^{+23\%}_{-31\%}$ \\
   $O_{qu}^{(8)}$ & -2.0$\pm 0.5\%^{+15\%}_{-15\%}$ & -6.9$\pm 0.4\%^{+12\%}_{-26\%}$ & -9.3$\cdot 10^{-1}\pm 0.5\%^{+20\%}_{-12\%}$ & -1.9$\cdot 10^{-1}\pm 0.5\%^{+21\%}_{-16\%}$ & -2.1$\cdot 10^2\pm 0.2\%^{+14\%}_{-24\%}$ \\
   $O_{qd}^{(1)}$ & $\times$ & $\times$ & 1.5$\cdot 10^{-2}\pm 1.3\%^{+20\%}_{-27\%}$ & 1.1$\cdot 10^{-3}\pm 9\%^{+270\%}_{-360\%}$ & 2.0$\pm 0.5\%^{+60\%}_{-60\%}$ \\
   $O_{qd}^{(8)}$ & -1.1$\pm 0.5\%^{+27\%}_{-15\%}$ & -2.4$\pm 0.5\%^{+42\%}_{-25\%}$ & -4.8$\cdot 10^{-1}\pm 0.6\%^{+21\%}_{-13\%}$ & -1.5$\cdot 10^{-1}\pm 0.5\%^{+27\%}_{-13\%}$ & -8.0$\cdot 10^1\pm 0.2\%^{+15\%}_{-24\%}$ \\
   \hline
   & \multicolumn{5}{c}{Partial $\mathcal{O}(1/\Lambda^4)$} \\ %\hline
   $O_{qq}^{(3)}$ & 1.0$\cdot 10^2 \pm 0.9\%^{+20\%}_{-10\%}$ & 1.9$\cdot 10^3 \pm 0.6\%^{+48\%}_{-20\%}$ & 9.8$\cdot 10^1 \pm 1.2\%^{+10\%}_{-16\%}$ & 1.8$\cdot 10^1 \pm 0.6\%^{+7\%}_{-11\%}$ & 6.2$\cdot 10^3 \pm 0.4\%^{+8\%}_{-8\%}$ \\
   \hline
\end{tabular}
}
\end{table}
%\end{sidewaystable}

\subsection{Results and distributions}
The total LO SM and linear cross sections for all the operators that contribute are summarised in Table \ref{tab:4lq_xsects}, with the relative numerical and scale-variation uncertainties, in the two $M_{jj}$ regions specified above. %By comparing them, it is possible to acknowledge a cross-section increase with the energy for all the 4LQ operators.

\begin{figure}
   \centering
   \caption{\footnotesize{Differential ({\it top}) and normalised ({\it bottom}) distributions of the exponential of the azimuthal distance among the two leading jets in multijet production, for the SM ({\it black}) and the contributing operators included in this analysis, with all the coefficients set to $C_i/\Lambda^2 = 1$ TeV${}^{-2}$. Two dijet invariant-mass regions are shown: [2.4, 3] TeV ({\it left}) and [6, 13] TeV ({\it right}). The numerical uncertainties are represented with shaded bands in both plots, while the scale variations are shown in hatched bands in both plots for the SM, and only on top for the 4LQ operators. Note that the cross-section unit is not the same in the two top plots. The experimental measurements are also included in the bottom plots}} \label{fig:4lq_2j_chijj}
   \includegraphics[width=.49\textwidth]{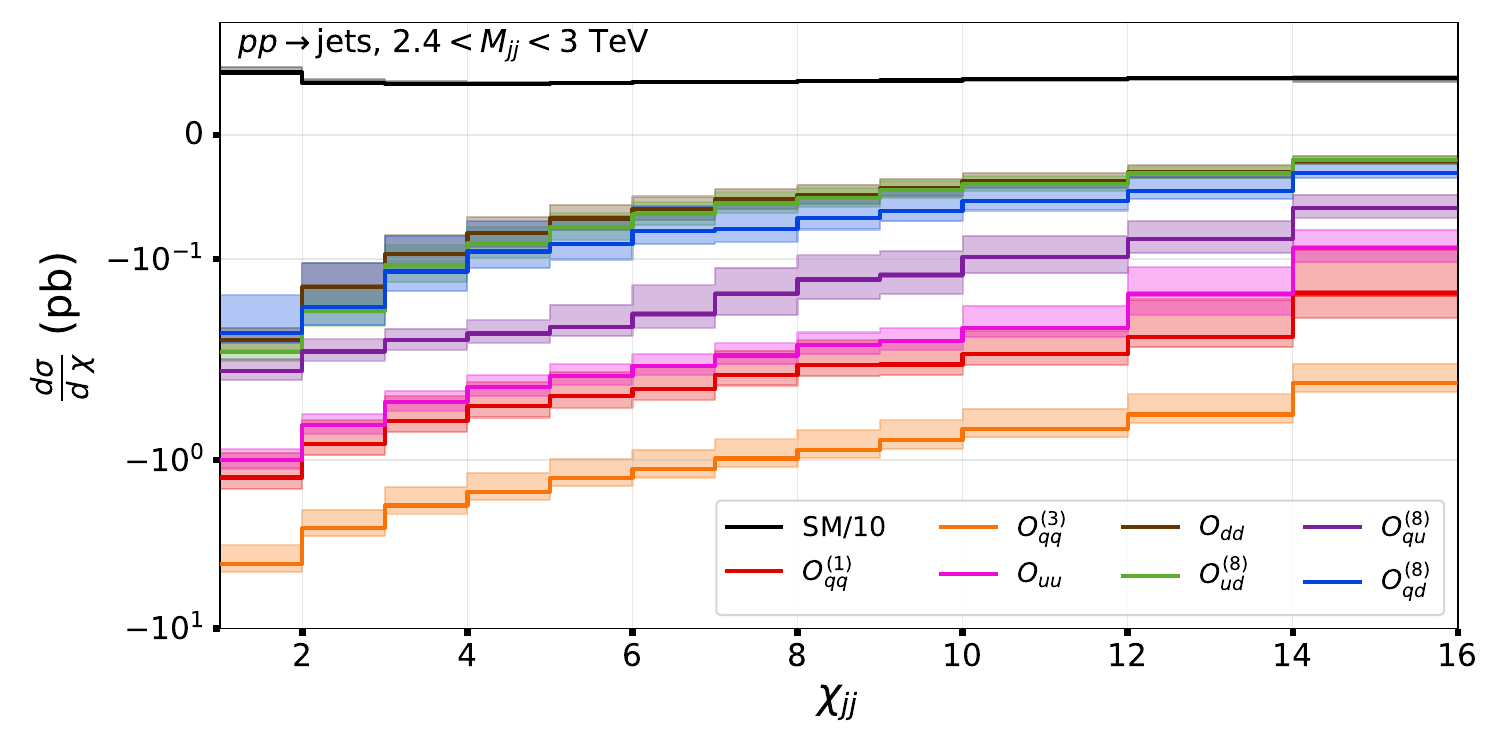}
   \includegraphics[width=.49\textwidth]{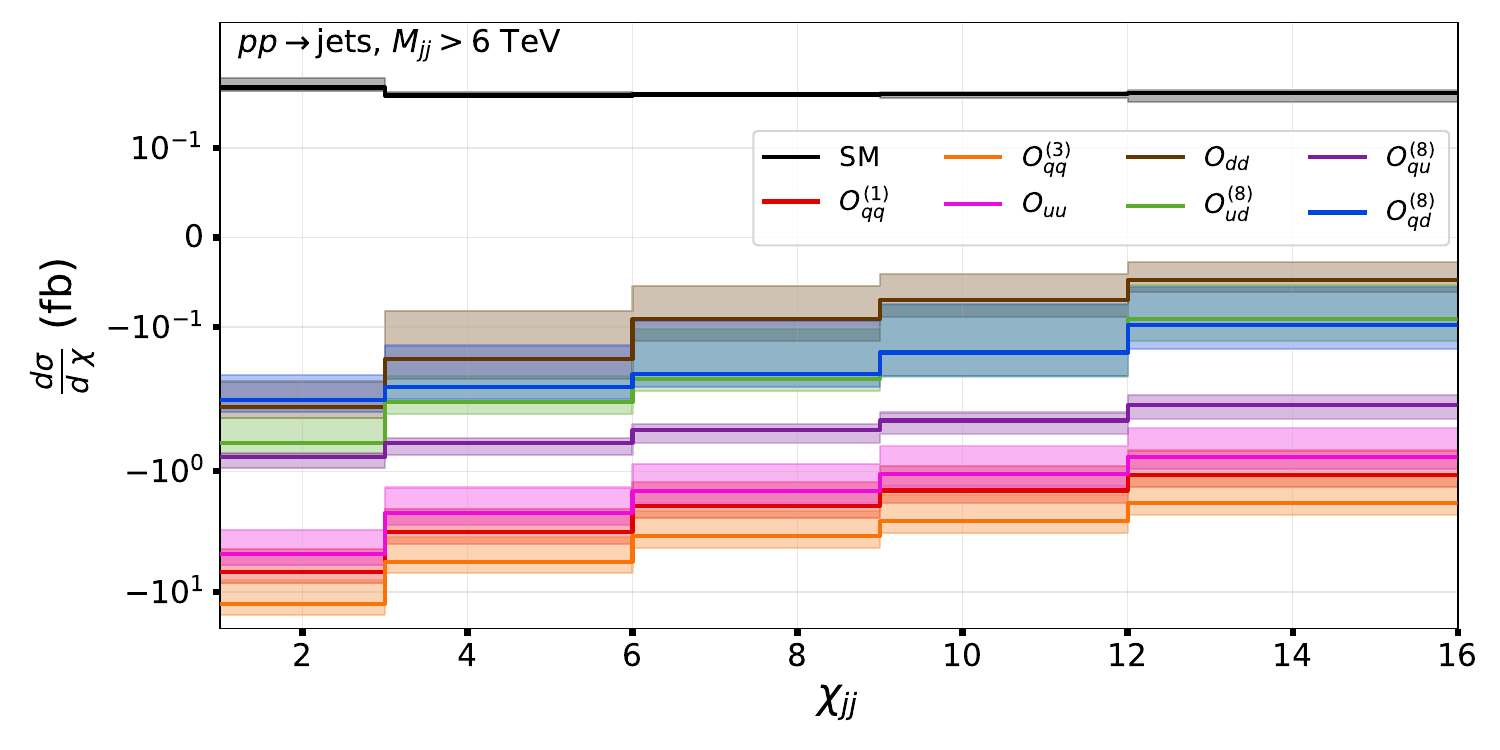}
   \includegraphics[width=.49\textwidth]{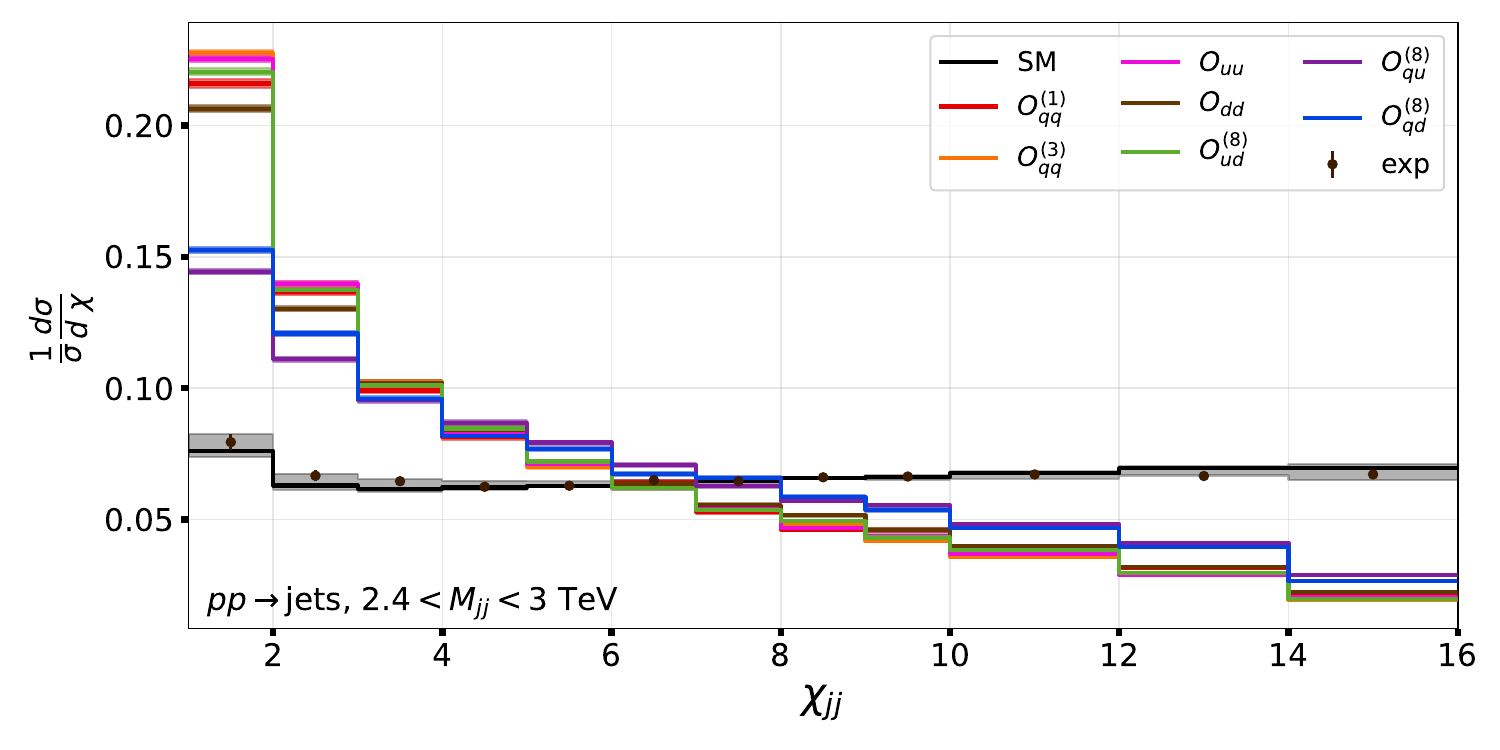}
   \includegraphics[width=.49\textwidth]{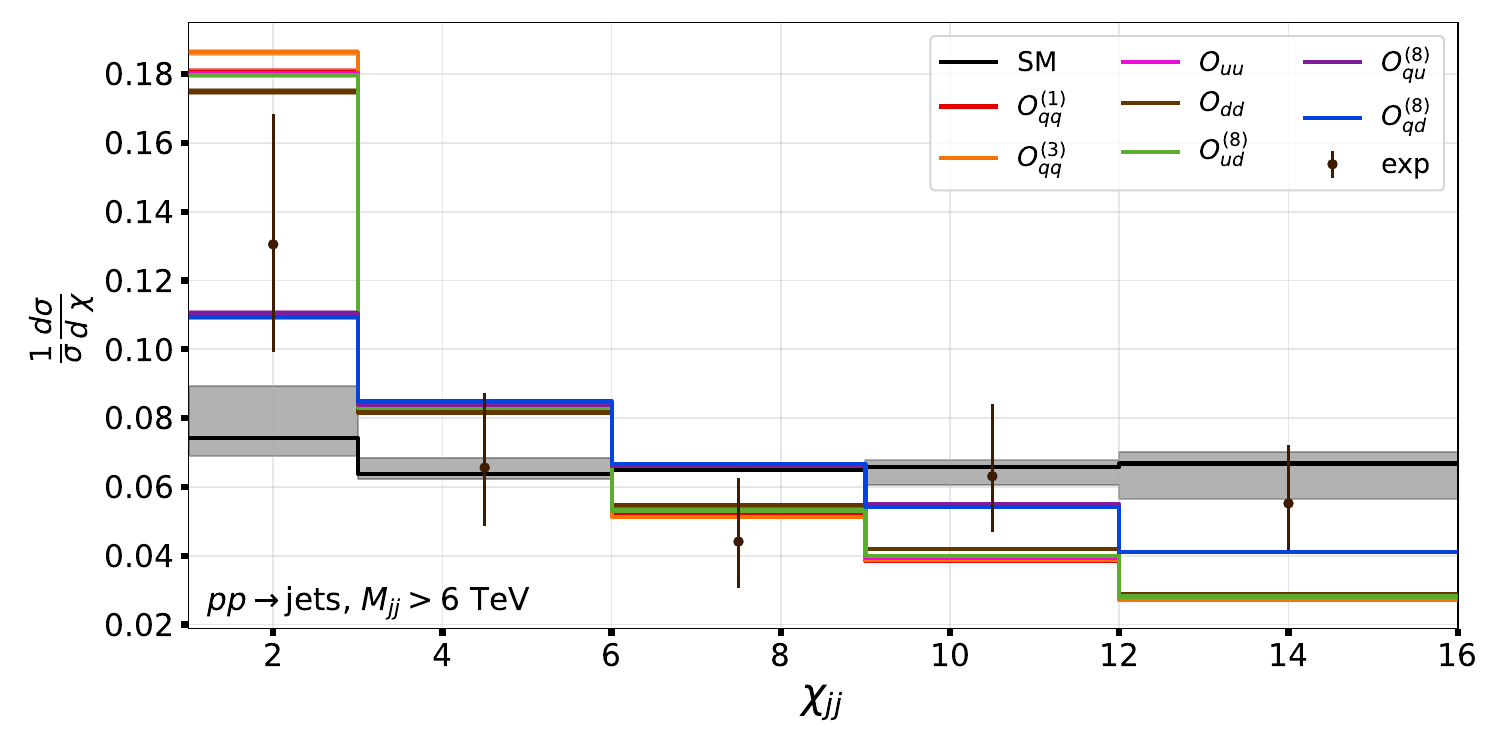}
   \caption{\footnotesize{Differential distributions of the transverse momentum of the $b$-tagged jets in the $b+$jets region for multijet production, for the SM ({\it black}) and the contributing operators at LO, with all the coefficients set to $C_i/\Lambda^2 = 1$ TeV${}^{-2}$. The numerical uncertainties are represented with shaded bands, while the scale variations are shown in hatched bands for the SM only. The last bin includes the overflow}} \label{fig:4lq_bj_pTb}
   \includegraphics[width=.7\textwidth]{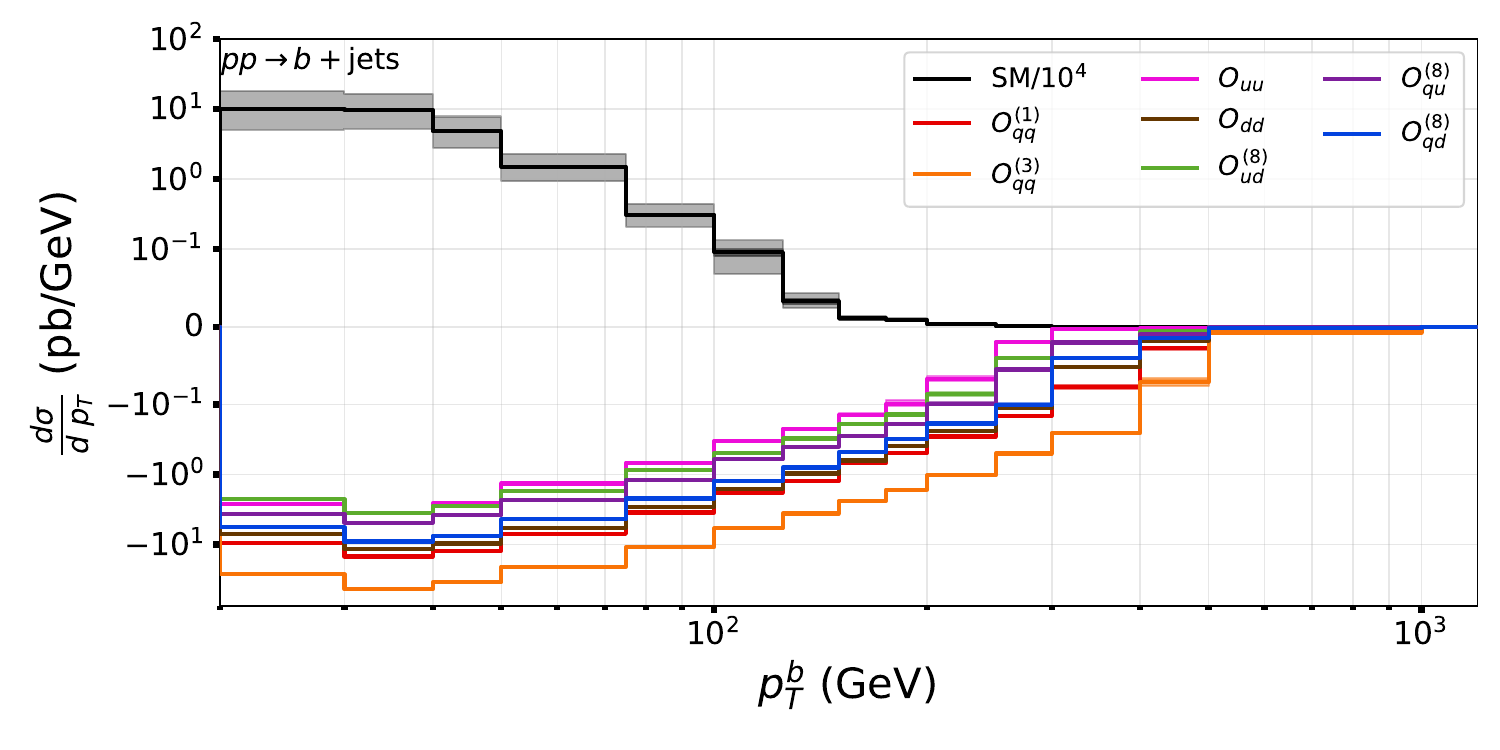}
\end{figure}

The normalised and differential distributions for $\chi_{jj}$ in the same $M_{jj}$ intervals are shown in Fig. \ref{fig:4lq_2j_chijj}. It can be seen that the two different shapes predicted at parton level for the main subprocess $uu \rightarrow uu$ are still visible: one more peaked at $\chi_{jj} \sim 1$ for the operators in \eqref{4lq_shape_a} and another more flat and closer to the SM for the \eqref{4lq_shape_b} group. This reinforces the statement above about PS effects and suggests that the consequence of PDF is only to favour different flavour mixes at different momentum fractions. Moreover, it shows that only two directions can be probed in the coefficient space by this process, independently of the number of bins.

The SM differential distributions we used are the ones generated through {\sc NLOJET++} in the experimental analysis \cite{nlojet}, rescaled to the total LO SM cross section from {\sc MadGraph}. The ones we obtained show lower tails than the normalised ones from the reference: this discrepancy might be bridged with a full NLO calculation, but we did not check if this is the case. The relative uncertainties vary up to 12\% and to 30\% in the two invariant-mass ranges, respectively.

For what concerns the experimental measurements, the errors on them lie between 1 and 8\% in the lowest-$M_{jj}$ region and between 50 and 70\% in the highest one.

If all the Wilson coefficients are equal, the largest correction to the SM comes from $O_{qq}^{(3)}$, that is the only operator that contributes to all subprocesses. Scale variations for its distribution lie around 35\% in the [2.4, 3] TeV interval and 60\% in the [6, 13] TeV one.

\subsection{Multijets with flavour-tagging}
Flavour-tagging of jets can increase the sensitivity to 4LQ operators containing only up- or down-like quarks. An example of the application of such algorithms to $Z+$jets production in the SM can be found in \cite{zjj_flavour_tag}.

We simulated the DL1r algorithm for $b$- and $c$-tagging of jets \cite{bc_tagging} and applied it to the simpler multijet case. If a reconstructed jet is within $\Delta R = 0.4$ from a MC-true $b$-parton, it is randomly tagged with an efficiency dependent on its $p_T$. These values were digitised from Figures 11 and 12 of the reference for the 20-250 and 250-3000 GeV ranges, in the 77\% working point: they slightly decrease with the $p_T$ up to 2 TeV, then drop. The pseudorapidity and luminosity were not taken into account, as multidifferential efficiencies are not available; in general, the latter might affect the efficiencies through pile-up, but the behaviour of this algorithm seems stable with respect to it. $c$-jets and light-flavour ones can be mistagged as $b$-jets depending on their $p_T$ and with rejection rates that we took from the same plots.

We followed a similar strategy for the $c$-tagging, but with a probability fixed at 30\% for all $p_T$ values. This procedure was applied only to jets with a transverse momentum between 20 and 250 GeV. Also $b$- and light-jets mistagging rates were constant, and we set them to the values in Fig. 16 of the reference for the 30\% $c$-tagging efficiency.

Simultaneous $b$- and $c$-tagging is possible for the same jet in this simulated method: since the full output of the DL1r net was not available, we checked the MC truth in those cases, or performed a random extraction in case of doubly-mistagged light jets.

Because of the different typical energies of $b$- and $c$-jets, we relaxed the cuts for this analysis with respect to the previous paragraphs: jets were included if they presented $p_T^j > 20$ GeV and $|y^j| < 2.5$. The operational windows of the two algorithms need to be taken into account, too: jets with $p_T > 250$ (3000) GeV cannot be $c$- ($b$-) tagged. The experimental data was supposed to follow the LO SM distribution that we generated.

Thanks to this set-up, we studied $b+$jets, $c+$jets and $bb+$jets productions, where at least one $b$-jet, one $c$-jet or two $b$-jets had to be identified, respectively. The total LO cross sections for the three cases are shown in Table \ref{tab:4lq_bc_xsects}: it can be seen that scale uncertainties are large and that the SM term is at least three orders of magnitude larger than the $O_{qq}^{(3)}$ one, that induces the largest linear correction if all coefficients are equal. This is a consequence of the relatively low $p_T$ scales where the tagging algorithms operate. These results show that the jet mistagging does not allow cancelling the contributions of operators affecting only up- or down-like quarks to these processes, like $O_{uu}$ when at least a $b$-jet is required.

\begin{table}
\centering
\caption{\footnotesize{Cross sections, in pb, for multijet production when at least one $b$-jet, at least one $c$-jet and at least two $b$-jets are tagged, at LO matched to PS. The SM and 4LQ interference values are reported. The first uncertainty source is numerical, while the following ones are from scale variation}} \label{tab:4lq_bc_xsects}
\begin{tabular}{c|ccc}
%\hline
 & $b+$jets & $c+$jets & $bb+$jets \\ \hline
SM & 2.9$\cdot 10^6 \pm 0.3\%^{+69\%}_{-44\%}$ & 1.9$\cdot 10^6 \pm 1.1\%^{+57\%}_{-42\%}$ & 2.5$\cdot 10^6 \pm 3\%^{+54\%}_{-35\%}$ \\
$O_{qq}^{(1)}$ & -8$\cdot 10^2 \pm 0.6\%^{+38\%}_{-62\%}$ & -4$\cdot 10^2 \pm 1.0\%^{+40\%}_{-50\%}$ & 1$\cdot 10^1 \pm 17\% ^{+400\%}_{-400\%}$ \\
$O_{qq}^{(3)}$ & -2.3$\cdot 10^3 \pm0.7\%^{+43\%}_{-65\%}$ & -1.4$\cdot 10^3 \pm 0.9\% ^{+43\%}_{-57\%}$ & 7$\cdot 10^1 \pm 7\%^{+142\%}_{-142\%}$ \\
$O_{uu}$ & -1.6$\cdot 10^2 \pm 1.1\%^{+44\%}_{-63\%}$ & -3.7$\cdot 10^2 \pm 0.8\%^{+41\%}_{-54\%}$ & -2$\pm 15\%^{+100\%}_{-100\%}$ \\
$O_{dd}$ & -6$\cdot 10^2 \pm 0.5\%^{+50\%}_{-67\%}$ & -3$\cdot 10^1 \pm 3\%^{+67\%}_{-67\%}$ & 1$\cdot 10^1 \pm 10\%^{+400\%}_{-400\%}$ \\
$O_{ud}^{(8)}$ & -1.9$\cdot 10^2 \pm 0.7\%^{+42\%}_{-63\%}$ & -1.3$\cdot 10^2 \pm 0.8\%^{+38\%}_{-62\%}$ & 7$\pm 6\%^{+114\%}_{-114\%}$ \\
$O_{qu}^{(8)}$ & -2.6$\cdot 10^2 \pm 0.8\%^{+42\%}_{-62\%}$ & -2.9$\cdot 10^2 \pm 0.7\%^{+41\%}_{-59\%}$ & 6 $\pm 10\%^{+150\%}_{-150\%}$ \\
$O_{qd}^{(8)}$ & -5$\cdot 10^2 \pm 0.6\%^{+20\%}_{-60\%}$ & -1.3$\cdot 10^2 \pm 1.2\%^{+46\%}_{-62\%}$ & 1$\cdot 10^1 \pm 10\%^{+300\%}_{-300\%}$ \\
\hline
\end{tabular}
\end{table}

The differential distributions for the transverse momentum $p_T^b$ of the $b$-tagged jets in the $b+$jets case is shown in Fig. \ref{fig:4lq_bj_pTb}. Scale variations are large for the predictions: they lie above 60\% for the SM and between 50 and 80\% for $O_{qq}^{(3)}$. In both cases, the uncertainties reach above 100\% in the first bins.

%\begin{figure}
%   \caption{\footnotesize{Differential distributions of the transverse momentum of the $b$-tagged jets in the $b+$jets region for multijet production, for the SM ({\it black}) and the contributing operators at LO, with all the coefficients set to $C_i/\Lambda^2 = 1$ TeV${}^{-2}$. The numerical uncertainties are represented with shaded bands, while the scale variations are shown in hatched bands for the SM only. The last bin includes the overflow}} \label{fig:4lq_bj_pTb}
%   \begin{center}
%   \includegraphics[width=.7\textwidth]{figures/bj_pTb_diff.pdf}
%   \end{center}
%\end{figure}

%%%%%%%%%%%%%%%%%%%%%%%%% Zjj %%%%%%%%%%%%%%%%%%%%%%%%%

\section{$Z+$jets production through VBF}
This is the same process that we analysed in Sect. \ref{sec:Zjj_oW} for the $O_W$ study: we implemented the same details about the generation and limited ourselves to the same phase-space cuts as in the experimental setup, even if they might not be the most suitable ones for the 4LQ operators. 
In this case, though, we did not include NLO corrections, so we could simulate all the LO contributions without having to require only some particles in the $t$-channel. Also, we did not run it at FO but asked for up to three jets that we merged and showered in \pyth : because of the discrepancies among different PS algorithms discussed in the previous chapter, we still considered for the SM the distributions generated in the experimental analysis \cite{atlas:2021Zjj} through \hw +{\sc Vbfnlo}.

As in the other study, we focused on the signed azimuthal distance between the two leading jets, $\Delta \phi_{jj}$, to extract limits on the 4LQ coefficients. We also investigated the dijets invariant mass $M_{jj}$, their rapidity difference $|\Delta y_{jj}|$ and the dilepton-system transverse momentum $p_T^{\ell\ell}$.

\begin{figure}
   \centering
   \caption{\footnotesize{Differential ({\it top}) and normalised ({\it bottom}) distributions of the azimuthal distance between the two leading jets in $\ell^+ \ell^- +$jets production, for the SM ({\it black}) and the interference of the ten operators included in this analysis at LO, with all the coefficients set to $C_i/\Lambda^2 = 1$ TeV${}^{-2}$. The numerical uncertainties are represented with shaded bands in both plots, while the scale variations are shown in hatched bands in both plots for the SM, and only on top for the 4LQ operators. The experimental measurements are also included}} \label{fig:4lq_zjj_Dphijj}
   \includegraphics[width=.7\textwidth]{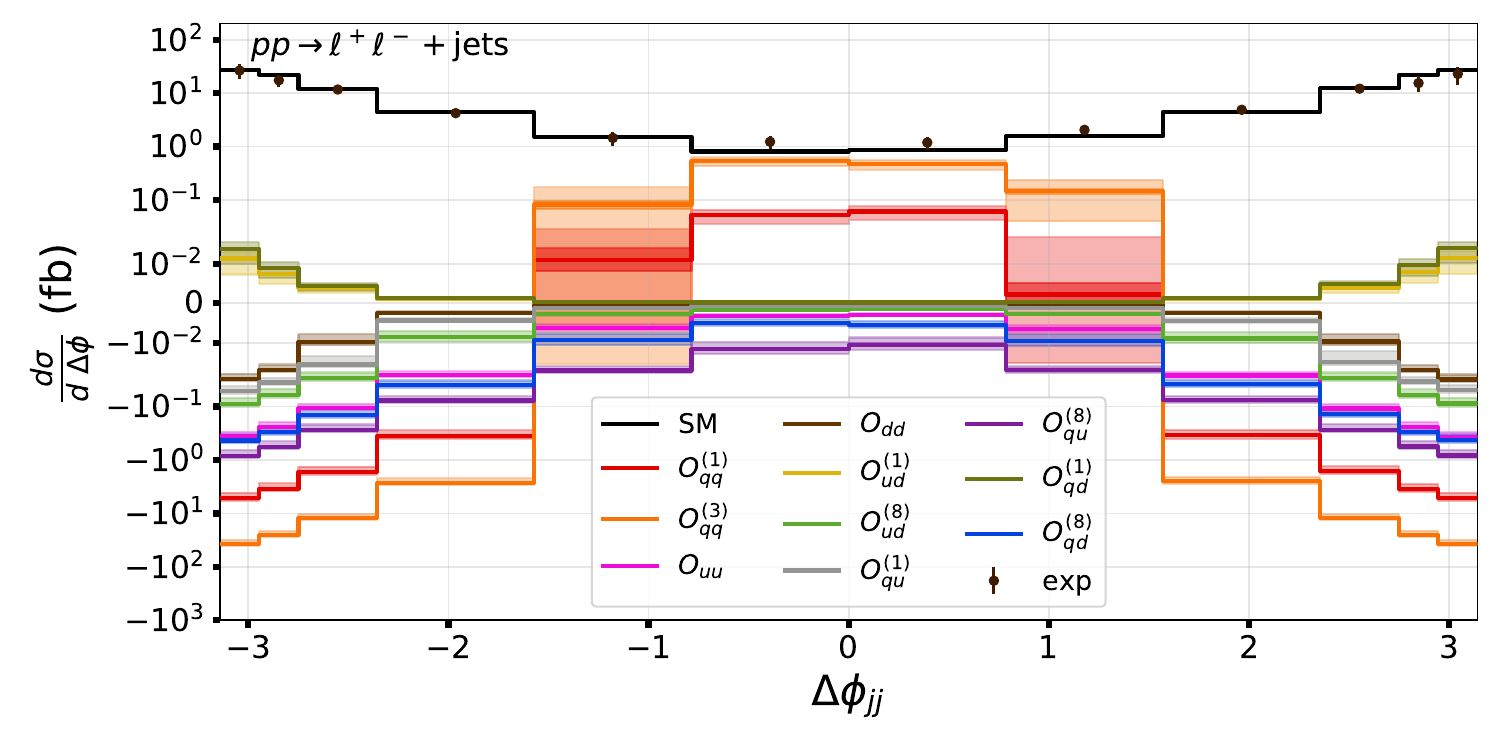}
   \includegraphics[width=.7\textwidth]{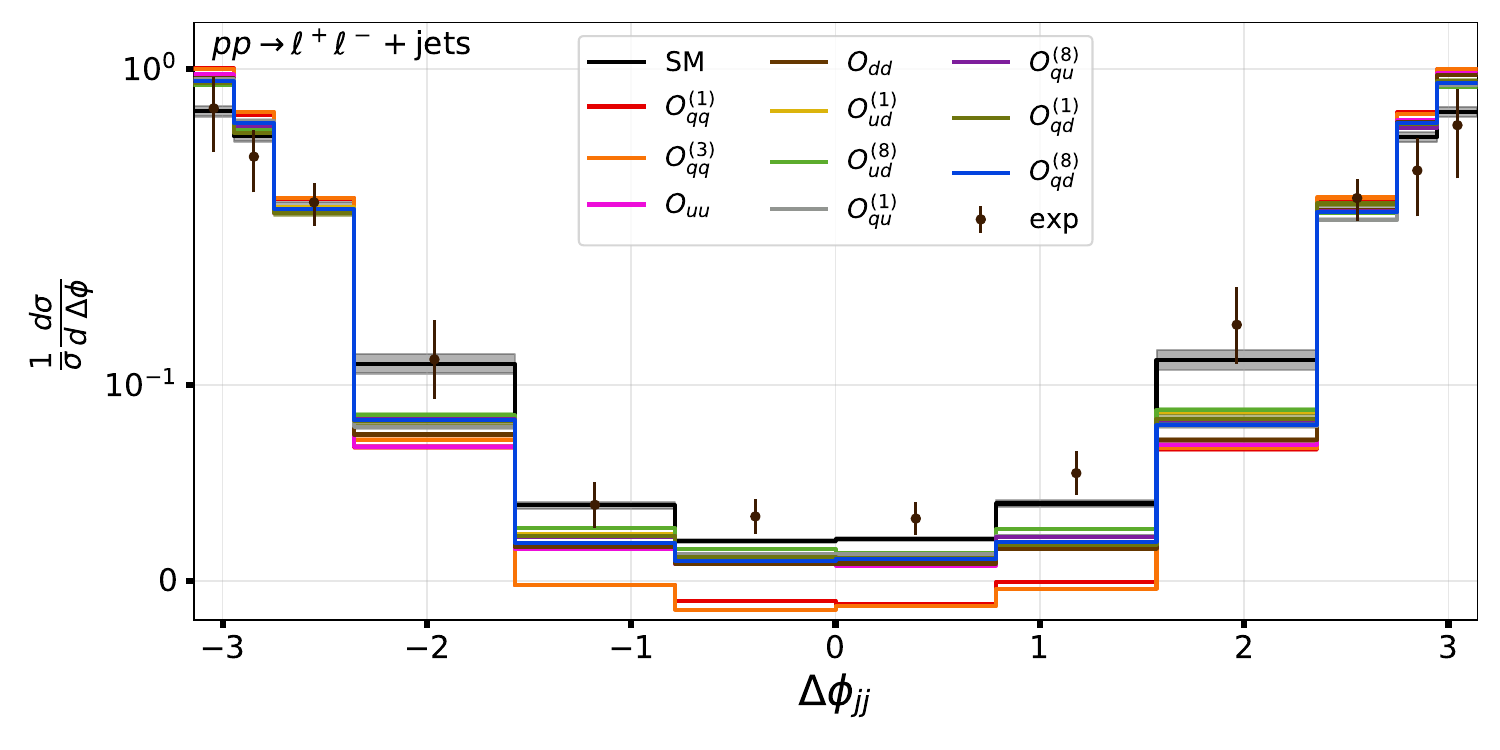}
\end{figure}

\subsection{Results and distributions}
The total cross sections for the SM and the 4LQ interference terms are summarised in Table \ref{tab:4lq_xsects}, together with the experimental measurement. As discussed in the previous chapter, the largest source of uncertainty in the latter is due to the differences in predictions from various MC generators, even for the SM distributions.

The differential and normalised histograms for the SM and linear 4LQ contributions to $\Delta \phi_{jj}$ are shown in Fig. \ref{fig:4lq_zjj_Dphijj}. As in multijet production, $O_{qq}^{(3)}$ induces the largest deviation from the SM. Most of the interference terms have the same sign if all coefficients are equal, but a flip occurs in the central bins for $O_{qq}^{(1)}$ and $O_{qq}^{(3)}$. The SM results from ATLAS present uncertainties around 9\% in all bins, while the experimental ones lie between 10\% in the central bins to 30\% in the external ones. Our $O_{qq}^{(3)}$ predictions show scale variations of order 25\% all over the variable range, except for the bins in which the sign change occurs, where they raise to more than 100\%.

%%%%%%%%%%%%%%%%%%%%%%%%% Wjj %%%%%%%%%%%%%%%%%%%%%%%%%

\section{$W+$jets production through VBF}
The VBF regime for $W$-boson production in association with jets can bring insights about the SM and parton radiation, but also anomalous triple-gauge couplings \cite{bsm_wjj_1,bsm_wjj_2,bsm_wjj_3}. In our study, we focused on the leptonic decay of the EW boson, $W^\pm \rightarrow \ell^\pm \nu$. As anticipated in the introduction to this chapter, because of the left-handed nature of the interaction, $O_{uu}$, $O_{dd}$, $O_{ud}^{(1)}$ and $O_{ud}^{(8)}$ do not contribute to this process.

We checked observables like $p_T^{j1}$, $p_T^{j2}$, $M_{jj}$, $|\Delta y_{jj}|$ and $\Delta \phi_{jj}$ of the two leading jets, the angular distances $\Delta R_{W,j1}$ and $\Delta R_{\ell, j1}$ of the leading jet with the reconstructed $W$-boson and lepton, the lepton transverse momentum $p_T^\ell$, the transverse mass $M_{T}^{\ell\nu}$ of the lepton-neutrino system, the triple product $\frac{(\vec{j}_1 \times \vec{j}_2)\cdot \vec{\ell}}{|\vec{j}_1 \times \vec{j}_2||\vec{\ell}|}$ and the azimuthal angle $\phi_W$ defined in Eq. \eqref{phiV_def}. As for $Z+$jets, we found that the best bounds come from $\Delta \phi_{jj}$, as the 4LQ interference effects over the SM ones are larger for it.

The angular distance $\Delta R_{W,j1}$ is also an interesting observable, as it is sensitive to the jet chiralities. The individual bounds it yields are similar to the ones from $\Delta\phi_{jj}$. The $W$-boson reconstruction from the lepton and neutrino, though, induces some smearing in the distribution trend. This can be partially fixed by considering $\Delta R_{\ell, j1}$, but both these quantities suffer from slightly larger uncertainties on the SM predictions and lower new-physics effects over the SM ones, compared to $\Delta\phi_{jj}$.

\subsection{Phase-space cuts}
We followed the experimental analysis \cite{cms:2020wjj}, that requires the events to contain exactly one lepton with $p_T^\ell > 25$ GeV and $|\eta^\ell|<2.4$, together with at least two jets satisfying $M_{jj}>200$ GeV and with $p_T > 50$ and 30 GeV respectively. Any additional jets with $p_T^j > 25$ GeV and $|y^j|<5$ are also included in the analysis. Both $p_T^\text{miss}$ and $M_T^{\ell\nu}$ need to be above 40 GeV. {\sc NNPDF3.0} is chosen as PDF set.

The neutrino is reconstructed as for $WZ$ and $W\gamma$ productions in the previous chapter, following the procedure depicted in Sect. \ref{sec:WZ_oW_cuts}.

Different classes of diagrams contribute to the EW $Wjj$ generation at LO SM, and some of those induce large cancellations when interfering with the VBF ones of interest \cite{cms:2020wjj}. Because of this, a Boosted Decision Tree (BDT) is employed in the experimental analysis to isolate the different contributions. Since it is not possible to simulate the output of such algorithm, we generated the LO SM via \amc\ and assumed the measurements to follow the same behaviour. We introduced an uncertainty on the experimental distributions equal to 10\%, similarly to the most precise bins in the SM $Z+$jets $\Delta \phi_{jj}$ predictions, as the $W+$jets cross section is larger and smaller errors can be expected.

To reduce the scale-variation uncertainties, the SM generation should be performed at NLO, but this would require to deal with the same pole-cancellation issues described in Sect. \ref{sec:Zjj_oW} for $Zjj$. More stringent cuts on the leading-jets invariant mass and $p_T$ than the ones employed in the measurements would increase the new-physics effect over the SM one, both at linear and quadratic orders.

\subsection{Results and distributions}
The total cross sections are reported in Table \ref{tab:4lq_xsects}, showing that $O_{qq}^{(3)}$ is again inducing the largest effects among all the 4LQ operators, if all their coefficients are equal.

\begin{figure}
   \centering
   \caption{\footnotesize{Same as Fig. \ref{fig:4lq_zjj_Dphijj}, but for $W+$jets production at LO}} \label{fig:4lq_wjj_Dphijj}
   \includegraphics[width=.7\textwidth]{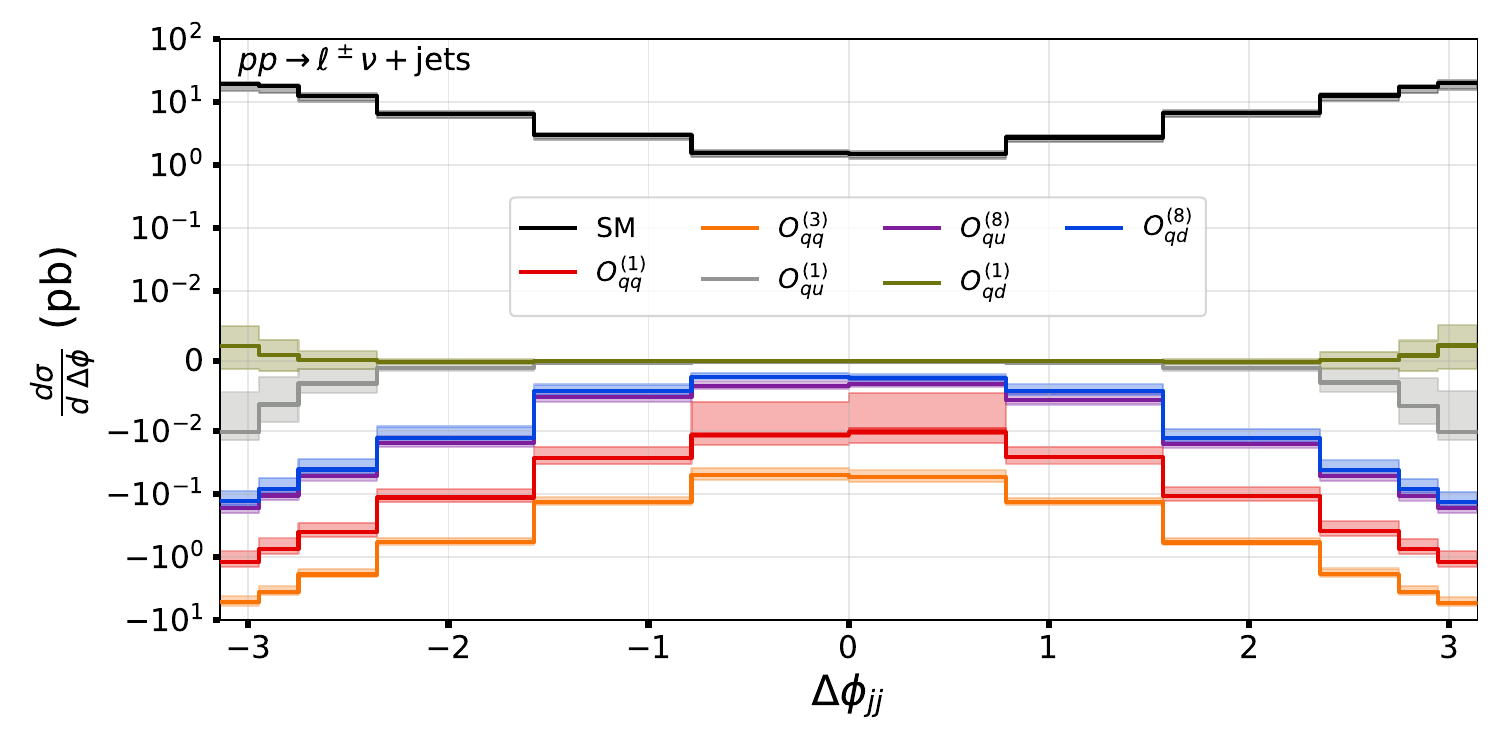}
   \includegraphics[width=.7\textwidth]{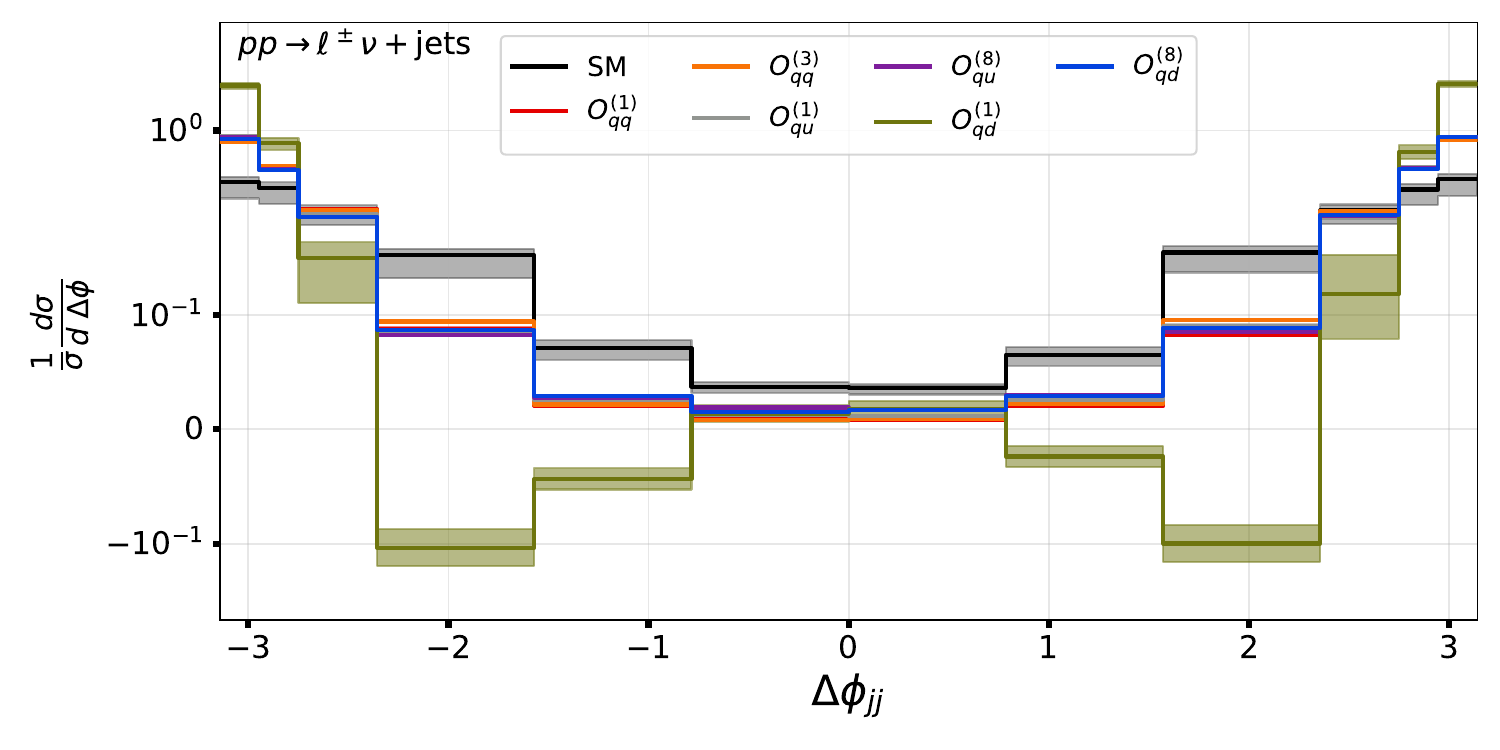}
\end{figure}

The differential and normalised distributions for $\Delta \phi_{jj}$ are shown in Fig. \ref{fig:4lq_wjj_Dphijj}, for the LO SM and the contributing operators at linear order. The scale-variation uncertainties for the SM lie between 25 and 30\%, while they spread between 25 and 50\% in the $O_{qq}^{(3)}$ distribution. $O_{qd}^{(1)}$ produces a different shape for this observable than the other operators, with sign changes occurring in some bins, but its effect is small compared to them.

%%%%%%%%%%%%%%%%%%%%%%%%% ajj %%%%%%%%%%%%%%%%%%%%%%%%%

\section{$\gamma+$jets production}
When a photon is produced in association with jets, it can be generated from the hard interaction or from the fragmentation of a high-$p_T$ parton. The first scenario is usually referred to as ``direct process'' and the second one as ``fragmentation process''; the latter can be distinguished from the decay into photons of a hadron inside the jets through isolation cuts \cite{Atlas:2020ajj}.

All the 4LQ operators can take part to this process, but $O_{ud}^{(1)}$, $O_{qu}^{(1)}$ and $O_{qd}^{(1)}$ do not interfere with the main SM contribution, the $t$-channel exchange of a gluon, so their effects are expected to be small \cite{smeft_global_mfv}.

We studied, over the whole phase space and for the two scenarios described above, the transverse momenta $p_T^\gamma$ and $p_T^j$, the jet rapidities $|y^j|$, the rapidity and azimuthal distances between the jets and the photon $|\Delta y_{jj}|$, $|\Delta \phi_{jj}|$, $|\Delta y_{\gamma j}|$ and $|\Delta \phi_{\gamma j}|$, and the invariant masses $M_{jj}$ and $M_{\gamma jj}$. Previous analyses \cite{smeft_global_mfv,Atlas:2024ajj} suggested the dijet invariant mass $M_{jj}$ to investigate this process, as it is more sensitive to the dynamics of the hard interaction. Our simulations showed instead that the photon transverse momentum presents larger 4LQ effects compared to the SM ones, by at least a factor four in the tail of its distribution. For this reason, we used $p_T^\gamma$ to extract bounds over the coefficients.

\subsection{Phase-space cuts}
We delimited the phase space as in the experimental analysis \cite{Atlas:2020ajj}: at least one photon with $p_T^\gamma > 150$ GeV and $|y^\gamma|< 2.37$ needed to be present, in combination with at least two jets with minimum $p_T^j$ of 100 GeV and $|\eta^j|<2.5$. The angular separation $\Delta R_{\gamma j}$ among the leading photon and the leading jets had to exceed 0.8. Other jets were included only if they showed $p_T > 20$ GeV and $|y|<2.5$. In addition to these cuts, that define the inclusive phase-space area, two more were used to investigate the two topologies described above: $p_T^\gamma > p_T^{j1}$ for a direct-enriched region and $p_T^\gamma < p_T^{j2}$ for a fragmentation-enriched one. {\sc NNPDF3.0} was used to model PDF effects.

The SM prediction that we employed are the ones presented in the reference, that are modelled with {\sc Sherpa} \cite{sherpa}.

\subsection{Results and distributions}
As for the previous processes, the total cross sections for the SM and linear terms are reported in Table \ref{tab:4lq_xsects}, and $O_{qq}^{(3)}$ leads the deviations from the SM.

\begin{figure}
   \centering
   \caption{\footnotesize{As in Fig. \ref{fig:4lq_zjj_Dphijj}, but for the photon transverse momentum in $\gamma+$jets production}} \label{fig:4lq_ajj_pTa}
      \includegraphics[width=.7\textwidth]{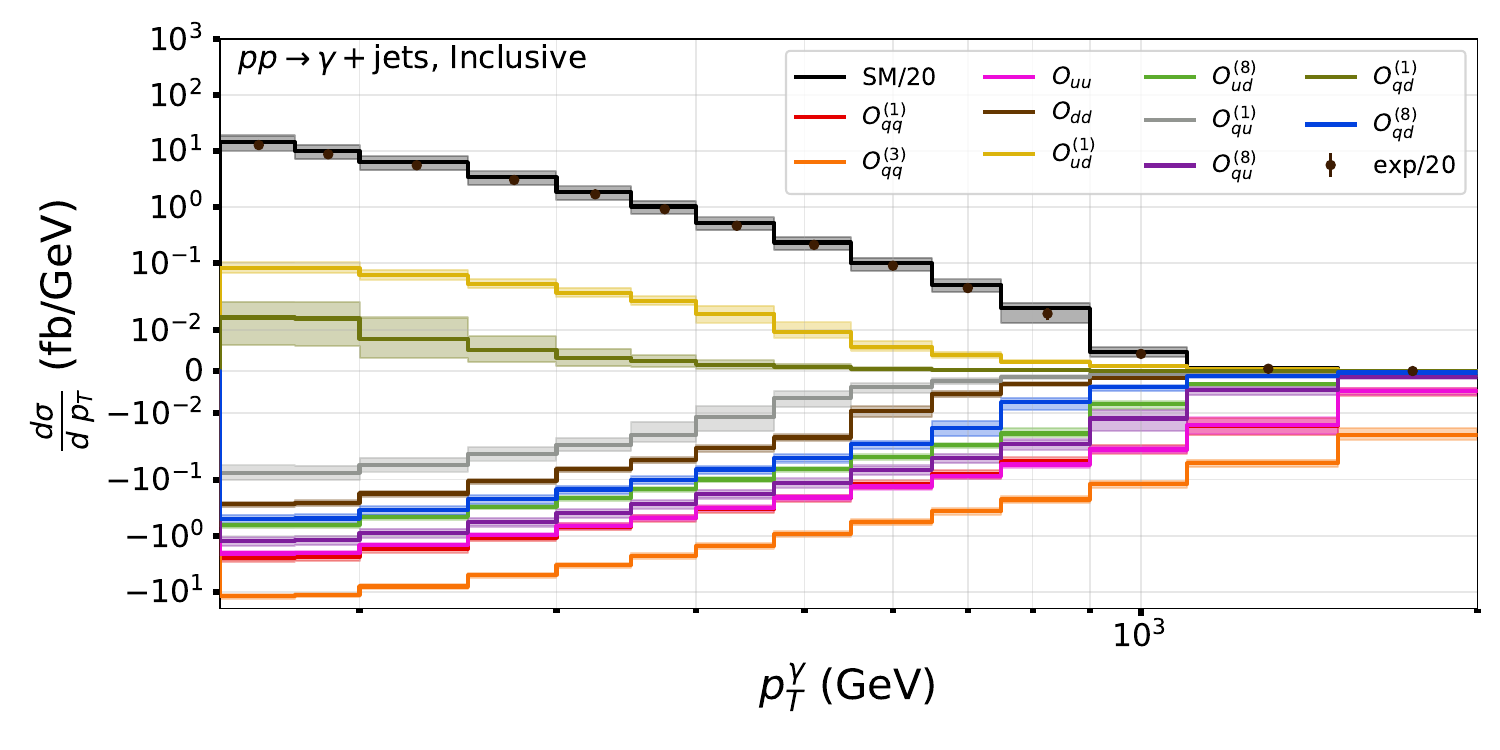}
      \includegraphics[width=.7\textwidth]{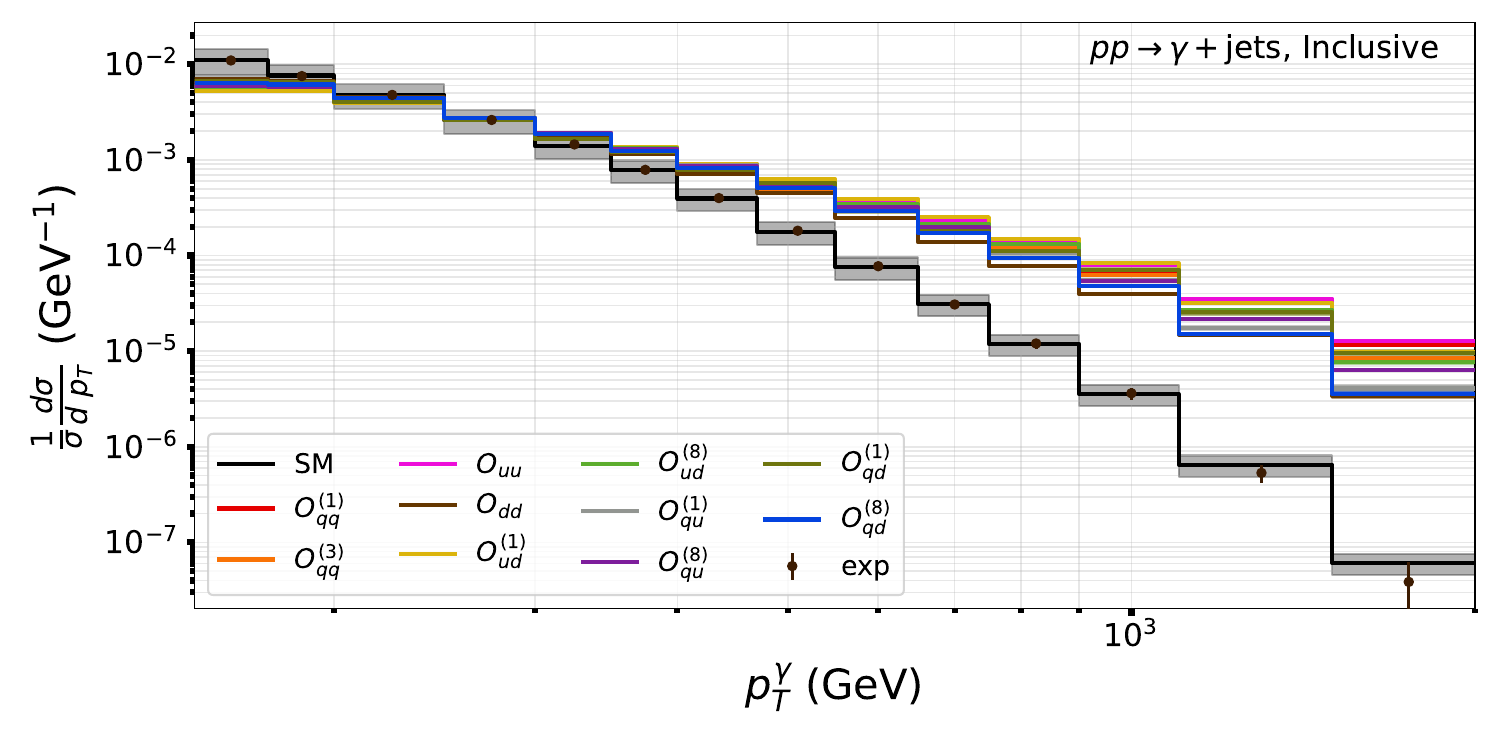}
   \caption{\footnotesize{As in Fig. \ref{fig:4lq_zjj_Dphijj}, but for the leading-jets invariant mass in $\gamma+$jets production}} \label{fig:4lq_ajj_mjj}
   \includegraphics[width=.7\textwidth]{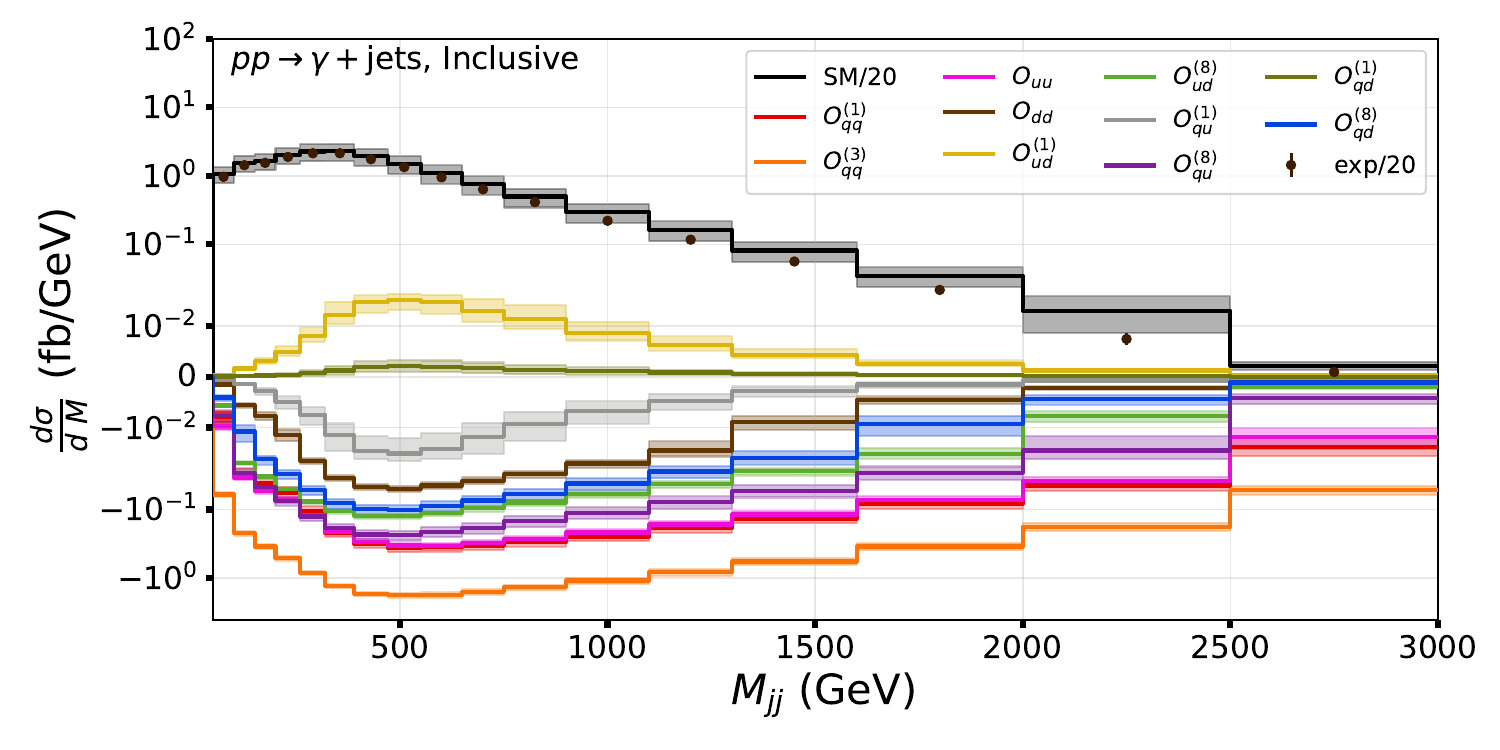}
   \includegraphics[width=.7\textwidth]{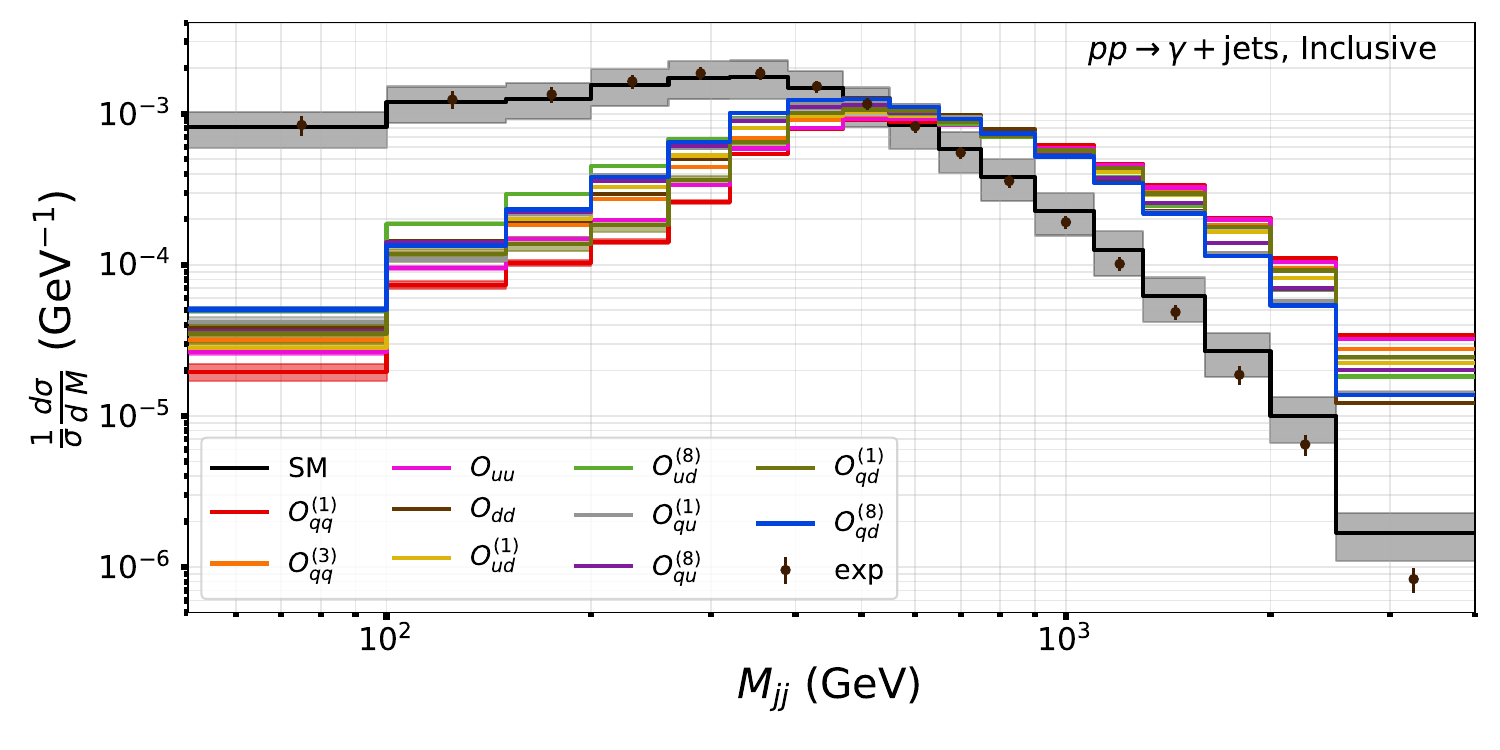}
\end{figure}

The differential and normalised distributions for $p_T^\gamma$ in the inclusive region are shown in Fig. \ref{fig:4lq_ajj_pTa}, where we applied the binning [150, 175, 200, 250, 300, 350, 400, 470, 550, 650, 750, 900, 1100, 1500, 2000] GeV with the last one including the overflow. The SM uncertainties are between 50 and 60\% in all bins, while the experimental ones are stable around 10\% almost everywhere but increase in the tail. For $O_{qq}^{(3)}$, the scale variations vary from 15 to 30\%.

The same distributions for $M_{jj}$ are also shown in Fig. \ref{fig:4lq_ajj_mjj}: the relative uncertainties are similar to the $p_T^\gamma$ ones. For both the observables, the 4LQ linear shapes are different from the SM ones, but no remarkable distinctions can be identified among the operators.

%\begin{figure}
%   \caption{\footnotesize{As in Fig. \ref{fig:4lq_zjj_Dphijj}, but for the leading-jets invariant mass in $\gamma+$jets production}} \label{fig:4lq_ajj_mjj}
%   \begin{center}
%   \includegraphics[width=.7\textwidth]{figures/ajj_mjj_diff.pdf}
%   \includegraphics[width=.7\textwidth]{figures/ajj_mjj_norm.pdf}
%   \end{center}
%\end{figure}

%%%%%%%%%%%%%%%%%%%%%%%%% Bounds %%%%%%%%%%%%%%%%%%%%%%%%%

\section{Constrained directions in the coefficient space}
For a given observable $X$, we built a $\chi^2$ function out of the experimental, SM and linear differential distributions through the expression
\begin{equation}
   \chi^2 = \sum_{i=1}^{N_\text{bins}} \frac{1}{\Delta_i^2} \left( \frac{d \sigma_i^\text{exp}}{d X} -\frac{d \sigma_i^\text{SM}}{d X} -\sum_j^\text{operators} \frac{c_j}{\Lambda^2} \frac{d\sigma^{1/\Lambda^2}_{j,i}}{d X} \right)^2. \label{4lq_chisq}
\end{equation}
Here above, the $c_j$ are components of $\vec{c}^{\hspace{.5mm}T} = \left( C_{qq}^{(1)}, C_{qq}^{(3)}, \ldots, C_{qd}^{(8)} \right)$, that is the vector of ten 4LQ coefficients. The $\Delta_i$ terms contained the numerical and scale uncertainties from the SM and experimental data only, summed in quadrature; the theoretical uncertainties over the 4LQ distributions were not included, as the scale variations are correlated among the various operators because of their dependence on $\alpha_s$. To get an estimate of the effect of the $\mathcal{O}(1/\Lambda^2)$ errors, we tried to include them at the numerator as in Eq. \eqref{chisq_uncNum}, namely
\begin{equation}
   \chi^2 = \sum_{i=1}^{N_\text{bins}} \frac{1}{\Delta_i^2} \left[ \frac{d \sigma_i^\text{exp}}{d X} -\frac{d \sigma_i^\text{SM}}{d X} -\sum_j^\text{operators} \frac{c_j}{\Lambda^2} \left( \frac{d\sigma^{1/\Lambda^2}_{j,i}}{d X} \pm \Delta_{j,i}^{1/\Lambda^2} \right) \right]^2. \label{4lq_chisq_1}
\end{equation}
The scale variations in the $\Delta^{1/\Lambda^2}$ terms were combined linearly to partially account for their correlation \cite{2j_tevatron}. The limits from this second expression were only considered for comparison.

We defined the vector of uncorrelated directions that the measurements are able to constrain as $\vec{c}^{\hspace{1mm}\prime} {}^T = \left( C_1, C_2, \ldots, C_{10} \right)$. It is related to $\vec{c}$ via a change of basis $\vec{c}^{\hspace{1mm}\prime} = \mathcal{R}^T \cdot \vec{c}$, where the columns of $\mathcal{R}$ are the eigenvectors of the matrix of coefficients of the quadratic terms in the $\chi^2$ expressions above. This happens because Eqs. \eqref{4lq_chisq} and \eqref{4lq_chisq_1} are quadratic polynomials in the Wilson coefficients; the inclusion of correlations among the 4LQ operators would spoil this argument \cite{andres_heavyquarks}. Each column of $\mathcal{R}$ represents a direction in the space of coefficients that the measurements are able to probe, and the larger the eigenvalue $\lambda$ associated to it, the better it can be constrained. The eigenvectors are defined with a unitary norm and showing a positive scalar product with $\left( 1,1,\ldots,1 \right)$.

Some directions among those are flat, meaning that no bounds can be placed along them. In principle, they would be associated to null eigenvalues, but the numerical nature of the computations makes them difficult to identify. Indeed, the ten $\lambda$ usually span over multiple orders of magnitude. To be able to estimate an uncertainty over each eigenvalue, we followed the procedure described in the reference \cite{ew_4f_smeft}: from each differential distribution, we generated various toy ones where the bin contents were replaced by random numbers, extracted from a Gaussian centred in the bin value and with standard deviation equal to the MC errors in each bin. We computed the eigenvalues from each toy plot and the standard deviations $\sigma$ of all the results were considered as uncertainties. After this, each eigenvector whose $\lambda$ turned out to be compatible with zero within $2\sigma$ was tagged as a flat direction.

We found that the second-best constrained directions for each observable we considered presented eigenvalues at least one order of magnitude smaller than the respective best ones, meaning that a much larger precision would be needed both at experiments and predictions to investigate them. The procedure described here above strengthens this hypothesis, as it showed that all the eigenvalues but the largest ones for every variable were compatible with zero and affected by strong numerical fluctuations, given the SM and experimental uncertainties included in the $\chi^2$ formula.

The best-constrained directions by each main observable for all the processes discussed in this chapter are shown in Table \ref{tab:4lq_dirs}, with their eigenvalues. These vectors mostly make use of the information from the total cross sections and not from the distribution shapes. All of the directions are much closer to $C_{qq}^{(3)}/\Lambda^2$ than to any other coefficient, suggesting that its operator would receive the most stringent bounds. For multijet production, moving to the highest $M_{jj}$ bin seems to increase the sensitivity to $O_{qq}^{(1)}$ and $O_{uu}$, even though the limits strength is lower because of larger experimental uncertainties. $W+$jets and $Z+$jets can probe very similar directions, at least for the operators that contribute to both, and the same happens for $\gamma+$jets and the low-$M_{jj}$ region in multijets; the processes involving EW bosons present quite small eigenvalues compared to the multijet ones. The application of flavour-tagging algorithms in $b+$jets production slightly moves the best-constrained direction towards $C_{dd}/\Lambda^2$, compared to the other untagged multijet cases; the related eigenvalue, though, is much smaller than the others, implying that the limits would be very loose. This is a consequence of the large SM cross section and scale uncertainties for that process and cuts. The best direction obtained from the sum of all the single-process $\chi^2$ polynomials in the table is shown in the last column: it is very close to the low-$M_{jj}$ region one, as expected from the large eigenvalue of the latter.

All the observables we checked from each process are only able to probe very close directions in the coefficient space, with variations of at most few percent in the $C_{qq}^{(3)}/\Lambda^2$ components. This confirms that the best-constrained axes are related to the total cross sections more than the shapes of the differential plots. The same happens when the $\gamma+$jet variables are computed in the direct- and fragmentation-enriched regions, compared to the inclusive one. Moreover, changing from Eq. \eqref{4lq_chisq} to \eqref{4lq_chisq_1} for the $\chi^2$ does not modify the directions significantly.

\begin{table}
\caption{\footnotesize{Best constrained directions, in the coefficient space, by the processes and main observables included in this study. The last column was obtained from the sum of all the other $\chi^2$ polynomials in the table. The eigenvalues $\lambda$ for each of them, with $1\sigma$ uncertainties, are also shown. The symbol $\times$ is used to mark the operators that do not contribute to a certain process, while zeros replace entries that are smaller than $10^{-2}$}} \label{tab:4lq_dirs}
\resizebox{\textwidth}{!}{
\begin{tabular}{c|ccccccc}
%\hline
\multicolumn{8}{c}{Best-constrained directions in coefficient space} \\ %\hline
 & \multicolumn{2}{c}{Multijets $\chi_{jj}$} & $b+$jets & $\ell^+ \ell^- +$jets & $\ell^\pm \nu +$jets & $\gamma+$jets & Combined \\
 & $2.4<M_{jj}<3$ TeV & $M_{jj}>6$ TeV & $p_T^b$ & $\Delta\phi_{jj}$ & $\Delta\phi_{jj}$ & $p_T^\gamma$ & \\ \hline
$C_{qq}^{(1)}$ & 0.41 & 0.52 & 0.39 & 0.25 & 0.36 & 0.38 & 0.43 \\
$C_{qq}^{(3)}$ & 0.83 & 0.74 & 0.83 & 0.97 & 0.93 & 0.84 & 0.82 \\
$C_{uu}$ & 0.32 & 0.40 & 0.02 & 0.02 & $\times$ & 0.37 & 0.33 \\
$C_{dd}$ & 0.07 & 0.03 & 0.28 & 0 & $\times$ & 0.01 & 0.07 \\
$C_{ud}^{(1)}$ & $\times$ & $\times$ & $\times$ & 0 & $\times$ & -0.01 & 0 \\
$C_{ud}^{(8)}$ & 0.08 & 0.05 & 0.10 & 0 & $\times$ & 0.07 & 0.07 \\
$C_{qu}^{(1)}$ & $\times$ & $\times$ & $\times$ & 0 & 0 & 0 & 0 \\
$C_{qu}^{(8)}$ & 0.17 & 0.10 & 0.12 & 0.04 & 0.05 & 0.11 & 0.16 \\
$C_{qd}^{(1)}$ & $\times$ & $\times$ & $\times$ & 0 & 0 & 0 & 0 \\
$C_{qd}^{(8)}$ & 0.10 & 0.04 & 0.22 & 0.02 & 0.04 & 0.03 & 0.09 \\ \hline
$\lambda$ & 1.1$\cdot 10^5$ & 2.1$\cdot 10^4$ & 1.2$\cdot 10^{-1}$ & 1.6$\cdot 10^2$ & 3.1 & 4.4$\cdot 10^2$ & $1.3\cdot 10^5$ \\
 & $\pm 50\%$ & $\pm 57\%$ & $\pm 67\%$ & $\pm 40\%$ & $\pm 52\%$ & $\pm 50\%$ & $\pm 54\%$ \\
\hline
\end{tabular}
}
\end{table}

\section{Individual and marginalised bounds on the coefficients}
We computed limits on the 4LQ coefficients through the $\chi^2$ \eqref{4lq_chisq}: for individual bounds, all the coefficients but one are set to zero, while for marginalised ones on a set of operators, all the other $C_i/\Lambda^2$ are fixed to the values that minimise the $\chi^2$ function along their direction.

\begin{figure}
   \centering
   \caption{\footnotesize{Individual limits, at 95\% CL, on the 4LQ operators from the processes and variables included in this study. For multijet production, the same observable is considered in two dijet invariant-mass regions separately}} \label{fig:4lq_ind_limits}
   \includegraphics[width=\textwidth]{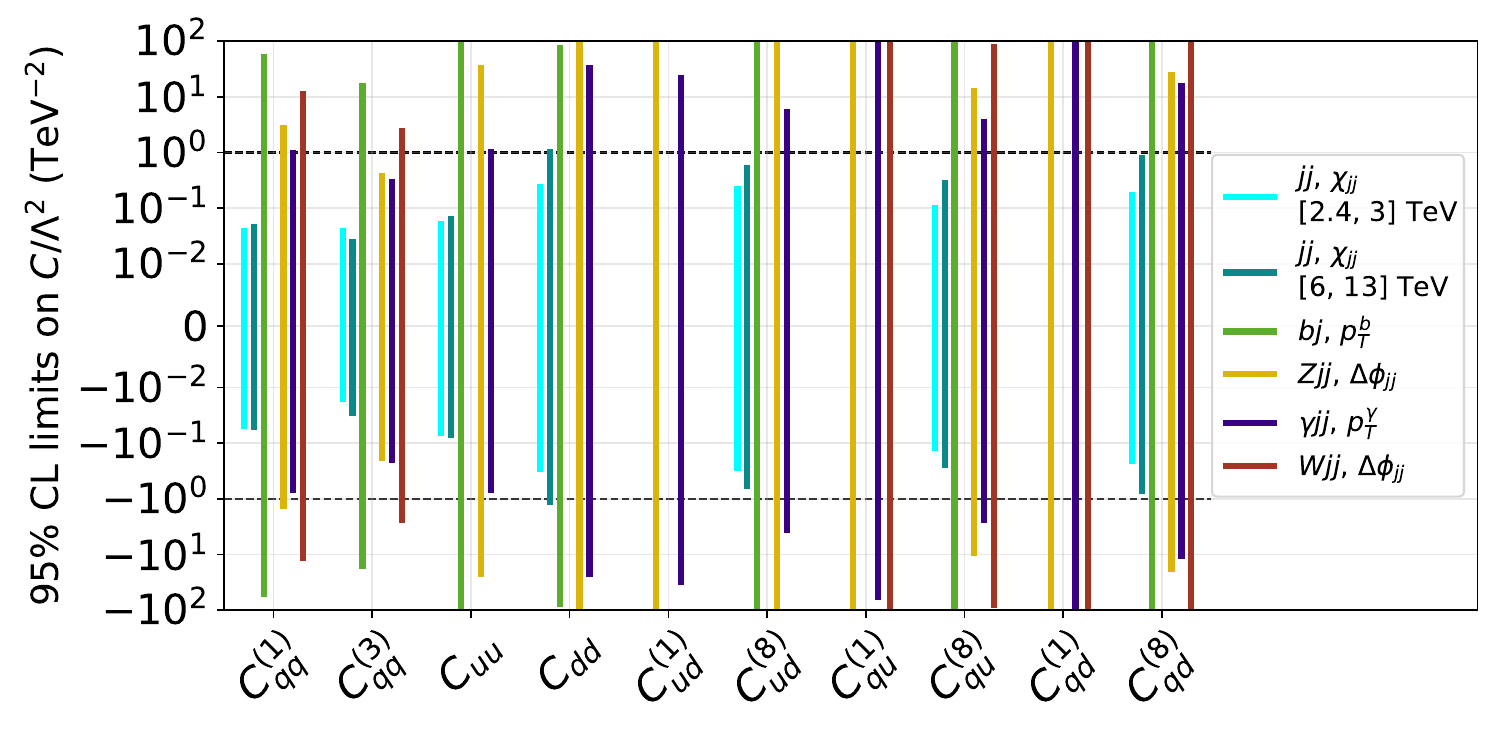}
\end{figure}

The individual bounds from the main observables described in this chapter are shown in Fig. \ref{fig:4lq_ind_limits}, and a summary of the experimental datasets used for each process is presented in Table \ref{tab:4lq_dataset}. The marginalised limits are not shown for single operators, as there are not enough different shapes, with respect to the SM and experimental uncertainties, to obtain valid results from a fit featuring all ten of them. The inclusion of more variables from the same processes is not always feasible because the correlations among bins are not always available.

The best bounds for all the 4LQ operators come from multijet production, and in particular from the [2.4, 3] TeV $M_{jj}$ region where the measurement uncertainties are lower. As a reminder, since the total cross sections for the SM and experiment are not public in the experimental analysis, we used the LO SM value to rescale the normalised distributions presented there. As expected from the previous paragraph, the bounds on $C_{qq}^{(3)}/\Lambda^2$ are the most stringent ones; indeed, its operator induces the largest deviations from the SM in all processes and contributes to all subprocesses in multijets. The three operators $O_{ud}^{(1)}$, $O_{qu}^{(1)}$ and $O_{qd}^{(1)}$, that do not contribute to multijet production and do not interfere with the main SM diagrams for other processes, are basically unconstrained.

\begin{table}
\centering
\caption{\footnotesize{List of experimental measurements used to obtain bounds on the 4LQ coefficients, for the processes studied in this chapter. For $W+$jets and $b+$jets, we assumed that the measurements follow the LO SM distribution. For multijets, we used the total LO SM cross section to multiply the normalised experimental distributions presented in the reference. The two $N_\text{data}$ values for multijets refer to the investigated low- and high-$M_{jj}$ regions, respectively}} \label{tab:4lq_dataset}
\begin{tabular}{cc|cccc}
   Proc. & Observable & $\sqrt{s}$, $\mathcal{L}$ & Final state & $N_\text{data}$ & Ref. \\ \hline
   Multijets & $d^2 \sigma / (d\chi_{jj} \hspace{1mm} d M_{jj})$ & 13 TeV, 35.9 fb${}^{-1}$ & jets & 12, 5 & \cite{2j_2018} \\
   $b+$jets & $d \sigma /d p_T^b$ & \multicolumn{4}{c}{Exp. data taken as LO SM} \\
   $Z+$jets & $d\sigma / d\Delta \phi_{jj}$ & 13 TeV, 139 fb${}^{-1}$ & $\ell^+ \ell^- +$jets, $\ell=e,\mu$ & 12 & \cite{atlas:2021Zjj} \\
   $W+$jets & $d\sigma / d\Delta \phi_{jj}$ & \multicolumn{4}{c}{Exp. data taken as LO SM} \\
   $\gamma +$jets & $d\sigma /d p_T^\gamma$ & 13 TeV, 36.1 fb${}^{-1}$ & $\gamma +$jets & 14 & \cite{Atlas:2020ajj} \\ \hline
\end{tabular}
\end{table}

For what concerns the uncertainties at the $\chi^2$ denominator, for multijets they are dominated by the SM ones in the lowest $M_{jj}$ range and by the experimental ones in the highest one. When the interference errors are included as in Eq. \eqref{4lq_chisq_1}, the individual limits worsen by factors from 1.3 to 1.7.

In $Z+$jets, the largest uncertainty contribution comes from the measurements, that also include the MC-generators discrepancies. Switching to the second $\chi^2$ formula widens the limits by 1.2 to 1.5 times.

For $W+$jets, since a BDT is employed in the CMS analysis we followed, we generated the LO SM distributions and assumed that the experimental ones had the same behaviour, with a 10\% relative error. Including the theoretical uncertainties over the linear terms increases the limit sizes by factors between 1.4 and 2.%, with the exception of the $O_{qd}^{(1)}$ and $O_{qu}^{(1)}$ ones that become 5 and 8 times larger.

In $\gamma+$jets, the errors are again dominated by the SM predictions. The second $\chi^2$ returns bounds that are between 1.2 and 2.3 times worse than the ones from the first formula.

We computed the partial $\mathcal{O}(1/\Lambda^4)$ corrections, with just the inclusion of the square of the amplitudes with up to one insertion of 4LQ operators: the results for $O_{qq}^{(3)}$ can be found in Table \ref{tab:4lq_xsects}, for a coefficient value of 1 TeV${}^{-1}$. As in the previous chapters, the interference among dimension-8 operators and the SM was not included, due to the large number of terms that should be taken into account. Moreover, incorporating the squared terms for all the operators would require to deal with the correlations among them, and this would complicate the analysis even more, especially in the computation of the constrained directions in the parameter space.
This check was performed because the squared contributions might dominate the linear ones at the high energies required by the cuts in this analysis. Indeed, a comparison of the linear and quadratic cross sections for $O_{qq}^{(3)}$ in the table shows that the latter are at least as big as the former for coefficient values that are similar to the ones obtained in the limit computation, with the only exception of the $2.4 < M_{jj} < 3$ TeV region in multijet production. For the unconstrained operators $O_{ud}^{(1)}$, $O_{qu}^{(1)}$ and $O_{qd}^{(1)}$, the quadratic contributions would probably always dominate.

\begin{figure}
   \caption{\footnotesize{Individual contours in the $C_{qq}^{(1)}$ {\it vs} $C_{qq}^{(3)}$ plane ({\it top left}), in the $C_{uu}$ {\it vs} $C_{qq}^{(3)}$ one ({\it top right}), in the $C_{dd}$ {\it vs} $C_{qq}^{(3)}$ one ({\it bottom left}) and in the $C_{qu}^{(8)}$ {\it vs} $C_{qq}^{(3)}$ one ({\it bottom right}). The $W+$jets and $b+$jets distributions are not shown  because they are larger than the axes limits. Note that the ranges on the two axes are different in each plot}} \label{fig:4lq_contours}
   \includegraphics[width=.49\textwidth]{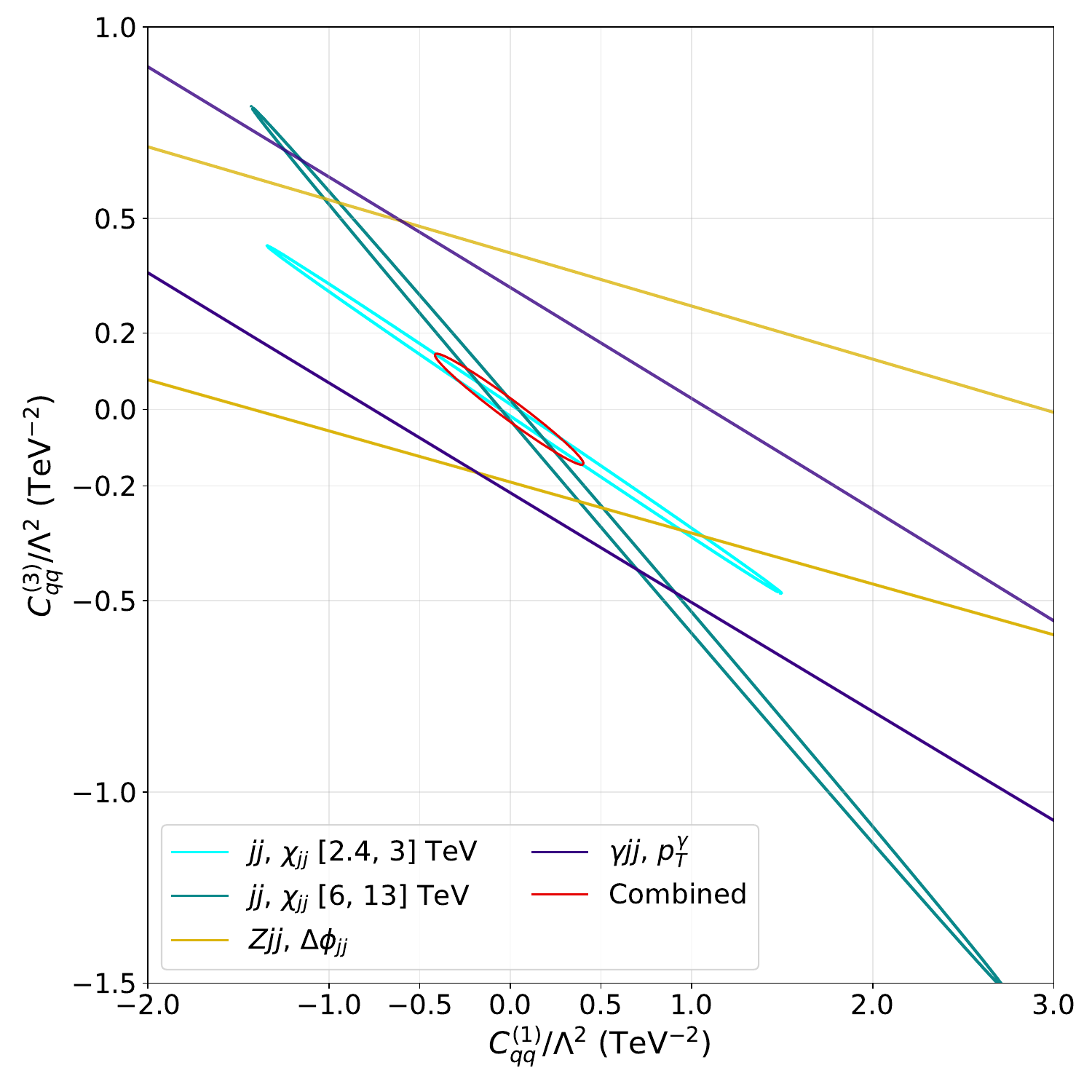}
   \includegraphics[width=.49\textwidth]{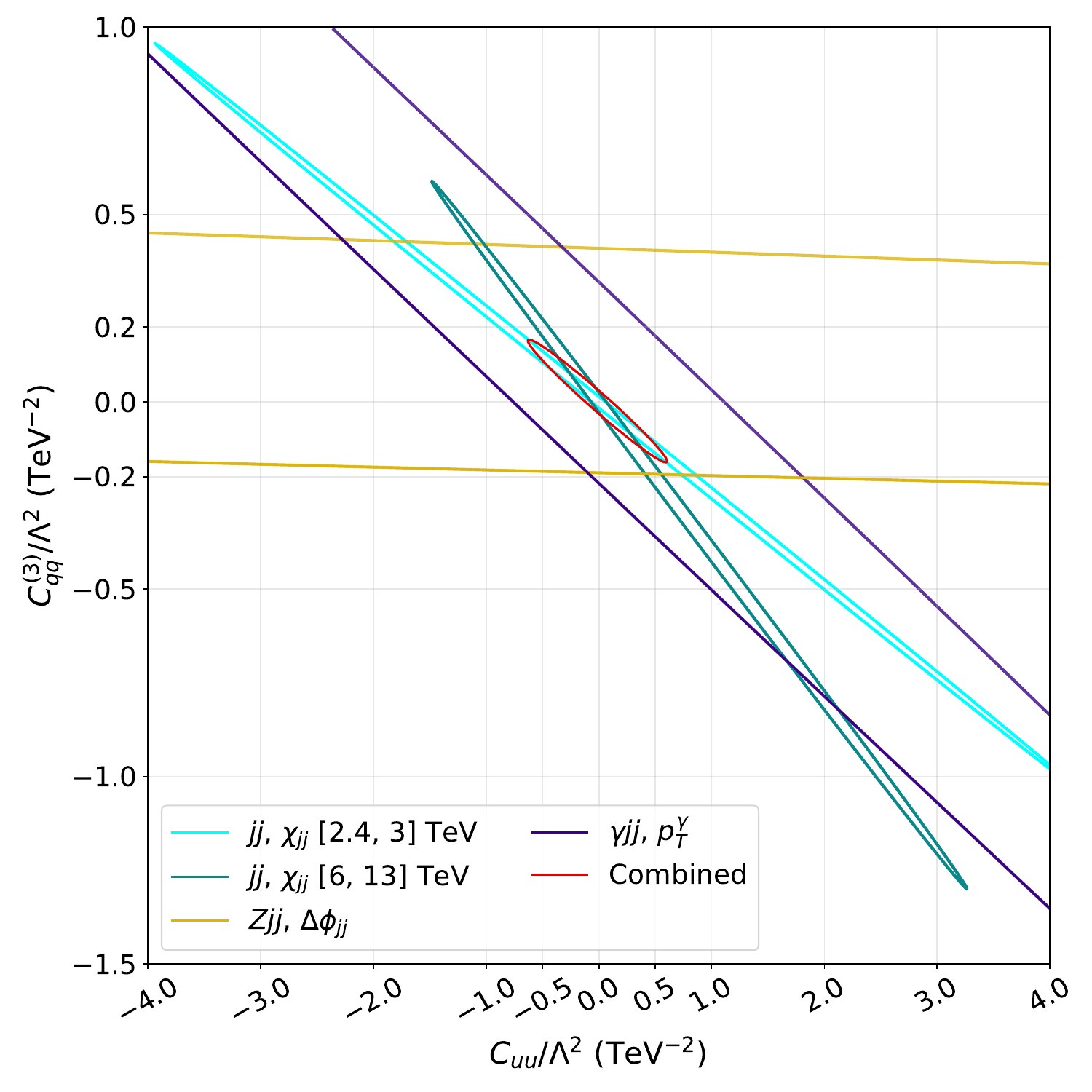}
   \includegraphics[width=.49\textwidth]{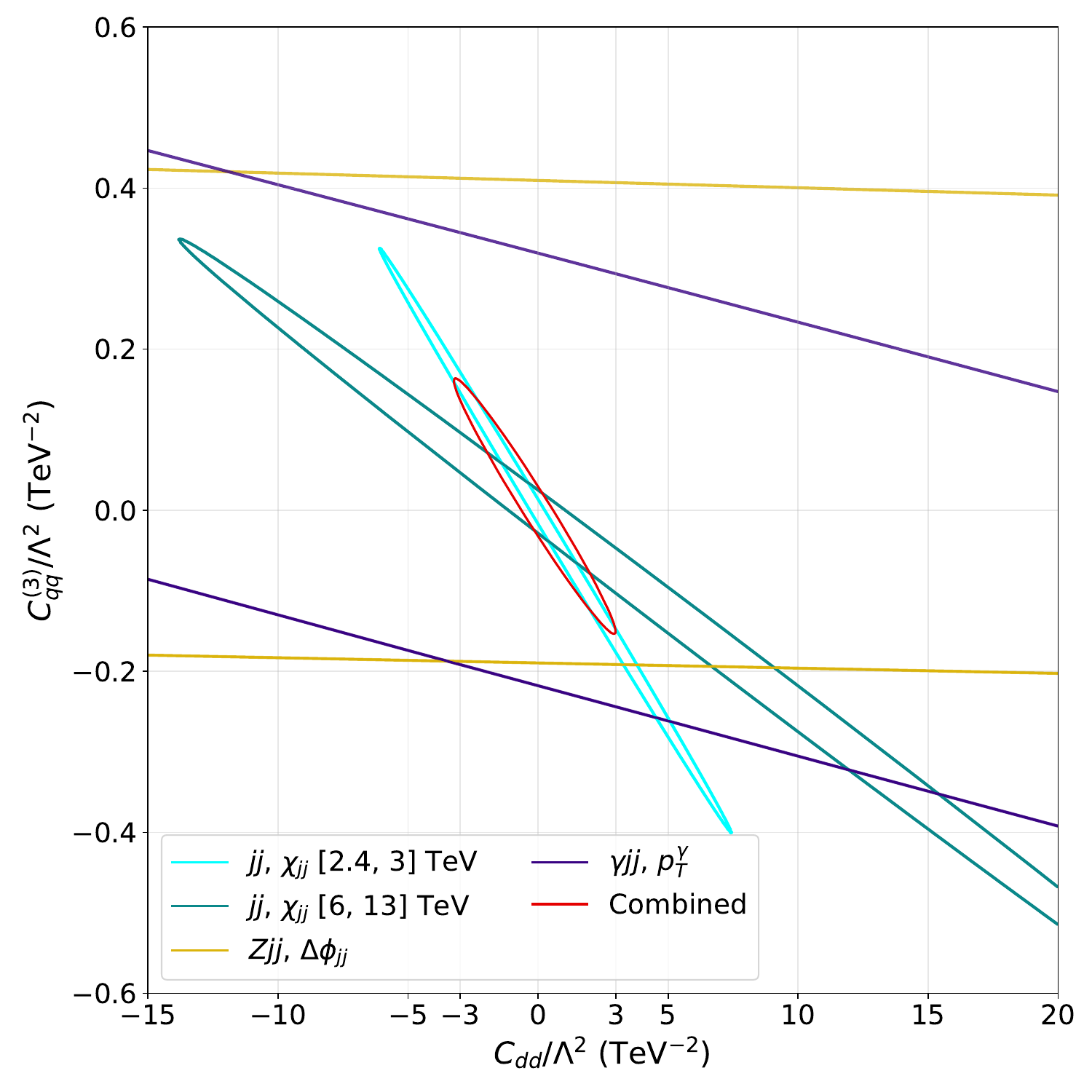}
   \includegraphics[width=.49\textwidth]{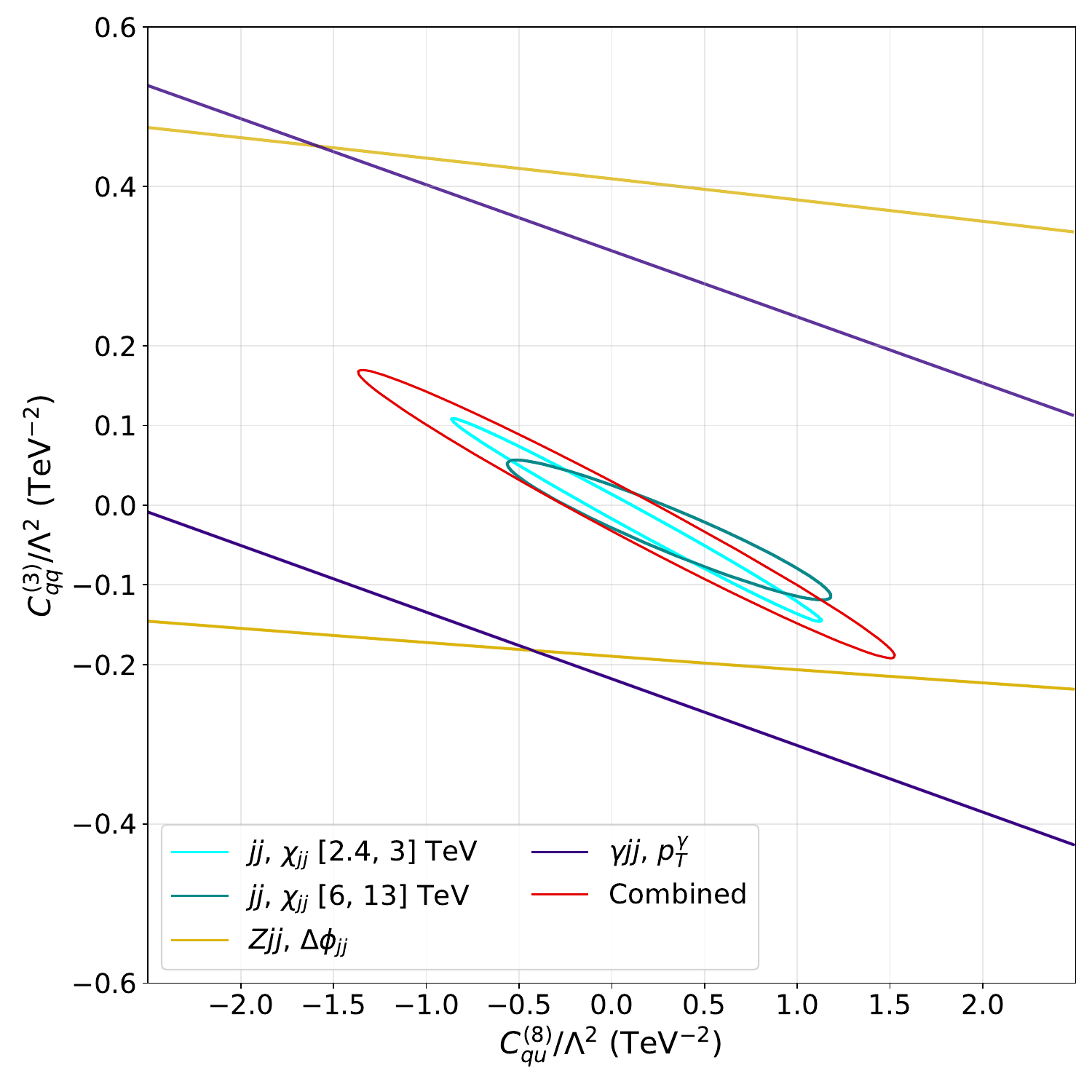}
\end{figure}

The individual contours in some two-dimensional coefficient planes are illustrated in Fig. \ref{fig:4lq_contours}. The two $M_{jj}$ regions for multijets are both included separately, as the correlations among their bins presented in \cite{2j_2018} are orders of magnitude smaller than the ones inside the same intervals. As discussed above, the most stringent contours come from multijets, as the two invariant-mass regions can probe slightly different directions in the coefficient space. The $Z+$jets and $W+$jets ellipses point along similar directions, coherently to what is reported in Table \ref{tab:4lq_dirs} for the best-constrained directions. Similarly happens for $\gamma+$jets and the low-$M_{jj}$ region in multijets. All the main axes of the contours lean towards the $C_{qq}^{(3)}/\Lambda^2$ one.

\begin{figure}
   \centering
   \caption{\footnotesize{Individual and combined marginalised contours in the $C_{qq}^{(1)}$ {\it vs} $C_{qq}^{(3)}$ plane ({\it top left}), in the $C_{uu}$ {\it vs} $C_{qq}^{(3)}$ one ({\it top right}) and in the $C_{qu}^{(8)}$ {\it vs} $C_{qq}^{(3)}$ one ({\it bottom}). Only these four operators are included in the fit, and only the predictions from the two $M_{jj}$ regions in multijets, $Z+$jets and $\gamma +$jets are used. Note that the ranges on the two axes are different in each plot}} \label{fig:4lq_contours_red}
   \includegraphics[width=.49\textwidth]{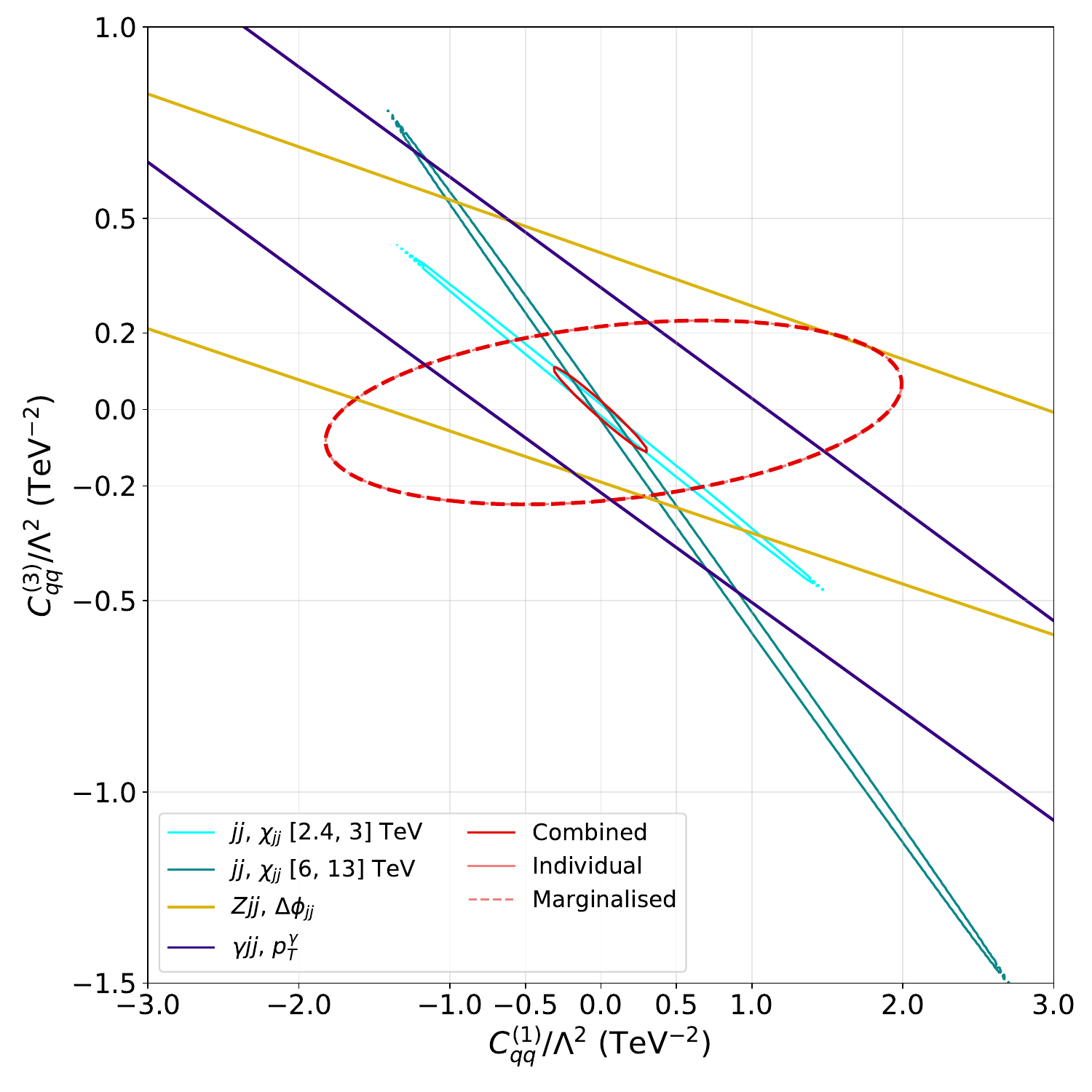}
   \includegraphics[width=.49\textwidth]{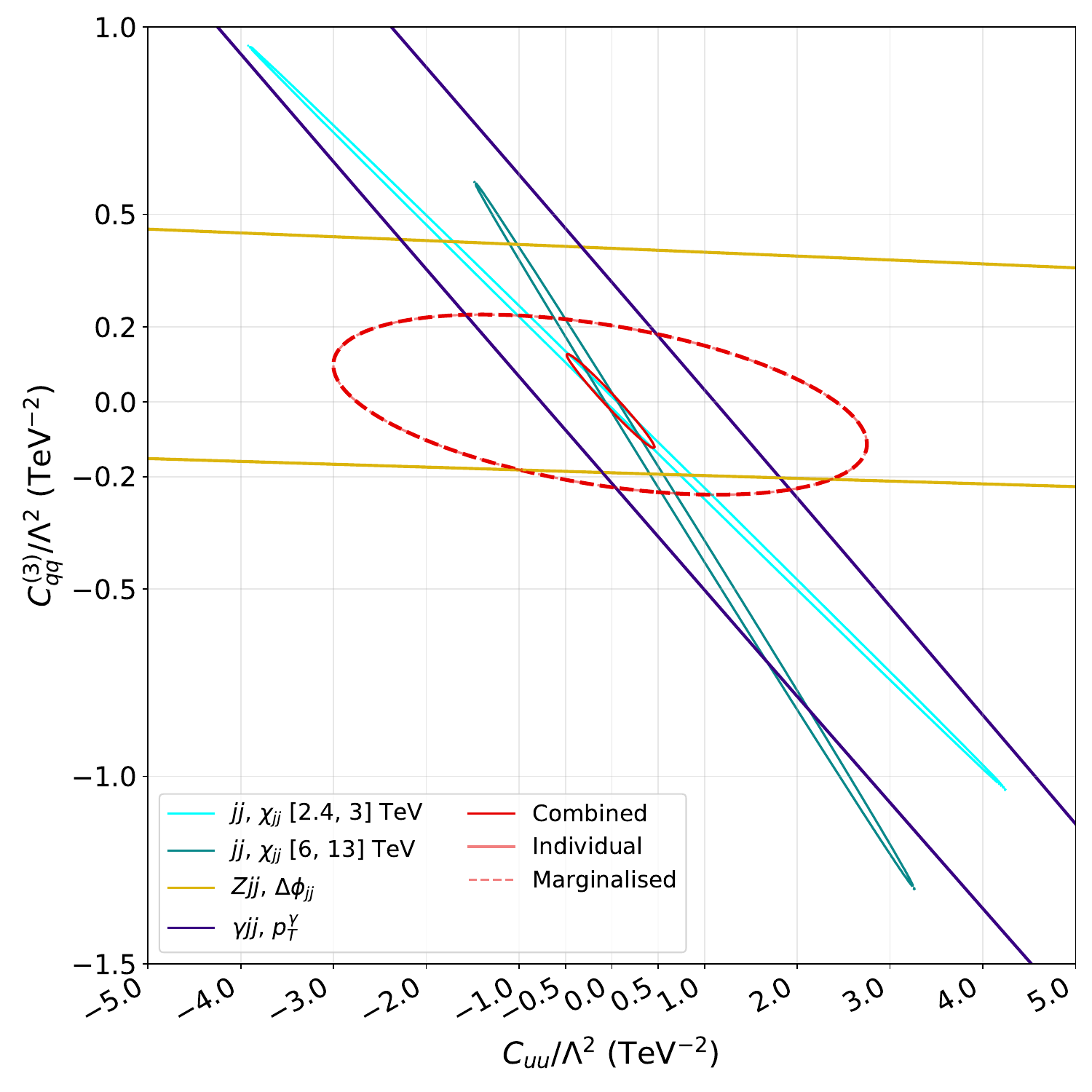}
   \includegraphics[width=.49\textwidth]{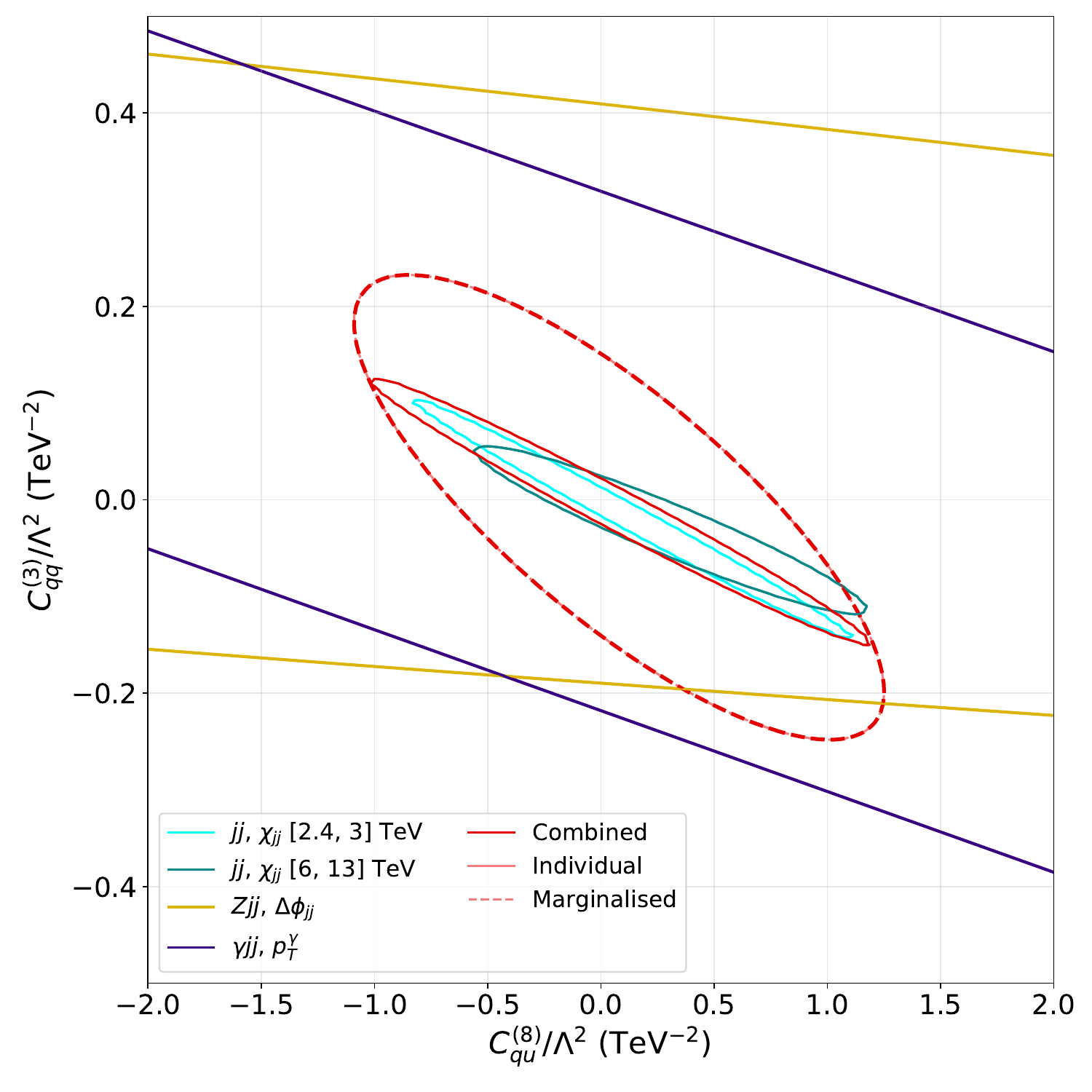}
\end{figure}

We then repeated the fit including only the four most constrained operators $O_{qq}^{(1)}$, $O_{qq}^{(3)}$, $O_{uu}$ and $O_{qu}^{(8)}$, using exclusively the $\chi_{jj}$, $\Delta \phi_{jj}$ and $p_T^\gamma$ differential distributions from the two $M_{jj}$ regions in multijets, $Z+$jets and $\gamma +$jets respectively, since they are the most constraining processes. For this scenario, both the individual and combined marginalised contours are shown in Fig. \ref{fig:4lq_contours_red}. It can be seen that the latter are tilted along the $C_{qq}^{(3)}/\Lambda^2$ axis as the individual ones, and that the related bounds on the $O_{qq}^{(3)}$ coefficient are better by at least a factor $\sim 5$ compared to the other operators. The weaker constraints on those three operators make them once again more vulnerable to validity issues. The marginalised limits in the $C_{qu}^{(8)}$ {\it vs} $C_{qq}^{(3)}$ plane are better than the ones in $C_{qq}^{(1)}$ {\it vs} $C_{qq}^{(3)}$, even though the combined individual bounds are worse; the marginalised contour in the second plane becomes more stringent than in the first one if $C_{uu}$ is set to zero. The same happens if $C_{uu}$ and $C_{qq}^{(1)}$ are swapped in the previous sentence: this suggests that $O_{qq}^{(1)}$ and $O_{uu}$ might partially cancel each other along the main direction probed by the combination of the datasets included in this fit.

\section{Conclusions and prospects}
This chapter described how we investigated multiple processes that might be affected by ten 4LQ operators at linear level. We included multijet production and cases where the jets are generated together with EW bosons; for each process, we checked which observable is able to provide the best bounds, finding that $O_{qq}^{(3)}$ is always the most constrained operator. This is in part a result of the normalisation chosen for it: if the $\tau^I = \sigma^I /2$ matrices had to be used instead of the Pauli ones as in \cite{2j_SMEFT}, it would produce similar or smaller contributions than $O_{qq}^{(1)}$ and $O_{uu}$ to the studied processes. It goes without saying that physics should not depend on the normalisation: the assumption of some UV models that could run into the 4LQ operators would remove this issue through the matching procedure. Furthermore, this might allow to neglect some of the operators depending on the model, as the renormalisation group would not mix all of them. Even when keeping model independence, though, using the $\tau^I$ matrices to define $O_{qq}^{(3)}$ would make the comparison among the 4LQ operators easier, especially when estimating the constrained directions in the coefficient space and the square of the dimension-6 amplitudes.

We also simulated a flavour-tagging algorithm and used it in multijets to increase the sensitivity to operators featuring $b$- and $c$-jets. This did not completely cancel the contributions of operators that do not include them, because of jet mistagging. Moreover, the $p_T$ intervals of operation for the algorithms do not allow imposing strong enough cuts to reduce the SM contribution, that remains at least three orders of magnitude larger than the main linear one.

We checked which directions in the coefficient space could be probed by each observable, and saw that only the ones that make use of the normalisation information are accessible by the data: all the others, that involve more details about the distribution shapes, seem to be out of the reach of current measurements. More precision in the measurements and predictions would be needed to investigate them, in particular for the SM ones where relative uncertainties are large and MC generators do not agree on the results. This is true both for the SM predictions that we took from the experimental collaborations and the ones that we obtained via \amc , like in the $W+$jets case.
As a consequence, all the variables from the same processes are only able to probe the same directions in the coefficient space, and all of them are very close to the $C_{qq}^{(3)}/\Lambda^2$ axis.

The partial $\mathcal{O}(1/\Lambda^4)$ corrections for $O_{qq}^{(3)}$ seem to be already as large as the linear ones for the coefficient values obtained in the limits. This is true for all processes and regions, exception made for multijets in the low-$M_{jj}$ interval: as expected, there is a sweet spot in the energy range where the new-physics effects are significant compared to the SM, while the experimental precision is still high enough and EFT uncertainties are under control.

The interference among the SM and dimension-8 operators was not included in this study. Additionally, the two different flavour-indices contractions for the operators with bilinears of same chirality should be split, and constraints on their coefficients should be obtained separately. This would probably yield less stringent bounds than the ones presented in this chapter.

% Conclusion
\clearpage
\addcontentsline{toc}{chapter}{Final remarks}
%!TEX root = main.tex

%%%%%%%%%%%%%%%%%%%%%%%%%%%%%%%%%%%%%%%%%%%%%%%%%%%%
%
%      Conclusions :
%
%
%%%%%%%%%%%%%%%%%%%%%%%%%%%%%%%%%%%%%%%%%%%%%%%%%%%

\chapter*{Final remarks}
\label{chap:conclusions}
\pagestyle{fancy}

%If someone only had time to read these few lines instead of the whole thesis, here are the key points that I would like them to take home.

The integrable and measurable cross sections, introduced in Chapter \ref{chap:interf_OG}, can be very useful in quantifying the suppression for the interference of the SM and the SMEFT operators that may affect some processes at certain colliders. If this is the result of a cancellation between two large contributions with opposite sign, some strategies might be found to lift it. Since the measurable cross section is usually quite costly to compute and model-dependent, we employed it to investigate the nature of the suppression and find simpler kinematic observables that can perform the same reviving job with a similar efficiency.

In the example of the $O_G$ operator, we checked that the interference is exactly zero for dijet production but not for three-jets, where the cancellation can be reduced if the pencil-like and isotropic event topologies are separated. For this reason, an event-shape variable called ``transverse sphericity'' is able to set constraints at linear level on the $O_G$ coefficient that are compatible with the ones from the following order in the expansion.

In Chapter \ref{chap:interf_OW} we used the same quantities to investigate the $O_W$ contribution at NLO to three different processes that involve EW bosons. In this case, azimuthal observables are quite sensitive to the suppression and their differential $K$-factors are more reasonable than the global ones, or they can be used to cut the phase space into regions where the opposite-sign contributions are separated. $\Delta\phi_{jj}$ in $Zjj$ production is able to set the best constraints on the operator.

The multiple sources of interference suppression that we found in these processes, together with the different strategies needed to lift them, suggest that a similar analysis should be repeated from scratch for other processes and operators that experience a cancellation of the same kind. Machine Learning might be employed to find kinematic observables with large reviving power, especially if the measurable cross section is used to pick the most suitable inputs to the networks, but uncertainties must be reduced in our predictions to appreciate improvements in the results for bounds and $K$-factors. Since NLO and parton-shower corrections can introduce negative weights in the Monte Carlo samples that cannot be distinguished from the interference ones, the measurable and integrable cross sections have to be computed at parton level with PDF. After suitable observables have been found to lift the suppression, the effects of NLO, parton shower and detectors need to be checked over their distributions to assess how much their reviving power is affected.

Chapter \ref{chap:fourLQ} describes a preliminary analysis on the four-light quark operators in the SMEFT, attempting to set bounds on their coefficients. These operators are quite tricky to investigate, but they might affect any process at NLO. Despite our simplistic approach, it is unlikely that competitive constraints would be placed on them unless higher precision is reached in the predictions, and especially in the SM ones, as uncertainties are very large and different Monte Carlo generators do not agree on the results.

It is needless to say that all these studies can be further improved, and detailed discussions about the issues and missing contributions can be found in each chapter. A common denominator to all of them is that the interference of dimension-8 operators and the SM is not included due to the large number of the former, even if it could induce large corrections at the $\mathcal{O}(1/\Lambda^4)$ order. This could either increase or reduce the constraining power of that level compared to the linear one.

\vspace{0.5cm}
\hfill
\begin{minipage}{\textwidth}
   \centering
   \includegraphics[width=.7\textwidth]{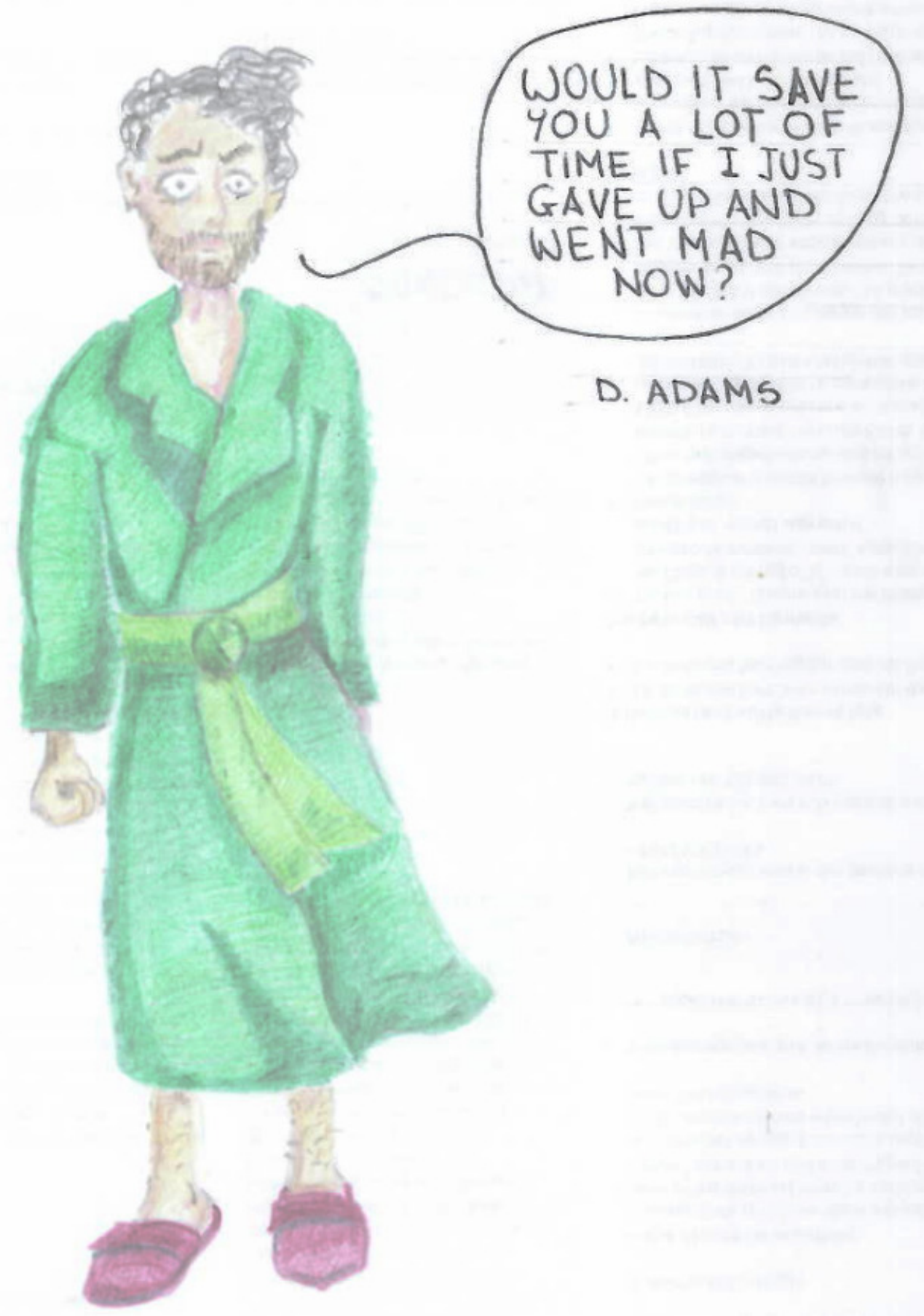}
\end{minipage}

\clearpage
\addcontentsline{toc}{chapter}{Bibliography}
\bibliography{refs.bib}
\bibliographystyle{ieeepes}

\printindex

\end{document}